\documentclass[3p,times]{elsarticle}
\usepackage[latin9]{inputenc}
\usepackage{verbatim}
\usepackage{lmodern}
\usepackage{graphicx}
\usepackage{amsmath,mathrsfs,amssymb}
\usepackage{amsfonts}
\usepackage{caption}
\usepackage{array}
\usepackage{longtable}
\usepackage{algpseudocode}
\usepackage{amsmath}
\usepackage{mathpazo}
\usepackage{float}
\usepackage{algorithm}
\graphicspath{ {images/} }
\usepackage{epstopdf}
\usepackage{enumitem}
\usepackage{booktabs}
\usepackage{subcaption}
\usepackage[unicode=true,
 bookmarks=false,
 breaklinks=false,pdfborder={0 0 1},colorlinks=false]
 {hyperref}
\hypersetup{
 colorlinks,bookmarksopen,bookmarksnumbered,citecolor=red,urlcolor=red}

\makeatletter

\pdfpageheight\paperheight
\pdfpagewidth\paperwidth
\usepackage[english]{babel}
\usepackage{lineno,hyperref}
\modulolinenumbers[5]

\newtheorem{defn}{Definition}

\newtheorem{rem}{Remark}

\newtheorem{assum}{Assumption}

\begin{document}

\begin{frontmatter}

\title{Attitude Determination and Estimation using Vector Observations: Review, Challenges and Comparative Results}

\author[HAH]{Hashim A.~Hashim}
\ead{h.a.hashiim@gmail.com, hhashim@tru.ca}

\address[HAH]{Department of Engineering and Applied Science, Thompson Rivers University, Kamloops, Britich Columbia, Canada, V2C-0C8}
\vspace{9cm}
\begin{abstract}
This paper concerns the problem of attitude determination and estimation. The early applications considered algebraic methods of attitude determination. Attitude determination algorithms were supplanted by the Gaussian attitude estimation filters (which continue to be widely used in commercial applications). However, the sensitivity of the Gaussian attitude filter to the measurement noise prompted the introduction of the nonlinear attitude filters which account for the nonlinear nature of the attitude dynamics problem and allow for a simpler filter derivation. This paper presents a survey of several types of attitude determination and estimation algorithms. Each category is detailed and illustrated with literature examples in both continuous and discrete form. A comparison between these algorithms is demonstrated in terms of transient and steady-state error through simulation results. The comparison is supplemented by statistical analysis of the error-related mean, infinity norm, and standard deviation of each algorithm in the steady-state.
\noindent\rule{16.5cm}{1pt}
\vspace{0.2cm}
\end{abstract}
\begin{keyword}
Comparative Study, Attitude, Determination, Estimation, Filter,
Adaptive Filter, Gaussian Filter, Nonlinear Filter, Overview, Rodrigues Vector, Special Orthogonal Group, Unit-quaternion, Angle-axis, Determinstic, Stochastic, Continuous, Discrete.
\end{keyword}
\end{frontmatter}{}


\vspace{2.4cm}

\noindent\rule{16.5cm}{2pt}
{\small
\textbf{\textcolor{red}{Bibtex formatted citatio}}\textcolor{red}{n}:\\

@article\{hashim2020attitude, 

title=\{Attitude Determination and Estimation using Vector Observations: Review, Challenges and Comparative Results\}, 

author=\{Hashim A. Hashim\}, 

journal=\{arXiv preprint arXiv:2001.03787\}, 

year=\{2020\} 

\}

}

\noindent\rule{16.5cm}{2pt}

\newpage

\tableofcontents{}

\noindent\makebox[1\linewidth]{%
	\rule{0.4\textwidth}{1.4pt}%
}

\listoftables

\noindent\makebox[1\linewidth]{%
	\rule{0.4\textwidth}{1.4pt}%
}

\listoffigures

\noindent\makebox[1\linewidth]{%
	\rule{0.4\textwidth}{1.4pt}%
}

\newpage

\section{Introduction}

Automated and semi-automated robotic applications such as unmanned
aerial vehicles (UAVs), autonomous underwater vehicles (AUVs), ground
vehicles, satellites, radars and others can be controlled to rotate
successfully in the three dimensional (3D) space if the orientation
of the rigid-body is accurately known. However, the true orientation
of a rigid-body, generally referred to as attitude, cannot be extracted
directly. Alternatively, the attitude can be determined using
\begin{enumerate}
	\item[1)] a set of measurements available in the body-frame and
	\item[2)] known observations in the inertial-frame. 
\end{enumerate}
In general, measurement units are corrupted with unknown bias and
noise components. However, the quality of measurement units has a
significant impact on the level of noise and bias components attached
to the measurements. The measurement units can be broadly divided
into two categories:
\begin{enumerate}
	\item[1)] high-cost or high-quality measurement units and 
	\item[2)] low-cost or low-quality inertial measurement units (IMUs).
\end{enumerate}
There are three main approaches to establishing the attitude:
\begin{enumerate}
	\item[1)] algebraic determination algorithms,
	\item[2)] vector-based filter dynamics, and
	\item[3)] filter dynamics that mimic the true nature of the attitude dynamics
	problem.
\end{enumerate}
As such, attitude determination or estimation problem is a fundamental
sub-task in the majority of robotic applications. The accurate knowledge
of the attitude is indispensable for the control process of most robotic
applications. This is especially true for the applications that require
fast maneuvering. Lack of accurate attitude information may result
into an unstable control process. In this paper, the terms ``filter''
and ``estimator'' are equivalent and will be used interchangeably.
Also, the term ``attitude'', ``orientation'' and ``rotational
matrix'' are equivalent and will be used interchangeably. The main
goals of this paper are as follows:
\begin{itemize}
	\item[1)] introducing the attitude dynamics problem,
	\item[2)] providing the assumptions necessary for the attitude determination
	and estimation problem,
	\item[3)] presenting a brief survey of different types of attitude determination
	algorithms and attitude filters,
	\item[4)] demonstrating several types of filter design in both continuous and
	discrete form as well as attitude determination algorithms,
	\item[5)] comparing the results between different categories of attitude determination
	and estimation algorithms.
\end{itemize}
The paper is organized as follows: Section \ref{sec:Comp_Math-Notations}
contains abbreviations, math and attitude notations, math identities
and attitude preliminaries. The attitude problem, inertial-frame observations,
body-frame measurements and basic assumptions are outlined in Section
\ref{sec:SO3_kinematics}. Section \ref{sec:Comp_Determination} gives
a brief overview of attitude determination algorithms and presents
a detailed description of the three most common algorithms. Section
\ref{sec:Comp_Determination} explains the structure of Gaussian attitude
filters and discusses two main algorithms of Gaussian attitude filters.
Section \ref{sec:Comp_Nonlinear} describes the structure of nonlinear
attitude filters as well as different types of nonlinear attitude
filters. Comparative results between the different categories of attitude
determination algorithms and filters are given in Section \ref{sec:Comp_Simulation}.
Finally, Section \ref{sec:Comp_Conclusion} summarizes the work.

\section{Notation and Preliminaries \label{sec:Comp_Math-Notations}}

Table \ref{tab:Table-of-acronyms} lists the abbreviations used throughout
the paper. Table \ref{tab:Table-of-Notations1} contains the important
math notation used throughout the paper. Table \ref{tab:Table-of-Notations2}
provides some important attitude-related definitions and notation. 

\begin{table}[H]
	\centering{}\caption{\label{tab:Table-of-acronyms}Abbreviations in order of appearance}
	\begin{tabular}{lll}
		\toprule 
		\addlinespace
		UAVs & :  & Unmanned aerial vehicles \tabularnewline
		\addlinespace
		AUVs & :  & autonomous underwater vehicles\tabularnewline
		\addlinespace
		IMU & :  & Inertial measurement unit\tabularnewline
		\midrule 
		\addlinespace
		QUEST & :  & Quaternion estimator\tabularnewline
		\addlinespace
		SVD & :  & Singular value decomposition\tabularnewline
		\addlinespace
		TRIAD & :  & Triaxial attitude determination\tabularnewline
		\midrule 
		\addlinespace
		KF & :  & Kalman filter\tabularnewline
		\addlinespace
		EKF & :  & Extended Kalman filter\tabularnewline
		\addlinespace
		MEKF & : & Multiplicative extended Kalman filter\tabularnewline
		\addlinespace
		GAMEF & : & Geometric Approximate Minimum-Energy Filter\tabularnewline
		\midrule 
		\addlinespace
		NDAF & : & Nonlinear deterministic attitude filter\tabularnewline
		\addlinespace
		CG-NDAF & :  & Constant gain NDAF\tabularnewline
		\addlinespace
		CGD-NDAF & :  & Constant gain direct NDAF\tabularnewline
		\addlinespace
		CGSd-NDAF & :  & Constant gain semi-direct NDAF\tabularnewline
		\addlinespace
		AG-NDAF & :  & Adaptive gain NDAF\tabularnewline
		\addlinespace
		GP-NDAF & :  & Guaranteed performance NDAF\tabularnewline
		\addlinespace
		GPSd-NDAF & :  & Guaranteed performance semi-direct NDAF\tabularnewline
		\addlinespace
		GPD-NDAF & :  & Guaranteed performance direct NDAF\tabularnewline
		\addlinespace
		NSAF & : & Nonlinear stochastic attitude filter\tabularnewline
		\addlinespace
		AGI-NSAF & : & Adaptive gain Ito NSAF\tabularnewline
		\addlinespace
		AGS-NSAF & : & Adaptive gain Stratonovich NSAF\tabularnewline
		\addlinespace
		GP-NSAF & :  & Guaranteed performance NSAF\tabularnewline
		\addlinespace
		GPSd-NSAF & :  & Guaranteed performance semi-direct NSAF\tabularnewline
		\addlinespace
		GPD-NSAF & :  & Guaranteed performance direct NSAF\tabularnewline
		\bottomrule
	\end{tabular}
\end{table}

\begin{table}[H]
	\centering{}\caption{\label{tab:Table-of-Notations1}Mathematical Notation}
	\begin{tabular}{lll}
		\toprule 
		\addlinespace
		$\mathbb{N}$  & :  & The set of integer numbers\tabularnewline
		\addlinespace
		$\mathbb{R}_{+}$  & :  & The set of nonnegative real numbers\tabularnewline
		\addlinespace
		$\mathbb{R}^{n}$  & :  & Real $n$-dimensional vector\tabularnewline
		\addlinespace
		$\mathbb{R}^{n\times m}$  & :  & Real $n\times m$ dimensional matrix\tabularnewline
		\addlinespace
		$\left\Vert \cdot\right\Vert $  & :  & Euclidean norm, for $x\in\mathbb{R}^{n}$, $\left\Vert x\right\Vert =\sqrt{x^{\top}x}$\tabularnewline
		\addlinespace
		$\mathbb{S}^{2}$  & :  & Two-sphere, $\mathbb{S}^{2}=\left\{ \left.x=\left[x_{1},x_{2},x_{3}\right]^{\top}\in\mathbb{R}^{3}\right|\left\Vert x\right\Vert =1\right\} $\tabularnewline
		\addlinespace
		$\mathbb{S}^{3}$  & :  & 3-sphere, $\mathbb{S}^{3}=\left\{ \left.x\in\mathbb{R}^{4}\right|\left\Vert x\right\Vert =1\right\} $ \tabularnewline
		\addlinespace
		$^{\top}$  & :  & Transpose of a component\tabularnewline
		\addlinespace
		$\times$ & : & Cross multiplication\tabularnewline
		\addlinespace
		$\left[\,\cdot\,\right]_{\times}$ & : & Skew-symmetric of a matrix\tabularnewline
		\addlinespace
		$\mathbf{I}_{n}$  & :  & Identity matrix with dimension $n$-by-$n$\tabularnewline
		\addlinespace
		${\rm det}\left(\,\cdot\,\right)$  & :  & Determinant of a component\tabularnewline
		\addlinespace
		${\rm Tr}\left\{ \,\cdot\,\right\} $ & :  & Trace of a component\tabularnewline
		\addlinespace
		${\rm exp}\left(\,\cdot\,\right)$ & :  & Exponential value of a component\tabularnewline
		\addlinespace
		$\lambda\left(\,\cdot\,\right)$ & :  & A group of eigenvalues of a matrix\tabularnewline
		\addlinespace
		$\underline{\lambda}\left(\,\cdot\,\right)$ & :  & Minimum eigenvalue of a matrix\tabularnewline
		\addlinespace
		$\mathbb{E}\left[\,\cdot\,\right]$ & :  & Expected value of a component\tabularnewline
		\addlinespace
		$\mathbb{P}\left\{ \,\cdot\,\right\} $ & :  & Probability of a component\tabularnewline
		\bottomrule
	\end{tabular}
\end{table}

\begin{table}[H]
	\centering{}\caption{\label{tab:Table-of-Notations2}Attitude Notation}
	\begin{tabular}{ll>{\raggedright}p{12cm}}
		\toprule 
		\addlinespace
		$\left\{ \mathcal{I}\right\} $ & :  & Inertial-frame of reference\tabularnewline
		\addlinespace
		$\left\{ \mathcal{B}\right\} $ & :  & Body-frame of reference\tabularnewline
		\addlinespace
		$\mathbb{SO}\left(3\right)$  & :  & Special Orthogonal Group\tabularnewline
		\addlinespace
		$\mathfrak{so}\left(3\right)$  & :  & The space of $3\times3$ skew-symmetric matrices, and Lie-algebra
		of $\mathbb{SO}\left(3\right)$\tabularnewline
		\addlinespace
		$\boldsymbol{\mathcal{P}}_{a}$  & :  & Anti-symmetric projection operator\tabularnewline
		\addlinespace
		$\boldsymbol{\mathcal{P}}_{s}$  & :  & Symmetric projection operator\tabularnewline
		\midrule
		\addlinespace
		$R$  & :  & True attitude/Rotational matrix/Orientation of a rigid-body, $R\in\mathbb{SO}\left(3\right)$\tabularnewline
		\addlinespace
		$\Omega$  & :  & Angular velocity vector with $\Omega=\left[\Omega_{x},\Omega_{y},\Omega_{z}\right]^{\top}\in\mathbb{R}^{3}$\tabularnewline
		\addlinespace
		$\Omega_{m}$  & :  & Angular velocity measurement vector\tabularnewline
		\addlinespace
		$b$  & :  & The bias associated with $\Omega_{m}$, $b\in\mathbb{R}^{3}$\tabularnewline
		\addlinespace
		$\omega$  & :  & The noise associated with $\Omega_{m}$, $\omega\in\mathbb{R}^{3}$\tabularnewline
		\addlinespace
		$\mathcal{Q}_{\omega}$ & : & Diagonal covariance matrix of the noise $\omega$\tabularnewline
		\addlinespace
		$\sigma$ & : & Upper bound of $\mathcal{Q}_{\omega}$, $\sigma\in\mathbb{R}^{3}$\tabularnewline
		\addlinespace
		${\rm v}_{i}^{\mathcal{I}}$ & :  & The $i$th vector in the inertial-frame, ${\rm v}_{i}^{\mathcal{I}}\in\mathbb{R}^{3}$\tabularnewline
		\addlinespace
		${\rm v}_{i}^{\mathcal{B}}$ & :  & The $i$th vector in the body-frame, ${\rm v}_{i}^{\mathcal{B}}\in\mathbb{R}^{3}$\tabularnewline
		\addlinespace
$\mathring{{\rm v}}_{i}^{\mathcal{B}}$ & :  & The true value of the $i$th vector in the body-frame, $\mathring{{\rm v}}_{i}^{\mathcal{B}}\in\mathbb{R}^{3}$\tabularnewline
		\addlinespace
		$b_{i}^{\mathcal{B}}$  & :  & The $i$th bias component of ${\rm v}_{i}^{\mathcal{B}}$, $b_{i}^{\mathcal{B}}\in\mathbb{R}^{3}$\tabularnewline
		\addlinespace
		$\omega_{i}^{\mathcal{B}}$ & :  & The $i$th noise component of ${\rm v}_{i}^{\mathcal{B}}$, $\omega_{i}^{\mathcal{B}}\in\mathbb{R}^{3}$\tabularnewline
		\addlinespace
		$s_{i}$ & :  & The confidence level of $i$th measurement, $s_{i}\in\mathbb{R}_{+}$\tabularnewline
		\addlinespace
		$\upsilon_{i}^{\mathcal{I}}$ & :  & Normalized value of ${\rm v}_{i}^{\mathcal{I}}$, $\upsilon_{i}^{\mathcal{I}}\in\mathbb{R}^{3}$\tabularnewline
		\addlinespace
		$\upsilon_{i}^{\mathcal{B}}$ & :  & Normalized value of ${\rm v}_{i}^{\mathcal{B}}$, $\upsilon_{i}^{\mathcal{B}}\in\mathbb{R}^{3}$\tabularnewline
		$\mathring{\upsilon}_{i}^{\mathcal{B}}$ & :  & Normalized value of $\mathring{{\rm v}}_{i}^{\mathcal{B}}$, $\mathring{\upsilon}_{i}^{\mathcal{B}}\in\mathbb{R}^{3}$\tabularnewline
		\midrule
		\addlinespace
		$\mathcal{R}_{Q}$  & :  & Attitude representation obtained using unit-quaternion vector, $\mathcal{R}_{Q}\in\mathbb{SO}\left(3\right)$\tabularnewline
		\addlinespace
		$Q$  & :  & True unit-quaternion vector, $Q=\left[q_{0},q^{\top}\right]^{\top}\in\mathbb{S}^{3}$\tabularnewline
		\addlinespace
		$Q^{*}$  & :  & Complex conjugate of unit-quaternion, $Q^{*}\in\mathbb{S}^{3}$\tabularnewline
		\addlinespace
		$\odot$  & :  & Multiplication operator of two unit-quaternion vectors\tabularnewline
		\midrule
		\addlinespace
		$\mathcal{R}_{\alpha}$  & :  & Attitude representation obtained using angle-axis parameterization,
		$\mathcal{R}_{\alpha}\in\mathbb{SO}\left(3\right)$\tabularnewline
		\addlinespace
		$\alpha$  & :  & Angle of rotation, $\alpha\in\mathbb{R}$\tabularnewline
		\addlinespace
		$u$  & :  & Unit vector, $u=\left[u_{1},u_{2},u_{3}\right]^{\top}\in\mathbb{S}^{2}$\tabularnewline
		\midrule
		\addlinespace
		$\mathcal{R}_{\rho}$  & :  & Attitude representation obtained using Rodriguez vector, $\mathcal{R}_{\rho}\in\mathbb{SO}\left(3\right)$\tabularnewline
		\addlinespace
		$\rho$  & :  & Rodriguez vector, $\rho=\left[\rho_{1},\rho_{2},\rho_{3}\right]^{\top}\in\mathbb{R}^{3}$\tabularnewline
		\midrule
		\addlinespace
		$R_{y}$ & : & Reconstructed attitude, $R_{y}\in\mathbb{SO}\left(3\right)$\tabularnewline
		\addlinespace
		$||R||_{I}$ & : & Normalized Euclidean distance of $R\in\mathbb{SO}\left(3\right)$\tabularnewline

	\end{tabular}
\end{table}
\begin{table}[H]
	\centering{}
	\begin{tabular}{ll>{\raggedright}p{12cm}}
		\addlinespace
$Q_{y}$ & : & Reconstructed unit-quaternion, $Q_{y}\in\mathbb{S}^{3}$\tabularnewline
		\addlinespace
		$Q_{{\rm opt}}$ & : & Optimal unit-quaternion, $Q_{{\rm opt}}\in\mathbb{S}^{3}$\tabularnewline
		\addlinespace
		$\hat{R}$ & : & Estimate of the true attitude, $\hat{R}\in\mathbb{SO}\left(3\right)$\tabularnewline
		\addlinespace
		$\hat{Q}$ & : & Estimate of the true unit-quaternion, $\hat{Q}\in\mathbb{S}^{3}$\tabularnewline
		\addlinespace
		$\hat{b}$ & : & Estimate of the true bias, $\hat{b}\in\mathbb{R}^{3}$\tabularnewline
		\addlinespace
		$\hat{\sigma}$ & : & Estimate of $\sigma$, $\hat{\sigma}\in\mathbb{R}^{3}$\tabularnewline
		\addlinespace
		$\tilde{R}$ & : & Attitude error, $\tilde{R}\in\mathbb{SO}\left(3\right)$\tabularnewline
		\addlinespace
		$\tilde{\rho}$ & : & Rodriguez vector error, $\tilde{\rho}\in\mathbb{R}^{3}$\tabularnewline
		\addlinespace
		$\tilde{\alpha}$ & : & Angle of rotation error, $\tilde{\alpha}\in\mathbb{R}$\tabularnewline
		\addlinespace
		$\tilde{b}$ & : & Bias error, $\tilde{b}\in\mathbb{R}^{3}$\tabularnewline
		\addlinespace
		$\tilde{\sigma}$ & : & Upper bound covariance error, $\tilde{\sigma}\in\mathbb{R}^{3}$\tabularnewline
		\addlinespace
		$\mathcal{E}$ & : & Unconstrained error or transformed error, $\mathcal{E}\in\mathbb{R}$\tabularnewline
		\addlinespace
		$\xi$ & : & Prescribed performance measure function, $\xi\in\mathbb{R}$\tabularnewline
		\addlinespace
		$\xi_{0}$ & : & Intial value of $\xi$ (upper bound), $\xi_{0}\in\mathbb{R}$\tabularnewline
		\addlinespace
		$\xi_{\infty}$ & : & Steady-state value of $\xi$ (lower bound), $\xi_{\infty}\in\mathbb{R}$\tabularnewline
		\addlinespace
		$\ell$ & : & Convergence factor of $\xi$ from $\xi_{0}$ to $\xi_{\infty}$, $\ell\in\mathbb{R}$\tabularnewline
		\bottomrule
	\end{tabular}
\end{table}

Let $\mathbb{SO}\left(3\right)$ denote the Special Orthogonal Group.
The relative orientation of a rigid-body in the body-frame $\left\{ \mathcal{B}\right\} $
with respect to the inertial frame $\left\{ \mathcal{I}\right\} $
is referred to as attitude or a rotational matrix $R$ and is given
by:
\[
\mathbb{SO}\left(3\right):=\left\{ \left.R\in\mathbb{R}^{3\times3}\right|R^{\top}R=RR^{\top}=\mathbf{I}_{3}\text{, }{\rm det}\left(R\right)=1\right\} 
\]
with ${\rm det\left(\cdot\right)}$ denoting a determinant of a matrix.
The Lie-algebra related to $\mathbb{SO}\left(3\right)$ is denoted
by $\mathfrak{so}\left(3\right)$ and is defined by
\[
\mathfrak{so}\left(3\right):=\left\{ \left.\mathcal{X}=\left[\begin{array}{ccc}
0 & -x_{3} & x_{2}\\
x_{3} & 0 & -x_{1}\\
-x_{2} & x_{1} & 0
\end{array}\right]\right|\mathcal{X}^{\top}=-\mathcal{X}\right\} 
\]
where $\mathcal{X}\in\mathbb{R}^{3\times3}$ is a skew-symmetric matrix.
The map $\left[\cdot\right]_{\times}:\mathbb{R}^{3}\rightarrow\mathfrak{so}\left(3\right)$
is given by
\begin{equation}
\mathcal{X}=\left[x\right]_{\times}=\left[\begin{array}{ccc}
0 & -x_{3} & x_{2}\\
x_{3} & 0 & -x_{1}\\
-x_{2} & x_{1} & 0
\end{array}\right],\hspace{1em}x=\left[\begin{array}{c}
x_{1}\\
x_{2}\\
x_{3}
\end{array}\right]\in\mathbb{R}^{3}\label{eq:Comp_skew}
\end{equation}
For $x,y\in\mathbb{R}^{3}$, one has
\[
\left[x\right]_{\times}y=x\times y
\]
where $\times$ is a cross product of the two given vectors. The mapping
of a skew-symmetric matrix $\left[\cdot\right]_{\times}$ to vector
form is defined by a vex operator $\mathbf{vex}:\mathfrak{so}\left(3\right)\rightarrow\mathbb{R}^{3}$
such that
\[
\mathbf{vex}\left(\mathcal{X}\right)=x
\]
with $x\in\mathbb{R}^{3}$ and $\mathcal{X}\in\mathfrak{so}\left(3\right)$
as defined in \eqref{eq:Comp_skew}. Let $\boldsymbol{\mathcal{P}}_{a}$
be the anti-symmetric projection operator on the Lie-algebra $\mathfrak{so}\left(3\right)$
\cite{murray1994mathematical}. The related mapping is given by $\boldsymbol{\mathcal{P}}_{a}:\mathbb{R}^{3\times3}\rightarrow\mathfrak{so}\left(3\right)$
\begin{equation}
\boldsymbol{\mathcal{P}}_{a}\left(\mathcal{Y}\right)=\frac{1}{2}\left(\mathcal{Y}-\mathcal{Y}^{\top}\right)\in\mathfrak{so}\left(3\right),\hspace{1em}\mathcal{Y}\in\mathbb{R}^{3\times3}\label{eq:Comp_Pa}
\end{equation}
The symmetric projection operator in the space of a square matrix
is given by
\begin{equation}
\boldsymbol{\mathcal{P}}_{s}\left(\mathcal{Y}\right)=\frac{1}{2}\left(\mathcal{Y}+\mathcal{Y}^{\top}\right),\hspace{1em}\mathcal{Y}\in\mathbb{R}^{3\times3}\label{eq:Comp_Ps}
\end{equation}
The normalized Euclidean distance of a rotational matrix $R\in\mathbb{SO}\left(3\right)$
can be represented as follows
\begin{equation}
||R||_{I}:=\frac{1}{4}{\rm Tr}\left\{ \mathbf{I}_{3}-R\right\} \label{eq:Comp_PER_Ecul_Dist}
\end{equation}
where ${\rm Tr}\left\{ \cdot\right\} $ denotes a trace of the matrix
and $||R||_{I}\in\left[0,1\right]$. The following identities will
prove useful in the subsequent derivations:
\begin{align}
||\alpha||^{2}= & {\rm Tr}\left\{ \alpha\alpha^{\top}\right\} \label{eq:Comp_Identity_Trace}\\
\left[\alpha\times\beta\right]_{\times}= & \beta\alpha^{\top}-\alpha\beta^{\top},\quad\alpha,\beta\in{\rm \mathbb{R}}^{3}\label{eq:Comp_Identity1}\\
\left[R\alpha\right]_{\times}= & R\left[\alpha\right]_{\times}R^{\top},\quad R\in\mathbb{SO}\left(3\right),\alpha\in\mathbb{R}^{3}\label{eq:Comp_Identity2}\\
\left[\alpha\right]_{\times}^{2}= & -||\alpha||^{2}\mathbf{I}_{3}+\alpha\alpha^{\top},\quad\alpha\in\mathbb{R}^{3}\label{eq:Comp_Identity3}\\
\left[A,B\right]= & AB-BA,\quad A,B\in\mathbb{R}^{3\times3}\label{eq:Comp_Identity8}\\
{\rm Tr}\left\{ \left[A,B\right]\right\} = & {\rm Tr}\left\{ AB-BA\right\} =0,\quad A,B\in\mathbb{R}^{3\times3}\label{eq:Comp_Identity5}\\
{\rm Tr}\left\{ B\left[\alpha\right]_{\times}\right\} = & 0,\quad B=B^{\top}\in\mathbb{R}^{3\times3},\alpha\in\mathbb{R}^{3}\label{eq:Comp_Identity6}\\
{\rm Tr}\left\{ A\left[\alpha\right]_{\times}\right\} = & {\rm Tr}\left\{ \boldsymbol{\mathcal{P}}_{a}\left(A\right)\left[\alpha\right]_{\times}\right\} =-2\mathbf{vex}\left(\boldsymbol{\mathcal{P}}_{a}\left(A\right)\right)^{\top}\alpha,\quad A\in\mathbb{R}^{3\times3},\alpha\in\mathbb{R}^{3}\label{eq:Comp_Identity7}\\
B\left[\alpha\right]_{\times}+\left[\alpha\right]_{\times}B= & {\rm Tr}\left\{ B\right\} \left[\alpha\right]_{\times}-\left[B\alpha\right]_{\times},\quad B=B^{\top}\in\mathbb{R}^{3\times3},\alpha\in\mathbb{R}^{3}\label{eq:Comp_Identity4}
\end{align}
The unit-quaternion is defined by
\[
Q=\left[\begin{array}{c}
q_{0}\\
q
\end{array}\right]\in\mathbb{S}^{3}
\]
where $q_{0}\in\mathbb{R}$ and $q=\left[q_{1},q_{2},q_{3}\right]^{\top}\in\mathbb{R}^{3}$
such that
\begin{equation}
\mathbb{S}^{3}=\left\{ \left.Q\in\mathbb{R}^{4}\right|\left\Vert Q\right\Vert =1\right\} \label{eq:Comp_PER_Q_1}
\end{equation}
Let $Q=\left[q_{0},q^{\top}\right]^{\top}\in\mathbb{S}^{3}$. Hence,
$Q^{*}=Q^{-1}\in\mathbb{S}^{3}$ can be defined as follows
\begin{equation}
Q^{*}=Q^{-1}=\left[\begin{array}{c}
q_{0}\\
-q
\end{array}\right]\in\mathbb{S}^{3}\label{eq:Comp_PER_Q_2}
\end{equation}
where $Q^{*}$ and $Q^{-1}$ are a complex conjugate and an inverse
of the unit-quaternion, respectively. For any $Q_{1},Q_{2}\in\mathbb{S}^{3}$,
the quaternion product between $Q_{1}$ and $Q_{2}$ can be found
in the following manner
\begin{align}
Q_{3}=Q_{1}\odot Q_{2} & =\left[\begin{array}{c}
q_{01}\\
q_{1}
\end{array}\right]\odot\left[\begin{array}{c}
q_{02}\\
q_{2}
\end{array}\right]\nonumber \\
& =\left[\begin{array}{c}
q_{01}q_{02}-q_{1}^{\top}q_{2}\\
q_{01}q_{2}+q_{02}q_{1}+\left[q_{1}\right]_{\times}q_{2}
\end{array}\right]\in\mathbb{S}^{3}\label{eq:Comp_PER_Q_Mul}
\end{align}
where $q_{01},q_{02}\in\mathbb{R}$ and $q_{1},q_{2}\in\mathbb{R}^{3}$.
The coordinates of a moving frame can be defined with respect to the
reference frame:
\begin{align}
\mathcal{R}_{Q}\left(Q\right) & =\left(q_{0}^{2}-\left\Vert q\right\Vert ^{2}\right)\mathbf{I}_{3}+2qq^{\top}+2q_{0}\left[q\right]_{\times}\nonumber \\
& =\mathbf{I}_{3}+2q_{0}\left[q\right]_{\times}+2\left[q\right]_{\times}^{2}\label{eq:Comp_Q_R}
\end{align}
The attitude of a rigid-body can be obtained given a unit-axis $u\in\mathbb{R}^{3}$
and an angle of rotation $\alpha\in\mathbb{R}$ in the 2-sphere $\mathbb{S}^{2}$
\cite{shuster1993survey,hashim2019AtiitudeSurvey}
\begin{align}
\mathcal{R}_{\alpha}\left(\alpha,u\right) & ={\rm exp}\left(\alpha\left[u\right]_{\times}\right)\nonumber \\
& =\mathbf{I}_{3}+\sin\left(\alpha\right)\left[u\right]_{\times}+\left(1-\cos\left(\alpha\right)\right)\left[u\right]_{\times}^{2}\label{eq:Comp_R_att_ang}
\end{align}
Also, the attitude can be established using Rodriguez parameters vector
$\rho=\left[\rho_{1},\rho_{2},\rho_{3}\right]^{\top}\in\mathbb{R}^{3}$
such that related map from vector form to $\mathbb{SO}\left(3\right)$
is
\begin{align}
\mathcal{R}_{\rho}= & \frac{1}{1+\left\Vert \rho\right\Vert ^{2}}\left(\left(1-\left\Vert \rho\right\Vert ^{2}\right)\mathbf{I}_{3}+2\rho\rho^{\top}+2\left[\rho\right]_{\times}\right)\label{eq:OVERVIEW_PER_ROD}
\end{align}
A more thorough overview of attitude mapping, important properties
and helpful notes can be found in \cite{hashim2019AtiitudeSurvey}.

\noindent\makebox[1\linewidth]{%
	\rule{0.8\textwidth}{1.4pt}%
}

\section{Attitude Dynamics and Measurements \label{sec:SO3_kinematics}}

\begin{figure}[h]
	\centering{}\includegraphics[scale=0.4]{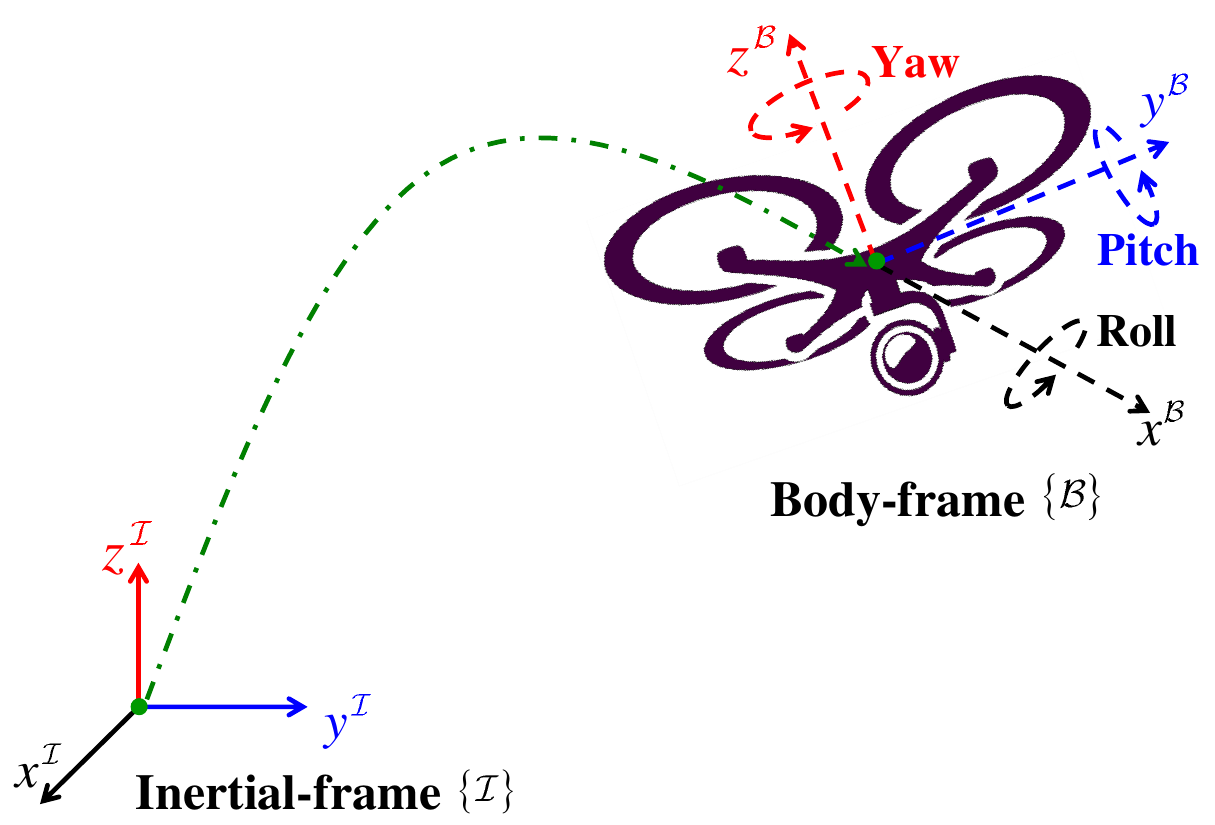}\caption{The orientation of a 3D rigid-body in body-frame relative to inertial-frame
		\cite{hashim2018SO3Stochastic}.}
	\label{fig:SO3_PPF_STCH_1}
\end{figure}

Let $R\in\mathbb{SO}\left(3\right)$ denote the attitude (rotational
matrix), which describes the relative orientation of the moving rigid-body
in the body-frame $\left\{ \mathcal{B}\right\} $ with respect to
the fixed inertial-frame $\left\{ \mathcal{I}\right\} $ as illustrated
in Figure \ref{fig:SO3_PPF_STCH_1}. 

The attitude can be extracted through $n$-known non-collinear vectors
in the inertial-frame and their measurements done relative to the
coordinate system fixed to the rigid-body. For simplicity, let the
superscripts $\mathcal{I}$ and $\mathcal{B}$ indicate that a vector
is associated with the inertial-frame and body-frame, respectively.
Let ${\rm v}_{i}^{\mathcal{I}}\in\mathbb{R}^{3}$ be a known vector
in the inertial-frame which is measured in the coordinate system fixed
to the rigid-body such that
\begin{equation}
{\rm v}_{i}^{\mathcal{B}}=R^{\top}{\rm v}_{i}^{\mathcal{I}}+{\rm b}_{i}^{\mathcal{B}}+\omega_{i}^{\mathcal{B}}\in\mathbb{R}^{3}\label{eq:Comp_Vect_True}
\end{equation}
where ${\rm b}_{i}^{\mathcal{B}}\in\mathbb{R}^{3}$ stands for the
bias component, and $\omega_{i}^{\mathcal{B}}\in\mathbb{R}^{3}$ denotes
the noise component attached to the $i$th body-frame measurement
for all $i=1,2,\ldots,n$. The measurement in \eqref{eq:Comp_Vect_True}
represents output of a typical measurement unit attached to a moving
body. However, the values of ${\rm b}_{i}^{\mathcal{B}}$ and $\omega_{i}^{\mathcal{B}}$
are heavily dependent on the quality of the measurement unit. Define
the following two sets
\begin{align}
{\rm v}^{\mathcal{I}} & =\left[{\rm v}_{1}^{\mathcal{I}},{\rm v}_{2}^{\mathcal{I}},\ldots,{\rm v}_{n}^{\mathcal{I}}\right]\in\mathbb{R}^{3\times n}\nonumber \\
{\rm v}^{\mathcal{B}} & =\left[{\rm v}_{1}^{\mathcal{B}},{\rm v}_{2}^{\mathcal{B}},\ldots,{\rm v}_{n}^{\mathcal{B}}\right]\in\mathbb{R}^{3\times n}\label{eq:Comp_Set}
\end{align}

\begin{rem}
	\label{rem:Rem1_attitude}The attitude can be extracted given the
	availability of at least two known non-collinear observations in the
	inertial-frame and their measurements in the body-frame (The sets
	in \eqref{eq:Comp_Set} are at least of rank two). In case when $n=2$,
	the third inertial-frame and body-frame vectors can be obtained by
	the cross product such that ${\rm v}_{3}^{\mathcal{I}}={\rm v}_{1}^{\mathcal{I}}\times{\rm v}_{2}^{\mathcal{I}}$
	and ${\rm v}_{3}^{\mathcal{B}}={\rm v}_{1}^{\mathcal{B}}\times{\rm v}_{2}^{\mathcal{B}}$,
	respectively, which ensures non-collinearity of the vectors ${\rm v}_{1}^{\mathcal{I}}$,
	${\rm v}_{2}^{\mathcal{I}}$, and ${\rm v}_{3}^{\mathcal{I}}$ as
	well as ${\rm v}_{1}^{\mathcal{B}}$, ${\rm v}_{2}^{\mathcal{B}}$,
	and ${\rm v}_{3}^{\mathcal{B}}$.
\end{rem}
The dynamics of the true attitude are described by
\begin{equation}
\dot{R}=R\left[\Omega\right]_{\times}\label{eq:Comp_R_dynam}
\end{equation}
where $\Omega\in\mathbb{R}^{3}$ is the true value of angular velocity.
Angular velocity of a moving body can be measured by the rate gyros,
and its typical measurement is equivalent to
\begin{equation}
\Omega_{m}=\Omega+b+\omega\in\left\{ \mathcal{B}\right\} \label{eq:Comp_Angular}
\end{equation}
where $b$ and $\omega$ denote the bias and noise components, respectively,
attached to the measurement of angular velocity for all $b,\omega\in\mathbb{R}^{3}$.
A low-cost module of an inertial measurement unit may consist of three
different measuring subunits:
\begin{itemize}
	\item[i)] 3-axis magnetometers which can be represented by
	\[
	{\rm v}_{1}^{\mathcal{B}}=R^{\top}{\rm v}_{1}^{\mathcal{I}}+{\rm b}_{1}^{\mathcal{B}}+\omega_{1}^{\mathcal{B}}
	\]
	with ${\rm v}_{1}^{\mathcal{I}}$ being the earth-magnetic field,
	and ${\rm b}_{1}^{\mathcal{B}}$ and $\omega_{1}^{\mathcal{B}}$ being
	the additive unknown bias and noise components, respectively.
	\item[ii)] 3-axis accelerometers that can be represented by
	\[
	{\rm v}_{2}^{\mathcal{B}}=R^{\top}\left(\dot{\mathcal{V}}-{\rm v}_{2}^{\mathcal{I}}\right)+{\rm b}_{2}^{\mathcal{B}}+\omega_{2}^{\mathcal{B}}
	\]
	with ${\rm v}_{2}^{\mathcal{I}}:=\left[0,0,g^{\mathcal{I}}\right]^{\top}\approx\left[0,0,9.8\right]^{\top}$
	being the gravitational acceleration field defined in $\left\{ \mathcal{I}\right\} $,
	$\dot{\mathcal{V}}\in\mathbb{R}^{3}$ denoting the linear acceleration
	in $\left\{ \mathcal{I}\right\} $, and ${\rm b}_{2}^{\mathcal{B}}$
	and $\omega_{2}^{\mathcal{B}}$ being the unknown bias and noise components
	added during the measurement process, respectively. At low frequency,
	$\left\Vert {\rm v}_{2}^{\mathcal{I}}\right\Vert >>||\dot{\mathcal{V}}||$
	which allow one to obtain
	\[
	{\rm v}_{2}^{\mathcal{B}}\approx-R^{\top}{\rm v}_{2}^{\mathcal{I}}+{\rm b}_{2}^{\mathcal{B}}+\omega_{2}^{\mathcal{B}}
	\]
	\item[iii)] 3-axis rate gyros record the angular velocity measurement which can
	be denoted by $\Omega_{m}$ as defined in \eqref{eq:Comp_Angular}.
\end{itemize}
Attitude determination or estimation may be utilized through normalized
values of the vectorial measurements. The $i$th inertial-frame and
body-frame vectors in \eqref{eq:Comp_Vect_True} are normalized in
the following manner:
\begin{equation}
\upsilon_{i}^{\mathcal{I}}=\frac{{\rm v}_{i}^{\mathcal{I}}}{||{\rm v}_{i}^{\mathcal{I}}||},\hspace{1em}\upsilon_{i}^{\mathcal{B}}=\frac{{\rm v}_{i}^{\mathcal{B}}}{||{\rm v}_{i}^{\mathcal{B}}||},\hspace{1em}\forall i=1,2,\cdots,n\label{eq:Comp_Vector_norm}
\end{equation}
As such, the sets of normalized values presented below get utilized
by the attitude determination or estimation algorithms:
\begin{align}
\upsilon^{\mathcal{I}} & =\left[\upsilon_{1}^{\mathcal{I}},\upsilon_{2}^{\mathcal{I}},\ldots,\upsilon_{n}^{\mathcal{I}}\right]\in\mathbb{R}^{3\times n}\nonumber \\
\upsilon^{\mathcal{B}} & =\left[\upsilon_{1}^{\mathcal{B}},\upsilon_{2}^{\mathcal{B}},\ldots,\upsilon_{n}^{\mathcal{B}}\right]\in\mathbb{R}^{3\times n}\label{eq:Comp_Set-1}
\end{align}
The exact integration of \eqref{eq:Comp_R_dynam} is equivalent to
\begin{equation}
R\left[k+1\right]=R\left[k\right]\exp\left(\left[\Omega\left[k\right]\right]_{\times}\Delta t\right)\label{eq:Comparison_R_discrete}
\end{equation}
where $\Delta t$ is a small time sample and $\left[k\right]$ associated
with a variable refers to its value at the $k$th sample for $k\in\mathbb{N}$.

On the other side, the vectorial measurements in \eqref{eq:Comp_Vect_True},
can be written in terms of unit-quaternion as
\begin{equation}
\left[\begin{array}{c}
0\\
{\rm v}_{i}^{\mathcal{B}}
\end{array}\right]=Q^{-1}\odot\left[\begin{array}{c}
0\\
{\rm v}_{i}^{\mathcal{I}}
\end{array}\right]\odot Q+\left[\begin{array}{c}
0\\
{\rm b}_{i}^{\mathcal{B}}
\end{array}\right]+\left[\begin{array}{c}
0\\
\omega_{i}^{\mathcal{B}}
\end{array}\right]\in\mathbb{R}^{4}\label{eq:Comp_Q_VM}
\end{equation}
where $Q\in\mathbb{S}^{3}$. In the same spirit, attitude dynamics
can be redefined in terms of unit-quaternion as
\begin{align}
\dot{Q} & =\frac{1}{2}\Gamma\left(\Omega\right)Q=\frac{1}{2}\left[\begin{array}{cc}
0 & -\Omega^{\top}\\
\Omega & -\left[\Omega\right]_{\times}
\end{array}\right]Q\label{eq:Comp_Q_dynam}
\end{align}
The exact integration of \eqref{eq:Comp_Q_dynam} results into
\begin{equation}
Q\left[k+1\right]=\exp\left(\frac{1}{2}\Gamma\left(\Omega\left[k\right]\right)\Delta t\right)Q\left[k\right]\label{eq:Comp_Q_discrete}
\end{equation}
For a comprehensive overview of attitude parameterization, mapping
and related useful properties visit \cite{hashim2019AtiitudeSurvey}.
\begin{defn}
	\label{def:Comp_unstable}Consider a forward invariant unstable set
	$\mathcal{U}_{s}\subseteq\mathbb{SO}\left(3\right)$ defined by
	\begin{equation}
	\mathcal{U}_{s}=\left\{ \left.R\in\mathbb{SO}\left(3\right)\right|{\rm Tr}\left\{ R\right\} =-1\right\} \label{eq:Comp_Us}
	\end{equation}
	where ${\rm Tr}\left\{ R\right\} =-1$ only at one of the following
	three orientations
	\begin{equation}
	\left\{ \begin{array}{c}
	R=\left[\begin{array}{ccc}
	-1 & 0 & 0\\
	0 & -1 & 0\\
	0 & 0 & 1
	\end{array}\right]\\
	\\
	R=\left[\begin{array}{ccc}
	-1 & 0 & 0\\
	0 & 1 & 0\\
	0 & 0 & -1
	\end{array}\right]\\
	\\
	R=\left[\begin{array}{ccc}
	1 & 0 & 0\\
	0 & -1 & 0\\
	0 & 0 & -1
	\end{array}\right]
	\end{array}\right.\label{eq:Comp_Us_R}
	\end{equation}
	Directly substituting $R$ value in \eqref{eq:Comp_Us_R} with its
	definition in \eqref{eq:Comp_PER_Ecul_Dist}, it becomes apparent
	that ${\rm Tr}\left\{ R\right\} =-1$ implies that $||R||_{I}=1$.
\end{defn}
\begin{assum}
	\label{Assum:Comp_b} (Uniform boundedness of unknown bias $b$ in
	\eqref{eq:Comp_Angular}) Let vector $b$ belong to a given compact
	set $\Delta_{b}$ where $b\in\Delta_{b}\subset\mathbb{R}^{3}$, and
	let $b$ be upper bounded by a scalar $\Gamma_{b}$ such that $\left\Vert \Delta_{b}\right\Vert \leq\Gamma_{b}<\infty$.
\end{assum}
\begin{assum}
	\label{Assum:Comp_s} (Uniform boundedness of unknown noise $\omega$
	in \eqref{eq:Comp_Angular}) Let vector $\omega$ belong to a given
	compact set $\Delta_{\omega}$ where $\omega\in\Delta_{\omega}\subset\mathbb{R}^{3}$,
	and let $\omega$ be upper bounded by a scalar $\Gamma_{\omega}$
	such that $\left\Vert \Delta_{\omega}\right\Vert \leq\Gamma_{\omega}<\infty$.
\end{assum}
\noindent\makebox[1\linewidth]{%
	\rule{0.8\textwidth}{1.4pt}%
}

\section{Attitude Determination\label{sec:Comp_Determination}}

As previously mentioned, attitude determination or estimation is an
essential sub-task in most robotics and control applications. The
attitude can be determined using a set of vector measurements made
in body-frame and their observations in the inertial-frame as it acts
as a linear transformation from one frame to the other \cite{hashim2018SO3Stochastic,hashim2019SO3Det}.
Attitude determination, in contrast to attitude estimation, takes
an algebraic approach to attitude reconstruction. Every attitude determination
and estimation algorithm holds minimization of the cost function as
its main objective. Wahba's Problem presents an example of such a
cost function \cite{wahba1965least}:
\begin{equation}
\mathcal{J}\left(R\right)=\frac{1}{2}\sum_{i=1}^{n}s_{i}\left\Vert {\rm v}_{i}^{\mathcal{B}}-R^{\top}{\rm v}_{i}^{\mathcal{I}}\right\Vert ^{2}\label{eq:Comparison_wahba}
\end{equation}
where $s_{i}\in\mathbb{R}_{+}$ is the confidence level of the $i$th
sensor measurement and at the same time it is a non-negative weight.
The work proposed by \cite{wahba1965least} was purely algebraic.
Over the following decades, a considerable effort was made in developing
attitude determination algorithms based on a set of simultaneous inertial
and body-frame vectors, for instance \cite{davenport1968vector,shuster1978approximate,shuster1981three,markley1988attitude,markley1993attitude,mortari2000second,lerner1978three,tanygin2007many}.
All the algorithms in \cite{davenport1968vector,shuster1978approximate,shuster1981three,markley1988attitude,markley1993attitude,mortari2000second,lerner1978three,tanygin2007many}
are applicable if and only if the statement in Remark \ref{rem:Rem1_attitude}
is met. In the subsections that follow, three common algebraic attitude
determination algorithms are detailed, namely, triaxial attitude determination
(TRIAD) \cite{black1964passive}, quaternion estimator (QUEST) \cite{shuster1981three},
and singular value decomposition (SVD) \cite{markley1988attitude}.

\noindent\makebox[1\linewidth]{%
	\rule{0.8\textwidth}{0.5pt}%
}

\subsection{Attitude Determination using TRIAD Algorithm \label{subsec:Comp_Attit_TRIAD}}

TRIaxial Attitude Determination (TRIAD) algorithm is one of the earliest
and the simplest methods of attitude determination \cite{black1964passive}.
The TRIAD algorithm has been commonly used as a tool of attitude determination
for almost two decades from the date invented until its replacement
by more advanced algorithms. The underlining assumption of the TRIAD
algorithm is the availability of two non-collinear vector observations
at each time instant. Also, these vectors have to be non-collinear.
The implementation of the TRIAD algorithm can be summarized in the
following three steps:
\begin{equation}
\begin{cases}
& \text{\textbf{Step1)} normalization}\\
& \upsilon_{1}^{\mathcal{I}}={\rm v}_{1}^{\mathcal{I}}\left/||{\rm v}_{1}^{\mathcal{I}}||\right.,\hspace{1em}\upsilon_{2}^{\mathcal{I}}={\rm v}_{2}^{\mathcal{I}}\left/||{\rm v}_{2}^{\mathcal{I}}||\right.\\
& \upsilon_{1}^{\mathcal{B}}={\rm v}_{1}^{\mathcal{B}}\left/||{\rm v}_{1}^{\mathcal{B}}||\right.,\hspace{1em}\upsilon_{2}^{\mathcal{B}}={\rm v}_{2}^{\mathcal{B}}\left/||{\rm v}_{2}^{\mathcal{B}}||\right.\\
& \text{\textbf{Step2)} collect three non-collinear vectors}\\
& \bar{\boldsymbol{\upsilon}}_{1}^{\mathcal{I}}=\upsilon_{1}^{\mathcal{I}},\hspace{1em}\hspace{1em}\hspace{1.4em}\bar{\boldsymbol{\upsilon}}_{1}^{\mathcal{B}}=\upsilon_{1}^{\mathcal{B}}\\
& \bar{\boldsymbol{\upsilon}}_{2}^{\mathcal{I}}=\bar{\boldsymbol{\upsilon}}_{1}^{\mathcal{I}}\times\upsilon_{2}^{\mathcal{I}},\hspace{1em}\bar{\boldsymbol{\upsilon}}_{2}^{\mathcal{B}}=\bar{\boldsymbol{\upsilon}}_{1}^{\mathcal{B}}\times\upsilon_{2}^{\mathcal{B}}\\
& \bar{\boldsymbol{\upsilon}}_{3}^{\mathcal{I}}=\bar{\boldsymbol{\upsilon}}_{1}^{\mathcal{I}}\times\bar{\boldsymbol{\upsilon}}_{2}^{\mathcal{I}},\hspace{1em}\bar{\boldsymbol{\upsilon}}_{3}^{\mathcal{B}}=\bar{\boldsymbol{\upsilon}}_{1}^{\mathcal{B}}\times\bar{\boldsymbol{\upsilon}}_{2}^{\mathcal{B}}\\
& \text{\textbf{Step3)} obtain }R_{y}\in\mathbb{SO}\left(3\right)\\
& \left[\begin{array}{ccc}
\bar{\boldsymbol{\upsilon}}_{1}^{\mathcal{B}} & \bar{\boldsymbol{\upsilon}}_{2}^{\mathcal{B}} & \bar{\boldsymbol{\upsilon}}_{3}^{\mathcal{B}}\end{array}\right]=R_{y}^{\top}\left[\begin{array}{ccc}
\bar{\boldsymbol{\upsilon}}_{1}^{\mathcal{I}} & \bar{\boldsymbol{\upsilon}}_{2}^{\mathcal{I}} & \bar{\boldsymbol{\upsilon}}_{3}^{\mathcal{I}}\end{array}\right]\\
& R_{y}=\left[\begin{array}{ccc}
\bar{\boldsymbol{\upsilon}}_{1}^{\mathcal{I}} & \bar{\boldsymbol{\upsilon}}_{2}^{\mathcal{I}} & \bar{\boldsymbol{\upsilon}}_{3}^{\mathcal{I}}\end{array}\right]\left[\begin{array}{ccc}
\bar{\boldsymbol{\upsilon}}_{1}^{\mathcal{B}} & \bar{\boldsymbol{\upsilon}}_{2}^{\mathcal{B}} & \bar{\boldsymbol{\upsilon}}_{3}^{\mathcal{B}}\end{array}\right]^{-1}\\
& \text{or}\\
& R_{y}=\left[\begin{array}{ccc}
\bar{\boldsymbol{\upsilon}}_{1}^{\mathcal{I}} & \bar{\boldsymbol{\upsilon}}_{2}^{\mathcal{I}} & \bar{\boldsymbol{\upsilon}}_{3}^{\mathcal{I}}\end{array}\right]\left[\begin{array}{ccc}
\bar{\boldsymbol{\upsilon}}_{1}^{\mathcal{B}} & \bar{\boldsymbol{\upsilon}}_{2}^{\mathcal{B}} & \bar{\boldsymbol{\upsilon}}_{3}^{\mathcal{B}}\end{array}\right]^{\top}
\end{cases}\label{eq:Comp_TRIAD}
\end{equation}
where $R_{y}\in\mathbb{SO}\left(3\right)$ denotes a reconstructed
attitude. The aim of the algorithm is to drive $R_{y}\rightarrow R$.
Later, a series of modifications of the basic TRIAD algorithm \cite{black1964passive}
were proposed, such as the symmetric TRIAD algorithm \cite{lerner1978three}
and optimal TRIAD algorithm \cite{tanygin2007many}. The main shortcoming
of the TRIAD algorithm in \cite{black1964passive} is that by design
it can use only two non-collinear observations. However, in spite
of the above-mentioned drawback, TRIAD is a pioneer algorithm that
served as a doorway to the more advanced methods and promoted the
growth of the attitude estimation and determination research. As such,
it is my believe that the earliest version of the TRIAD algorithm
is brilliant in its simplicity and can be considered a predecessor
of all the algorithms proposed after.

\noindent\makebox[1\linewidth]{%
	\rule{0.8\textwidth}{0.5pt}%
}

\subsection{Attitude Determination using QUEST\label{subsec:Comp_Attit_QUEST}}

TRIAD was displaced by QUaternion ESTimator (QUEST) \cite{shuster1981three},
since QUEST allowed for attitude determination when two or more non-collinear
observations are available ($n\geq2$), in consistence with Remark
\ref{rem:Rem1_attitude}. QUEST algorithm is able to find an optimal
solution to Wahba's problem \cite{wahba1965least} in \eqref{eq:Comparison_wahba}
given $n$ observations. QUEST algorithm is a modification of its
precursor, Davenport $q$-method \cite{davenport1968vector}, which
provided an early solution to Wahba's problem \cite{wahba1965least}.
Let us represent the attitude with respect to the true unit-quaternion
$Q=\left[q_{0},q^{\top}\right]^{\top}\in\mathbb{S}^{3}$ \cite{shuster1993survey,hashim2019AtiitudeSurvey}
\begin{equation}
\mathcal{R}_{Q}=\left(q_{0}^{2}-\left\Vert q\right\Vert ^{2}\right)\mathbf{I}_{3}+2qq^{\top}+2q_{0}\left[q\right]_{\times}\label{eq:Comp_Attit_Q}
\end{equation}
as defined in \eqref{eq:Comp_Q_R}. Consider the following weighting
scheme
\[
w_{i}=s_{i}\left/\sum_{i=1}^{n}s_{i}\right.
\]
with
\begin{equation}
B=\sum_{i=1}^{n}w_{i}\upsilon_{i}^{\mathcal{B}}\left(\upsilon_{i}^{\mathcal{I}}\right)^{\top}\label{eq:Comp_Attit_B}
\end{equation}
Since the attitude in \eqref{eq:Comp_Attit_Q} represents a homogeneous
quadratic function with respect to $Q$, one may obtain
\[
{\rm Tr}\left\{ \mathcal{R}_{Q}B^{\top}\right\} =Q^{\top}\mathcal{M}Q
\]
where $\mathcal{M}$ is a symmetric matrix equivalent to
\begin{equation}
\mathcal{M}=\left[\begin{array}{cc}
{\rm Tr}\left\{ B\right\}  & \left(\sum_{i=1}^{n}w_{i}\upsilon_{i}^{\mathcal{B}}\times\upsilon_{i}^{\mathcal{I}}\right)^{\top}\\
\sum_{i=1}^{n}w_{i}\upsilon_{i}^{\mathcal{B}}\times\upsilon_{i}^{\mathcal{I}} & B+B^{\top}-{\rm Tr}\left\{ B\right\} \mathbf{I}_{3}
\end{array}\right]\label{eq:Comp_Attit_M}
\end{equation}
Thus, it can be shown that the optimal unit-quaternion $Q_{{\rm opt}}\in\mathbb{S}^{3}$
satisfies
\[
\mathcal{M}Q_{{\rm opt}}=\max\left\{ \lambda\left(\mathcal{M}\right)\right\} Q_{{\rm opt}}
\]
Accordingly, the complete QUEST algorithm can be given as follows:
\begin{equation}
\begin{cases}
w_{i} & =s_{i}\left/\sum_{i=1}^{n}s_{i}\right.\\
B & =\sum_{i=1}^{n}w_{i}\upsilon_{i}^{\mathcal{B}}\left(\upsilon_{i}^{\mathcal{I}}\right)^{\top}\\
S & =B+B^{\top}\\
z & =\left[\begin{array}{c}
B_{23}-B_{32}\\
B_{31}-B_{13}\\
B_{12}-B_{21}
\end{array}\right]=\sum_{i=1}^{n}w_{i}\upsilon_{i}^{\mathcal{B}}\times\upsilon_{i}^{\mathcal{I}}\\
\mathcal{M} & =\left[\begin{array}{cc}
{\rm Tr}\left\{ B\right\}  & z^{\top}\\
z & S-{\rm Tr}\left\{ B\right\} \mathbf{I}_{3}
\end{array}\right]\\
\lambda_{{\rm max}} & =\max\left\{ \lambda\left(\mathcal{M}\right)\right\} \\
\beta_{1} & =\lambda_{{\rm max}}^{2}-{\rm Tr}\left\{ B\right\} ^{2}+{\rm Tr}\left\{ {\rm adj}\left(S\right)\right\} \\
\beta_{2} & =\lambda_{{\rm max}}-{\rm Tr}\left\{ B\right\} \\
x_{0} & =\det\left(\left(\lambda_{{\rm max}}+{\rm Tr}\left\{ B\right\} \right)\mathbf{I}_{3}-S\right)\\
x & =\left(\beta_{1}\mathbf{I}_{3}+\beta_{2}S+S^{2}\right)z\\
Q_{y} & =\frac{1}{\sqrt{x_{0}^{2}+\left\Vert x\right\Vert ^{2}}}\left[\begin{array}{c}
x_{0}\\
x
\end{array}\right]
\end{cases}\label{eq:Comp_QUEST}
\end{equation}
where $Q_{y}\left[q_{0y},q_{y}^{\top}\right]^{\top}\in\mathbb{S}^{3}$
denotes a reconstructed unit-quaternion with $q_{0y}\in\mathbb{R}$
and $q_{y}\in\mathbb{R}^{3}$, $\lambda\left(\mathcal{M}\right)$
represents a set of eigenvalues of matrix $\mathcal{M}$, and $\lambda_{{\rm max}}$
stands for the maximum value of $\lambda\left(\mathcal{M}\right)$.
Also, ${\rm adj}\left(S\right)$ is an adjoint or adjugate of the
square matrix $S$. It is obvious that the QUEST algorithm aims to
drive $Q_{y}\rightarrow Q$. Up to the current moment QUEST remains
one of the most widely used algorithms for solving Wahba's problem
\cite{markley1993attitude}.

\noindent\makebox[1\linewidth]{%
	\rule{0.8\textwidth}{0.5pt}%
}

\subsection{Attitude Determination using SVD\label{subsec:Comp_Attit_SVD}}

Singular Value Decomposition (SVD) is another commonly used method
of attitude determination. In consistence with Remark \ref{rem:Rem1_attitude},
it is able to use two or more non-collinear observations ($n\geq2$).
Considering the loss function in \cite{farrell1966least} and using
SVD algorithm, attitude can be determined through the following series
of steps \cite{markley1988attitude}:
\begin{equation}
\begin{cases}
& \text{\textbf{Step1)} normalize weights}\\
& w_{i}=s_{i}\left/\sum_{i=1}^{n}s_{i}\right.,\hspace{1em}i=1,2,\ldots,n\\
& \text{\textbf{Step2)} minimize loss function}\\
& \mathcal{J}\left(R\right)=1-\sum_{i=1}^{n}w_{i}\left(\upsilon_{i}^{\mathcal{B}}\right)^{\top}R^{\top}\upsilon_{i}^{\mathcal{I}}\\
& \hspace{3em}=1-{\rm Tr}\left\{ R^{\top}B^{\top}\right\} \\
& \text{where}\\
& B=\sum_{i=1}^{n}w_{i}\upsilon_{i}^{\mathcal{B}}\left(\upsilon_{i}^{\mathcal{I}}\right)^{\top}=USV^{\top}\\
& \text{\textbf{Step3)} solve for }U\text{, }S,\text{ and }V\text{ using SVD}\\
& \text{\textbf{Step4)} obtain }U_{+}\text{ and \ensuremath{V_{+}}}\\
& U_{+}=U\left[\begin{array}{ccc}
1 & 0 & 0\\
0 & 1 & 0\\
0 & 0 & {\rm det}\left(U\right)
\end{array}\right]\\
& V_{+}=V\left[\begin{array}{ccc}
1 & 0 & 0\\
0 & 1 & 0\\
0 & 0 & {\rm det}\left(V\right)
\end{array}\right]\\
& \text{\textbf{Step5)} obtain }R_{y}\in\mathbb{SO}\left(3\right)\\
& R_{y}=V_{+}U_{+}^{\top}
\end{cases}\label{eq:Comp_SVD}
\end{equation}
where $R_{y}\in\mathbb{SO}\left(3\right)$ denotes a reconstructed
attitude and SVD aims to drive $R_{y}\rightarrow R$. The solution
obtained by SVD is equivalent to the solution proposed in \cite{farrell1966least},
with the only difference being the necessity to compute the SVD. In
fact, SVD is one of the most robust numerical algorithms \cite{horn1990matrix}.

TRIAD, SVD, and QUEST algorithms outlined above along with \cite{davenport1968vector,shuster1978approximate,shuster1981three,markley1988attitude,markley1993attitude,mortari2000second,lerner1978three,tanygin2007many}
are among several other algebraic algorithms proposed for attitude
determination. Body-frame vector measurement are uncertain and are
subject to significant bias and noise, which is not accounted for
by the above-mentioned algebraic algorithms. In spite of their simplicity,
the category of attitude determination algorithms in \cite{davenport1968vector,shuster1978approximate,shuster1981three,markley1988attitude,markley1993attitude,mortari2000second,lerner1978three,tanygin2007many}
produce poor results in comparison with Gaussian and nonlinear attitude
filters as will be illustrated in Section \ref{sec:Comp_Simulation}.
Therefore, the attitude observation problem is best addressed using
Gaussian and nonlinear filters.

\noindent\makebox[1\linewidth]{%
	\rule{0.8\textwidth}{1.4pt}%
}

\section{Gaussian Attitude Filters\label{sec:Comp_Gaussian}}

Define $\Omega=\left[\Omega_{x},\Omega_{y},\Omega_{z}\right]^{\top}\in\mathbb{R}^{3}$
and let $Q=\left[q_{0},q^{\top}\right]^{\top}\in\mathbb{S}^{3}$ denote
a unit-quaternion vector that satisfies \eqref{eq:Comp_PER_Q_1}.
Referring to the notation above define the following set of equations:{\small{}
	\[
	\begin{cases}
	\bar{\Omega} & =\left[\begin{array}{c}
	0\\
	\Omega
	\end{array}\right]\\
	\Gamma\left(\Omega\right) & =\left[\begin{array}{cc}
	0 & -\Omega^{\top}\\
	\Omega & -\left[\Omega\right]_{\times}
	\end{array}\right]=\left[\begin{array}{cccc}
	0 & -\Omega_{x} & -\Omega_{y} & -\Omega_{z}\\
	\Omega_{x} & 0 & \Omega_{z} & -\Omega_{y}\\
	\Omega_{y} & -\Omega_{z} & 0 & \Omega_{x}\\
	\Omega_{z} & \Omega_{y} & -\Omega_{x} & 0
	\end{array}\right]\\
	\Xi\left(Q\right) & =\left[\begin{array}{c}
	-q^{\top}\\
	q_{0}\mathbf{I}_{3}+\left[q\right]_{\times}
	\end{array}\right]
	\end{cases}
	\]
}Recall the true attitude dynamics in unit-quaternion form $\dot{Q}=\frac{1}{2}\Gamma\left(\Omega\right)Q$
in \eqref{eq:Comp_Q_dynam}. Let $\hat{Q}=\left[\hat{q}_{0},\hat{q}^{\top}\right]^{\top}\in\mathbb{S}^{3}$
denote the estimate of the true unit-quaternion vector $Q$, where
$\hat{q}_{0}\in\mathbb{R}$ and $\hat{q}\in\mathbb{R}^{3}$. Gaussian
attitude filters aim to drive $\hat{Q}\rightarrow Q$. The general
design of a Gaussian attitude filter can be described with respect
to unit-quaternion vector as follows:
\begin{equation}
\dot{\hat{Q}}=\frac{1}{2}\Gamma\left(\hat{\Omega}\right)\hat{Q},\hspace{1em}\hat{Q}\in\mathbb{S}^{3}\text{ and }\hat{\Omega}\in\mathbb{R}^{3}\label{eq:Comp_Qdot_hat}
\end{equation}
where $\hat{\Omega}\in\mathbb{R}^{3}$ is to be designed subsequently.
Gaussian filters can be easily utilized for the moving vehicles given
the availability of:
\begin{itemize}
	\item two or more non-collinear vectorial measurements in accordance with
	Remark \ref{rem:Rem1_attitude} as well as,
	\item a rate gyroscope measurement ($\Omega_{m}$).
\end{itemize}
Let us modify the angular velocity measurements in \eqref{eq:Comp_Angular}
by adding a noise term:
\[
\Omega_{m}=\Omega+b+\mathcal{Q}_{\omega}\omega
\]
with $\mathcal{Q}_{\omega}\in\mathbb{R}^{3\times3}$ being a nonzero
diagonal weighting matrix associated with the angular velocity measurements
whose covariance is $\mathcal{\bar{Q}}_{\omega}=\mathcal{Q}_{\omega}\mathcal{Q}_{\omega}^{\top}$.
Consider the bias $b$ attached to angular velocity measurements to
be unknown and slowly time-varying such that
\[
\dot{b}=\mathcal{Q}_{b}\nu\left(t\right)
\]
where $\mathcal{Q}_{b}\in\mathbb{R}^{3\times3}$ is a nonzero diagonal
weighting matrix and $\mathcal{\bar{Q}}_{b}=\mathcal{Q}_{b}\mathcal{Q}_{b}^{\top}$.
Slightly modifying the true body-frame measurements defined in \eqref{eq:Comp_Q_VM}
we obtain
\begin{equation}
\left[\begin{array}{c}
0\\
{\rm v}_{i}^{\mathcal{B}}
\end{array}\right]=Q^{-1}\odot\left[\begin{array}{c}
0\\
{\rm v}_{i}^{\mathcal{I}}
\end{array}\right]\odot Q+\left[\begin{array}{c}
0\\
\mathcal{Q}_{v\left(i\right)}\omega_{i}^{\mathcal{B}}
\end{array}\right]\label{eq:Comp_Q_VM_New}
\end{equation}
where $\mathcal{Q}_{v\left(i\right)}\in\mathbb{R}^{3\times3}$ is
a nonzero diagonal weighting matrix such that the covariance associated
with the body-frame measurements is $\mathcal{\bar{Q}}_{v\left(i\right)}=\mathcal{Q}_{v\left(i\right)}\mathcal{Q}_{v\left(i\right)}^{\top}$
for all $i=1,2,\ldots,n$. It can be noticed that $\mathcal{\bar{Q}}_{\omega}$,
$\mathcal{\bar{Q}}_{b}$, and $\mathcal{\bar{Q}}_{v\left(i\right)}$
are positive definite matrices.

Over the past few decades, several Gaussian attitude filters have
been proposed with the aim of improving the estimation process. The
majority of the attitude filters within the Gaussian family formulate
the attitude problem with respect to unit-quaternion \cite{hashim2018SO3Stochastic}.
The benefit of using unit-quaternion is the fact that it provides
a nonsingular solution to the attitude parameterization. However,
its main drawback is non-uniqueness in representation \cite{shuster1993survey,hashim2019AtiitudeSurvey}.
The unit-quaternion attitude dynamics offer three main advantages,
namely the dynamics in \eqref{eq:Comp_Q_dynam} are characterized
by
\begin{enumerate}
	\item[1)] vector form representation,
	\item[2)] linearity, and
	\item[3)] dependence on the quaternion state.
\end{enumerate}
In consistence with the fact that the orientation of a rigid-body
in the 3-dimensional space can be described by a 4-dimensional vector,
the covariance matrix associated with noise has dimensions $4\times4$
and a rank of 3. One of the earliest attitude filters is the extended
Kalman filter (EKF) proposed in \cite{lefferts1982kalman}. EKF was
followed by several Gaussian filters before the novel Kalman filter
(KF) proposed in \cite{choukroun2006novel} which outperformed its
predecessor. Multiplicative extended Kalman filter, which is a modification
of EKF, is the state-of-the-art technology and an industry standard
in the area of attitude estimation \cite{crassidis2007survey,zamani2013minimum,zamani2015nonlinear}.
The other members of the Gaussian filter family include a modification
of the EKF an invariant extended Kalman filter (IEKF); right IEKF
which models the error in the inertial-frame \cite{bonnable2009invariant};
left IEKF that is analogous to MEKF; and a Geometric approximate minimum
energy filter (GAMEF) \cite{zamani2013minimum} that is developed
based on the Mortensen's approach \cite{mortensen1968maximum}. When
comparing the aforementioned Gaussian attitude filter, the following
points should be taken into consideration \cite{hashim2018SO3Stochastic,mohamed2019filters}:
\begin{enumerate}
	\item[1)] KF, EKF, IEKF, and MEKF are quaternion-based, while GAMEF is developed
	on $\mathbb{SO}\left(3\right)$.
	\item[2)] KF, EKF, and IEKF are based on optimal minimum-energy which is first
	order, while MEKF and GAMEF are based on optimal minimum-energy which
	is second order.
	\item[3)] KF, EKF, and IEKF require less computational cost when compared to
	MEKF and GAMEF.
	\item[4)] MEKF and GAMEF demonstrate better tracking performance when compared
	to KF, EKF, and IEKF.
\end{enumerate}
Unscented Kalman filter (UKF) follows the Gaussian assumptions and
has a structure analogous to KF. The only difference is that UKF uses
a set of sigma points to improve the probability distribution \cite{vandyke2004unscented,de2012uav,hashim2018SO3Stochastic,mohamed2019filters}.
In comparison, one can find that
\begin{enumerate}
	\item[1)] UKF outperforms KF and EKF in terms of tracking performance.
	\item[2)] UKF requires more computational cost than both KF and EKF.
	\item[3)] The use of sigma might add complexity to the estimation process.
\end{enumerate}
Particle filters (PFs), despite being classified as stochastic filters,
do not follow the Gaussian assumption \cite{arulampalam2002tutorial,oshman2006attitude}.
In comparison, it can noted that \cite{hashim2018SO3Stochastic,mohamed2019filters}
\begin{enumerate}
	\item[1)] PFs outperform UKF in terms of tracking performance.
	\item[2)] PFs computational cost is higher than UKF.
	\item[3)] PFs are not an optimal fit for small scale vehicles.
	\item[4)] PFs do not have a clear measure of how close the obtained solution
	is to the optimal one.
\end{enumerate}
In this Section, three of the most common continuous Gaussian attitude
filters are presented, namely KF, MEKF and GAMEF. The discrete form
of KF, MEKF and GAMEF can be found in the \nameref{sec:Appendix}.

\noindent\makebox[1\linewidth]{%
	\rule{0.8\textwidth}{0.5pt}%
}

\subsection{Kalman Filter\label{subsec:KF}}

The normalized vectors of the inertial-frame observations and body-frame
measurements defined in unit-quaternion form in \eqref{eq:Comp_Q_VM_New}
are as follows:
\[
\upsilon_{i}^{\mathcal{I}}=\frac{{\rm v}_{i}^{\mathcal{I}}}{||{\rm v}_{i}^{\mathcal{I}}||},\hspace{1em}\upsilon_{i}^{\mathcal{B}}=\frac{{\rm v}_{i}^{\mathcal{B}}}{||{\rm v}_{i}^{\mathcal{B}}||},\hspace{1em}\forall i=1,2,\ldots,n
\]
Define the true body-frame vector and its normalized values, respectively,
by
\begin{equation}
\begin{cases}
\left[\begin{array}{c}
0\\
\mathring{{\rm v}}_{i}^{\mathcal{B}}
\end{array}\right] & =Q^{-1}\odot\left[\begin{array}{c}
0\\
{\rm v}_{i}^{\mathcal{I}}
\end{array}\right]\odot Q\\
\mathring{\upsilon}_{i}^{\mathcal{B}} & =\mathring{{\rm v}}_{i}^{\mathcal{B}}\left/||\mathring{{\rm v}}_{i}^{\mathcal{B}}||\right.,\hspace{1em}\forall i=1,2,\ldots,n
\end{cases}\label{eq:Comp_Q_V_True}
\end{equation}
One could rewrite \eqref{eq:Comp_Q_V_True} as
\begin{align*}
Q\odot\left[\begin{array}{c}
0\\
\mathring{\upsilon}_{i}^{\mathcal{B}}
\end{array}\right] & =\left[\begin{array}{c}
0\\
\upsilon_{i}^{\mathcal{I}}
\end{array}\right]\odot Q\\
\left[\begin{array}{cc}
0 & -\left(\mathring{\upsilon}_{i}^{\mathcal{B}}\right)^{\top}\\
\mathring{\upsilon}_{i}^{\mathcal{B}} & -\left[\mathring{\upsilon}_{i}^{\mathcal{B}}\right]_{\times}
\end{array}\right]Q & =\left[\begin{array}{cc}
0 & -\left(\upsilon_{i}^{\mathcal{I}}\right)^{\top}\\
\upsilon_{i}^{\mathcal{I}} & \left[\upsilon_{i}^{\mathcal{I}}\right]_{\times}
\end{array}\right]Q
\end{align*}
Consequently,
\[
\mathring{\mathcal{Y}}=\mathbf{0}_{4\times1}=\sum_{i}^{n}\left[\begin{array}{cc}
0 & -\left(\mathring{\upsilon}_{i}^{\mathcal{B}}-\upsilon_{i}^{\mathcal{I}}\right)^{\top}\\
\mathring{\upsilon}_{i}^{\mathcal{B}}-\upsilon_{i}^{\mathcal{I}} & -\left[\mathring{\upsilon}_{i}^{\mathcal{B}}+\upsilon_{i}^{\mathcal{I}}\right]_{\times}
\end{array}\right]Q
\]
where $\mathring{\mathcal{Y}}$ denotes an ideal output signal. Accordingly,
the true attitude problem can be represented as a linear time-variant
state-space problem
\begin{equation}
\begin{cases}
\dot{Q} & =\frac{1}{2}\Gamma\left(\Omega\right)Q\\
\mathring{\mathcal{Y}} & =\mathbf{0}_{4\times1}
\end{cases}\label{eq:Comp_Q_SS_True}
\end{equation}
Unfortunately, the measuring unit cannot provide the true body-frame
vector ($\mathring{\upsilon}_{i}^{\mathcal{B}}$). From \eqref{eq:Comp_Q_VM_New},
it can be found that
\[
Q\odot\left[\begin{array}{c}
0\\
{\rm v}_{i}^{\mathcal{B}}
\end{array}\right]=\left[\begin{array}{c}
0\\
{\rm v}_{i}^{\mathcal{I}}
\end{array}\right]\odot Q+\Gamma\left(\mathcal{Q}_{v\left(i\right)}\omega_{i}^{\mathcal{B}}\right)Q
\]
that is
\begin{align*}
\mathcal{Y}= & \sum_{i}^{n}\left[\begin{array}{cc}
0 & -\left(\mathring{\upsilon}_{i}^{\mathcal{B}}-\upsilon_{i}^{\mathcal{I}}\right)^{\top}\\
\mathring{\upsilon}_{i}^{\mathcal{B}}-\upsilon_{i}^{\mathcal{I}} & -\left[\mathring{\upsilon}_{i}^{\mathcal{B}}+\upsilon_{i}^{\mathcal{I}}\right]_{\times}
\end{array}\right]Q+\sum_{i}^{n}\Gamma\left(\mathcal{Q}_{v\left(i\right)}\omega_{i}^{\mathcal{B}}\right)Q\\
= & \frac{1}{2}\sum_{i}^{n}\Xi\left(Q\right)\mathcal{Q}_{v\left(i\right)}\omega_{i}^{\mathcal{B}}
\end{align*}
where $\mathcal{Q}_{v\left(i\right)}\omega_{i}^{\mathcal{B}}$ is
to be readjusted after normalization. For $\Omega_{m}=\Omega+\mathcal{Q}_{\omega}\omega$,
the attitude problem becomes
\begin{equation}
\begin{cases}
\dot{Q} & =\frac{1}{2}\Gamma\left(\Omega_{m}-\mathcal{Q}_{\omega}\omega\right)Q\\
\mathcal{Y} & =\frac{1}{2}\sum_{i}^{n}\Xi\left(Q\right)\mathcal{Q}_{v\left(i\right)}\omega_{i}^{\mathcal{B}}
\end{cases}\label{eq:Comp_Q_SS_Noise}
\end{equation}
The basic Kalman filter of the problem in \eqref{eq:Comp_Q_SS_Noise}
and the novel Kalman filter proposed in \cite{choukroun2006novel}
in their discrete form can be found in the \nameref{sec:Appendix}.

\noindent\makebox[1\linewidth]{%
	\rule{0.8\textwidth}{0.5pt}%
}

\subsection{Multiplicative Extended Kalman Filter\label{subsec:MEKF}}

The MEKF and GAMEF are second order filters driven with respect to
a cost function. For $\mathcal{J}=\mathcal{J}\left(t;X_{0},\left.\omega\right|_{\left[0,t\right]},\left.\nu\right|_{\left[0,t\right]},\left.\omega_{i}^{\mathcal{B}}\right|_{\left[0,t\right]}\right)$,
consider the following cost function \cite{zamani2013minimum}
\begin{align*}
\mathcal{J}= & \frac{1}{2}{\rm Tr}\left\{ \left(\mathbf{I}_{3}-R\left(0\right)\right)K_{R\left(0\right)}^{\top}\left(\mathbf{I}_{3}-R\left(0\right)\right)^{\top}\right\} +\frac{1}{2}\int_{0}^{T}\left(\omega^{\top}\omega+\nu^{\top}\nu+\sum_{i=1}^{n}\left(\omega_{i}^{\mathcal{B}}\right)^{\top}\omega_{i}^{\mathcal{B}}\right)d\tau
\end{align*}
The optimal control problem of the cost function above can be approached
in terms of the pre-Hamiltonian ($\mathcal{H}^{-}$). Next, let us
define a value function that is subject to minimization
\[
V\left(R,t\right)=\underset{\left.\omega\right|_{\left[0,t\right]}}{{\rm min}}\mathcal{J}
\]
Applying the principle of dynamic programming in \cite{athans1966optimal}
yields a Hamilton-Jacobi-Bellman (HJB) equation
\[
\mathcal{H}-\frac{\partial}{\partial t}V\left(R,t\right)=0
\]
Resorting to the Mortensen's approach \cite{mortensen1968maximum}
allows to obtain an explicit, recursive solution. The complete steps
of the MEKF and GAMEF derivation can be found in \cite{markley2003attitude,zamani2015nonlinear}
and \cite{zamani2013minimum}, respectively.

Multiplicative extended Kalman filter (MEKF) \cite{markley2003attitude}
is a standard in the industry of recursive attitude filtering applications
\cite{crassidis2007survey,zamani2013minimum,zamani2015nonlinear,mohamed2019filters}.
The structure of MEKF is as follows \cite{markley2003attitude,zamani2015nonlinear}
\begin{equation}
\begin{cases}
\left[\begin{array}{c}
0\\
\hat{\upsilon}_{i}^{\mathcal{B}}
\end{array}\right] & =\hat{Q}^{-1}\odot\left[\begin{array}{c}
0\\
\upsilon_{i}^{\mathcal{I}}
\end{array}\right]\odot\hat{Q}\\
\hspace{1em}\dot{\hat{Q}} & =\frac{1}{2}\Gamma\left(\Omega_{m}-\hat{b}+P_{a}W\right)\hat{Q}\\
\hspace{1em}W & =\sum_{i=1}^{n}\hat{\upsilon}_{i}^{\mathcal{B}}\times\mathcal{\bar{Q}}_{v\left(i\right)}^{-1}\left(\hat{\upsilon}_{i}^{\mathcal{B}}-\upsilon_{i}^{\mathcal{B}}\right)
\end{cases}\label{eq:MEKF}
\end{equation}
where $\hat{Q}\in\mathbb{S}^{3}$ is an estimate of the true unit-quaternion,
$\odot$ is a quaternion multiplication operator, $\upsilon_{i}^{\mathcal{I}}\in\mathbb{R}^{3}$
is the $i$th vectorial measurement in the inertial-frame, $\hat{\upsilon}_{i}^{\mathcal{B}}\in\mathbb{R}^{3}$
is the $i$th body-frame vectorial estimate. Additionally,
\begin{equation}
\begin{cases}
\dot{\hat{b}} & =P_{c}^{\top}W\\
S & =\sum_{i=1}^{n}\left[\hat{\upsilon}_{i}^{\mathcal{B}}\right]_{\times}\mathcal{\bar{Q}}_{v\left(i\right)}^{-1}\left[\hat{\upsilon}_{i}^{\mathcal{B}}\right]_{\times}\\
\dot{P}_{a} & =\mathcal{\bar{Q}}_{\omega}+2\boldsymbol{\mathcal{P}}_{s}\left(P_{a}\left[\Omega_{m}-\hat{b}\right]_{\times}-P_{c}\right)-P_{a}SP_{a}\\
\dot{P}_{b} & =\mathcal{\bar{Q}}_{b}-P_{c}SP_{c}\\
\dot{P}_{c} & =-\left[\Omega_{m}-\hat{b}\right]_{\times}P_{c}-P_{a}SP_{c}-P_{b}
\end{cases}\label{eq:MEKF-1}
\end{equation}
with $\mathcal{\bar{Q}}_{v\left(i\right)},\mathcal{\bar{Q}}_{\omega},\mathcal{\bar{Q}}_{b}\in\mathbb{R}^{3\times3}$
being covariance matrices, for all $i=1,2,\ldots,n$.

\noindent\makebox[1\linewidth]{%
	\rule{0.8\textwidth}{0.5pt}%
}

\subsection{Geometric Approximate Minimum-Energy Filter\label{subsec:GAMEF}}

GAMEF is one of the recent Gaussian attitude filters \cite{zamani2013minimum}.
Its structure is similar to the MEKF and can be presented as follows
\cite{zamani2013minimum}:
\begin{equation}
\begin{cases}
\left[\begin{array}{c}
0\\
\hat{\upsilon}_{i}^{\mathcal{B}}
\end{array}\right] & =\hat{Q}^{-1}\odot\left[\begin{array}{c}
0\\
\upsilon_{i}^{\mathcal{I}}
\end{array}\right]\odot\hat{Q}\\
\hspace{1em}\dot{\hat{Q}} & =\frac{1}{2}\Gamma\left(\Omega_{m}-\hat{b}+P_{a}W\right)\hat{Q}\\
\hspace{1em}W & =\sum_{i=1}^{n}\hat{\upsilon}_{i}^{\mathcal{B}}\times\mathcal{\bar{Q}}_{v\left(i\right)}^{-1}\left(\hat{\upsilon}_{i}^{\mathcal{B}}-\upsilon_{i}^{\mathcal{B}}\right)
\end{cases}\label{eq:GAMEF1}
\end{equation}
with $\hat{Q}\in\mathbb{S}^{3}$ being the estimate of the true unit-quaternion,
$\odot$ being a quaternion multiplication operator, $\upsilon_{i}^{\mathcal{I}}\in\mathbb{R}^{3}$
being the $i$th vectorial measurement in the inertial-frame, and
$\hat{\upsilon}_{i}^{\mathcal{B}}\in\mathbb{R}^{3}$ being the $i$th
body-frame vectorial estimate. Additionally
\begin{equation}
\begin{cases}
\dot{\hat{b}} & =P_{c}^{\top}W\\
S & =\sum_{i=1}^{n}\left[\hat{\upsilon}_{i}^{\mathcal{B}}\right]_{\times}\mathcal{\bar{Q}}_{v\left(i\right)}^{-1}\left[\hat{\upsilon}_{i}^{\mathcal{B}}\right]_{\times}\\
C & =\sum_{i=1}^{n}\boldsymbol{\mathcal{P}}_{s}\left(\mathcal{\bar{Q}}_{v\left(i\right)}^{-1}\left(\hat{\upsilon}_{i}^{\mathcal{B}}-\upsilon_{i}^{\mathcal{B}}\right)\left(\hat{\upsilon}_{i}^{\mathcal{B}}\right)^{\top}\right)\\
E & ={\rm Tr}\left\{ C\right\} \mathbf{I}_{3}-C\\
\dot{P}_{a} & =\mathcal{\bar{Q}}_{\omega}+2\boldsymbol{\mathcal{P}}_{s}\left(P_{a}\left[\Omega_{m}-\hat{b}-\frac{1}{2}P_{a}W\right]_{\times}-P_{c}\right)+P_{a}\left(E-S\right)P_{a}\\
\dot{P}_{b} & =\mathcal{\bar{Q}}_{b}+P_{c}\left(E-S\right)P_{c}\\
\dot{P}_{c} & =-\left[\Omega_{m}-\hat{b}-\frac{1}{2}P_{a}W\right]_{\times}P_{c}+P_{a}\left(E-S\right)P_{c}-P_{b}
\end{cases}\label{eq:GAMEF2}
\end{equation}
where $\mathcal{\bar{Q}}_{v\left(i\right)},\mathcal{\bar{Q}}_{\omega},\mathcal{\bar{Q}}_{b}\in\mathbb{R}^{3\times3}$
are covariance matrices, for all $i=1,2,\ldots,n$.

\noindent\makebox[1\linewidth]{%
	\rule{0.8\textwidth}{1.4pt}%
}

\section{Nonlinear Attitude Filters\label{sec:Comp_Nonlinear}}

This section presents different categories of nonlinear attitude filters
in continuous form, while the discrete representation can be found
in the \nameref{sec:Appendix}. Recall the true attitude dynamics
$\dot{R}=R\left[\Omega\right]_{\times}$ in \eqref{eq:Comp_R_dynam}.
Let $\hat{R}\in\mathbb{SO}\left(3\right)$ denote the estimate of
the true attitude $R$. The goal of nonlinear attitude filters is
to drive $\hat{R}\rightarrow R$. Due to the fact that the true attitude
dynamics
\begin{enumerate}
	\item[1)] modeled on the Lie group of SO(3) and
	\item[2)] naturally nonlinear,
\end{enumerate}
nonlinear attitude filter design generally has the following structure
\begin{equation}
\dot{\hat{R}}=\hat{R}\left[\hat{\Omega}\right]_{\times},\hspace{1em}\hat{R}\in\mathbb{SO}\left(3\right)\text{ and }\hat{\Omega}\in\mathbb{R}^{3}\label{eq:Comp_Rdot_hat}
\end{equation}
Such filter design \eqref{eq:Comp_Rdot_hat} is a perfect fit for
the attitude kinematics as it is modeled on the Lie group of SO(3)
and accounts for their nonlinear nature. $\hat{\Omega}$ is to be
defined in the subsequent subsection. The need for nonlinear attitude
filters that would be robust against uncertainty in sensor measurements
has grown dramatically over the past two decades, in particular with
the advancement of low-cost IMUs technology \cite{mahony2008nonlinear,hamel2006attitude,zlotnik2017nonlinear,grip2012attitude,hashim2018SO3Stochastic,hashim2018Conf1}.
The nonlinear filter design presented above can be implemented given
\begin{itemize}
	\item two or more non-collinear vectorial measurements in accordance with
	Remark \ref{rem:Rem1_attitude}, as well as
	\item a rate gyroscope measurement ($\Omega_{m}$).
\end{itemize}
The above-mentioned measurements can be obtained, for example, by
a low-cost IMU module as explained in Section \ref{sec:SO3_kinematics}.
It is worth noting that high quality sensors are not an optimal fit
for small vehicles due to the fact that they are normally 
\begin{enumerate}
	\item[1)] large in size, 
	\item[2)] heavy in weight, and
	\item[3)] expensive. 
\end{enumerate}
In contrast, a typical low-cost IMU module has the following three
merits:
\begin{enumerate}
	\item[1)] small size,
	\item[2)] low weight, and
	\item[3)] low price.
\end{enumerate}
However, the main challenge of working with the low-cost IMU modules
is the fact that they are subject to high levels of noise and bias
components \cite{hashim2018SO3Stochastic,mohamed2019filters}. First
and higher orders of Gaussian attitude filters provide reasonable
estimates if the rigid-body is equipped with high quality sensors.
Whereas, if the rigid-body is fitted with a low-cost IMU module, first
order Gaussian attitude filter produce poor results. Thus, in that
case, the user has to resort to either a nonlinear attitude filter
or a high order Gaussian attitude filter. Nonlinear attitude filters
have the following three advantages:
\begin{enumerate}
	\item[1)] better tracking performance,
	\item[2)] simplicity of filter derivation, and
	\item[3)] less computational power requirements
\end{enumerate}
when compared with Gaussian attitude filters \cite{hashim2018SO3Stochastic,hashim2019SO3Det,mahony2008nonlinear,mohamed2019filters,crassidis2007survey}.
Therefore, nonlinear attitude filters have received considerable attention
over the last few decades, for example \cite{crassidis2007survey,mahony2008nonlinear,mahony2005complementary,zlotnik2017nonlinear,grip2012attitude,hashim2018Conf1,hashim2018SO3Stochastic,hashim2019SO3Det,mohamed2019filters}.

The family of nonlinear attitude filters can be further subdivided
into two distinct categories:
\begin{enumerate}
	\item[1)] \textbf{Nonlinear deterministic attitude filters}, which consider
	the measurements of angular velocity in \eqref{eq:Comp_Angular} to
	be corrupted with unknown constant bias such as
	\[
	\Omega_{m}=\Omega+b\in\left\{ \mathcal{B}\right\} ,\hspace{1em}\left(\omega=0\right)
	\]
	However, they disregard the noise attached to $\Omega_{m}$ in both
	filter derivation and the stability analysis, for example \cite{mahony2008nonlinear,mahony2005complementary,zlotnik2017nonlinear,grip2012attitude,hashim2019SO3Det}.
	\item[2)] \textbf{Nonlinear stochastic attitude filters}, that consider angular
	velocity measurements in \eqref{eq:Comp_Angular} to be
	\[
	\Omega_{m}=\Omega+b+\omega\in\left\{ \mathcal{B}\right\} ,\hspace{1em}\left(\omega\neq0\right)
	\]
	This way, both the unknown bias and the unknown noise attached to
	$\Omega_{m}$ are accounted for in the process of filter derivation
	and the stability analysis, for instance, \cite{hashim2018Conf1,hashim2018SO3Stochastic,hashim2018SE3Stochastic,hashim2019SO3Wiley}.
\end{enumerate}
\noindent\makebox[1\linewidth]{%
	\rule{0.8\textwidth}{0.5pt}%
}

\subsection{Error Criteria, Filter Structure and Setup}

Let $\hat{R}\in\mathbb{SO}\left(3\right)$ be the estimate of the
true body-fixed rotation matrix. Let the error from the body-fixed
frame to the estimator frame be given as
\begin{equation}
\tilde{R}=R^{\top}\hat{R}\label{eq:Comp_R_error}
\end{equation}
Recall \eqref{eq:Comp_Rdot_hat} and consider the estimate of the
attitude dynamics to be defined as
\begin{equation}
\dot{\hat{R}}=\hat{R}\left[\Omega_{m}-\hat{b}-k_{w}W\right]_{\times}\label{eq:Comp_Rdot_hat_all}
\end{equation}
where $\Omega_{m}$ is a gyro measurement as in \eqref{eq:Comp_Angular},
$\hat{b}$ is an estimate of the true bias $b$ associated with angular
velocity measurement, and $W$ is a correction factor. The design
of $\dot{\hat{b}}$ and $W$ will vary based on the type of filter,
and therefore, will be defined separately in each of the following
Subsections: \ref{subsec:CGNDAF}, \ref{subsec:AGNDAF}, \ref{subsec:GPNDAF},
\ref{subsec:AGNDSF}, and \ref{subsec:GPNDSF}. It is worth noting
that the structure of the filter dynamics in \eqref{eq:Comp_Rdot_hat_all}
or a little bit of variation ($\dot{\hat{R}}=[\hat{\Omega}]_{\times}\hat{R}$)
is common when designing a nonlinear attitude filter, for example
\cite{crassidis2007survey,mahony2008nonlinear,mahony2005complementary,zlotnik2017nonlinear,grip2012attitude,hashim2018Conf1,hashim2018SO3Stochastic,hashim2019SO3Det,mohamed2019filters}.
Define the error between the true and the estimated bias as 
\begin{align}
\tilde{b} & =b-\hat{b}\label{eq:Comp_b_error}
\end{align}
The difference between various nonlinear filters consists mainly in
the design of $\dot{\hat{b}}$ and the correction factor $W$, which
in turn depend on
\begin{enumerate}
	\item[1)] the error function selection and
	\item[2)] the type of the nonlinear attitude filter (deterministic or stochastic).
\end{enumerate}
\noindent\makebox[1\linewidth]{%
	\rule{0.8\textwidth}{0.5pt}%
}

\subsubsection{Direct Filter Setup}

From \eqref{eq:Comp_Vect_True} and \eqref{eq:Comp_Vector_norm},
recall that $\upsilon_{i}^{\mathcal{I}}\in\left\{ \mathcal{I}\right\} $
and $\upsilon_{i}^{\mathcal{B}}\in\left\{ \mathcal{B}\right\} $ for
$i=1,2,\ldots,n$. Define
\begin{align}
M^{\mathcal{I}} & =\left(M^{\mathcal{I}}\right)^{\top}=\sum_{i=1}^{n}s_{i}\upsilon_{i}^{\mathcal{I}}\left(\upsilon_{i}^{\mathcal{I}}\right)^{\top}\nonumber \\
M^{\mathcal{B}} & =\left(M^{\mathcal{B}}\right)^{\top}=\sum_{i=1}^{n}s_{i}\upsilon_{i}^{\mathcal{B}}\left(\upsilon_{i}^{\mathcal{B}}\right)^{\top}\nonumber \\
& =R^{\top}M^{\mathcal{I}}R\label{eq:Comp_MB_MI}
\end{align}
where $s_{i}>0$ indicates the confidence level of the $i$th sensor
measurement for all $i=1,2,\ldots,n$. Since $\left\Vert \upsilon_{i}^{\mathcal{I}}\right\Vert ^{2}=\left\Vert \upsilon_{i}^{\mathcal{B}}\right\Vert ^{2}=1$
and in accordance with property \eqref{eq:Comp_Identity_Trace}, one
obtains
\begin{equation}
{\rm Tr}\left\{ M^{\mathcal{B}}\right\} ={\rm Tr}\left\{ M^{\mathcal{I}}\right\} =\sum_{i=1}^{n}s_{i}\label{eq:Comp_sensor}
\end{equation}
Define 
\begin{equation}
\hat{\upsilon}_{i}^{\mathcal{B}}=\hat{R}^{\top}\upsilon_{i}^{\mathcal{I}}\label{eq:Comp_v_hat}
\end{equation}
such that $\hat{\upsilon}_{i}^{\mathcal{B}}$ is the estimate of $\upsilon_{i}^{\mathcal{B}}$
for all $i=1,2,\ldots,n$. From \eqref{eq:Comp_MB_MI} and with the
aid of the identity in \eqref{eq:Comp_Identity8}, one obtains
\begin{align}
\dot{M}^{\mathcal{B}} & =\dot{R}^{\top}M^{\mathcal{I}}R+R^{\top}M^{\mathcal{I}}\dot{R}\nonumber \\
& =-\left[\Omega\right]_{\times}R^{\top}M^{\mathcal{I}}R+R^{\top}M^{\mathcal{I}}R\left[\Omega\right]_{\times}\nonumber \\
& =-\left[\Omega\right]_{\times}M^{\mathcal{B}}+M^{\mathcal{B}}\left[\Omega\right]_{\times}\nonumber \\
& =\left[M^{\mathcal{B}},\left[\Omega\right]_{\times}\right]\label{eq:Comp_MB_dot}
\end{align}
where $M^{\mathcal{B}}=R^{\top}M^{\mathcal{I}}R$ as in \eqref{eq:Comp_MB_MI}.
Also, $\dot{M}^{\mathcal{I}}=\mathbf{0}_{3\times3}$ due to the fact
that $\upsilon_{i}^{\mathcal{I}}\in\left\{ \mathcal{I}\right\} $
denotes a fixed observation. The next stage is the introduction of
the three auxiliary variables in terms of vectorial measurements,
namely, $\mathbf{vex}\left(\boldsymbol{\mathcal{P}}_{a}\left(M^{\mathcal{B}}\tilde{R}\right)\right)$,
$||M^{\mathcal{B}}\tilde{R}||_{I}$, and $\boldsymbol{\Upsilon}\left(M^{\mathcal{B}},\tilde{R}\right)$.
From identity \eqref{eq:Comp_Identity1}, one finds

\begin{align*}
\left[\sum_{i=1}^{n}\frac{s_{i}}{2}\hat{\upsilon}_{i}^{\mathcal{B}}\times\upsilon_{i}^{\mathcal{B}}\right]_{\times} & =\sum_{i=1}^{n}\frac{s_{i}}{2}\left(\upsilon_{i}^{\mathcal{B}}\left(\hat{\upsilon}_{i}^{\mathcal{B}}\right)^{\top}-\hat{\upsilon}_{i}^{\mathcal{B}}\left(\upsilon_{i}^{\mathcal{B}}\right)^{\top}\right)\\
& =\frac{1}{2}R^{\top}M^{\mathcal{I}}R\tilde{R}-\frac{1}{2}\tilde{R}^{\top}R^{\top}M^{\mathcal{I}}R\\
& =\boldsymbol{\mathcal{P}}_{a}\left(M^{\mathcal{B}}\tilde{R}\right)
\end{align*}
such that
\begin{equation}
\mathbf{vex}\left(\boldsymbol{\mathcal{P}}_{a}\left(M^{\mathcal{B}}\tilde{R}\right)\right)=\sum_{i=1}^{n}\frac{s_{i}}{2}\hat{\upsilon}_{i}^{\mathcal{B}}\times\upsilon_{i}^{\mathcal{B}}\label{eq:Comp_VEX_VM}
\end{equation}
where $s_{i}>0$ for all $i=1,2,\ldots,n$. In the light of \eqref{eq:Comp_PER_Ecul_Dist},
the normalized Euclidean distance of $M^{\mathcal{B}}\tilde{R}$ is
equivalent to
\begin{align}
||M^{\mathcal{B}}\tilde{R}||_{I} & =\frac{1}{4}{\rm Tr}\left\{ \mathbf{I}_{3}-M^{\mathcal{B}}\tilde{R}\right\} \nonumber \\
& =\frac{1}{4}{\rm Tr}\left\{ \mathbf{I}_{3}-\sum_{i=1}^{n}s_{i}\upsilon_{i}^{\mathcal{B}}\left(\hat{\upsilon}_{i}^{\mathcal{B}}\right)^{\top}\right\} \nonumber \\
& =\frac{1}{4}\sum_{i=1}^{n}s_{i}\left(1-\left(\hat{\upsilon}_{i}^{\mathcal{B}}\right)^{\top}\upsilon_{i}^{\mathcal{B}}\right)\label{eq:Comp_RI_VM}
\end{align}
Let us introduce the following variable
\begin{align}
\boldsymbol{\Upsilon}\left(M^{\mathcal{B}},\tilde{R}\right) & ={\rm Tr}\left\{ \left(M^{\mathcal{B}}\right)^{-1}M^{\mathcal{B}}\tilde{R}\right\} \nonumber \\
& ={\rm Tr}\left\{ \left(\sum_{i=1}^{n}s_{i}\upsilon_{i}^{\mathcal{B}}\left(\upsilon_{i}^{\mathcal{B}}\right)^{\top}\right)^{-1}\sum_{i=1}^{n}s_{i}\upsilon_{i}^{\mathcal{B}}\left(\hat{\upsilon}_{i}^{\mathcal{B}}\right)^{\top}\right\} \label{eq:Comp_Gamma_VM}
\end{align}

\noindent\makebox[1\linewidth]{%
	\rule{0.8\textwidth}{0.5pt}%
}

\subsubsection{Stochastic Filter Setup}

The nonlinear stochastic attitude filters presented in \cite{hashim2018Conf1,hashim2018SO3Stochastic,mohamed2019filters,hashim2019SO3Wiley}
consider the angular velocity measurements to be
\[
\Omega_{m}=\Omega+b+\omega
\]
where $\omega$ is a zero-mean Gaussian noise vector which is bounded
and therefore follows Assumption \ref{Assum:Comp_s}. Due to the fact
that the derivative of any Gaussian process results in a Gaussian
process \cite{khasminskii1980stochastic,deng2001stabilization,jazwinski2007stochastic,hashim2018SO3Stochastic},
the vector $\omega$ can be redefined as a function of Brownian motion
process vector 
\begin{equation}
\omega=\mathcal{Q}_{\omega}\frac{d\beta}{dt}\label{eq:Comp_omega}
\end{equation}
where $\mathcal{Q}_{\omega}\in\mathbb{R}_{+}^{3\times3}$ is a nonnegative
real matrix whose diagonal consists of unknown time-variant nonnegative
components while the off-diagonal components are zeros or, more simply
put, 
\[
\mathcal{Q}_{\omega}=\left[\begin{array}{ccc}
\mathcal{Q}_{\omega\left(1,1\right)} & 0 & 0\\
0 & \mathcal{Q}_{\omega\left(2,2\right)} & 0\\
0 & 0 & \mathcal{Q}_{\omega\left(3,3\right)}
\end{array}\right]
\]
The covariance of the noise vector $\omega$ is given by $\mathcal{Q}_{\omega}^{2}=\mathcal{Q}_{\omega}\mathcal{Q}_{\omega}^{\top}$.
Also, the properties of the Brownian motion process are given as follows
\cite{khasminskii1980stochastic,deng2001stabilization,jazwinski2007stochastic}
\[
\mathbb{P}\left\{ \beta\left(0\right)=0\right\} =1,\hspace{1em}\mathbb{E}\left[d\beta/dt\right]=0,\hspace{1em}\mathbb{E}\left[\beta\right]=0
\]
Nonlinear stochastic attitude filters aim to achieve adaptive stabilization
for the case of unknown bias and unknown time-variant covariance matrix.
Therefore, let us define a new variable $\sigma\in\mathbb{R}^{3}$
which denotes the upper bound of the covariance matrix $\mathcal{Q}_{\omega}^{2}$
\cite{hashim2018Conf1,hashim2018SO3Stochastic,mohamed2019filters,hashim2019SO3Wiley}
\begin{equation}
\sigma=\left[{\rm max}\left\{ \mathcal{Q}_{\omega\left(1,1\right)}^{2}\right\} ,{\rm max}\left\{ \mathcal{Q}_{\omega\left(2,2\right)}^{2}\right\} ,{\rm max}\left\{ \mathcal{Q}_{\omega\left(3,3\right)}^{2}\right\} \right]^{\top}\label{eq:Comp_s_factor}
\end{equation}
with ${\rm max}\left\{ \cdot\right\} $ being the maximum value of
a component. According to \eqref{eq:Comp_s_factor}, $\sigma$ is
a constant vector that refers to the upper bound of the diagonal of
the covariance matrix $\mathcal{Q}_{\omega}^{2}$. Let $\hat{\sigma}\in\mathbb{R}^{3}$
denote the estimate of $\sigma$, and define the error between $\sigma$
and $\hat{\sigma}$ by
\begin{align}
\tilde{\sigma} & =\sigma-\hat{\sigma}\label{eq:Comp_s_error}
\end{align}

\noindent\makebox[1\linewidth]{%
	\rule{0.8\textwidth}{0.5pt}%
}

\subsubsection{Error Dynamics and Error Function Criteria}

From \eqref{eq:Comp_R_dynam} and \eqref{eq:Comp_Rdot_hat_all}, the
dynamics of the error in \eqref{eq:Comp_R_error} are equivalent to
\begin{align}
\dot{\tilde{R}} & =R^{\top}\dot{\hat{R}}+\dot{R}^{\top}\hat{R}\nonumber \\
& =R^{\top}\hat{R}\left[\Omega+\tilde{b}-k_{w}W\right]_{\times}+\left[\Omega\right]_{\times}^{\top}R^{\top}\hat{R}\nonumber \\
& =\tilde{R}\left[\Omega\right]_{\times}-\left[\Omega\right]_{\times}\tilde{R}+\tilde{R}\left[\tilde{b}-k_{w}W\right]_{\times}\nonumber \\
& =\left[\tilde{R},\left[\Omega\right]_{\times}\right]+\tilde{R}\left[\tilde{b}-k_{w}W\right]_{\times}\label{eq:Comp_Rdot_error}
\end{align}
where $\left[\Omega\right]_{\times}^{\top}=-\left[\Omega\right]_{\times}$
and the Lie bracket $\left[\tilde{R},\left[\Omega\right]_{\times}\right]=\tilde{R}\left[\Omega\right]_{\times}-\left[\Omega\right]_{\times}\tilde{R}$
as in \eqref{eq:Comp_Identity8}.

In general terms, the most important component of designing a new
nonlinear attitude filter is a careful selection of an error function.
The attitude error function presented in \cite{mahony2005complementary}
has been one of the most commonly used error function over the last
few years. Multiple attempts have been made to improve the error function
in \cite{mahony2005complementary} through minor modifications \cite{mahony2008nonlinear,hamel2006attitude,grip2012attitude}.
However, the performance did not see significant improvement. The
critical weakness of the error function in \cite{mahony2005complementary,mahony2008nonlinear,hamel2006attitude,grip2012attitude}
consists in the slow convergence of attitude error, in particular
when faced with large error in attitude initialization. A new form
of an error function introduced in \cite{zlotnik2017nonlinear,lee2012exponential,hashim2018SO3Stochastic,hashim2018Conf1}
provides faster convergence of attitude error to the stable equilibrium
point or to its close neighborhood. Nonetheless, the error functions
proposed in \cite{zlotnik2017nonlinear,lee2012exponential,hashim2018SO3Stochastic,hashim2018Conf1}
offer no systematic convergence in transient and steady-state performance.
In simple terms, the transient performance of the error function in
\cite{zlotnik2017nonlinear,lee2012exponential,hashim2018SO3Stochastic,hashim2018Conf1}
does not follow predefined dynamically reducing boundaries of transient
and steady-state error. Therefore, the prediction of transient and
steady-state performance of attitude error in \cite{mahony2005complementary,mahony2008nonlinear,hamel2006attitude,grip2012attitude,zlotnik2017nonlinear,lee2012exponential,hashim2018SO3Stochastic,hashim2018Conf1}
is almost impossible. Aiming to provide fast and guaranteed transient
and steady-state performance, new solutions are proposed in \cite{hashim2019SO3Det,hashim2019SO3Wiley}.
The solution offered in \cite{hashim2019SO3Det} is a nonlinear deterministic
filter, while the solution in \cite{hashim2019SO3Wiley} is a nonlinear
stochastic filter. 

Before we proceed further, it is important to define $b$ as an unknown
constant bias bounded in accordance with Assumption \ref{Assum:Comp_b}.
Similarly, $\sigma$ is an unknown constant vector defined in \eqref{eq:Comp_s_factor}
and bounded in consistent with Assumption \ref{Assum:Comp_b} and
\ref{Assum:Comp_s}. Let us introduce the following unstable set which
is similar to Definition \ref{def:Comp_unstable} and includes three
unstable equilibrium points
\begin{equation}
\mathcal{U}_{s}=\left\{ \left.\tilde{R}\left(0\right)\in\mathbb{SO}\left(3\right)\right|{\rm Tr}\left\{ \tilde{R}\left(0\right)\right\} =-1\right\} \label{eq:Comp_Us_error}
\end{equation}

\noindent\makebox[1\linewidth]{%
	\rule{0.8\textwidth}{0.5pt}%
}

\subsection{Constant Gain Nonlinear Deterministic Attitude Filter\label{subsec:CGNDAF}}

\subsubsection{Semi-direct Filter}

Consider the error function defined in \cite{mahony2008nonlinear}
\begin{align*}
E_{cgs} & =\frac{1}{4}\left\Vert \mathbf{I}_{3}-\tilde{R}\right\Vert ^{2}\\
& =\frac{1}{4}{\rm Tr}\left\{ \left(\mathbf{I}_{3}-\tilde{R}\right)^{\top}\left(\mathbf{I}_{3}-\tilde{R}\right)\right\} \\
& =\frac{1}{2}{\rm Tr}\left\{ \mathbf{I}_{3}-\tilde{R}\right\} 
\end{align*}
Consider the following constant gain semi-direct nonlinear deterministic
attitude filter (CGSd-NDAF) \cite{mahony2008nonlinear}
\begin{equation}
\begin{cases}
\dot{\hat{R}} & =\hat{R}\left[\Omega_{m}-\hat{b}-k_{w}W\right]_{\times}\\
W & =\mathbf{vex}\left(\boldsymbol{\mathcal{P}}_{a}\left(\tilde{R}\right)\right),\hspace{1em}\tilde{R}=R_{y}^{\top}\hat{R}\\
\dot{\hat{b}} & =\gamma W
\end{cases}\label{eq:Comp_Non_CGSd_NDAF}
\end{equation}
where $\gamma,k_{w}\in\mathbb{R}_{+}$ are positive constants, $W$
is a correction factor, $\hat{b}$ is the estimate of the true bias,
and $R_{y}$ is a reconstructed attitude obtained by one of the algorithms
in \eqref{eq:Comp_TRIAD}, \eqref{eq:Comp_QUEST}, \eqref{eq:Comp_SVD},
or any other method of attitude determination. Consider the Lyapunov
function candidate
\[
V=2E_{cgs}+\frac{1}{\gamma}\left\Vert \tilde{b}\right\Vert ^{2}={\rm Tr}\left\{ \mathbf{I}_{3}-\tilde{R}\right\} +\frac{1}{\gamma}\tilde{b}^{\top}\tilde{b}
\]
provided that $\tilde{R}\left(0\right)\notin\mathcal{U}_{s}$ in \eqref{eq:Comp_Us_error}.
Differentiating $V$, considering $\dot{\tilde{R}}$ in \eqref{eq:Comp_Rdot_error},
and directly substituting $W$ and $\dot{\hat{b}}$ with their definitions
in \eqref{eq:Comp_Non_CGSd_NDAF}, one obtains
\begin{align*}
\dot{V}= & -{\rm Tr}\left\{ \tilde{R}\left[\tilde{b}-k_{w}W\right]_{\times}\right\} -\frac{2}{\gamma}\tilde{b}^{\top}\dot{\hat{b}}\\
= & 2\mathbf{vex}\left(\boldsymbol{\mathcal{P}}_{a}\left(\tilde{R}\right)\right)^{\top}\left(\tilde{b}-k_{w}W\right)-\frac{2}{\gamma}\tilde{b}^{\top}\dot{\hat{b}}\\
= & -2k_{w}\left\Vert \mathbf{vex}\left(\boldsymbol{\mathcal{P}}_{a}\left(\tilde{R}\right)\right)\right\Vert ^{2}
\end{align*}
where $-{\rm Tr}\left\{ \tilde{R}\left[\tilde{b}\right]_{\times}\right\} =2\mathbf{vex}\left(\boldsymbol{\mathcal{P}}_{a}(\tilde{R})\right)^{\top}\tilde{b}$
as defined in identity \eqref{eq:Comp_Identity7}. As stated by Barbalat's
lemma, $\left\Vert \mathbf{vex}\left(\boldsymbol{\mathcal{P}}_{a}(\tilde{R})\right)\right\Vert ^{2}$
converges to zero and $\tilde{R}\rightarrow\mathbf{I}_{3}$ as $t\rightarrow\infty$.

\noindent\makebox[1\linewidth]{%
	\rule{0.4\textwidth}{0.5pt}%
}

\subsubsection{Direct Filter}

From \eqref{eq:Comp_MB_MI} and \eqref{eq:Comp_v_hat}, consider the
error function below \cite{mahony2008nonlinear}
\begin{align*}
E_{cgd} & =\sum_{i=1}^{n}s_{i}\left(1-{\rm Tr}\left\{ \upsilon_{i}^{\mathcal{B}}\left(\hat{\upsilon}_{i}^{\mathcal{B}}\right)^{\top}\right\} \right)\\
& =\sum_{i=1}^{n}s_{i}\left(1-{\rm Tr}\left\{ R^{\top}\upsilon_{i}^{\mathcal{I}}\left(\upsilon_{i}^{\mathcal{I}}\right)^{\top}\hat{R}\right\} \right)\\
& =\sum_{i=1}^{n}s_{i}-{\rm Tr}\left\{ M^{\mathcal{B}}\tilde{R}\right\} 
\end{align*}
Consider the following constant gain direct nonlinear deterministic
attitude filter (CGD-NDAF) \cite{mahony2008nonlinear}
\begin{equation}
\begin{cases}
\dot{\hat{R}} & =\hat{R}\left[\Omega_{m}-\hat{b}-k_{w}W\right]_{\times}\\
W & =\mathbf{vex}\left(\boldsymbol{\mathcal{P}}_{a}\left(M^{\mathcal{B}}\tilde{R}\right)\right)\\
\dot{\hat{b}} & =\gamma W
\end{cases}\label{eq:Comp_Non_CGD_NDAF}
\end{equation}
with $\gamma,k_{w}\in\mathbb{R}_{+}$ being positive constants, $W$
being a correction factor, $\hat{b}$ being an estimate of the true
bias, and $\mathbf{vex}(\boldsymbol{\mathcal{P}}_{a}(M^{\mathcal{B}}\tilde{R}))$
being obtained through vectorial measurements as in \eqref{eq:Comp_VEX_VM}.
Define the following Lyapunov function candidate
\[
V=E_{cgd}+\frac{1}{\gamma}\left\Vert \tilde{b}\right\Vert ^{2}=\sum_{i=1}^{n}s_{i}-{\rm Tr}\left\{ M^{\mathcal{B}}\tilde{R}\right\} +\frac{1}{\gamma}\tilde{b}^{\top}\tilde{b}
\]
provided that $\tilde{R}\left(0\right)\notin\mathcal{U}_{s}$ in \eqref{eq:Comp_Us_error}.
Differentiating $V$, considering $\dot{\tilde{R}}$ in \eqref{eq:Comp_Rdot_error},
and directly substituting $W$ and $\dot{\hat{b}}$ with their definitions
in \eqref{eq:Comp_Non_CGSd_NDAF}, one finds
\begin{align*}
\dot{V}= & -{\rm Tr}\left\{ \left[M^{\mathcal{B}}\tilde{R},\left[\Omega\right]_{\times}\right]\right\} -{\rm Tr}\left\{ M^{\mathcal{B}}\tilde{R}\left[\tilde{b}-k_{w}W\right]_{\times}\right\} -\frac{2}{\gamma}\tilde{b}^{\top}\dot{\hat{b}}\\
= & 2\mathbf{vex}\left(\boldsymbol{\mathcal{P}}_{a}\left(M^{\mathcal{B}}\tilde{R}\right)\right)^{\top}\left(\tilde{b}-k_{w}W\right)-\frac{2}{\gamma}\tilde{b}^{\top}\dot{\hat{b}}\\
= & -2k_{w}\left\Vert \mathbf{vex}\left(\boldsymbol{\mathcal{P}}_{a}\left(M^{\mathcal{B}}\tilde{R}\right)\right)\right\Vert ^{2}
\end{align*}
where ${\rm Tr}\left\{ \left[M^{\mathcal{B}}\tilde{R},\left[\Omega\right]_{\times}\right]\right\} =0$
as in identity \eqref{eq:Comp_Identity5} and $-{\rm Tr}\left\{ M^{\mathcal{B}}\tilde{R}\left[\tilde{b}\right]_{\times}\right\} =2\mathbf{vex}\left(\boldsymbol{\mathcal{P}}_{a}\left(M^{\mathcal{B}}\tilde{R}\right)\right)^{\top}\tilde{b}$
as defined in the identity in \eqref{eq:Comp_Identity7}. Also, Barbalat's
lemma could be invoked to illustrate that $\left\Vert \mathbf{vex}(\boldsymbol{\mathcal{P}}_{a}(M^{\mathcal{B}}\tilde{R}))\right\Vert ^{2}$
converges to zero and $\tilde{R}\rightarrow\mathbf{I}_{3}$ as $t\rightarrow\infty$.

\noindent\makebox[1\linewidth]{%
	\rule{0.8\textwidth}{0.5pt}%
}

\subsection{Adaptive Gain Nonlinear Deterministic Attitude Filter\label{subsec:AGNDAF}}

The nonlinear deterministic filter proposed in Subsection \ref{subsec:CGNDAF}
is characterized by slow convergence of attitude error. Aiming to
address this shortcoming, several solutions designed with attitude
error-based adaptive gain have been proposed, for instance \cite{grip2012attitude,zlotnik2017nonlinear,hashim2018SO3Stochastic,hashim2018Conf1,hashim2019SO3Det,hashim2019SO3Wiley}.
This Subsection presents an adaptive gain nonlinear deterministic
attitude filter (AG-NDAF) proposed in \cite{zlotnik2017nonlinear}
which is semi-direct (requires attitude reconstruction). Consider
the following error function
\begin{align*}
E_{ag} & =\frac{1}{1+{\rm Tr}\{\tilde{R}\}}\mathbf{vex}\left(\boldsymbol{\mathcal{P}}_{a}\left(\tilde{R}\right)\right),\hspace{1em}\tilde{R}=R_{y}^{\top}\hat{R}
\end{align*}
Based on the error function given above, the AG-NDAF is designed as
follows
\begin{equation}
\begin{cases}
\dot{\hat{R}} & =\hat{R}\left[\Omega_{m}-\hat{b}-k_{w}W\right]_{\times}\\
W & =\frac{1}{1+{\rm Tr}\{\tilde{R}\}}\mathbf{vex}\left(\boldsymbol{\mathcal{P}}_{a}\left(\tilde{R}\right)\right),\hspace{1em}\tilde{R}=R_{y}^{\top}\hat{R}\\
\dot{\hat{b}} & =\gamma W
\end{cases}\label{eq:Comp_Non_AGNDAF}
\end{equation}
where $\gamma,k_{w}\in\mathbb{R}_{+}$ are positive constants, $W$
is a correction factor, $\hat{b}$ is an estimate of the true bias,
and $R_{y}$ is the reconstructed attitude obtained by one of the
algorithms in \eqref{eq:Comp_TRIAD}, \eqref{eq:Comp_QUEST}, \eqref{eq:Comp_SVD}
or any other method of attitude determination. It can be easily noticed
that $1/\left(1+{\rm Tr}\{\tilde{R}\}\right)$ is an adaptive gain
whose value becomes increasingly aggressive as ${\rm Tr}\{\tilde{R}\}\rightarrow-1$.
Define the following Lyapunov function candidate

\[
V=\ln\left(2\right)-\frac{1}{2}\ln\left(1+{\rm Tr}\{\tilde{R}\}\right)+\frac{1}{\gamma}\tilde{b}^{\top}\tilde{b}
\]
For $\tilde{R}\left(0\right)\notin\mathcal{U}_{s}$ in \eqref{eq:Comp_Us_error},
differentiating $V$, considering $\dot{\tilde{R}}$ in \eqref{eq:Comp_Rdot_error},
and directly substituting $W$ and $\dot{\hat{b}}$ in \eqref{eq:Comp_Non_AGNDAF},
one obtains\textbf{
	\begin{align*}
	\dot{V}= & -k_{w}\left\Vert E_{ag}\right\Vert ^{2}
	\end{align*}
}Hence, in the light of Barbalat's lemma, $\left\Vert E_{ag}\right\Vert ^{2}$
converges to zero and $\tilde{R}\rightarrow\mathbf{I}_{3}$ as $t\rightarrow\infty$.

\noindent\makebox[1\linewidth]{%
	\rule{0.8\textwidth}{0.5pt}%
}

\subsection{Guaranteed Performance Nonlinear Deterministic Attitude Filter\label{subsec:GPNDAF}}

The filter proposed in Subsection \ref{subsec:AGNDAF} tackles the
weakness of problem convergence of attitude error. However, it is
not characterized by guaranteed measures of transient and steady-state
performance of attitude error convergence \cite{hashim2019SO3Det,hashim2019SO3Wiley}.
This Subsection presents guaranteed performance nonlinear deterministic
attitude filters (GP-NDAF) introduced in \cite{hashim2019SO3Det}.
GP-NDAF achieves guaranteed performance though the following steps:

\textbf{Step 1)}: Define an attitude error function, for example,
in terms of normalized Euclidean distance
\begin{equation}
||\tilde{R}\left(t\right)||_{I}=\frac{1}{4}{\rm Tr}\left\{ \mathbf{I}_{3}-\tilde{R}\right\} \label{eq:Comp_Non_DPPF_error1}
\end{equation}
in accordance with \eqref{eq:Comp_PER_Ecul_Dist}. In order to achieve
guaranteed measures of transient and steady-state performance of the
error function in \eqref{eq:Comp_Non_DPPF_error1}, it is necessary
to constrain $||\tilde{R}\left(t\right)||_{I}$ to initially start
within a large set and reduce systematically and smoothly to settle
within a narrow set. Thus, the next step is the definition of the
dynamically reducing boundaries.

\textbf{Step 2)}: Define a dynamic reducing boundaries as
\[
\xi\left(t\right)=\left(\xi_{0}-\xi_{\infty}\right)\exp\left(-\ell t\right)+\xi_{\infty}
\]
with $\xi_{0}=\xi\left(0\right)$ being the upper bound of the predefined
large set, $\xi_{\infty}$ being the upper bound of the narrow set,
and $\ell$ being a positive constant refers to the convergence rate
of $\xi\left(t\right)$. Next, $||\tilde{R}\left(t\right)||_{I}$
should be defined as a function of the dynamically reducing boundaries
$\xi\left(t\right)$.

\textbf{Step 3)}: Redefine the error function
\[
||\tilde{R}\left(t\right)||_{I}=\xi\left(t\right)\mathcal{Z}\left(\mathcal{E}\right)
\]
such that $\mathcal{Z}\left(\mathcal{E}\right)$ is a smooth function
to be defined, for instance
\[
\mathcal{Z}\left(\mathcal{E}\right)=\frac{\bar{\delta}\exp\left(\mathcal{E}\right)-\underline{\delta}\exp\left(-\mathcal{E}\right)}{\exp\left(\mathcal{E}\right)+\exp\left(-\mathcal{E}\right)}
\]
where $\bar{\delta}$ and $\underline{\delta}$ are positive constants
selected to satisfy $-\underline{\delta}<\mathcal{Z}\left(\mathcal{E}\right)<\bar{\delta},{\rm \text{for}}||\tilde{R}\left(0\right)||_{I}\geq0$.
Since the error $||\tilde{R}\left(t\right)||_{I}$ is constrained
by $\xi\left(t\right)$, let us define the unconstrained error $\mathcal{E}$.

\textbf{Step 4)}: Obtain the unconstrained error
\begin{equation}
\begin{aligned}\mathcal{E}= & \frac{1}{2}\text{ln}\frac{\underline{\delta}+||\tilde{R}||_{I}/\xi}{\bar{\delta}-||\tilde{R}||_{I}/\xi}\end{aligned}
\label{eq:Comp_Non_PPF_eps}
\end{equation}
with the following unconstrained error dynamics
\begin{equation}
\dot{\mathcal{E}}=\mu\left(\frac{d}{dt}||\tilde{R}||_{I}-\frac{\dot{\xi}}{\xi}||\tilde{R}||_{I}\right)\label{eq:Comp_Non_PPF_eps_dot}
\end{equation}
and 
\[
\mu=\frac{1/2}{\underline{\delta}\xi+||\tilde{R}||_{I}}+\frac{1/2}{\bar{\delta}\xi-||\tilde{R}||_{I}}
\]
From \eqref{eq:Comp_Rdot_error}, one finds
\begin{equation}
\frac{d}{dt}||\tilde{R}||_{I}=\frac{1}{2}\mathbf{vex}\left(\boldsymbol{\mathcal{P}}_{a}\left(\tilde{R}\right)\right)^{\top}\left(\tilde{b}-W\right)\label{eq:Comp_Non_DPPF_error_dot}
\end{equation}

\noindent\makebox[1\linewidth]{%
	\rule{0.4\textwidth}{0.5pt}%
}

\subsubsection{Semi-direct Filter}

Consider the following design of a guaranteed performance semi-direct
nonlinear deterministic attitude filter (GPSd-NDAF) \cite{hashim2019SO3Det}
\begin{equation}
\begin{cases}
\dot{\hat{R}} & =\hat{R}\left[\Omega_{m}-\hat{b}-W\right]_{\times}\\
W & =2\frac{k_{w}\mu\mathcal{E}-\dot{\xi}/4\xi}{1-||\tilde{R}||_{I}}\mathbf{vex}\left(\boldsymbol{\mathcal{P}}_{a}\left(\tilde{R}\right)\right)\\
\dot{\hat{b}} & =\frac{\gamma}{2}\mu\mathcal{E}\mathbf{vex}\left(\boldsymbol{\mathcal{P}}_{a}\left(\tilde{R}\right)\right),\quad\tilde{R}=R_{y}^{\top}\hat{R}
\end{cases}\label{eq:Comp_Non_GPSd_NDAF}
\end{equation}
where $\gamma,k_{w}\in\mathbb{R}_{+}$ are positive constants, $W$
is a correction factor, $\hat{b}$ is the estimate of the true bias,
and $R_{y}$ is a reconstructed attitude obtained by one of the algorithms
in \eqref{eq:Comp_TRIAD}, \eqref{eq:Comp_QUEST}, \eqref{eq:Comp_SVD},
or any other method of attitude determination. From \eqref{eq:Comp_Non_GPSd_NDAF},
it becomes apparent that the term multiplied by $\mathbf{vex}(\boldsymbol{\mathcal{P}}_{a}(\tilde{R}))$
becomes increasingly aggressive as $||\tilde{R}||_{I}\rightarrow+1$.
Moreover, it forces the observer to obey the predefined transient
and steady-state measures. Consider the following Lyapunov function
candidate

\[
V=\frac{1}{2}\mathcal{E}^{2}+\frac{1}{2\gamma}\tilde{b}^{\top}\tilde{b}
\]
for any $\tilde{R}\left(0\right)\notin\mathcal{U}_{s}$ in \eqref{eq:Comp_Us_error}.
Differentiating $V$, considering $\dot{\mathcal{E}}$ in \eqref{eq:Comp_Non_PPF_eps_dot},
and directly substituting $W$ and $\dot{\hat{b}}$ with their definitions
in \eqref{eq:Comp_Non_GPSd_NDAF}, it can be found that \textbf{
	\begin{align*}
	\dot{V}= & -4k_{w}||\tilde{R}||_{I}\mu^{2}\mathcal{E}^{2}
	\end{align*}
}On the basis of Barbalat's lemma, $\dot{V}\rightarrow0$ as $t\rightarrow\infty$.
Also, according to Proposition 1 in \cite{hashim2019SO3Det}, $||\mathcal{E}||\rightarrow0$
implies that $||\tilde{R}||_{I}\rightarrow0$ and vice versa. In addition,
$\mu$ is positive for all $t\geq0$. As such, $\tilde{R}\rightarrow\mathbf{I}_{3}$
as $t\rightarrow\infty$ with guaranteed measures of transient and
steady-state performance \cite{hashim2019SO3Det}.

\noindent\makebox[1\linewidth]{%
	\rule{0.4\textwidth}{0.5pt}%
}

\subsubsection{Direct Filter}

Let us modify the error function in \eqref{eq:Comp_Non_DPPF_error1}
to
\begin{equation}
||M^{\mathcal{B}}\tilde{R}||_{I}=\frac{1}{4}{\rm Tr}\left\{ \mathbf{I}_{3}-M^{\mathcal{B}}\tilde{R}\right\} \label{eq:Comp_Non_DPPF_error2}
\end{equation}
where $||M^{\mathcal{B}}\tilde{R}||_{I}$ is defined in terms of vectorial
measurements as in \eqref{eq:Comp_RI_VM}. Thus, with the aid of \eqref{eq:Comp_Rdot_error},
the following equations can be easily obtained
\begin{equation}
\begin{cases}
\mathcal{E} & =\frac{1}{2}\text{ln}\frac{\underline{\delta}+||M^{\mathcal{B}}\tilde{R}||_{I}/\xi}{\bar{\delta}-||M^{\mathcal{B}}\tilde{R}||_{I}/\xi}\\
\mu & =\frac{1/2}{\underline{\delta}\xi+||M^{\mathcal{B}}\tilde{R}||_{I}}+\frac{1/2}{\bar{\delta}\xi-||M^{\mathcal{B}}\tilde{R}||_{I}}\\
\frac{d}{dt}||M^{\mathcal{B}}\tilde{R}||_{I} & =\frac{1}{2}\mathbf{vex}\left(\boldsymbol{\mathcal{P}}_{a}\left(M^{\mathcal{B}}\tilde{R}\right)\right)^{\top}\left(\tilde{b}-W\right)\\
\dot{\mathcal{E}} & =\mu\left(\frac{d}{dt}||M^{\mathcal{B}}\tilde{R}||_{I}-\frac{\dot{\xi}}{\xi}||M^{\mathcal{B}}\tilde{R}||_{I}\right)
\end{cases}\label{eq:Comp_Non_PPF_NDAF}
\end{equation}
Consider the following design of guaranteed performance direct nonlinear
deterministic attitude filter (GPD-NDAF) \cite{hashim2019SO3Det}

\begin{equation}
\begin{cases}
\dot{\hat{R}} & =\hat{R}\left[\Omega_{m}-\hat{b}-W\right]_{\times}\\
\dot{\hat{b}} & =\frac{\gamma}{2}\mu\mathcal{E}\mathbf{vex}\left(\boldsymbol{\mathcal{P}}_{a}\left(M^{\mathcal{B}}\tilde{R}\right)\right)\\
W & =\frac{4}{\underline{\lambda}}\frac{k_{w}\mu\mathcal{E}-\dot{\xi}/\xi}{1+\boldsymbol{\Upsilon}\left(M^{\mathcal{B}},\tilde{R}\right)}\mathbf{vex}\left(\boldsymbol{\mathcal{P}}_{a}\left(M^{\mathcal{B}}\tilde{R}\right)\right)
\end{cases}\label{eq:Comp_Non_GPD_NDAF}
\end{equation}
with $\gamma,k_{w}\in\mathbb{R}_{+}$ being positive constants, $W$
being a correction factor, $\hat{b}$ being the estimate of the true
bias, and $\mathbf{vex}(\boldsymbol{\mathcal{P}}_{a}(M^{\mathcal{B}}\tilde{R}))$
and $\boldsymbol{\Upsilon}(M^{\mathcal{B}},\tilde{R})$ being obtained
through vectorial measurements as in \eqref{eq:Comp_VEX_VM} and \eqref{eq:Comp_Gamma_VM},
respectively. Also, $\underline{\lambda}:=\underline{\lambda}\left({\rm Tr}\left\{ M^{\mathcal{B}}\right\} \mathbf{I}_{3}-M^{\mathcal{B}}\right)$
and denotes the minimum eigenvalue of the matrix. From \eqref{eq:Comp_Non_GPD_NDAF},
it can be noticed that the term multiplied by $\mathbf{vex}(\boldsymbol{\mathcal{P}}_{a}(M^{\mathcal{B}}\tilde{R}))$
becomes increasingly aggressive as $||\tilde{R}||_{I}\rightarrow+1$.
In addition, the above-mentioned term forces the observer to follow
the predefined measures of transient and steady-state. Define the
following Lyapunov function candidate

\[
V=\frac{1}{2}\mathcal{E}^{2}+\frac{1}{2\gamma}\tilde{b}^{\top}\tilde{b}
\]
for any $\tilde{R}\left(0\right)\notin\mathcal{U}_{s}$ in \eqref{eq:Comp_Us_error}.
Differentiating $V$, considering $\dot{\mathcal{E}}$ in \eqref{eq:Comp_Non_PPF_NDAF},
and directly substituting $W$ and $\dot{\hat{b}}$ in \eqref{eq:Comp_Non_GPD_NDAF},
one obtains\textbf{
	\begin{align*}
	\dot{V}\leq & -k_{w}\mu^{2}\mathcal{E}^{2}\left\Vert M^{\mathcal{B}}\tilde{R}\right\Vert _{I}
	\end{align*}
}Consistent with Barbalat's lemma, $\dot{V}\rightarrow0$ as $t\rightarrow\infty$.
Also, according to Proposition 1 in \cite{hashim2019SO3Det}, $||\mathcal{E}||\rightarrow0$
signifies that $||\tilde{R}||_{I}\rightarrow0$ and vice versa. Moreover,
$\mu$ is positive for all $t\geq0$. Thus, $\tilde{R}\rightarrow\mathbf{I}_{3}$
as $t\rightarrow\infty$ with guaranteed measures of transient and
steady-state performance \cite{hashim2019SO3Det}.

\noindent\makebox[1\linewidth]{%
	\rule{0.8\textwidth}{0.5pt}%
}

\subsection{Adaptive Gain Nonlinear Stochastic Attitude Filter\label{subsec:AGNDSF}}

The filters introduced in this Subsection were first proposed in \cite{hashim2018SO3Stochastic,hashim2018Conf1}.
Although they share the nonlinear structure of the filters in Subsections
\ref{subsec:CGNDAF} and \ref{subsec:AGNDAF}, their main advantage
is the stochastic design. One of the stochastic filters is developed
in the sense of Ito, while the other one is developed in the sense
of Stratonovich. The work in \cite{hashim2018SO3Stochastic} gives
a comparison between Ito and Stratonovich in terms of 
\begin{itemize}
	\item[1)] effectiveness of filtering out white and colored noise, and
	\item[2)] computational cost.
\end{itemize}
\noindent\makebox[1\linewidth]{%
	\rule{0.8\textwidth}{0.5pt}%
}

\subsubsection{Ito Filter}

Define the noise attached to angular velocity measurements by $\omega=\mathcal{Q}_{\omega}d\beta/dt$
as introduced in \eqref{eq:Comp_omega}. Define $\boldsymbol{\Psi}(\tilde{R})=\mathbf{vex}(\boldsymbol{\mathcal{P}}_{a}(\tilde{R}))$
and consider the design of adaptive gain Ito nonlinear stochastic
attitude filter (AGI-NSAF) 
\begin{equation}
\begin{cases}
\dot{\hat{R}} & =\hat{R}\left[\Omega_{m}-\hat{b}-W\right]_{\times}\\
\dot{\hat{b}} & =\gamma_{1}||\tilde{R}||_{I}\boldsymbol{\Psi}(\tilde{R})-\gamma_{1}k_{b}\hat{b}\\
\dot{\hat{\sigma}} & =k_{w}\gamma_{2}||\tilde{R}||_{I}\mathcal{D}_{\Psi}^{\top}\boldsymbol{\Psi}(\tilde{R})-\gamma_{2}k_{\sigma}\hat{\sigma}\\
W & =\frac{k_{w}}{\varepsilon}\frac{2-||\tilde{R}||_{I}}{1-||\tilde{R}||_{I}}\boldsymbol{\Psi}(\tilde{R})+k_{2}\mathcal{D}_{\Psi}\hat{\sigma}
\end{cases}\label{eq:Comp_Non_AGI_NDSF}
\end{equation}
where $\gamma_{1},\gamma_{2},k_{w},k_{2},k_{b},k_{\sigma}>0$ are
positive constants, $W$ is a correction factor, $\hat{b}$ is the
estimate of the true bias, $\hat{\sigma}$ is the estimate of the
true upper bound of the covariance $\sigma$, $\mathcal{D}_{\Psi}=\left[\boldsymbol{\Psi}(\tilde{R}),\boldsymbol{\Psi}(\tilde{R}),\boldsymbol{\Psi}(\tilde{R})\right]$,
$\tilde{R}=R_{y}^{\top}\hat{R}$, and $R_{y}$ is a reconstructed
attitude obtained by one of the algorithms in \eqref{eq:Comp_TRIAD},
\eqref{eq:Comp_QUEST}, \eqref{eq:Comp_SVD}, or any other method
of attitude determination. From \eqref{eq:Comp_R_dynam}, \eqref{eq:Comp_Angular},
\eqref{eq:Comp_omega}, and \eqref{eq:Comp_Non_AGI_NDSF}, the attitude
error dynamics of \eqref{eq:Comp_R_error} can be written in an incremental
form as follows
\begin{equation}
d\tilde{R}=\tilde{R}\left[\Omega-\tilde{R}^{\top}\Omega+\tilde{b}-W\right]_{\times}dt+\tilde{R}\left[\mathcal{Q}_{\omega}d\beta\right]_{\times}\label{eq:Comp_Non_NSAF_error_dot}
\end{equation}
Consider the attitude representation with respect to Rodriguez vector
\eqref{eq:OVERVIEW_PER_ROD}. The true attitude dynamics in terms
of Rodriguez vector in incremental form is \cite{hashim2018SO3Stochastic,shuster1993survey}
\begin{align*}
d\rho & =\frac{1}{2}\left(\mathbf{I}_{3}+\left[\rho\right]_{\times}+\rho\rho^{\top}\right)\Omega dt
\end{align*}
The error dynamics in \eqref{eq:Comp_Non_NSAF_error_dot} can be expressed
in terms of Rodriguez error vector as 
\[
d\tilde{\rho}=\tilde{f}dt+\tilde{g}\mathcal{Q}_{\omega}d\beta
\]
with
\begin{align*}
\tilde{g} & =\frac{1}{2}\left(\mathbf{I}_{3}+\left[\tilde{\rho}\right]_{\times}+\tilde{\rho}\tilde{\rho}^{\top}\right)\\
\tilde{f} & =\tilde{g}\left(\Omega-\mathcal{R}_{\tilde{\rho}}^{\top}\Omega+\tilde{b}-W\right)
\end{align*}
such that

\[
\mathcal{R}_{\tilde{\rho}}=\frac{1}{1+\left\Vert \tilde{\rho}\right\Vert ^{2}}\left(\left(1-\left\Vert \tilde{\rho}\right\Vert ^{2}\right)\mathbf{I}_{3}+2\tilde{\rho}\tilde{\rho}^{\top}+2\left[\tilde{\rho}\right]_{\times}\right)
\]
For more information on attitude mapping visit \cite{hashim2018SO3Stochastic,mohamed2019filters,hashim2019AtiitudeSurvey}.
In accordance with (Subsection IV.A \cite{hashim2018SO3Stochastic}),
the Lyapunov function candidate should be obtained as a function of
$\tilde{\rho}$ and it should be twice differential. Accordingly,
consider the following Lyapunov function candidate

\begin{equation}
V\left(\tilde{\rho},\tilde{b},\tilde{\sigma}\right)=\left(\frac{\left\Vert \tilde{\rho}\right\Vert ^{2}}{1+\left\Vert \tilde{\rho}\right\Vert ^{2}}\right)^{2}+\frac{1}{2\gamma_{1}}\tilde{b}^{\top}\tilde{b}+\frac{1}{2\gamma_{2}}\tilde{\sigma}^{\top}\tilde{\sigma}\label{eq:Comp_Non_NSAF_V}
\end{equation}
The first and second partial derivatives of the equation above \eqref{eq:Comp_Non_NSAF_V}
with respect to $\tilde{\rho}$ are
\begin{equation}
\begin{cases}
V_{\tilde{\rho}}=\frac{\partial V}{\partial\tilde{\rho}} & =4\frac{\left\Vert \tilde{\rho}\right\Vert ^{2}}{\left(1+\left\Vert \tilde{\rho}\right\Vert ^{2}\right)^{3}}\tilde{\rho}\\
V_{\tilde{\rho}\tilde{\rho}}=\frac{\partial^{2}V}{\partial\tilde{\rho}^{2}} & =4\frac{\left(1+\left\Vert \tilde{\rho}\right\Vert ^{2}\right)\left\Vert \tilde{\rho}\right\Vert ^{2}\mathbf{I}_{3}+\left(2-4\left\Vert \tilde{\rho}\right\Vert ^{2}\right)\tilde{\rho}\tilde{\rho}^{\top}}{\left(1+\left\Vert \tilde{\rho}\right\Vert ^{2}\right)^{4}}
\end{cases}\label{eq:Comp_Non_ito_NDSF_Lyap}
\end{equation}
For any $\tilde{R}\left(0\right)\notin\mathcal{U}_{s}$ in \eqref{eq:Comp_Us_error}
and with direct substitution of $W$, $\dot{\hat{b}}$, and $\dot{\hat{\sigma}}$
in \eqref{eq:Comp_Non_AGI_NDSF}, one has
\begin{align*}
\mathcal{L}V\leq & -\underline{\lambda}\left(\mathcal{H}\right)V+c_{2}
\end{align*}
where
\[
\mathcal{H}=\left[\begin{array}{ccc}
4k_{w}/\varepsilon & \underline{\mathbf{0}}_{3}^{\top} & \underline{\mathbf{0}}_{3}^{\top}\\
\underline{\mathbf{0}}_{3} & \gamma_{1}k_{b}\mathbf{I}_{3} & \mathbf{0}_{3\times3}\\
\underline{\mathbf{0}}_{3} & \mathbf{0}_{3\times3} & \gamma_{2}k_{\sigma}\mathbf{I}_{3}
\end{array}\right]\in\mathbb{R}^{7\times7}
\]
such that 
\[
0\leq\mathbb{E}\left[V\left(t\right)\right]\leq V\left(0\right){\rm exp}\left(-\underline{\lambda}\left(\mathcal{H}\right)t\right)+\frac{c_{2}}{\underline{\lambda}\left(\mathcal{H}\right)},\,\forall t\geq0
\]
with $\underline{\lambda}\left(\mathcal{H}\right)$ being the minimum
eigenvalue of $\mathcal{H}$. As such, the error vector $\left[\tilde{\rho}^{\top},\tilde{b}^{\top},\tilde{\sigma}^{\top}\right]^{\top}\in\mathbb{R}^{9}$
is semi-globally uniformly ultimately bounded \cite{hashim2018SO3Stochastic}.
The nonlinear stochastic filter proposed in this subsection has been
presented in terms of vectorial measurements in \cite{hashim2018Conf1}.

\noindent\makebox[1\linewidth]{%
	\rule{0.4\textwidth}{0.5pt}%
}

\subsubsection{Stratonovich Filter}

Define $\boldsymbol{\Psi}(\tilde{R})=\mathbf{vex}(\boldsymbol{\mathcal{P}}_{a}(\tilde{R}))$
and consider the following design of an adaptive gain Stratonovich
nonlinear stochastic attitude filter (AGS-NSAF) 
\begin{equation}
\begin{cases}
\dot{\hat{R}} & =\hat{R}\left[\Omega_{m}-\hat{b}-\frac{1}{2}\frac{{\rm diag}\left(\boldsymbol{\Psi}(\tilde{R})\right)}{1-||\tilde{R}||_{I}}\hat{\sigma}-W\right]_{\times}\\
\dot{\hat{b}} & =\gamma_{1}||\tilde{R}||_{I}\boldsymbol{\Psi}(\tilde{R})-\gamma_{1}k_{b}\hat{b}\\
\dot{\hat{\sigma}} & =\gamma_{2}||\tilde{R}||_{I}\left(k_{w}\mathcal{D}_{\Psi}^{\top}+\frac{1}{2}\frac{{\rm diag}\left(\boldsymbol{\Psi}(\tilde{R})\right)}{1-||\tilde{R}||_{I}}\right)\boldsymbol{\Psi}(\tilde{R})-\gamma_{2}k_{\sigma}\hat{\sigma}\\
W & =\frac{k_{w}}{\varepsilon}\frac{2-||\tilde{R}||_{I}}{1-||\tilde{R}||_{I}}\boldsymbol{\Psi}(\tilde{R})+k_{2}\mathcal{D}_{\Psi}\hat{\sigma}
\end{cases}\label{eq:Comp_Non_AGS_NDSF}
\end{equation}
where $\gamma_{1},\gamma_{2},k_{w},k_{2},k_{b},k_{\sigma}>0$ are
positive constants, $W$ is a correction factor, $\hat{b}$ is the
estimate of the true bias, $\hat{\sigma}$ is the estimate of the
true upper bound of the covariance $\sigma$, $\mathcal{D}_{\Psi}=\left[\boldsymbol{\Psi}(\tilde{R}),\boldsymbol{\Psi}(\tilde{R}),\boldsymbol{\Psi}(\tilde{R})\right]$,
$\tilde{R}=R_{y}^{\top}\hat{R}$, and $R_{y}$ is a reconstructed
attitude obtained by one of the algorithms in \eqref{eq:Comp_TRIAD},
\eqref{eq:Comp_QUEST}, \eqref{eq:Comp_SVD}, or any other method
of attitude determination. From \eqref{eq:Comp_R_dynam}, \eqref{eq:Comp_Angular},
\eqref{eq:Comp_omega}, and \eqref{eq:Comp_Non_AGS_NDSF}, the attitude
error dynamics of \eqref{eq:Comp_R_error} can be written in an incremental
form as follows
\begin{align}
d\tilde{R} & =\tilde{R}\left[\Omega-\tilde{R}^{\top}\Omega+\tilde{b}-\frac{1}{2}{\rm diag}\left(\tilde{\rho}\right)\hat{\sigma}-W\right]_{\times}dt+\tilde{R}\left[\mathcal{Q}_{\omega}d\beta\right]_{\times}\label{eq:Comp_Non_NSAF_error_dot2}
\end{align}
The error dynamics in \eqref{eq:Comp_Non_NSAF_error_dot2} can be
expressed in terms of Rodriguez error vector as 
\[
d\tilde{\rho}=\tilde{\mathcal{F}}dt+\tilde{g}\mathcal{Q}_{\omega}d\beta
\]
with 
\begin{align*}
\tilde{g} & =\frac{1}{2}\left(\mathbf{I}_{3}+\left[\tilde{\rho}\right]_{\times}+\tilde{\rho}\tilde{\rho}^{\top}\right)\\
\tilde{\mathcal{F}} & =\tilde{g}\left(\Omega-\tilde{R}^{\top}\Omega+\tilde{b}-\frac{1}{2}{\rm diag}\left(\tilde{\rho}\right)\hat{\sigma}-W\right)+\boldsymbol{\mathcal{W}}\left(\tilde{\rho}\right)\\
\boldsymbol{\mathcal{W}}\left(\tilde{\rho}\right) & =\frac{1}{4}\left(\mathbf{I}_{3}+\left[\tilde{\rho}\right]_{\times}+\tilde{\rho}\tilde{\rho}^{\top}\right)\mathcal{Q}_{\omega}^{2}\tilde{\rho}
\end{align*}
where $\boldsymbol{\mathcal{W}}(\tilde{\rho})$ is the Wong-Zakai
factor \cite{hashim2018SO3Stochastic,mohamed2019filters}. Consider
the following Lyapunov function candidate

\[
V\left(\tilde{\rho},\tilde{b},\tilde{\sigma}\right)=\left(\frac{\left\Vert \tilde{\rho}\right\Vert ^{2}}{1+\left\Vert \tilde{\rho}\right\Vert ^{2}}\right)^{2}+\frac{1}{2\gamma_{1}}\tilde{b}^{\top}\tilde{b}+\frac{1}{2\gamma_{2}}\tilde{\sigma}^{\top}\tilde{\sigma}
\]
The first and second partial derivatives of the equation above with
respect to $\tilde{\rho}$ are similar to \eqref{eq:Comp_Non_ito_NDSF_Lyap}.
For $\tilde{R}\left(0\right)\notin\mathcal{U}_{s}$ in \eqref{eq:Comp_Us_error},
and with directly substituting $W$, $\dot{\hat{b}}$, and $\dot{\hat{\sigma}}$
in \eqref{eq:Comp_Non_AGS_NDSF}, one obtains
\begin{align*}
\mathcal{L}V\leq & -\underline{\lambda}\left(\mathcal{H}\right)V+c_{2}
\end{align*}
where
\[
\mathcal{H}=\left[\begin{array}{ccc}
4k_{w}/\varepsilon & \underline{\mathbf{0}}_{3}^{\top} & \underline{\mathbf{0}}_{3}^{\top}\\
\underline{\mathbf{0}}_{3} & \gamma_{1}k_{b}\mathbf{I}_{3} & \mathbf{0}_{3\times3}\\
\underline{\mathbf{0}}_{3} & \mathbf{0}_{3\times3} & \gamma_{2}k_{\sigma}\mathbf{I}_{3}
\end{array}\right]\in\mathbb{R}^{7\times7}
\]
such that 
\[
0\leq\mathbb{E}\left[V\left(t\right)\right]\leq V\left(0\right){\rm exp}\left(-\underline{\lambda}\left(\mathcal{H}\right)t\right)+\frac{c_{2}}{\underline{\lambda}\left(\mathcal{H}\right)},\,\forall t\geq0
\]
where $\underline{\lambda}\left(\mathcal{H}\right)$ is the minimum
eigenvalue of matrix $\mathcal{H}$. Thus, the error vector $\left[\tilde{\rho}^{\top},\tilde{b}^{\top},\tilde{\sigma}^{\top}\right]^{\top}\in\mathbb{R}^{9}$
is proven to be semi-globally uniformly ultimately bounded \cite{hashim2018SO3Stochastic}.

\noindent\makebox[1\linewidth]{%
	\rule{0.8\textwidth}{0.5pt}%
}

\subsection{Guaranteed Performance Nonlinear Stochastic Attitude Filter\label{subsec:GPNDSF}}

The filters described in this Subsection were first proposed in \cite{hashim2019SO3Wiley}.
Despite sharing the nonlinear structure of the filters in Subsection
\ref{subsec:GPNDAF}, their main advantage is the stochastic design.
Both stochastic filters presented below are driven in the sense of
Stratonovich.

\noindent\makebox[1\linewidth]{%
	\rule{0.8\textwidth}{0.5pt}%
}

\subsubsection{Semi-direct Filter}

Given $\omega=\mathcal{Q}d\beta/dt$ as defined in \eqref{eq:Comp_omega},
the normalized Euclidean distance of attitude error dynamics in \eqref{eq:Comp_Non_DPPF_error1}
can be rewritten in an incremental form as follows
\begin{equation}
d||\tilde{R}||_{I}=\frac{1}{2}\mathbf{vex}\left(\boldsymbol{\mathcal{P}}_{a}\left(\tilde{R}\right)\right)^{\top}\left(\left(\Omega_{m}-b\right)dt-\mathcal{Q}_{\omega}d\beta\right)\label{eq:Comp_Non_SPPF_error_dot}
\end{equation}
Consider \eqref{eq:Comp_Non_SPPF_error_dot} and recall the following
set of equations
\begin{equation}
\begin{cases}
\mathcal{E} & =\frac{1}{2}\text{ln}\frac{\underline{\delta}+||\tilde{R}||_{I}/\xi}{\bar{\delta}-||\tilde{R}||_{I}/\xi}\\
\mu & =\frac{\exp\left(2\mathcal{E}\right)+\exp\left(-2\mathcal{E}\right)+2}{8\xi\bar{\delta}}\\
d||\tilde{R}||_{I} & =\frac{1}{2}\mathbf{vex}\left(\boldsymbol{\mathcal{P}}_{a}\left(\tilde{R}\right)\right)^{\top}\left(\left(\Omega_{m}-b\right)dt-\mathcal{Q}_{\omega}d\beta\right)\\
d\mathcal{E} & =2\mu\left(d||\tilde{R}||_{I}-\frac{\dot{\xi}}{\xi}||\tilde{R}||_{I}dt\right)
\end{cases}\label{eq:Comp_Non_PPF_NDSF1}
\end{equation}
Define $\boldsymbol{\Psi}(\tilde{R})=\mathbf{vex}(\boldsymbol{\mathcal{P}}_{a}(\tilde{R}))$
and consider the following design of a guaranteed performance semi-direct
nonlinear stochastic attitude filter (GPSd-NSAF) \cite{hashim2019SO3Wiley}
\begin{equation}
\begin{cases}
\dot{\hat{R}} & =\hat{R}\left[\Omega_{m}-\hat{b}-W\right]_{\times}\\
\dot{\hat{b}} & =\gamma_{1}\left(\mathcal{E}+1\right)\exp\left(\mathcal{E}\right)\mu\boldsymbol{\Psi}(\tilde{R})\\
\dot{\hat{\sigma}} & =\gamma_{2}\left(\mathcal{E}+2\right)\exp\left(\mathcal{E}\right)\mu^{2}{\rm diag}\left(\boldsymbol{\Psi}(\tilde{R})\right)\boldsymbol{\Psi}(\tilde{R})\\
W & =2\frac{\mathcal{E}+2}{\mathcal{E}+1}\mu{\rm diag}\left(\boldsymbol{\Psi}(\tilde{R})\right)\hat{\sigma}+2\frac{k_{w}\left(\mathcal{E}+1\right)\mu-\dot{\xi}/4\xi}{1-||\tilde{R}||_{I}}\boldsymbol{\boldsymbol{\Psi}}(\tilde{R})
\end{cases}\label{eq:Comp_Non_GPSd_NDSF}
\end{equation}
where $\gamma_{1},\gamma_{2},k_{w}\in\mathbb{R}_{+}$ are positive
constants, $W$ is a correction factor, $\hat{b}$ is the estimate
of the true bias, $\hat{\sigma}$ is the estimate of the true upper
bound of the covariance $\sigma$, $\tilde{R}=R_{y}^{\top}\hat{R}$,
and $R_{y}$ is a reconstructed attitude obtained by one of the algorithms
in \eqref{eq:Comp_TRIAD}, \eqref{eq:Comp_QUEST}, \eqref{eq:Comp_SVD},
or any other method of attitude determination. From \eqref{eq:Comp_Non_GPSd_NDSF},
it can be observed that the term multiplied by $\boldsymbol{\Psi}(\tilde{R})=\mathbf{vex}(\boldsymbol{\mathcal{P}}_{a}(\tilde{R}))$
becomes increasingly aggressive as $||\tilde{R}||_{I}\rightarrow+1$.
Additionally, the above-mentioned term forces the filter to obey the
predefined transient and steady-state measures. When selecting the
Lyapunov function candidate the two important considerations are:
it should be a function of $\mathcal{E}$ and it should be twice differentiable.
In the light of the above considerations, let us define Lyapunov function
candidate as follows:

\begin{equation}
V(\mathcal{E},\tilde{b},\tilde{\sigma})=\mathcal{E}\exp\left(\mathcal{E}\right)+\frac{1}{2\gamma_{1}}||\tilde{b}||^{2}+\frac{1}{\gamma_{2}}||\tilde{\sigma}||^{2}\label{eq:Comp_Non_GPSd_NDSF_V}
\end{equation}
The first and second partial derivatives of the equation above \eqref{eq:Comp_Non_GPSd_NDSF_V}
with respect to $\mathcal{E}$ are
\begin{equation}
\begin{cases}
V_{\mathcal{E}}=\frac{\partial V}{\partial\mathcal{E}} & =\left(\mathcal{E}+1\right)\exp\left(\mathcal{E}\right)\\
V_{\mathcal{E}\mathcal{E}}=\frac{\partial^{2}V}{\partial\mathcal{E}^{2}} & =\left(\mathcal{E}+2\right)\exp\left(\mathcal{E}\right)
\end{cases}\label{eq:Comp_Non_GPSd_NDSF_Lyap}
\end{equation}
For any $\tilde{R}\left(0\right)\notin\mathcal{U}_{s}$ in \eqref{eq:Comp_Us_error},
considering $\dot{\mathcal{E}}$ in \eqref{eq:Comp_Non_PPF_NDSF1},
and directly substituting $W$, $\dot{\hat{b}}$, and $\dot{\hat{\sigma}}$
with their definitions in \eqref{eq:Comp_Non_GPSd_NDSF}, one obtains\textbf{
	\begin{align*}
	\mathcal{L}V\leq & -4\bar{\delta}k_{w}\xi\mu^{2}\left(\mathcal{E}+1\right)^{2}\frac{\exp\left(\mathcal{E}\right)-\exp\left(-\mathcal{E}\right)}{\exp\left(\mathcal{E}\right)+\exp\left(-\mathcal{E}\right)}\exp\left(\mathcal{E}\right)
	\end{align*}
} According to the fact that $\mathcal{L}V$ is bounded and $V$ is
radially unbounded for any $\tilde{R}\left(0\right)\notin\mathcal{U}_{s}$
and $\mathcal{E}\left(0\right)\in\mathbb{R}$, it can be concluded
that a unique strong solution to the stochastic system in \eqref{eq:Comp_Non_SPPF_error_dot}
exists with a probability of one \cite{khasminskii1980stochastic}.
Thus, $\mathcal{E}\left(t\right)$ is regulated asymptotically to
the origin in probability of 1 for all $\tilde{R}\left(0\right)\notin\mathcal{U}_{s}$
and $\mathcal{E}\left(0\right)\in\mathbb{R}$ implying that $\mathbb{P}\{\lim_{t\rightarrow\infty}\tilde{R}=\mathbf{I}_{3}\}=1$
for all $\tilde{R}\left(0\right)\notin\mathcal{U}_{s}$ and $\mathcal{E}\left(0\right)\in\mathbb{R}$
\cite{hashim2019SO3Wiley}.

\noindent\makebox[1\linewidth]{%
	\rule{0.4\textwidth}{0.5pt}%
}

\subsubsection{Direct Filter}

Consider modifying the error function in \eqref{eq:Comp_Non_DPPF_error1}
to
\begin{equation}
||M^{\mathcal{B}}\tilde{R}||_{I}=\frac{1}{4}{\rm Tr}\left\{ \mathbf{I}_{3}-M^{\mathcal{B}}\tilde{R}\right\} \label{eq:Comp_Non_SPPF_error2}
\end{equation}
where $||M^{\mathcal{B}}\tilde{R}||_{I}$ is given with respect to
vectorial measurements as in \eqref{eq:Comp_RI_VM}. Hence, the error
function in \eqref{eq:Comp_Non_SPPF_error2} can be expressed in an
incremental form as follows
\begin{equation}
d||M^{\mathcal{B}}\tilde{R}||_{I}=\frac{1}{2}\mathbf{vex}\left(\boldsymbol{\mathcal{P}}_{a}\left(M^{\mathcal{B}}\tilde{R}\right)\right)^{\top}\left(\left(\Omega_{m}-b\right)dt-\mathcal{Q}_{\omega}d\beta\right)\label{eq:Comp_Non_SPPF_error_dot2}
\end{equation}
Accordingly, one can arrive at the following set of equations: 
\begin{equation}
\begin{cases}
\mathcal{E} & =\frac{1}{2}\text{ln}\frac{\underline{\delta}+||M^{\mathcal{B}}\tilde{R}||_{I}/\xi}{\bar{\delta}-||M^{\mathcal{B}}\tilde{R}||_{I}/\xi}\\
\mu & =\frac{\exp\left(2\mathcal{E}\right)+\exp\left(-2\mathcal{E}\right)+2}{8\xi\bar{\delta}}\\
d||M^{\mathcal{B}}\tilde{R}||_{I} & =\frac{1}{2}\mathbf{vex}\left(\boldsymbol{\mathcal{P}}_{a}\left(M^{\mathcal{B}}\tilde{R}\right)\right)^{\top}\left(\left(\Omega_{m}-b\right)dt-\mathcal{Q}_{\omega}d\beta\right)\\
d\mathcal{E} & =2\mu\left(d||M^{\mathcal{B}}\tilde{R}||_{I}-\frac{\dot{\xi}}{\xi}||M^{\mathcal{B}}\tilde{R}||_{I}dt\right)
\end{cases}\label{eq:Comp_Non_PPF_NDSF2}
\end{equation}
Define $\boldsymbol{\Psi}(M^{\mathcal{B}}\tilde{R})=\mathbf{vex}(\boldsymbol{\mathcal{P}}_{a}(M^{\mathcal{B}}\tilde{R}))$
and consider the following design of a guaranteed performance direct
nonlinear stochastic attitude filter (GPD-NSAF) \cite{hashim2019SO3Det}

\begin{equation}
\begin{cases}
\dot{\hat{R}} & =\hat{R}\left[\Omega_{m}-\hat{b}-W\right]_{\times}\\
\dot{\hat{b}} & =\gamma_{1}\left(\mathcal{E}+1\right)\exp\left(\mathcal{E}\right)\mu\boldsymbol{\Psi}(M^{\mathcal{B}}\tilde{R})\\
\dot{\hat{\sigma}} & =\gamma_{2}\left(\mathcal{E}+2\right)\exp\left(\mathcal{E}\right)\mu^{2}{\rm diag}\left(\boldsymbol{\Psi}(M^{\mathcal{B}}\tilde{R})\right)\boldsymbol{\Psi}(M^{\mathcal{B}}\tilde{R})\\
W & =2\frac{\mathcal{E}+2}{\mathcal{E}+1}\mu{\rm diag}\left(\boldsymbol{\Psi}(M^{\mathcal{B}}\tilde{R})\right)\hat{\sigma}+\frac{4}{\underline{\lambda}}\frac{k_{w}\mu\mathcal{E}-\dot{\xi}/\xi}{1+\boldsymbol{\Upsilon}\left(M^{\mathcal{B}},\tilde{R}\right)}\boldsymbol{\Psi}(M^{\mathcal{B}}\tilde{R})
\end{cases}\label{eq:Comp_Non_GPD_NDSF}
\end{equation}
with $\gamma_{1},\gamma_{2},k_{w}\in\mathbb{R}_{+}$ being positive
constants, $W$ being a correction factor, $\hat{b}$ being the estimate
of the true bias, $\hat{\sigma}$ being the estimate of the true upper
bound of the covariance, and $\mathbf{vex}(\boldsymbol{\mathcal{P}}_{a}(M^{\mathcal{B}}\tilde{R}))$
and $\boldsymbol{\Upsilon}(M^{\mathcal{B}},\tilde{R})$ being obtained
through vectorial measurements as in \eqref{eq:Comp_VEX_VM} and \eqref{eq:Comp_Gamma_VM},
respectively. Also, $\underline{\lambda}:=\underline{\lambda}\left({\rm Tr}\left\{ M^{\mathcal{B}}\right\} \mathbf{I}_{3}-M^{\mathcal{B}}\right)$
denotes the minimum eigenvalue. Consider the below Lyapunov function
candidate 

\[
V(\mathcal{E},\tilde{b},\tilde{\sigma})=\mathcal{E}\exp\left(\mathcal{E}\right)+\frac{1}{2\gamma_{1}}||\tilde{b}||^{2}+\frac{1}{\gamma_{2}}||\tilde{\sigma}||^{2}
\]
The first and second partial derivatives of the equation above are
similar to \eqref{eq:Comp_Non_GPSd_NDSF_Lyap}. For any $\tilde{R}\left(0\right)\notin\mathcal{U}_{s}$
in \eqref{eq:Comp_Us_error}, considering $\dot{\mathcal{E}}$ in
\eqref{eq:Comp_Non_PPF_NDSF1}, and directly substituting $W$, $\dot{\hat{b}}$,
and $\dot{\hat{\sigma}}$ in \eqref{eq:Comp_Non_GPD_NDSF}, one has

\textbf{
	\begin{align*}
	\mathcal{L}V\leq & -\bar{\delta}k_{w}\xi\mu^{2}\left(\mathcal{E}+1\right)^{2}\frac{\exp\left(\mathcal{E}\right)-\exp\left(-\mathcal{E}\right)}{\exp\left(\mathcal{E}\right)+\exp\left(-\mathcal{E}\right)}\exp\left(\mathcal{E}\right)
	\end{align*}
}Since $\mathcal{L}V$ is bounded and $V$ is radially unbounded for
any $\tilde{R}\left(0\right)\notin\mathcal{U}_{s}$ and $\mathcal{E}\left(0\right)\in\mathbb{R}$,
there exists a unique strong solution to the stochastic system in
\eqref{eq:Comp_Non_SPPF_error_dot2} with a probability of one \cite{khasminskii1980stochastic}.
Therefore, $\mathcal{E}\left(t\right)$ is regulated asymptotically
to the origin in probability for all $\tilde{R}\left(0\right)\notin\mathcal{U}_{s}$
and $\mathcal{E}\left(0\right)\in\mathbb{R}$ which, in turn, implies
that $\mathbb{P}\{\lim_{t\rightarrow\infty}\tilde{R}=\mathbf{I}_{3}\}=1$
for all $\tilde{R}\left(0\right)\notin\mathcal{U}_{s}$ and $\mathcal{E}\left(0\right)\in\mathbb{R}$
\cite{hashim2019SO3Wiley}.

\noindent\makebox[1\linewidth]{%
	\rule{0.8\textwidth}{1.4pt}%
}

\section{Simulation and Comparative Results\label{sec:Comp_Simulation}}

\subsection{Continuous Attitude Filters}

This Subsection provides comparative results in accordance with Table
\ref{tab:Simu1}.

\begin{table}[h!]
	\caption{\label{tab:Simu1}Attitude determination and estimation algorithms
		in comparison}
	
	\centering{}%
	\begin{tabular}{>{\raggedright}p{5cm}|>{\raggedright}p{7cm}}
		\hline 
		Category & Type\tabularnewline
		\hline 
		\hline 
		Attitude Determination & TRIAD (Equation \eqref{eq:Comp_TRIAD}), \vspace{0.1cm}
		
		QUEST (Equation \eqref{eq:Comp_QUEST}) and \vspace{0.1cm}
		
		SVD (Equation \eqref{eq:Comp_SVD})\tabularnewline
		\hline 
		Gaussian Attitude Filters & MEKF (Equation \eqref{eq:MEKF}) and\vspace{0.1cm}
		
		GAMEF (Equation \eqref{eq:GAMEF1})\tabularnewline
		\hline 
		Nonlinear Attitude Filters & CG-NDAF (Equation \eqref{eq:Comp_Non_CGSd_NDAF} and \eqref{eq:Comp_Non_CGD_NDAF}),\vspace{0.1cm}
		
		AG-NDAF (Equation \eqref{eq:Comp_Non_AGNDAF}), \vspace{0.1cm}
		
		GP-NDAF (Equation \eqref{eq:Comp_Non_GPSd_NDAF} and \eqref{eq:Comp_Non_GPD_NDAF}),
		\vspace{0.1cm}
		
		AG-NSAF (Equation \eqref{eq:Comp_Non_AGI_NDSF} and \eqref{eq:Comp_Non_AGS_NDSF}),
		and \vspace{0.1cm}
		
		GP-NSAF (Equation \eqref{eq:Comp_Non_GPSd_NDSF} and \eqref{eq:Comp_Non_GPD_NDSF})\tabularnewline
		\hline 
	\end{tabular}
\end{table}

According to the discussion given in the above Sections, AG-NDAF,
GP-NDAF, AG-NSAF, and GP-NSAF are adaptively tuned. Thus, to ensure
fair comparison, different scenarios have been considered for CG-NDAF,
MEKF, and GAMEF. Since KF demonstrates reasonable performance only
if the sensor measurements are free of high level uncertainties, KF
is not included in the comparison. The algorithms are implemented
and the results are obtained using ${\rm MATLAB}{}^{\circledR}$.
The filter performance will be tested on angular velocity and body-frame
vectorial measurements subject to
\begin{enumerate}
	\item[1)] constant bias, and
	\item[2)] noise that is normally distributed with a zero mean and a nonzero
	standard deviation (STD).
\end{enumerate}
\noindent\makebox[1\linewidth]{%
	\rule{0.8\textwidth}{0.5pt}%
}

\subsubsection{True Values and Measurements}

\begin{figure}[h!]
	\includegraphics[scale=0.2]{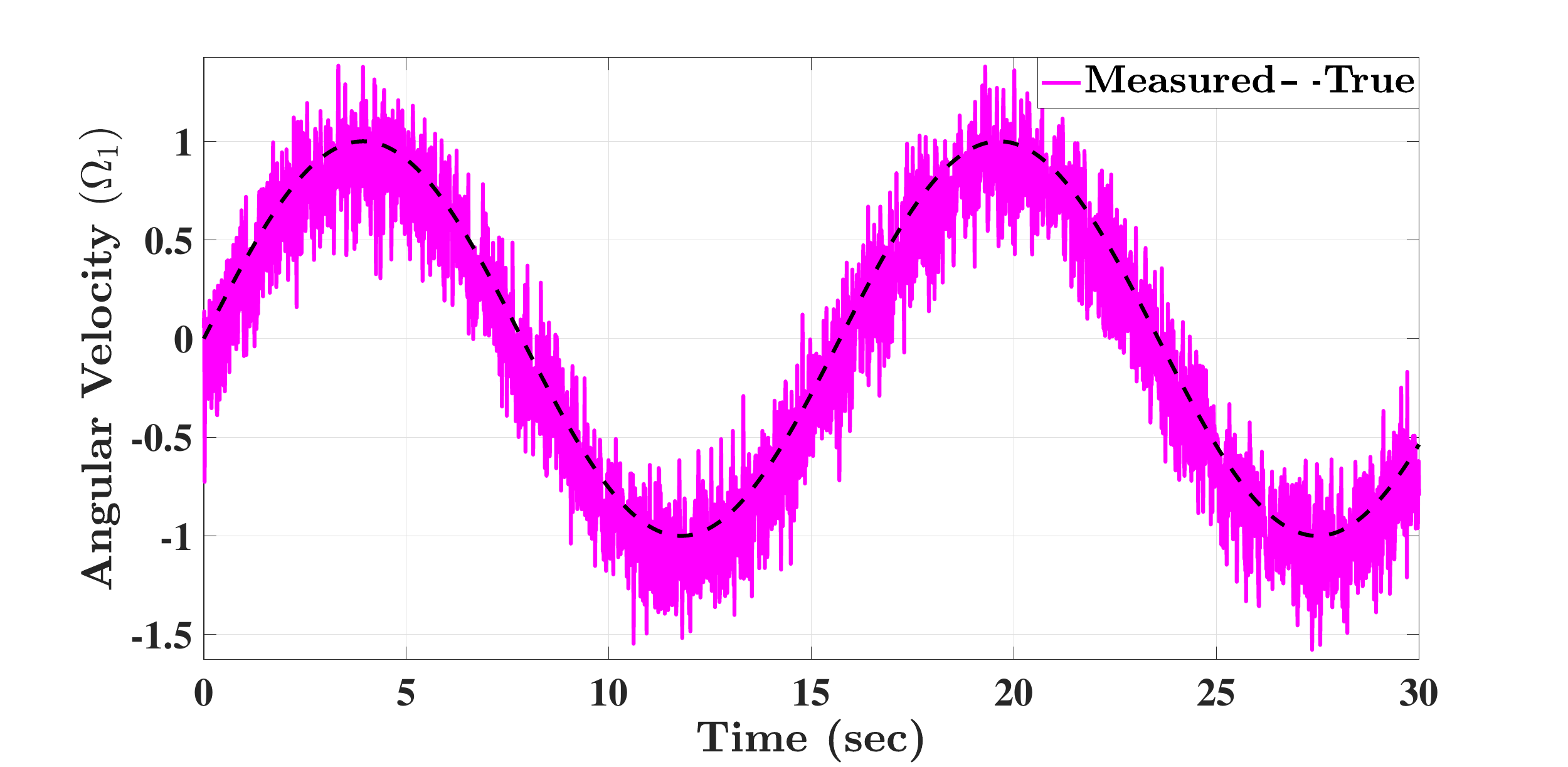}\includegraphics[scale=0.2]{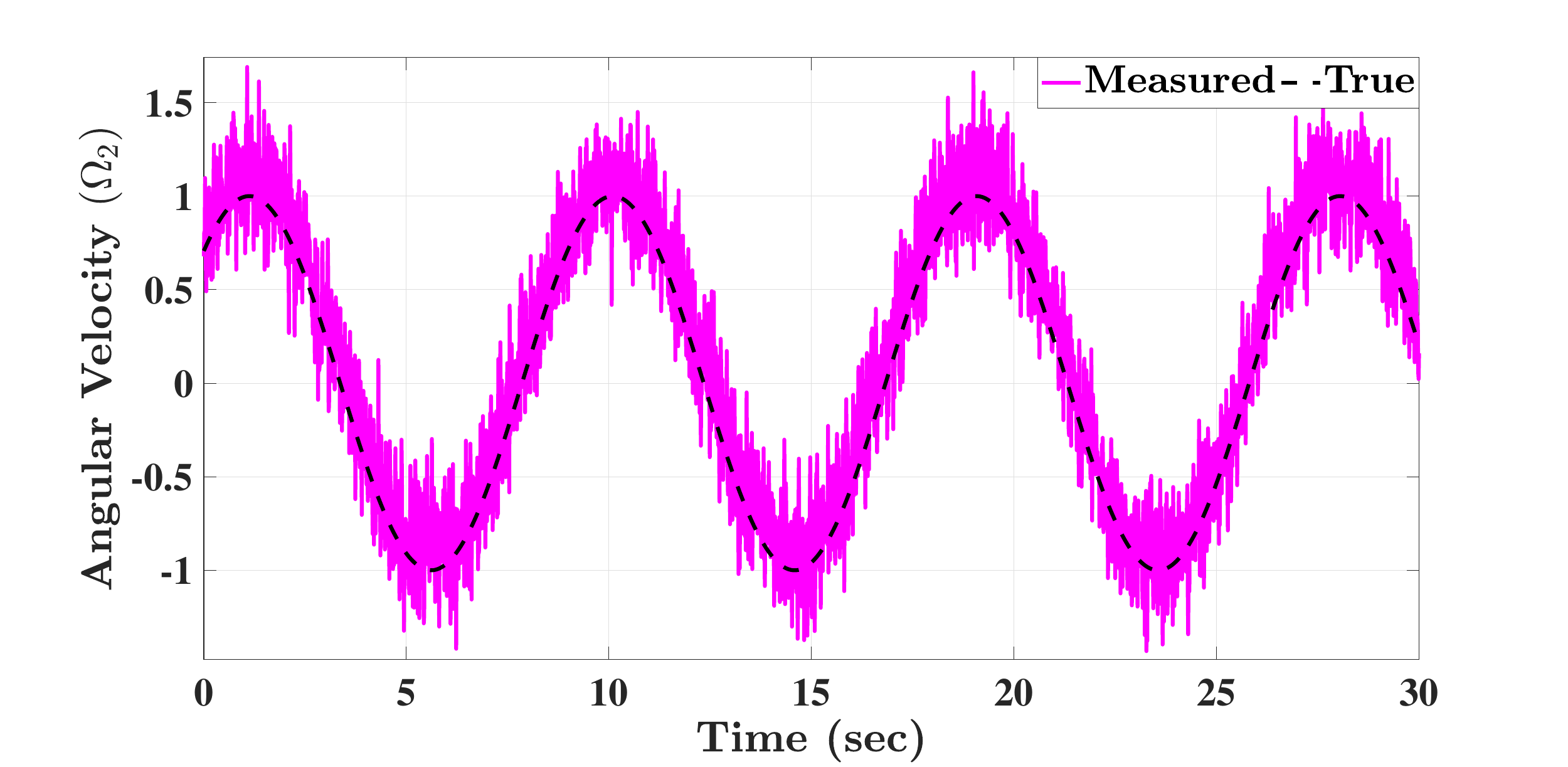}
	
	\includegraphics[scale=0.2]{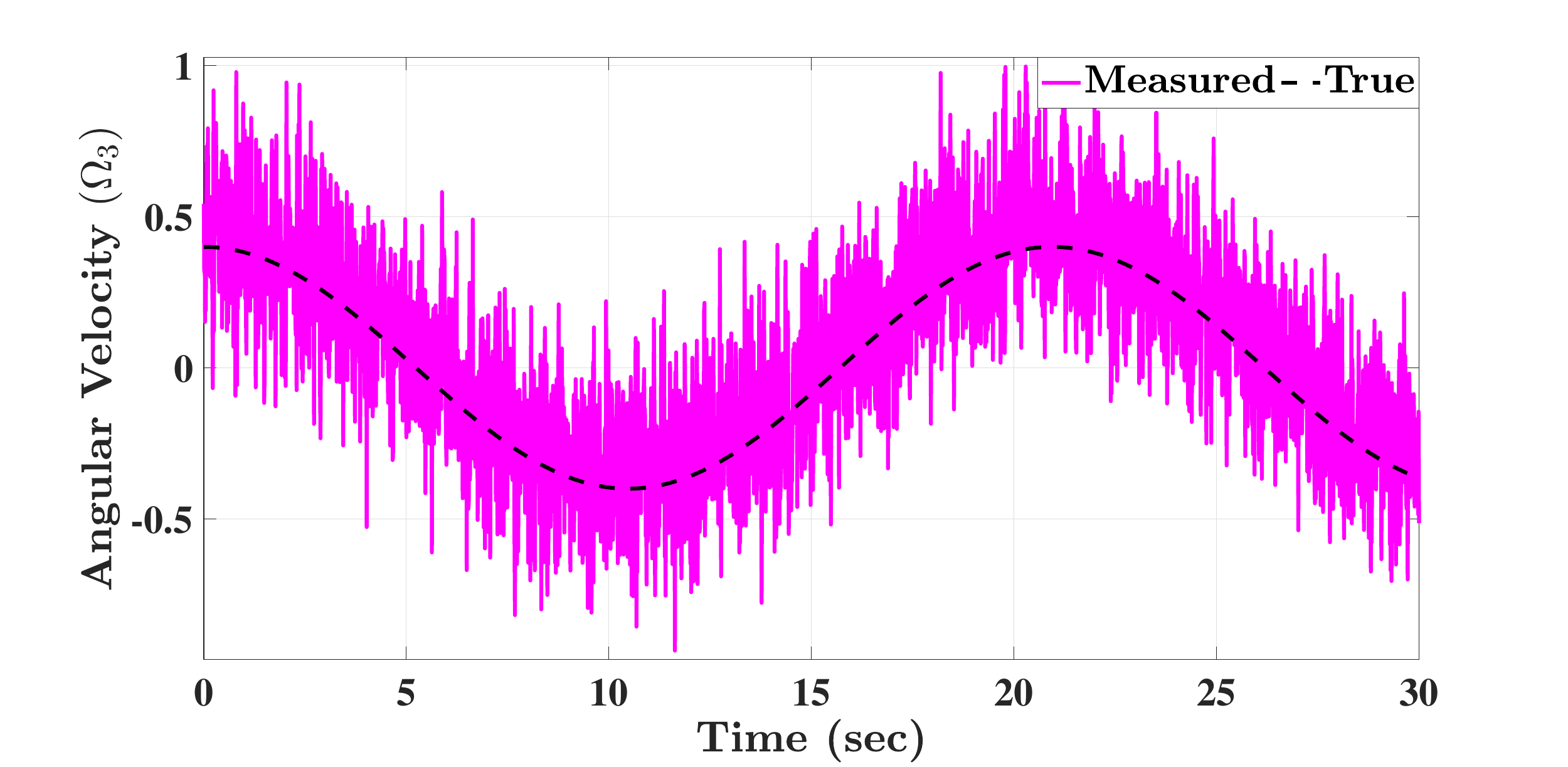}\includegraphics[scale=0.2]{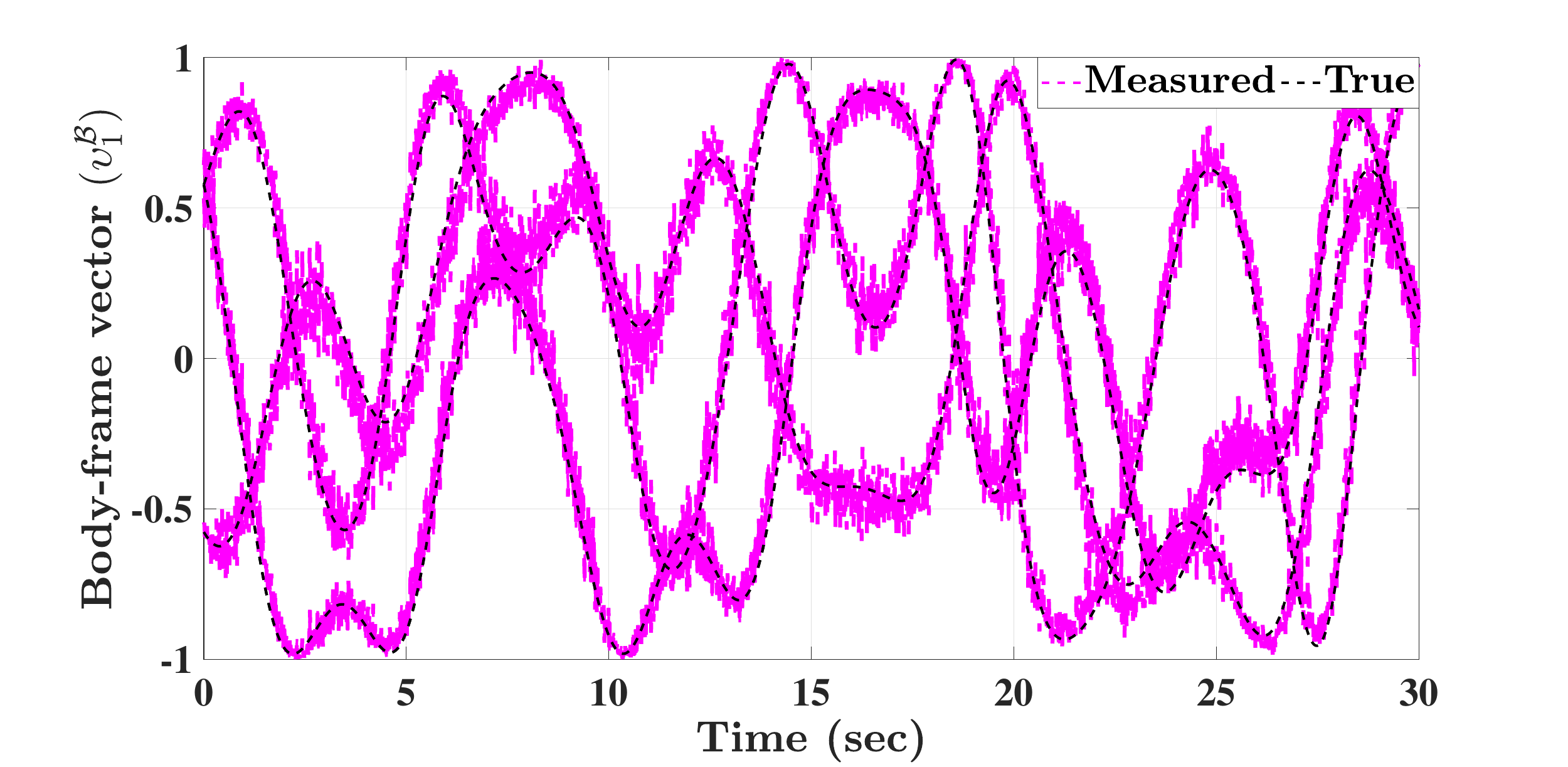}
	
	\includegraphics[scale=0.2]{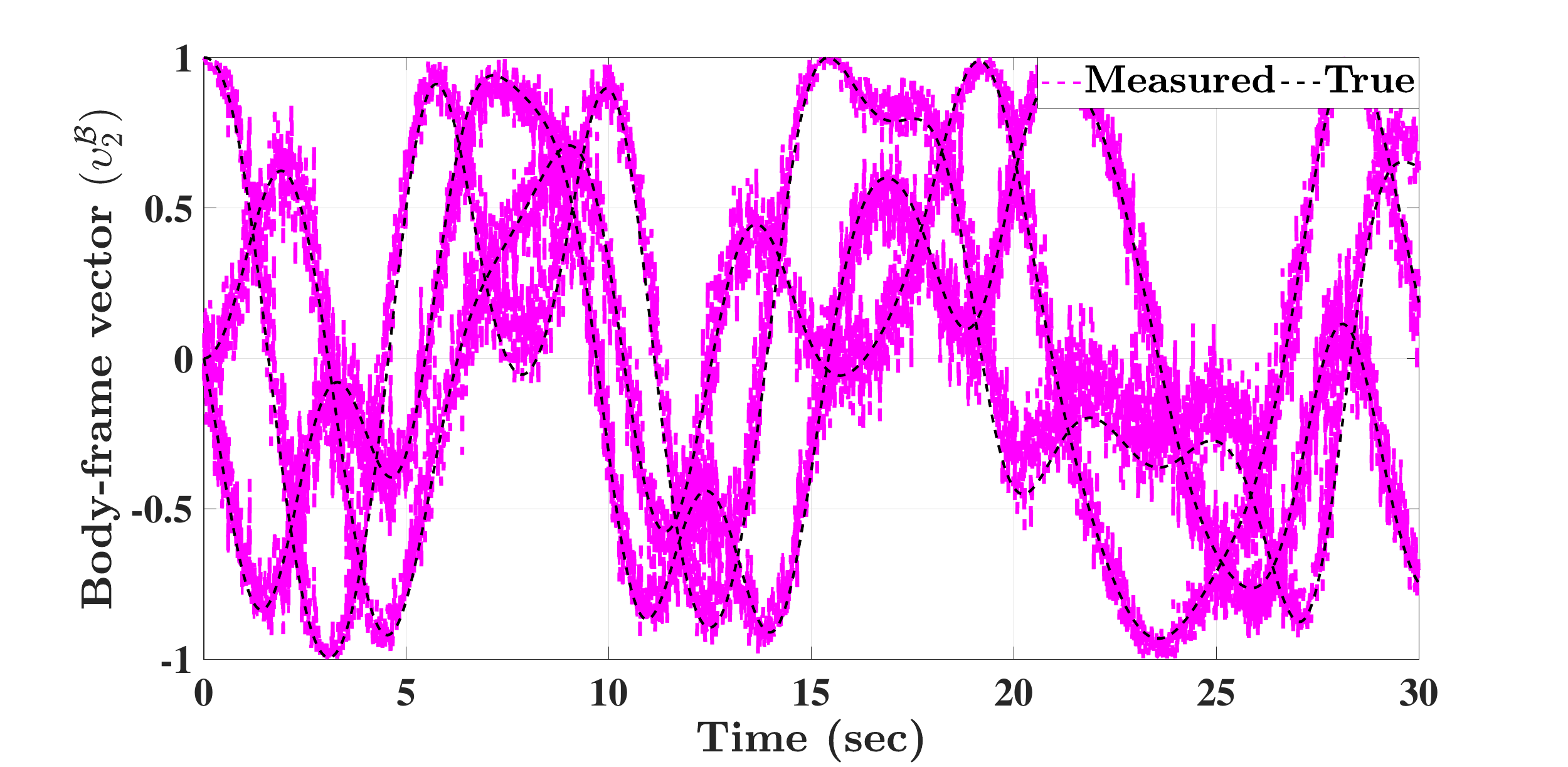}\includegraphics[scale=0.2]{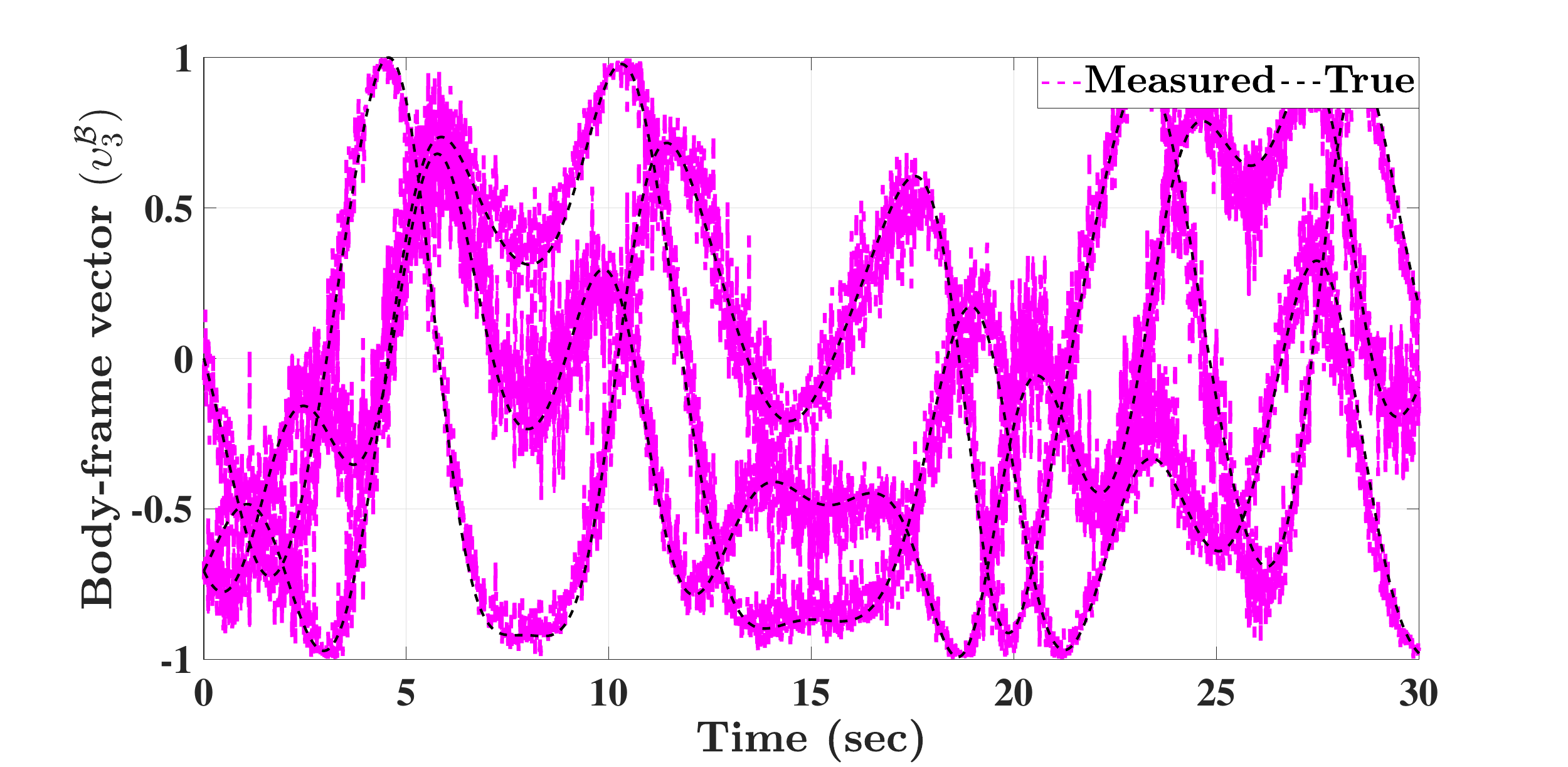}\caption{Angular velocity and body-frame vectors: Measured and true.}
	\label{fig:Comp_Simu_Ang_Vec}
\end{figure}

Consider the attitude dynamics in equation \eqref{eq:Comp_R_dynam},
true angular velocity input signal ($\Omega$), and initial attitude
($R\left(0\right)$) to be given as
\[
\begin{cases}
\dot{R} & =R\left[\Omega\right]_{\times}\\
\Omega & =\left[\begin{array}{c}
{\rm sin}\left(0.4t\right)\\
{\rm sin}\left(0.7t+\frac{\pi}{4}\right)\\
0.4{\rm sin}\left(0.3t+\frac{\pi}{2}\right)
\end{array}\right]\left({\rm rad/sec}\right)\\
R\left(0\right) & =\left[\begin{array}{ccc}
1 & 0 & 0\\
0 & 1 & 0\\
0 & 0 & 1
\end{array}\right]\\
T & =30\text{ sec},\hspace{1em}\text{(Total simulation time)}
\end{cases}
\]
Let the measurements of the true angular velocity ($\Omega_{m}$)
be corrupted with unknown random noise and constant bias such that
\begin{equation}
\begin{cases}
\Omega_{m} & =\Omega+b+\omega\\
b & =\left[-0.1,0.1,0.05\right]^{\top}\\
\omega & =\mathcal{N}\sim\left(0,0.2\right)
\end{cases}\label{eq:Comp_Simu_Ang}
\end{equation}
where $\omega=\mathcal{N}\sim\left(0,0.2\right)$ is a short-hand
notation indicating that the noise ($\omega$) is normally distributed
with zero mean ($\mathbb{E}\left[\omega\right]=0$) and ${\rm STD}=0.2$.
To implement $\omega=\mathcal{N}\sim\left(0,0.2\right)$ at instant
$t$ in MATLAB use the following command: $\omega\left(t\right)=0.2\times\text{randn}\left(3,1\right)$.
Consider the following two non-collinear inertial-frame vectors
\[
\begin{cases}
{\rm v}_{1}^{\mathcal{I}} & =\left[1,-1,1\right]^{\top}\\
{\rm v}_{2}^{\mathcal{I}} & =\left[0,0,1\right]^{\top}
\end{cases}
\]
The associated body-frame measurements are obtained as follows
\begin{equation}
\begin{cases}
{\rm v}_{1}^{\mathcal{B}} & =R^{\top}{\rm v}_{1}^{\mathcal{I}}+{\rm b}_{1}^{\mathcal{B}}+\omega_{1}^{\mathcal{B}}\\
{\rm v}_{2}^{\mathcal{B}} & =R^{\top}{\rm v}_{2}^{\mathcal{I}}+{\rm b}_{2}^{\mathcal{B}}+\omega_{2}^{\mathcal{B}}
\end{cases}\label{eq:Comp_Simu_Vec}
\end{equation}
with
\[
\begin{cases}
{\rm b}_{1}^{\mathcal{B}} & =\left[0.13,-0.13,0.13\right]^{\top}\\
{\rm b}_{2}^{\mathcal{B}} & =\left[0,0,0.13\right]^{\top}\\
\omega_{1}^{\mathcal{B}} & =\mathcal{N}\sim\left(0,0.13\right)\\
\omega_{2}^{\mathcal{B}} & =\mathcal{N}\sim\left(0,0.13\right)
\end{cases}
\]
The third inertial-frame and body-frame vectors are obtained as a
cross product as follows
\[
\begin{cases}
{\rm v}_{3}^{\mathcal{I}} & ={\rm v}_{1}^{\mathcal{I}}\times{\rm v}_{2}^{\mathcal{I}}\\
{\rm v}_{3}^{\mathcal{B}} & ={\rm v}_{1}^{\mathcal{B}}\times{\rm v}_{2}^{\mathcal{B}}
\end{cases}
\]
Next step is normalization performed according to \eqref{eq:Comp_Vector_norm}:
\begin{equation}
\upsilon_{i}^{\mathcal{I}}=\frac{{\rm v}_{i}^{\mathcal{I}}}{||{\rm v}_{i}^{\mathcal{I}}||},\hspace{1em}\upsilon_{i}^{\mathcal{B}}=\frac{{\rm v}_{i}^{\mathcal{B}}}{||{\rm v}_{i}^{\mathcal{B}}||},\hspace{1em}\forall i=1,2,3\label{eq:Comp_Simu_Vec2}
\end{equation}
Consider the measurements of angular velocity in \eqref{eq:Comp_Simu_Ang},
the body-frame measurements in \eqref{eq:Comp_Simu_Vec}, and the
normalized values of the body-frame measurements in \eqref{eq:Comp_Simu_Vec2}.
The true angular velocity and the normalized values of body-frame
vectors are plotted against angular velocity measurements and the
normalized values of body-frame vectorial measurements in Figure \ref{fig:Comp_Simu_Ang_Vec},
respectively. It can be observed in Figure \ref{fig:Comp_Simu_Ang_Vec}
that high values of noise and bias components corrupted the measurement
process of the three categories of attitude determination and estimation
algorithms listed in Table \ref{tab:Simu1}.

\noindent\makebox[1\linewidth]{%
	\rule{0.8\textwidth}{0.5pt}%
}

\subsubsection{Initialization and Design Parameters}

For the semi-direct filters in \eqref{eq:Comp_Non_CGSd_NDAF}, \eqref{eq:Comp_Non_AGNDAF},
\eqref{eq:Comp_Non_GPSd_NDAF}, and \eqref{eq:Comp_Non_GPSd_NDSF},
$R_{y}$ is reconstructed with the aid of SVD in \eqref{eq:Comp_SVD}.

For Gaussian and nonlinear attitude filters, the initial attitude
estimate is given with respect to angle-axis parameterization in \eqref{eq:Comp_R_att_ang}
such that
\[
\begin{cases}
\hat{R}\left(0\right) & =\mathcal{R}_{\alpha}\left(\alpha,u/\left\Vert u\right\Vert \right)\\
\alpha & =178\left({\rm deg}\right)\\
u & =\left[8,7,4\right]^{\top}
\end{cases}
\]
or more simply put
\[
\hat{R}\left(0\right)=\left[\begin{array}{ccc}
-0.0074 & 0.8557 & 0.5175\\
0.8802 & -0.2399 & 0.4094\\
0.4745 & 0.4586 & -0.7514
\end{array}\right]
\]
where $||\tilde{R}\left(0\right)||_{I}=0.9997$ initiated very close
to the unstable equilibria ($+1$). Initial estimates used for all
filters are as follows
\begin{align*}
\hat{b}\left(0\right) & =\left[0,0,0\right]^{\top}\\
\hat{\sigma}\left(0\right) & =\left[0,0,0\right]^{\top}
\end{align*}
The design parameters of the filters are summarized in Table \ref{tab:Simu2}.
Since MEKF, GAMEF, and CG-NDAF are not characterized with adaptive
gains,three cases of the design parameters are considered for each
of the above-mentioned filters to ensure fair comparison. The comparison
between the filtering methods in this section examines the transient
and steady-state performance of the attitude error in terms of
\begin{enumerate}
	\item[1)] normalized Euclidean distance of the attitude error
	\[
	||\tilde{R}||_{I}=\frac{1}{4}{\rm Tr}\left\{ \mathbf{I}_{3}-R^{\top}\hat{R}\right\} 
	\]
	\item[2)] the error in rotation angle about the unit axis \cite{hashim2019AtiitudeSurvey}
	\[
	\tilde{\alpha}={\rm cos}^{-1}\left(\frac{{\rm Tr}\{R^{\top}\hat{R}\}-1}{2}\right)
	\]
\end{enumerate}
In all the simulations, the output values of $||\tilde{R}||_{I}$
and $\tilde{\alpha}$ are recorded every 0.01 seconds with the infinity
norm $\left\Vert ||\tilde{R}||_{I}\right\Vert _{\infty}:=\max_{t}\left(||\tilde{R}(t)||_{I}\right)$
and $\left\Vert \tilde{\alpha}\right\Vert _{\infty}:=\max_{t}\left|\tilde{\alpha}(t)\right|$.

\begin{table}[h!]
	\caption{\label{tab:Simu2}Design parameters}
	
	\centering{}%
	\begin{tabular}{>{\raggedright}p{2cm}|>{\raggedright}p{11cm}}
		\hline 
		Filter & Design parameters\tabularnewline
		\hline 
		\hline 
		\noalign{\vskip\doublerulesep}
		MEKF & Case 1:$\mathcal{\bar{Q}}_{v\left(i\right)}=\mathbf{I}_{3}$, $\mathcal{\bar{Q}}_{\omega}=\mathbf{I}_{3}$,
		and $\mathcal{\bar{Q}}_{b}=\mathbf{I}_{3}$ 
		
		Case 2:$\mathcal{\bar{Q}}_{v\left(i\right)}=0.1\mathbf{I}_{3}$, $\mathcal{\bar{Q}}_{\omega}=10\mathbf{I}_{3}$,
		and $\mathcal{\bar{Q}}_{b}=10\mathbf{I}_{3}$
		
		Case 3:$\mathcal{\bar{Q}}_{v\left(i\right)}=0.01\mathbf{I}_{3}$,
		$\mathcal{\bar{Q}}_{\omega}=100\mathbf{I}_{3}$, and $\mathcal{\bar{Q}}_{b}=100\mathbf{I}_{3}$\tabularnewline
		\hline 
		\noalign{\vskip\doublerulesep}
		GAMEF & Case 1:$\mathcal{\bar{Q}}_{v\left(i\right)}=\mathbf{I}_{3}$, $\mathcal{\bar{Q}}_{\omega}=\mathbf{I}_{3}$,
		and $\mathcal{\bar{Q}}_{b}=\mathbf{I}_{3}$ 
		
		Case 2:$\mathcal{\bar{Q}}_{v\left(i\right)}=0.1\mathbf{I}_{3}$, $\mathcal{\bar{Q}}_{\omega}=10\mathbf{I}_{3}$,
		and $\mathcal{\bar{Q}}_{b}=10\mathbf{I}_{3}$
		
		Case 3:$\mathcal{\bar{Q}}_{v\left(i\right)}=0.01\mathbf{I}_{3}$,
		$\mathcal{\bar{Q}}_{\omega}=100\mathbf{I}_{3}$, and $\mathcal{\bar{Q}}_{b}=100\mathbf{I}_{3}$\tabularnewline
		\hline 
		CG-NDAF & Case 1: $k_{w}=1$
		
		Case 2: $k_{w}=10$
		
		Case 3: $k_{w}=100$\tabularnewline
		\hline 
		AG-NDAF & $k_{w}=8$\tabularnewline
		\hline 
		\noalign{\vskip\doublerulesep}
		GP-NDAF & $k_{w}=2$, $\bar{\delta}=1.7$, $\underline{\delta}=1.7$ $\xi_{0}=1.7$,
		$\xi_{\infty}=0.08$, $\ell=4$ and $\gamma=1$\tabularnewline
		\hline 
		AG-NSAF & $\gamma_{1}=1$, $\gamma_{2}=1$, $k_{b}=0.01$, $k_{\sigma}=0.01$,
		$k_{w}=2$, $k_{2}=0.5$ and $\varepsilon=0.1$\tabularnewline
		\hline 
		\noalign{\vskip\doublerulesep}
		GP-NSAF & $k_{w}=2$, $\bar{\delta}=1.7$, $\underline{\delta}=1.7$ $\xi_{0}=1.7$,
		$\xi_{\infty}=0.08$, $\ell=4$, $\gamma_{1}=1$ and $\gamma_{2}=0.1$\tabularnewline
		\hline 
	\end{tabular}
\end{table}

\noindent\makebox[1\linewidth]{%
	\rule{0.8\textwidth}{0.5pt}%
}

\subsubsection{Attitude Determination Results}

\begin{figure}[h!]
	\includegraphics[scale=0.2]{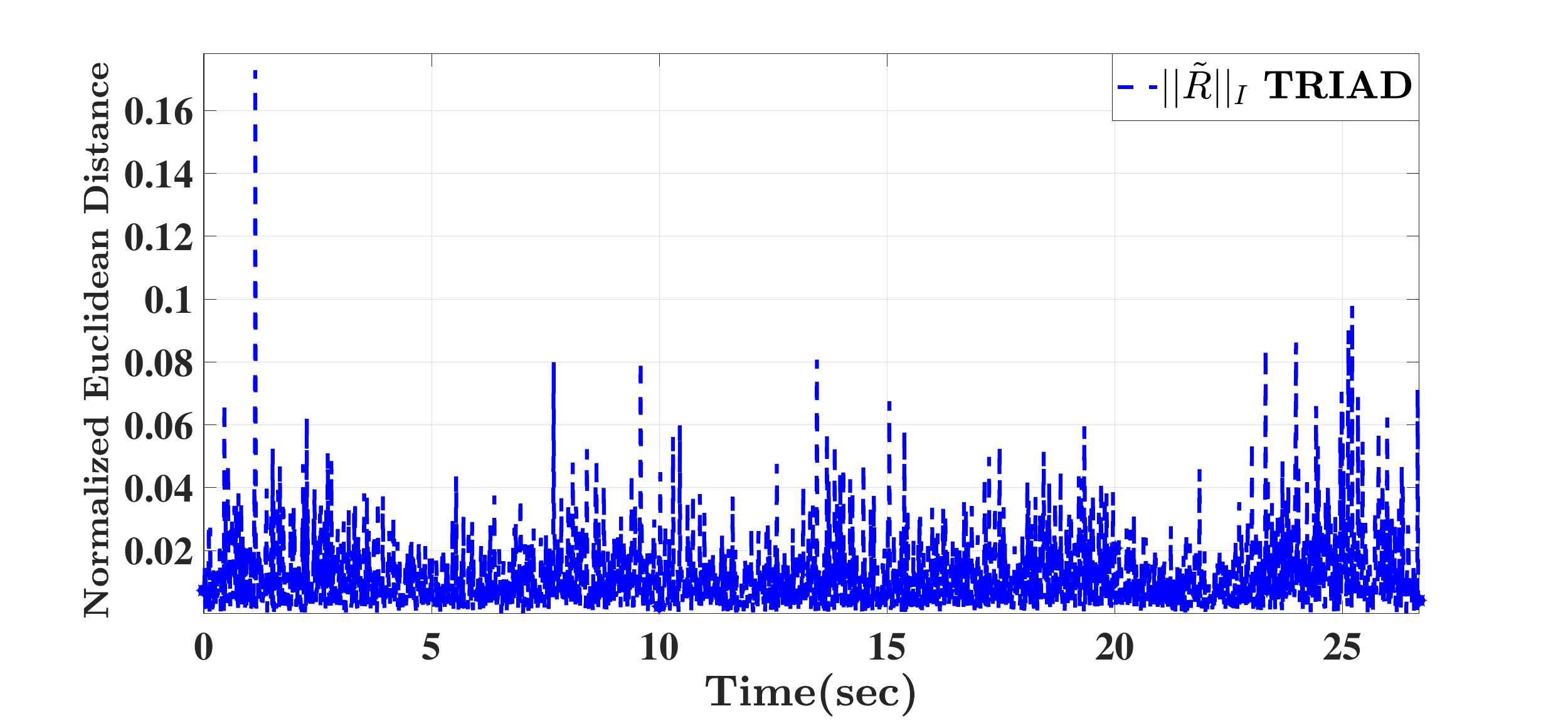}\includegraphics[scale=0.2]{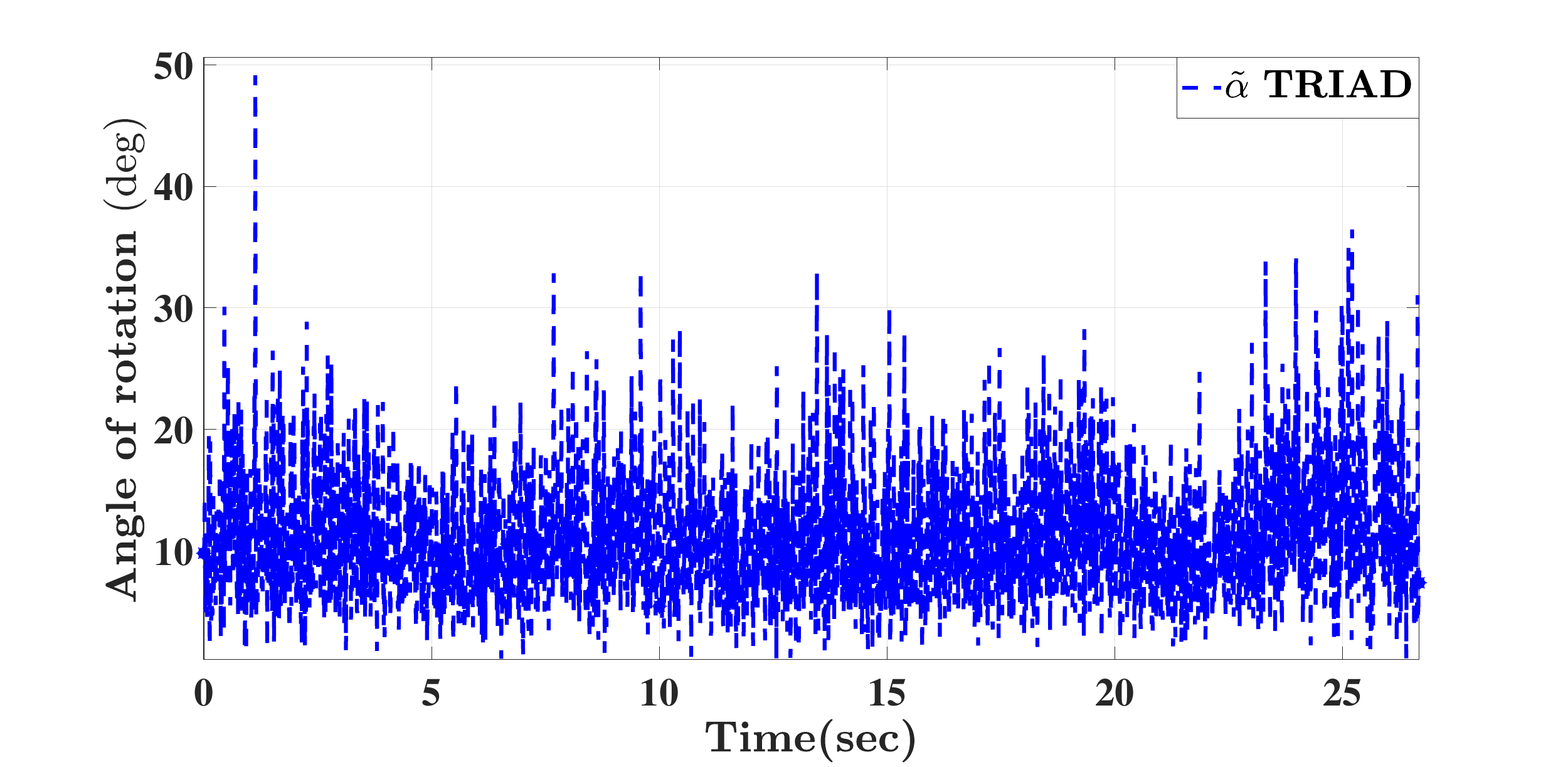}
	
	\includegraphics[scale=0.2]{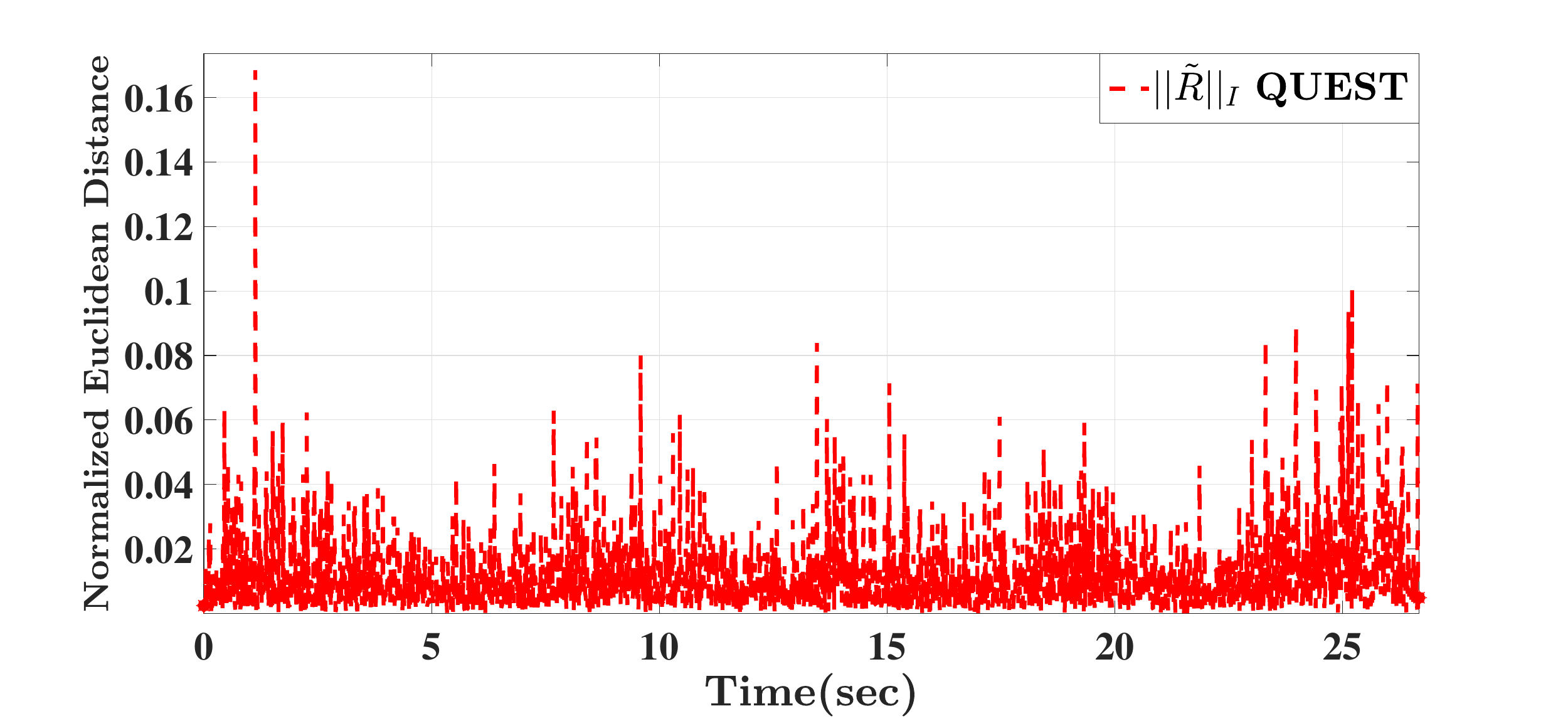}\includegraphics[scale=0.2]{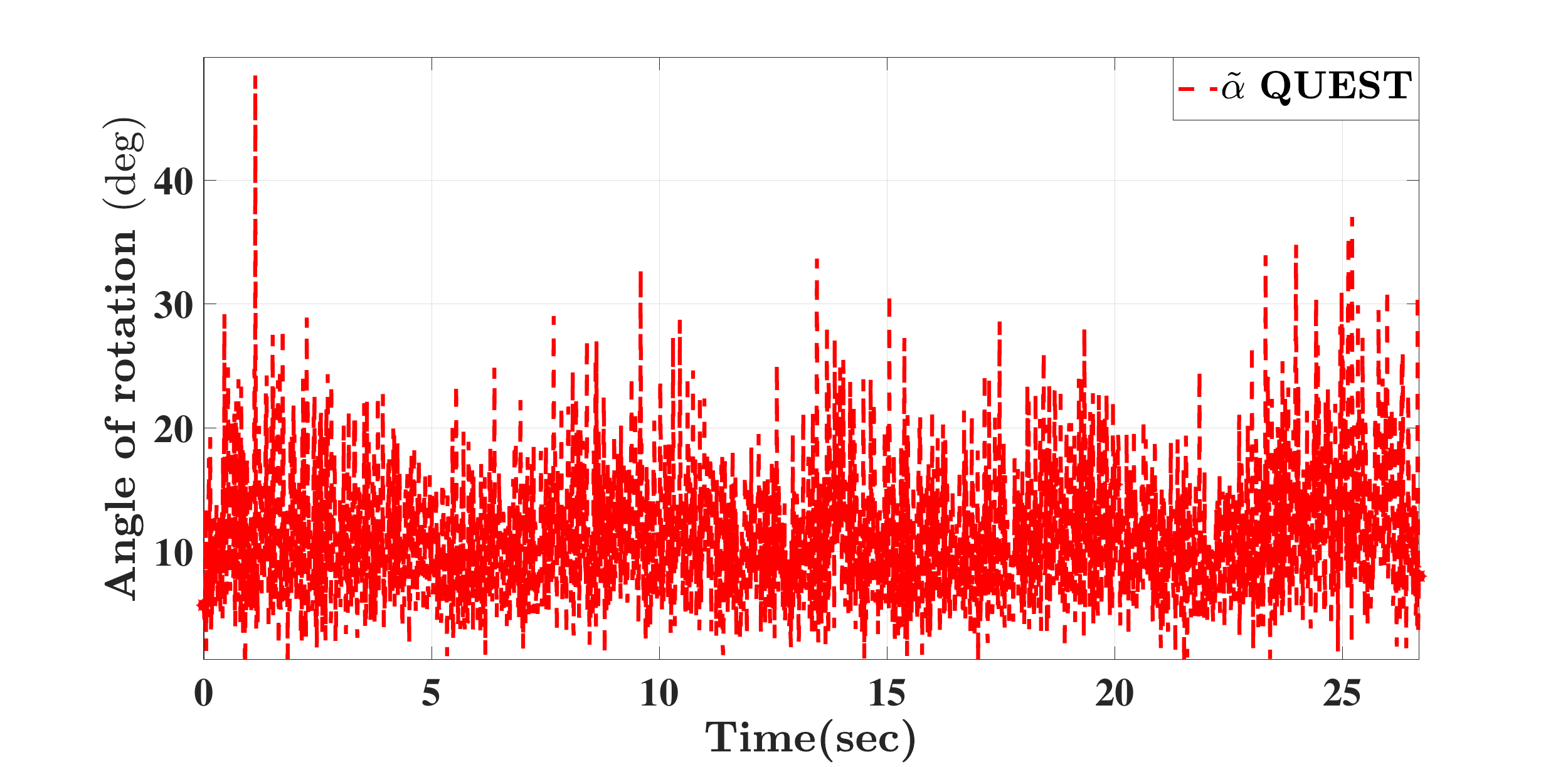}
	
	\includegraphics[scale=0.2]{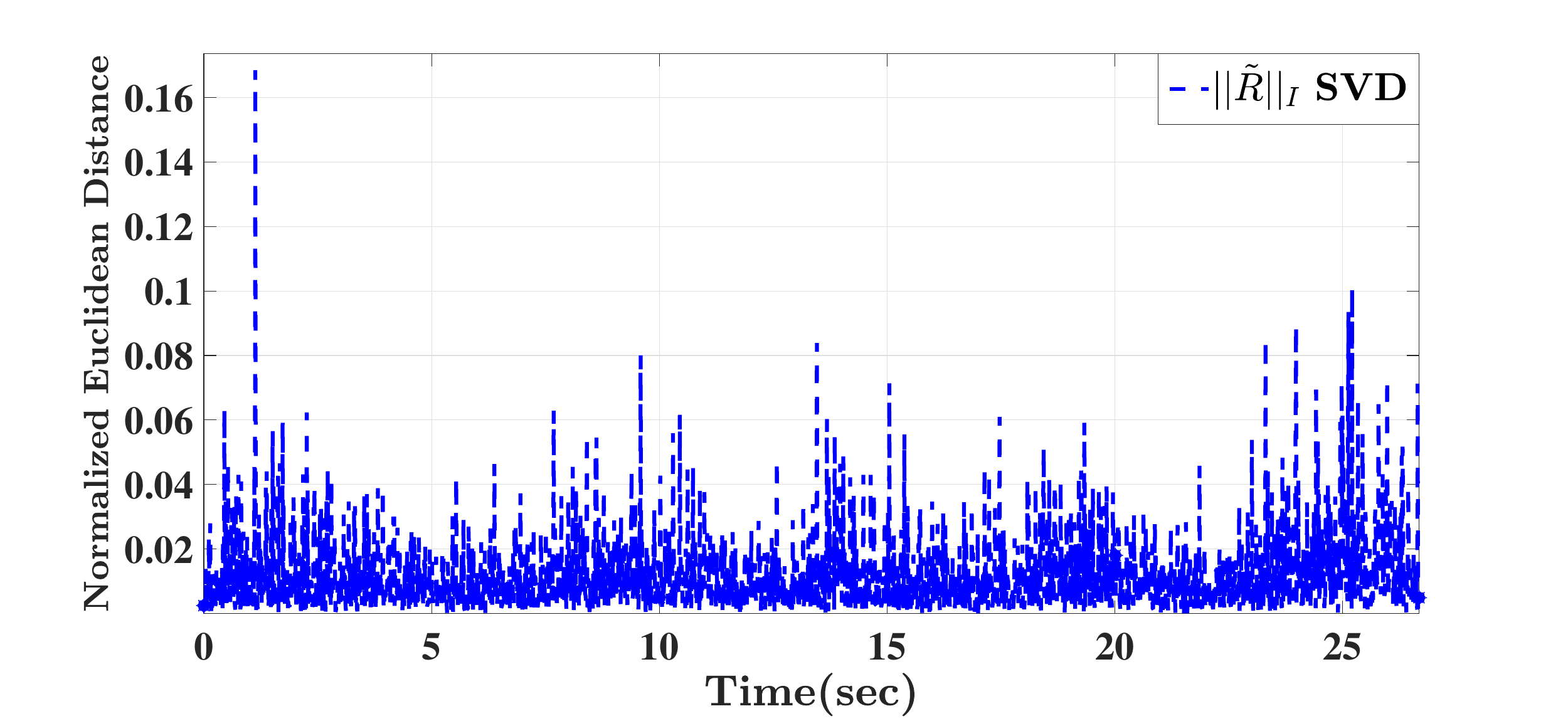}\includegraphics[scale=0.2]{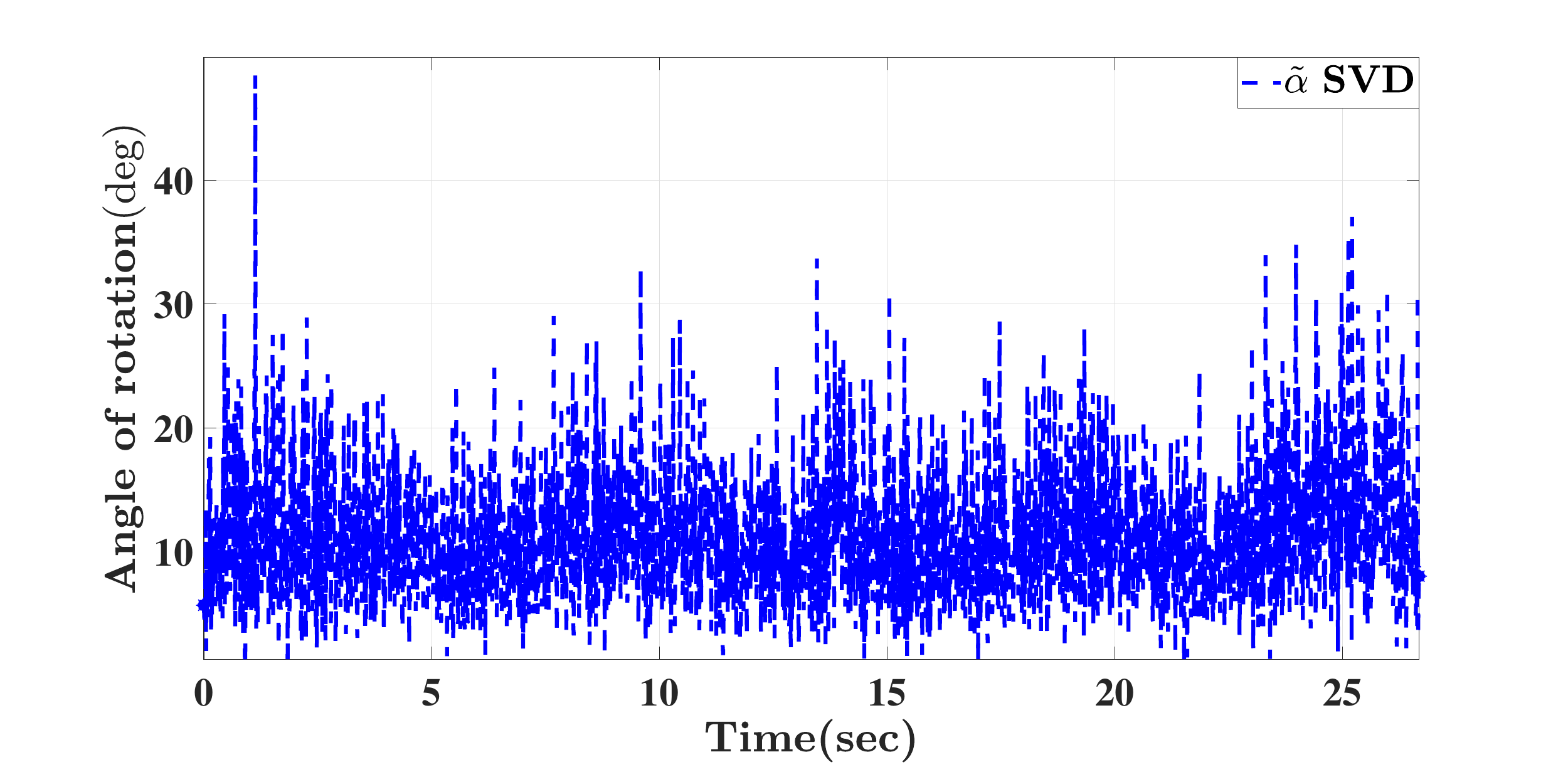}\caption{Tracking error of TRIAD, QUEST and SVD: $||\tilde{R}||_{I}$ and $\tilde{\alpha}$.}
	\label{fig:Comp_Simu_Determination} 
\end{figure}

\begin{figure}[h!]
	\includegraphics[scale=0.13]{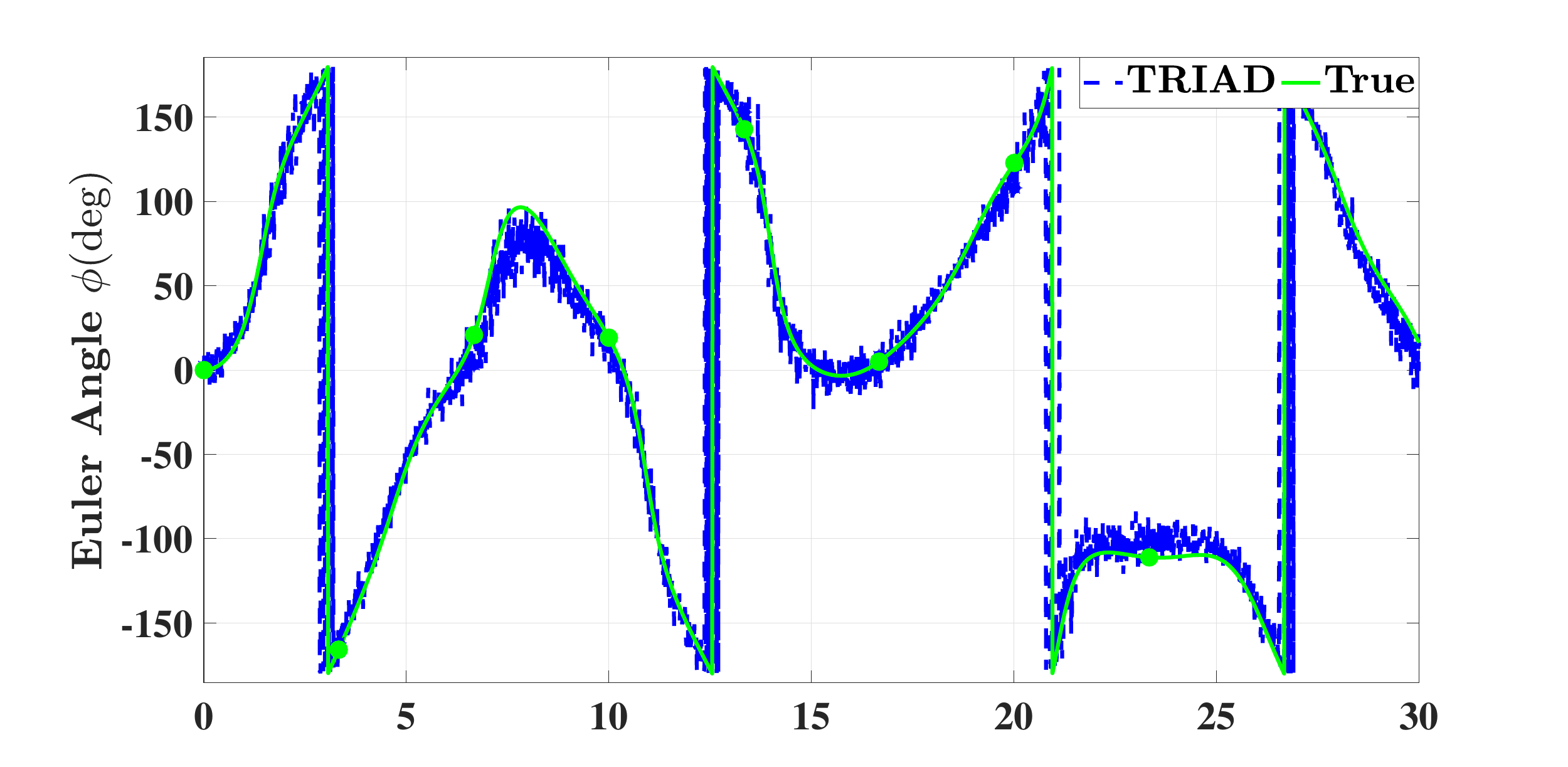}\includegraphics[scale=0.13]{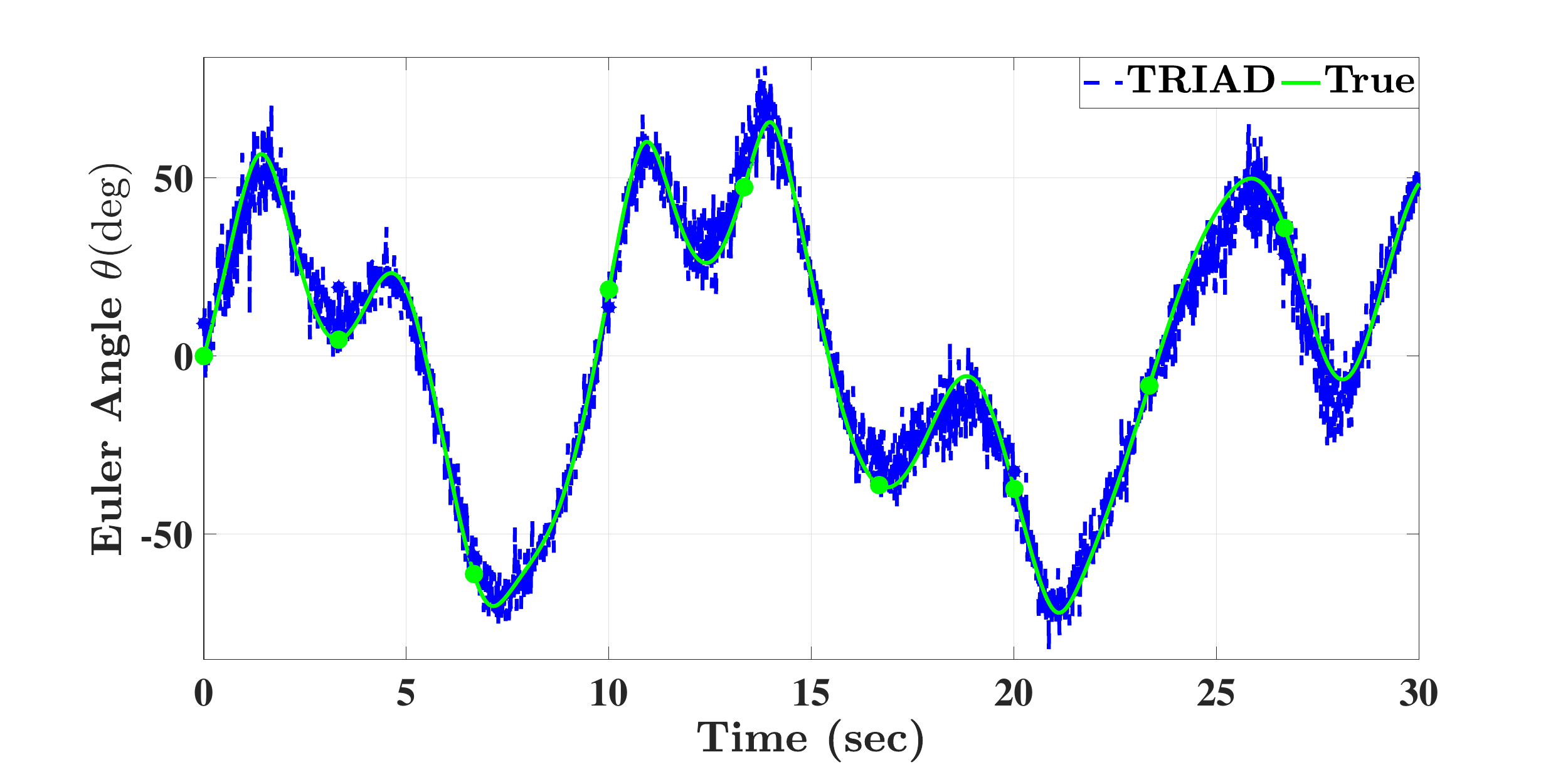}\includegraphics[scale=0.13]{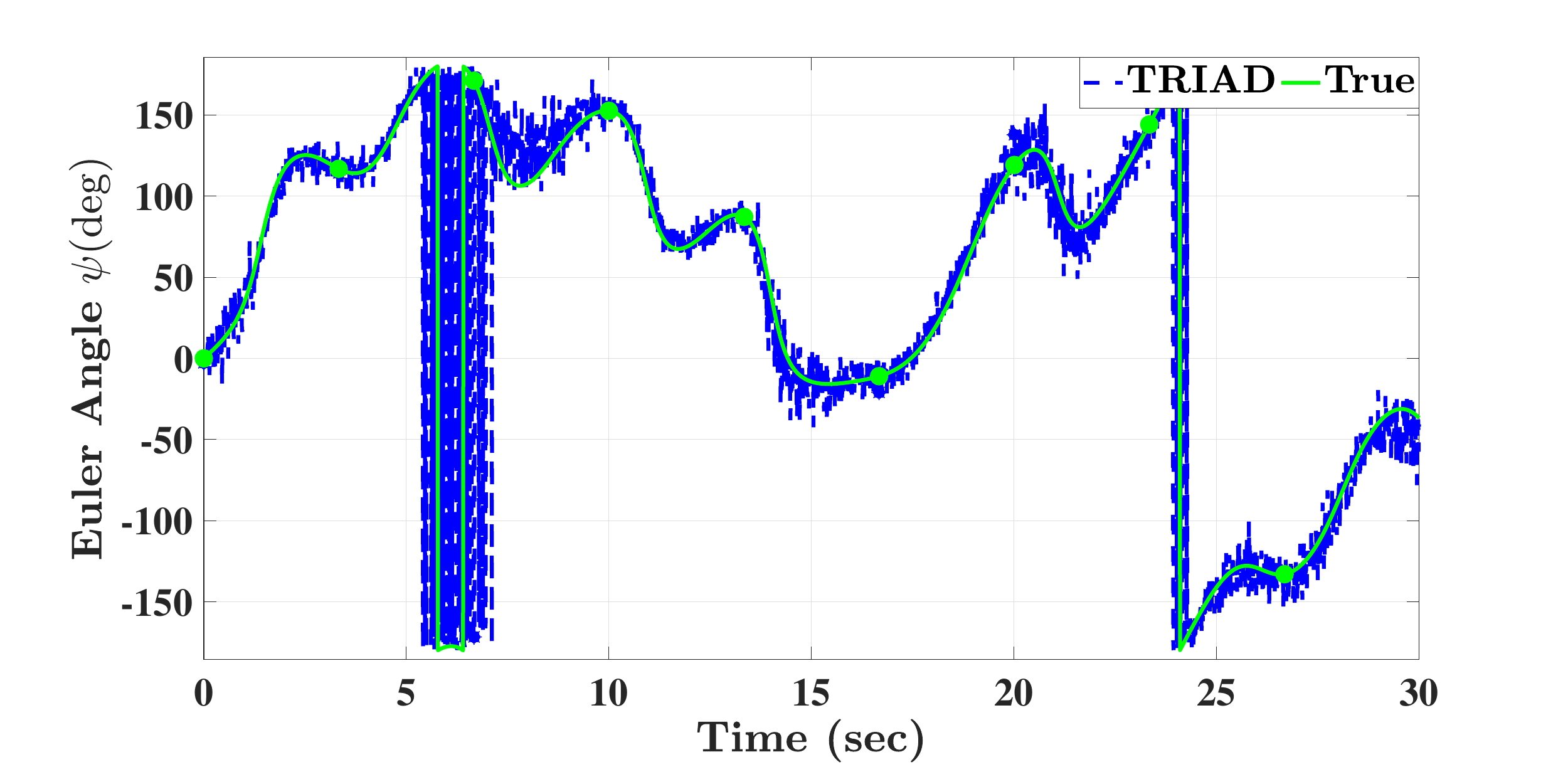}
	
	\includegraphics[scale=0.13]{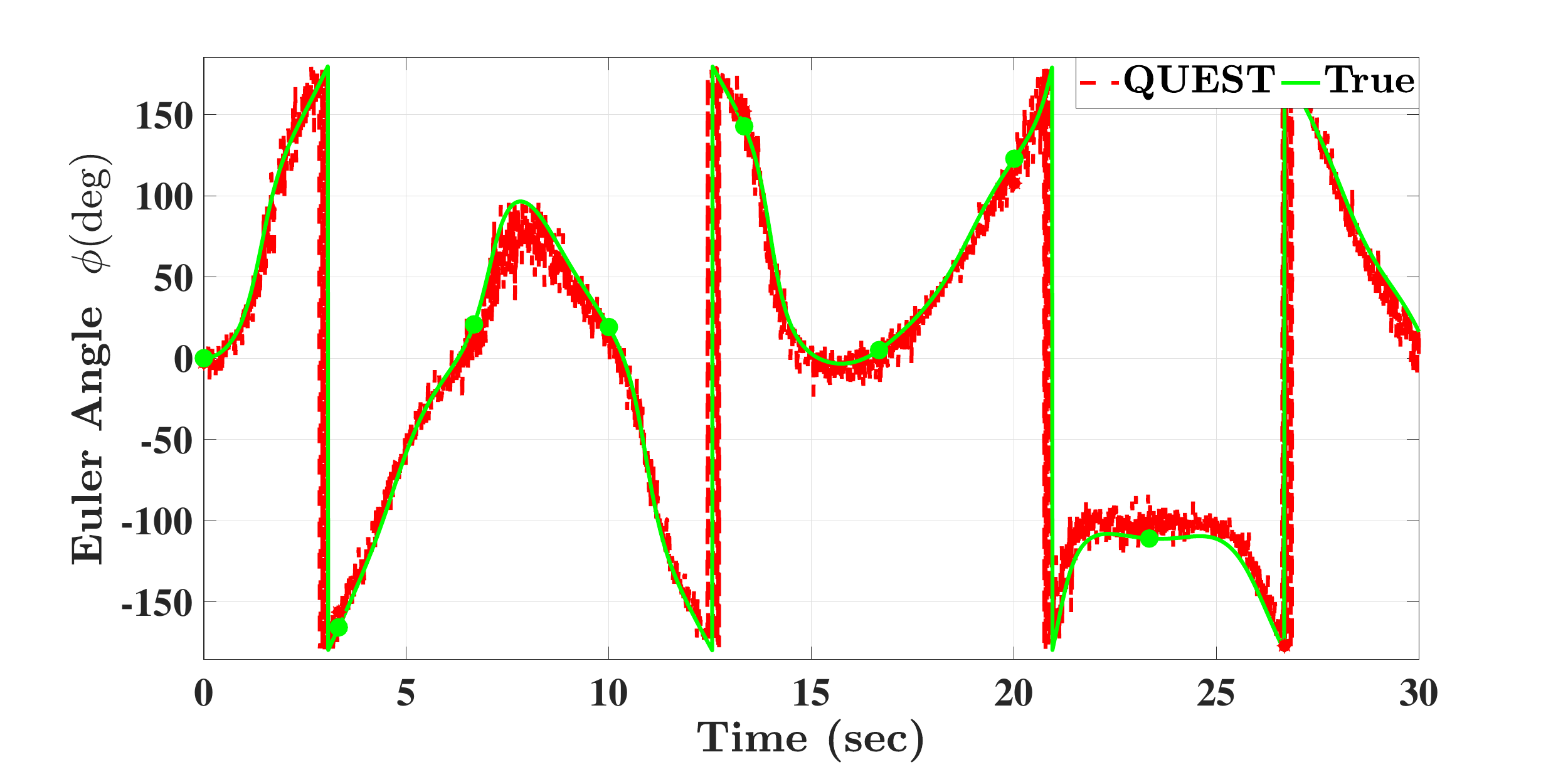}\includegraphics[scale=0.13]{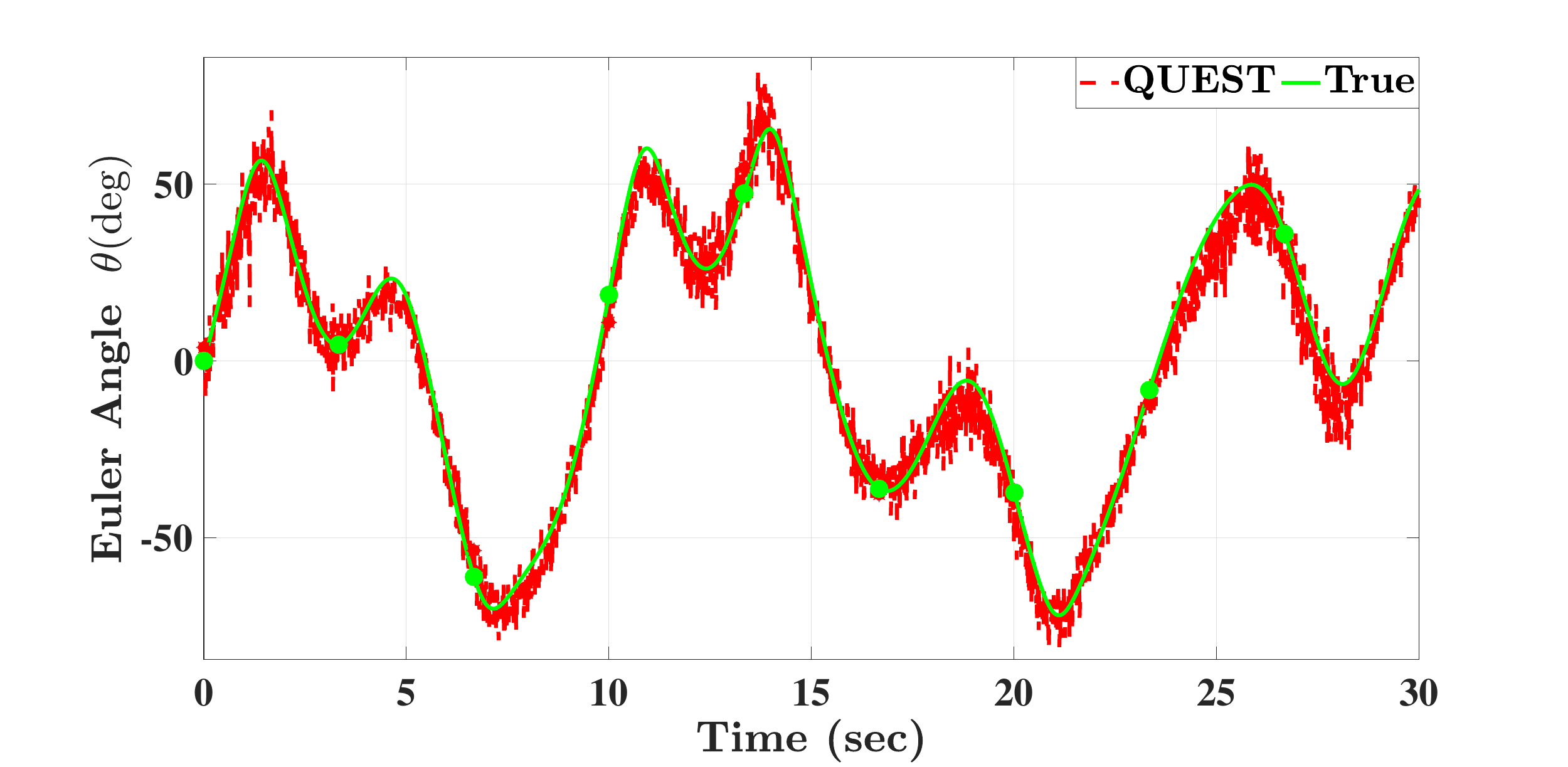}\includegraphics[scale=0.13]{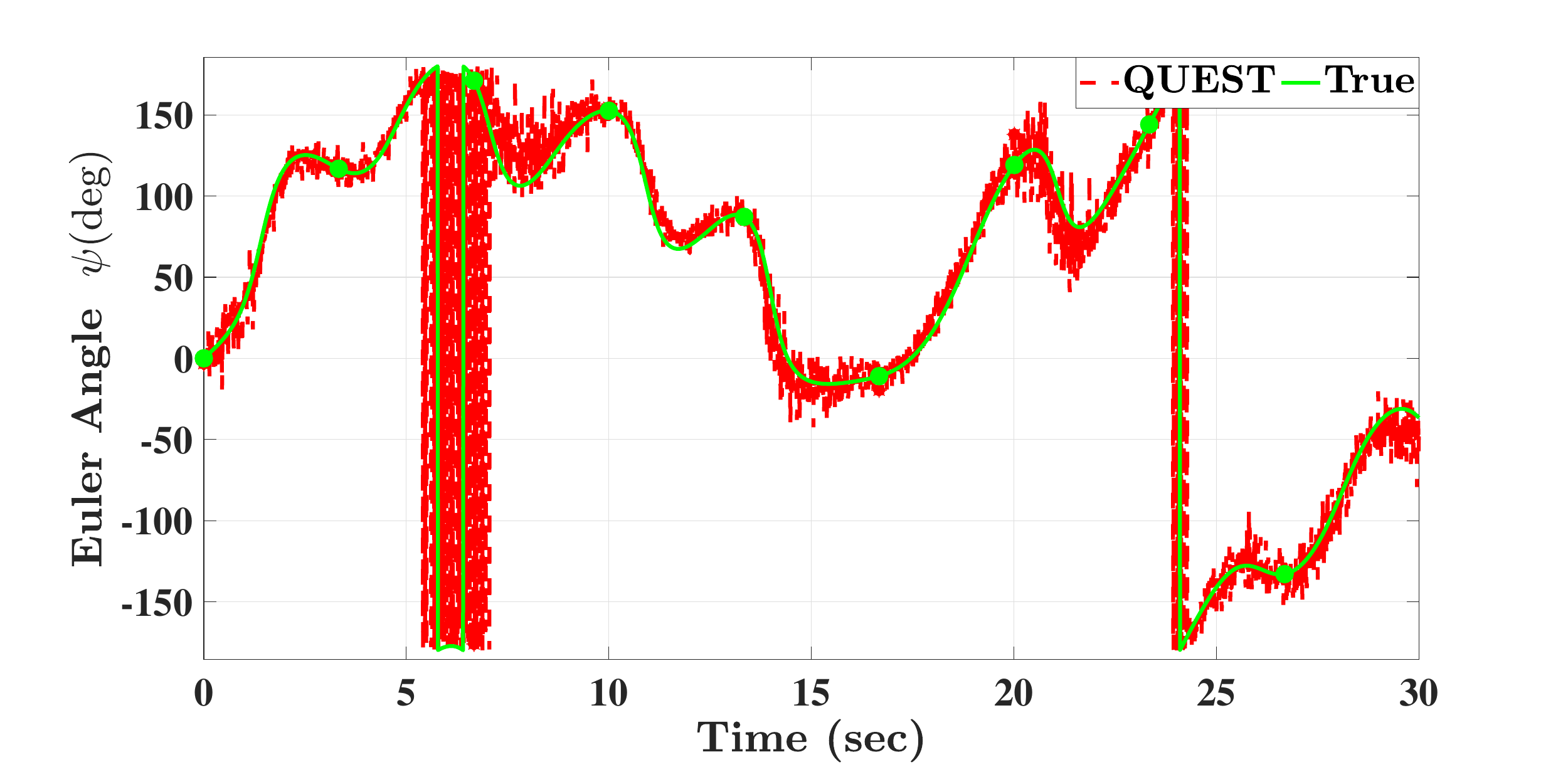}
	
	\includegraphics[scale=0.13]{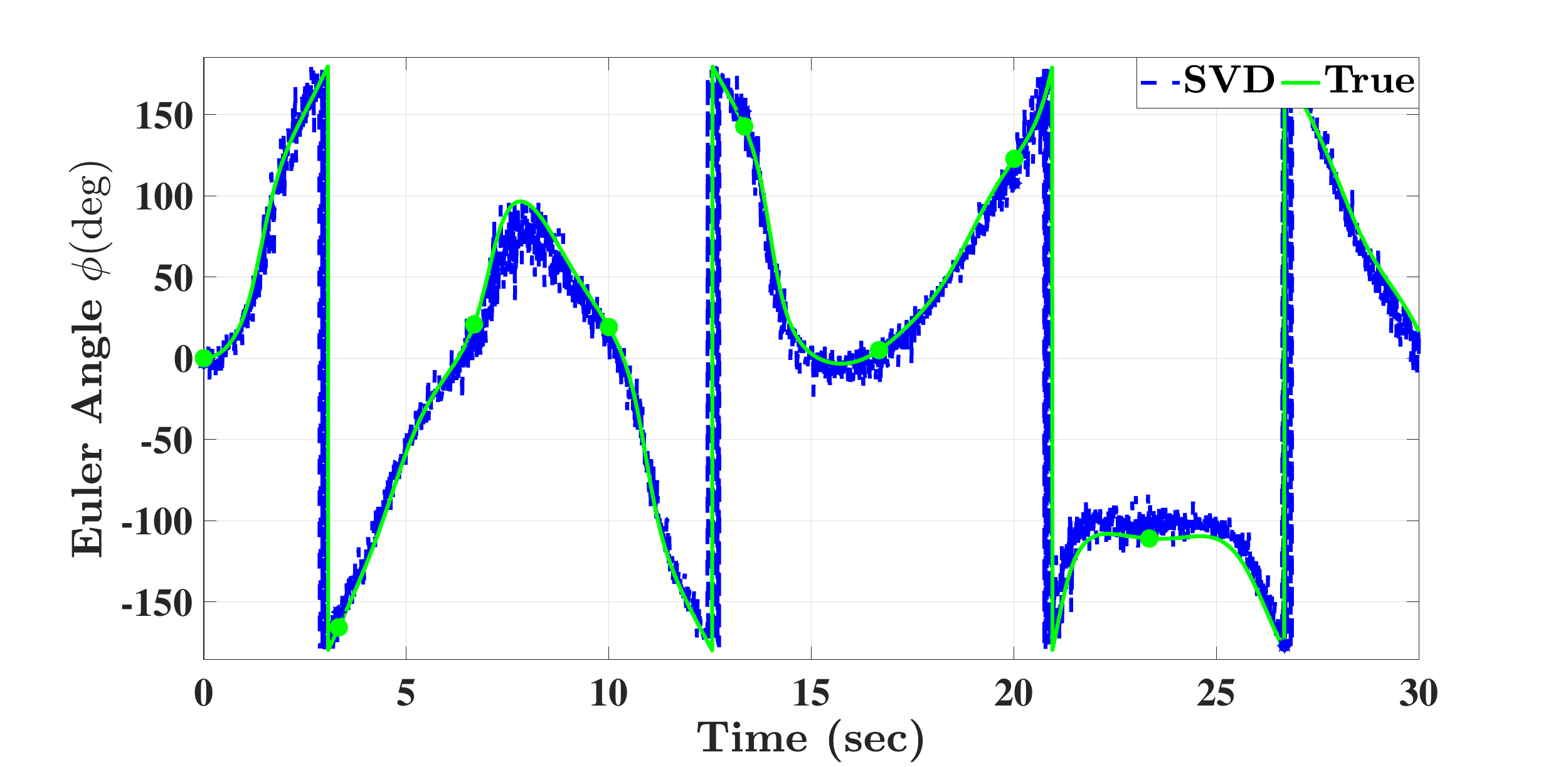}\includegraphics[scale=0.13]{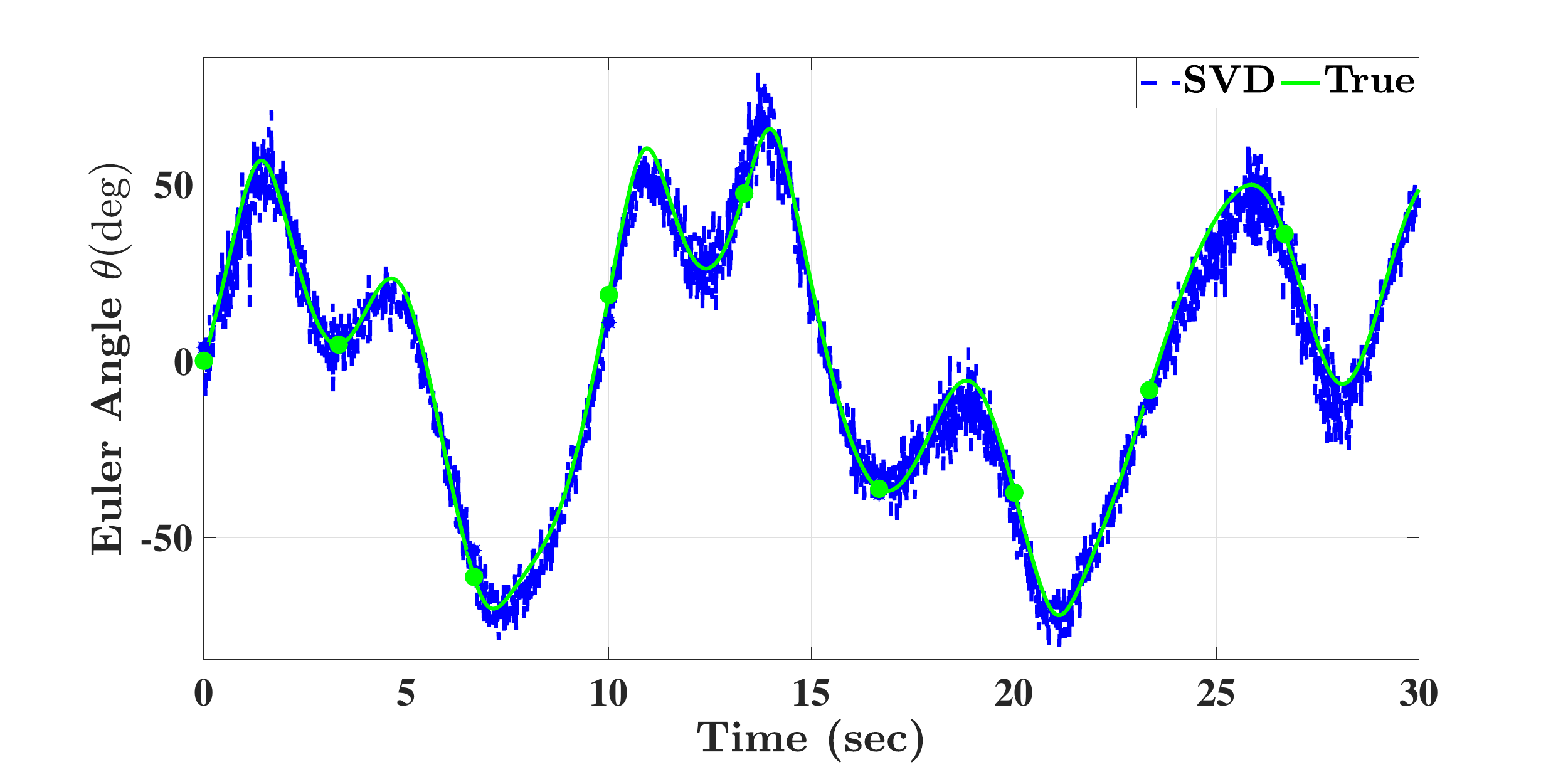}\includegraphics[scale=0.13]{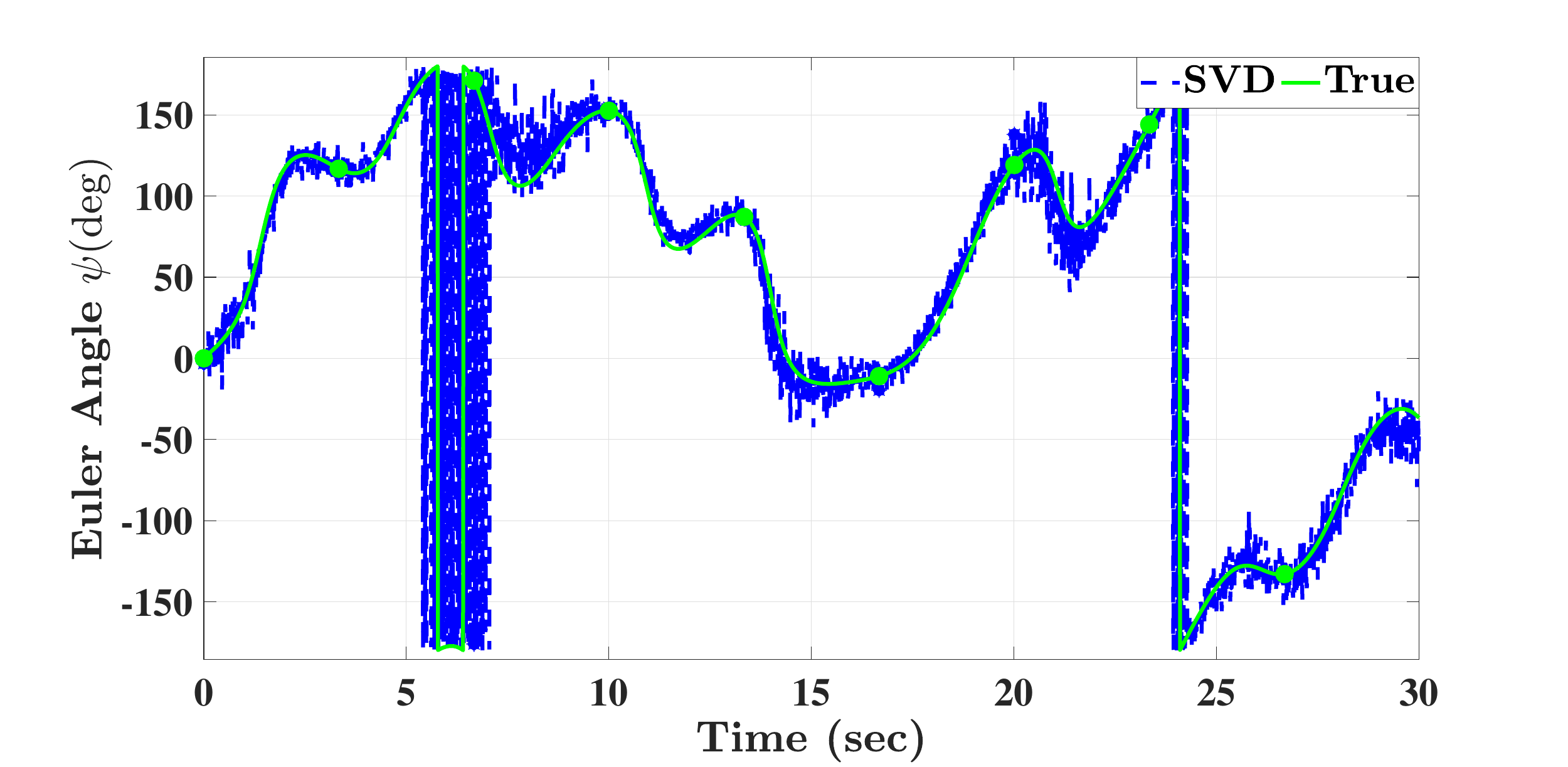}
	
	\caption{Tracking Euler angles ($\phi$, $\theta$ and $\psi$) of TRIAD, QUEST
		and SVD vs true angles.}
	\label{fig:Comp_Simu_Determination2}
\end{figure}

Figure \ref{fig:Comp_Simu_Determination} illustrates high sensitivity
of algebraic attitude determination algorithms to bias and noise present
in measurements. The poor performance observed in Figure \ref{fig:Comp_Simu_Determination}
is reinforced by the oscillatory behavior of the constructed Euler
angles when compared to the true Euler angles depicted in Figure \ref{fig:Comp_Simu_Determination2}.
Table \ref{tab:Comp_Simu_Determination} containing statistical results
of the mean, STD and $\left\Vert \,\cdot\,\right\Vert _{\infty}$
of $||\tilde{R}||_{I}$ and $\tilde{\alpha}$ provides additional
evidence of the poor performance of the algebraic attitude determination
algorithms: TRIAD, QUEST, and SVD when faced with biased and noisy
measurements. 

\begin{table}[h!]
	\caption{\textcolor{blue}{\label{tab:Comp_Simu_Determination} }Statistical
		analysis of $||\tilde{R}||_{I}$ and $\tilde{\alpha}$ of TRIAD, QUEST
		and SVD.}
	
	\centering{}%
	\begin{tabular}{l|c|c|c|c|c|c}
		\hline 
		\multicolumn{7}{c}{Output data of $||\tilde{R}||_{I}$ and $\tilde{\alpha}$ over the
			period (0-30 sec)}\tabularnewline
		\hline 
		\hline 
		\textbf{Filter} & Mean ($||\tilde{R}||_{I}$) & STD ($||\tilde{R}||_{I}$) & $\left\Vert ||\tilde{R}||_{I}\right\Vert _{\infty}$ & Mean ($\tilde{\alpha}$) & STD ($\tilde{\alpha}$) & $\left\Vert \tilde{\alpha}\right\Vert _{\infty}$\tabularnewline
		\hline 
		TRIAD & $0.0120$ & $0.0124$ & $0.1728$ & $11.2999$ & $5.6085$ & $49.130$\tabularnewline
		\hline 
		QUEST & $0.0124$ & $0.0123$ & $0.1685$ & $11.5080$ & $5.5839$ & $48.464$\tabularnewline
		\hline 
		SVD & $0.0124$ & $0.0123$ & $0.1685$ & $11.5080$ & $5.5839$ & $48.464$\tabularnewline
		\hline 
	\end{tabular}
\end{table}

\noindent\makebox[1\linewidth]{%
	\rule{0.8\textwidth}{0.5pt}%
}

\newpage

\subsubsection{Gaussian and Nonlinear Attitude Filters Results}

\begin{figure}[h!]
	\includegraphics[scale=0.2]{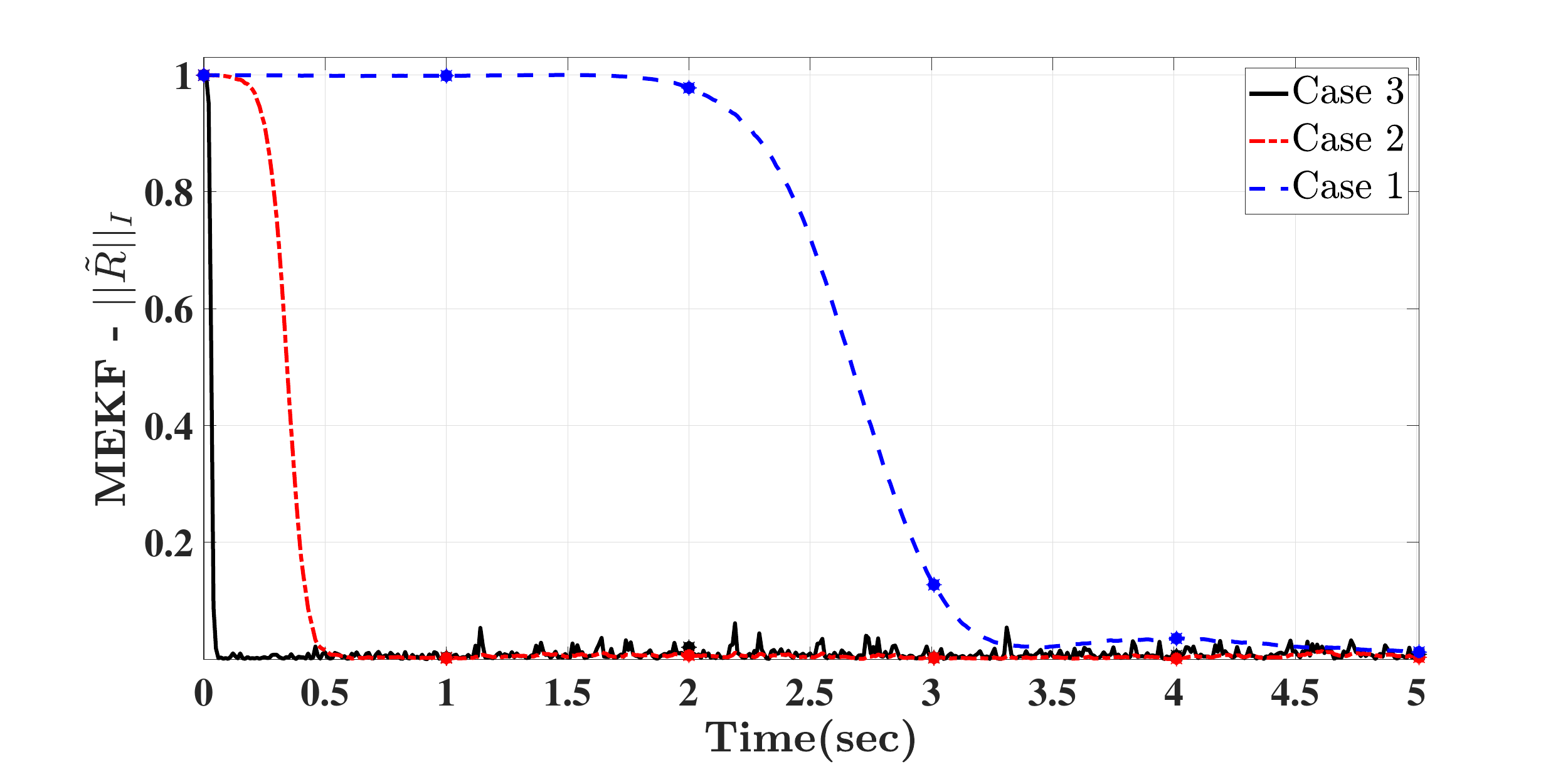}\includegraphics[scale=0.2]{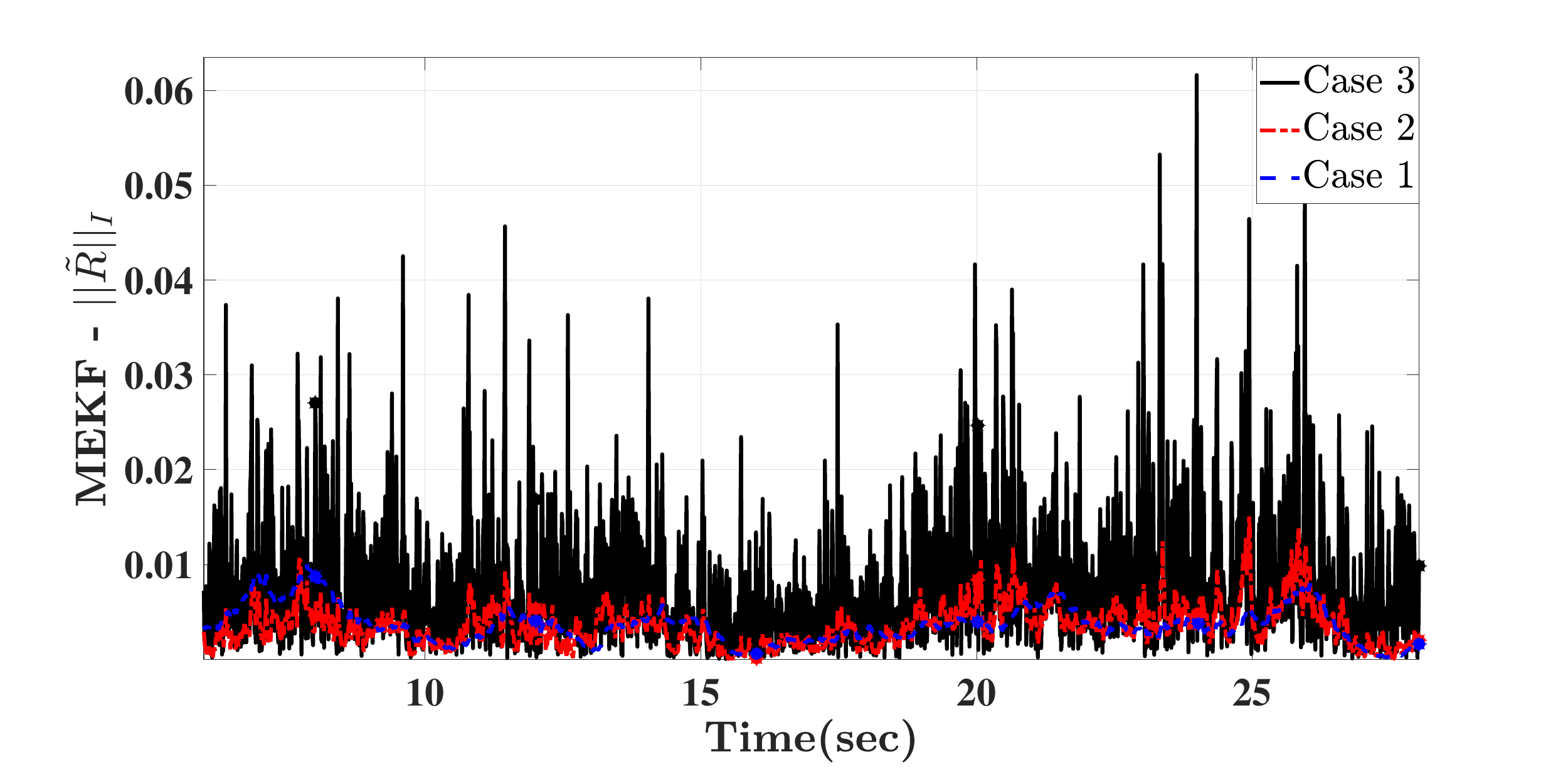}
	
	\includegraphics[scale=0.2]{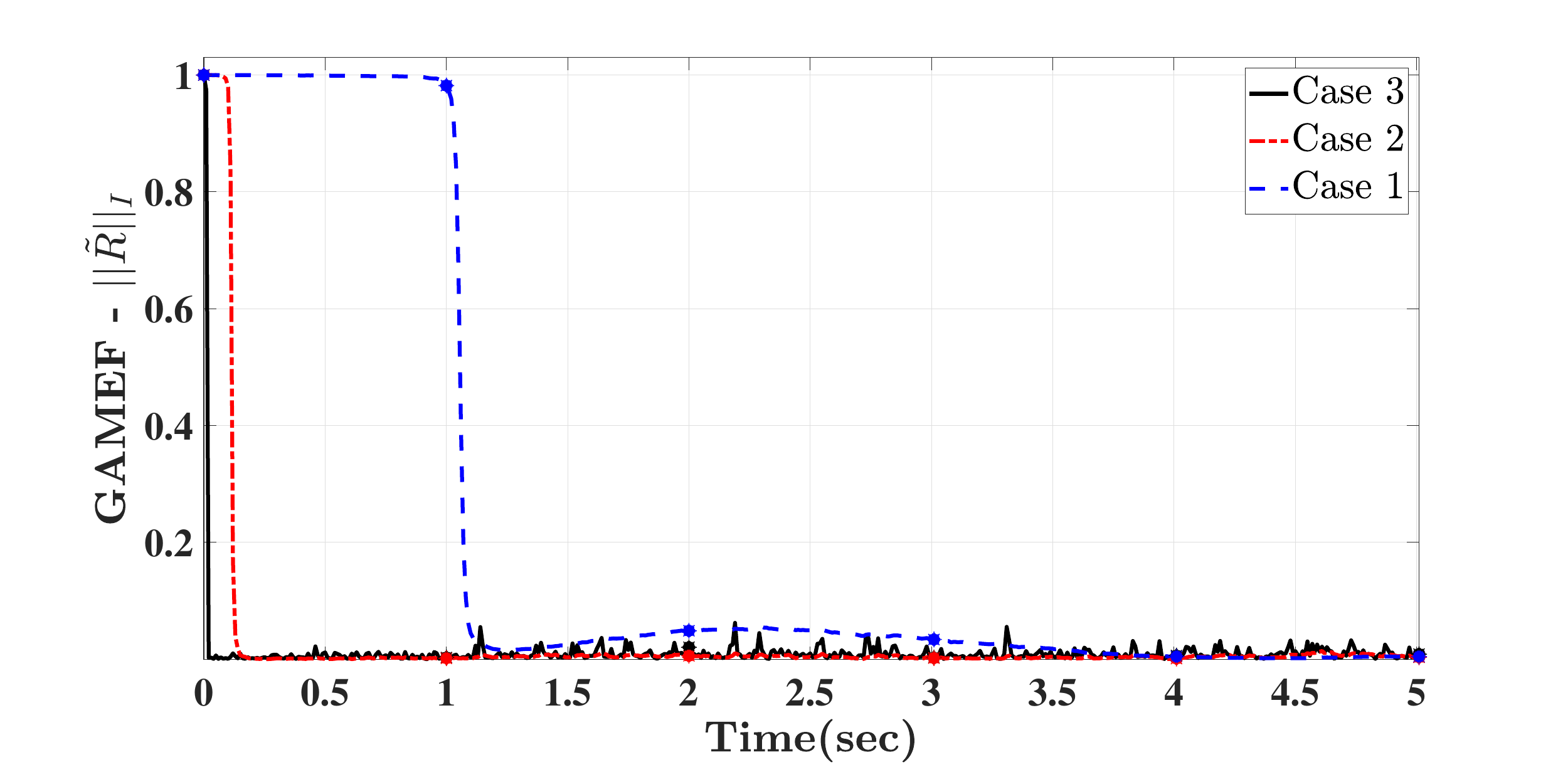}\includegraphics[scale=0.2]{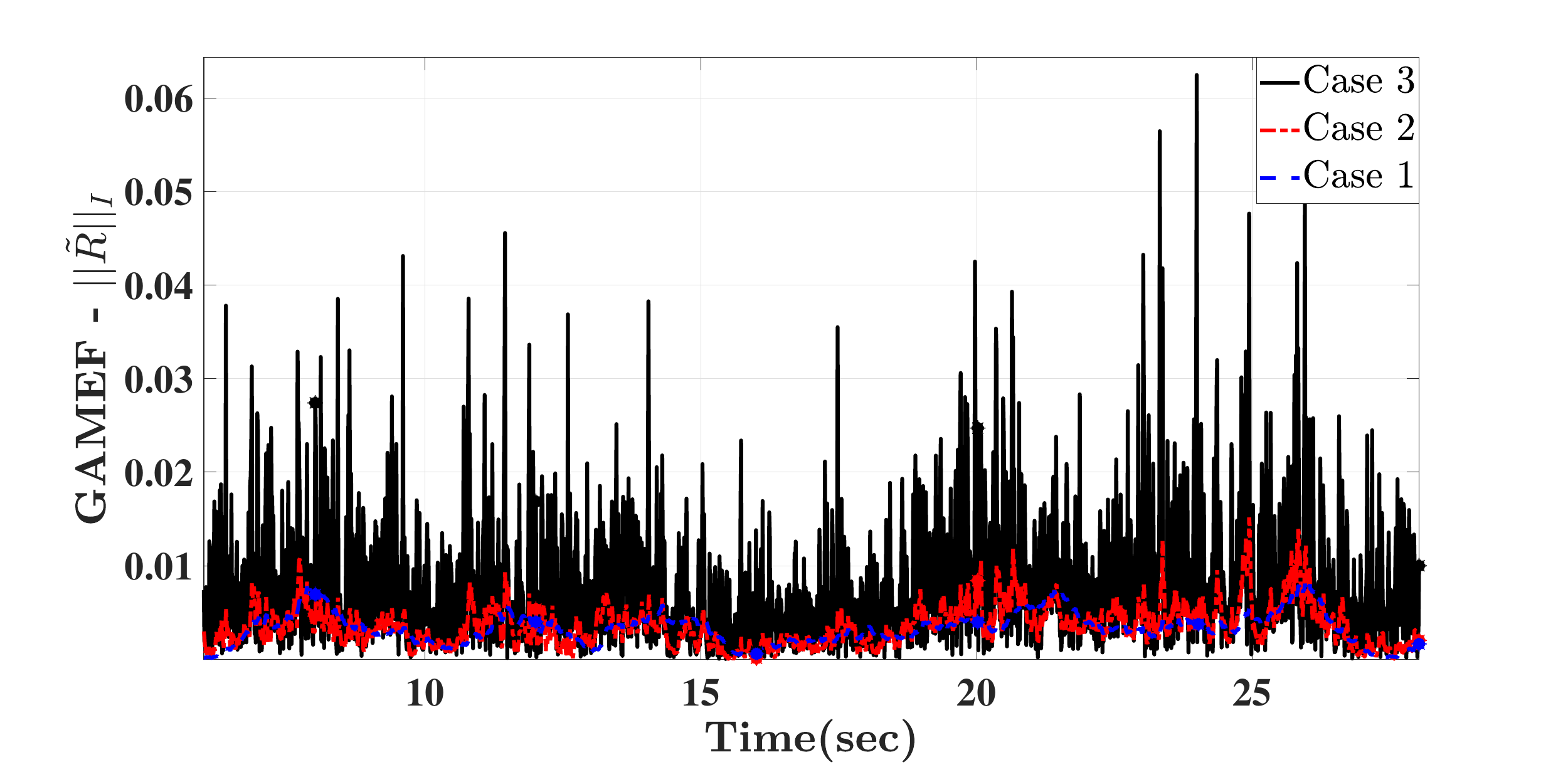}
	\begin{centering}
		\includegraphics[scale=0.3]{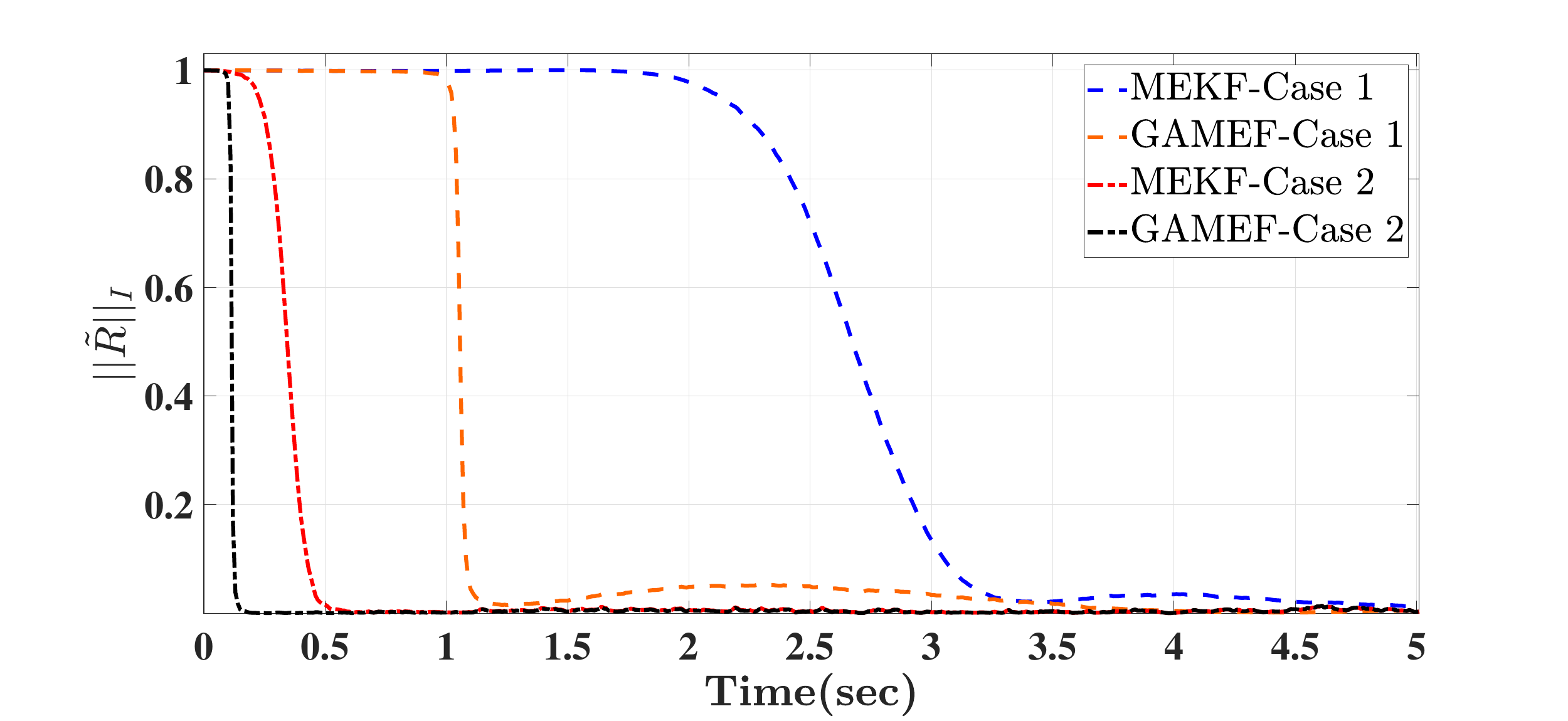}
		\par\end{centering}
	\centering{}\includegraphics[scale=0.3]{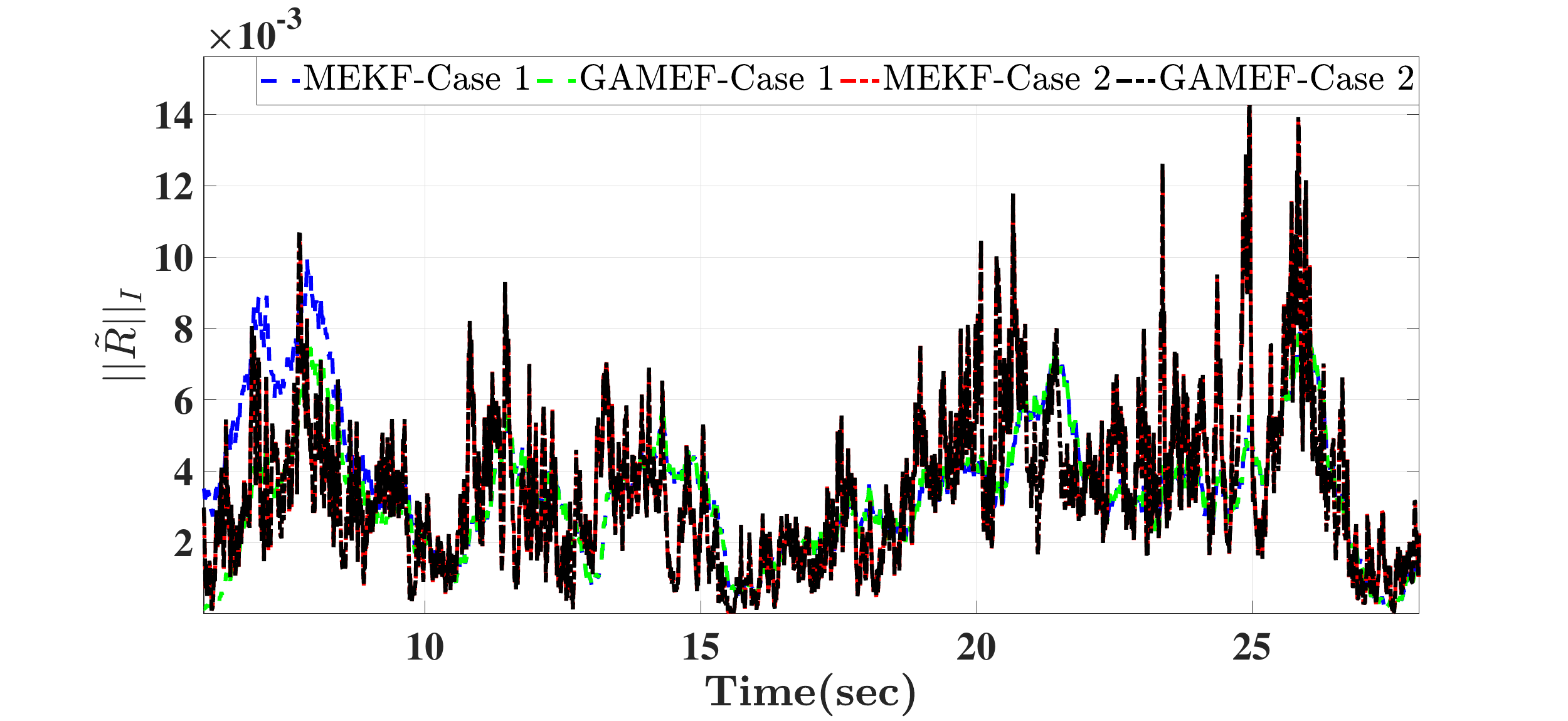}\caption{Tracking error ($||\tilde{R}||_{I}$) of Gaussian attitude filters:
		MEKF and GAMEF.}
	\label{fig:Comp_Simu_Gaus1} 
\end{figure}

\begin{figure}[h!]
	\includegraphics[scale=0.2]{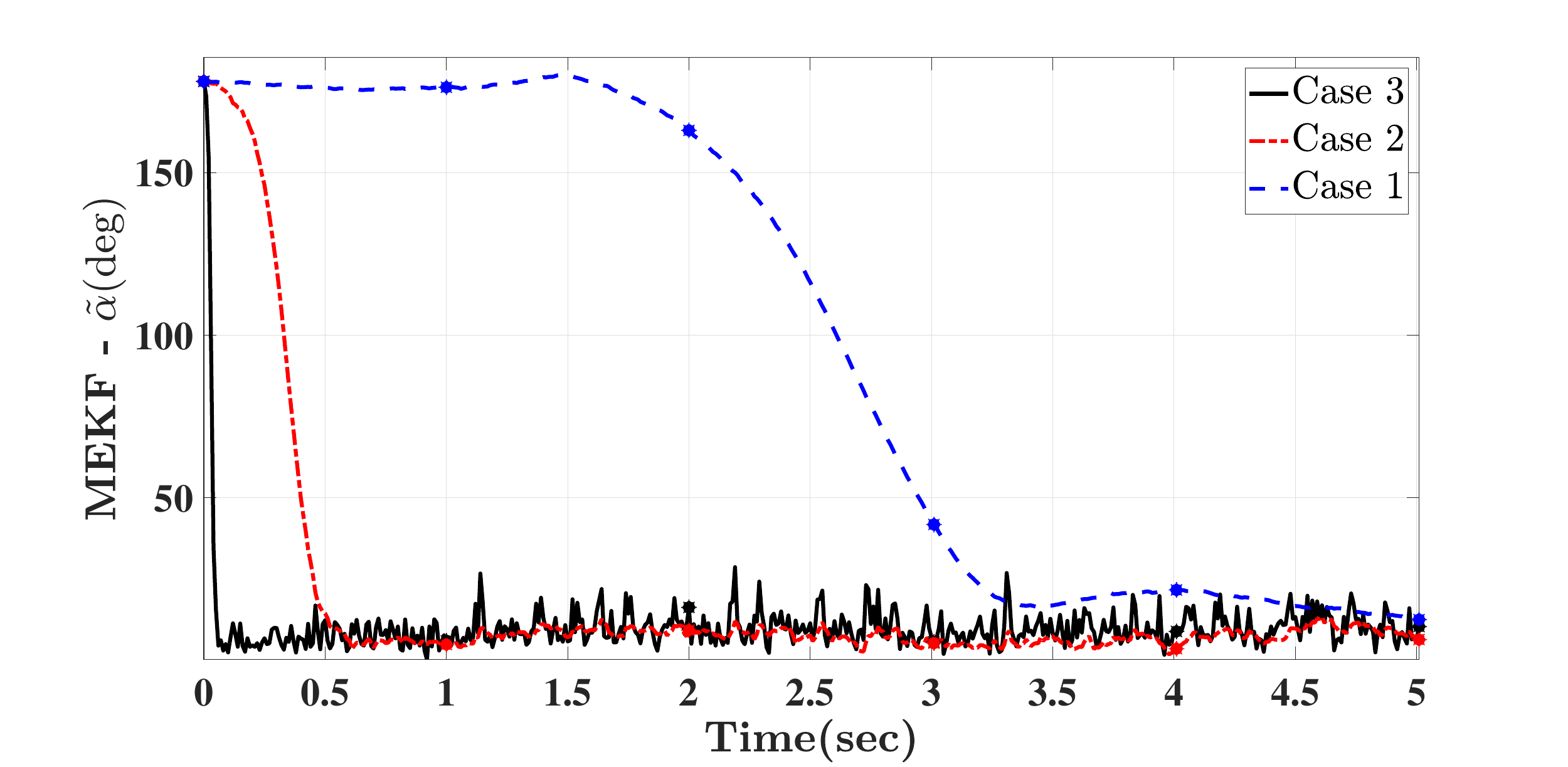}\includegraphics[scale=0.2]{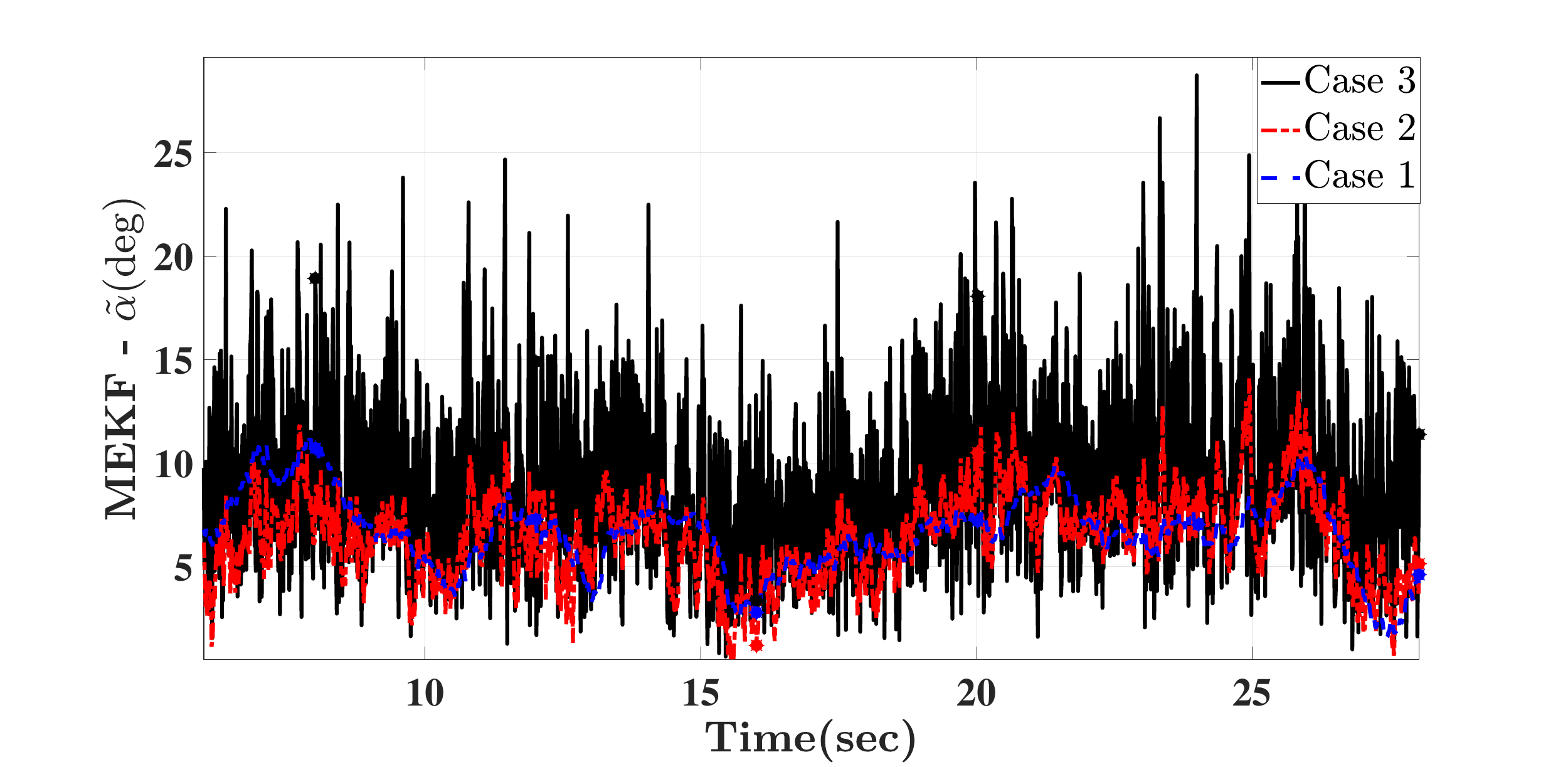}
	
	\includegraphics[scale=0.2]{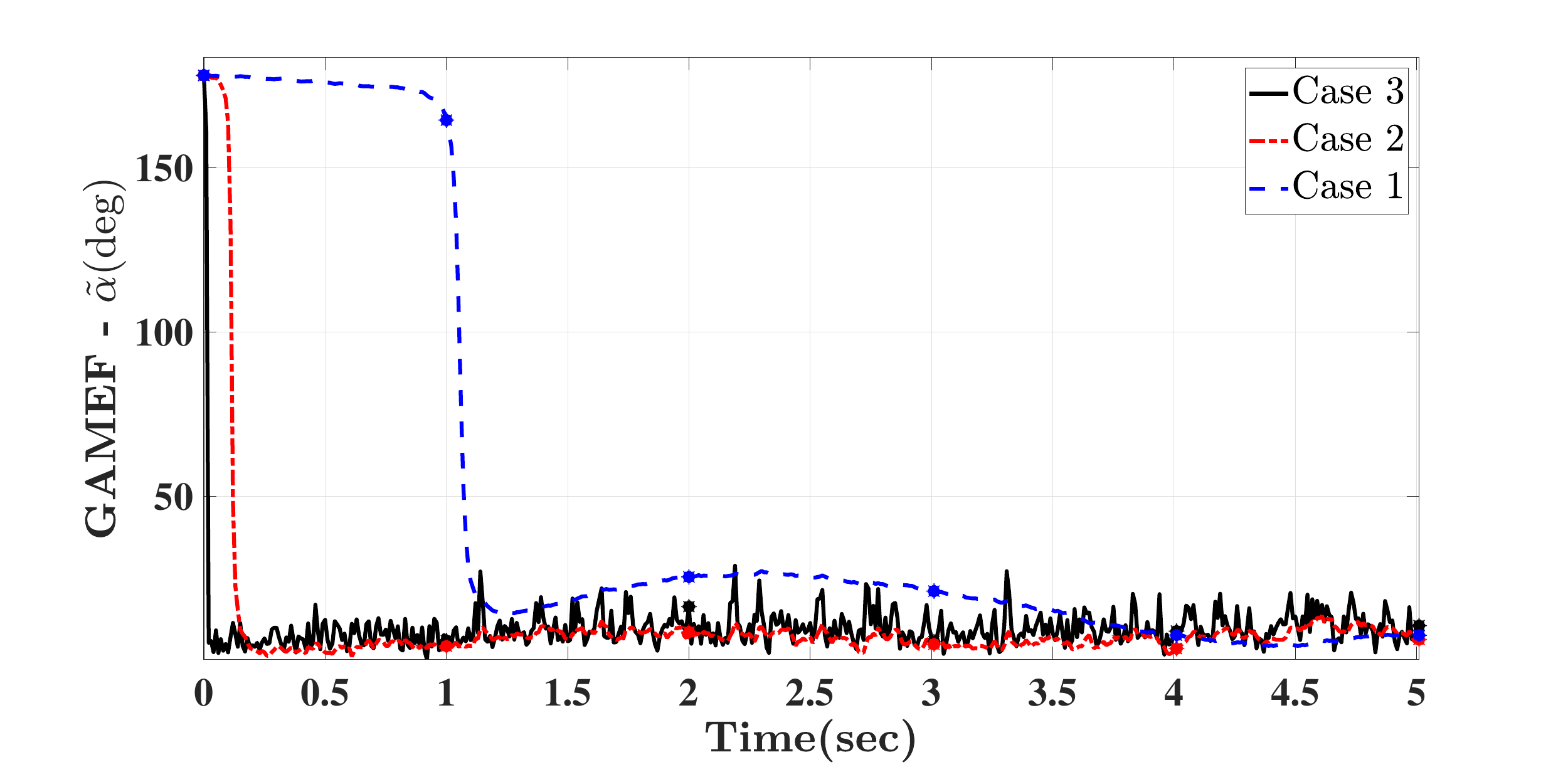}\includegraphics[scale=0.2]{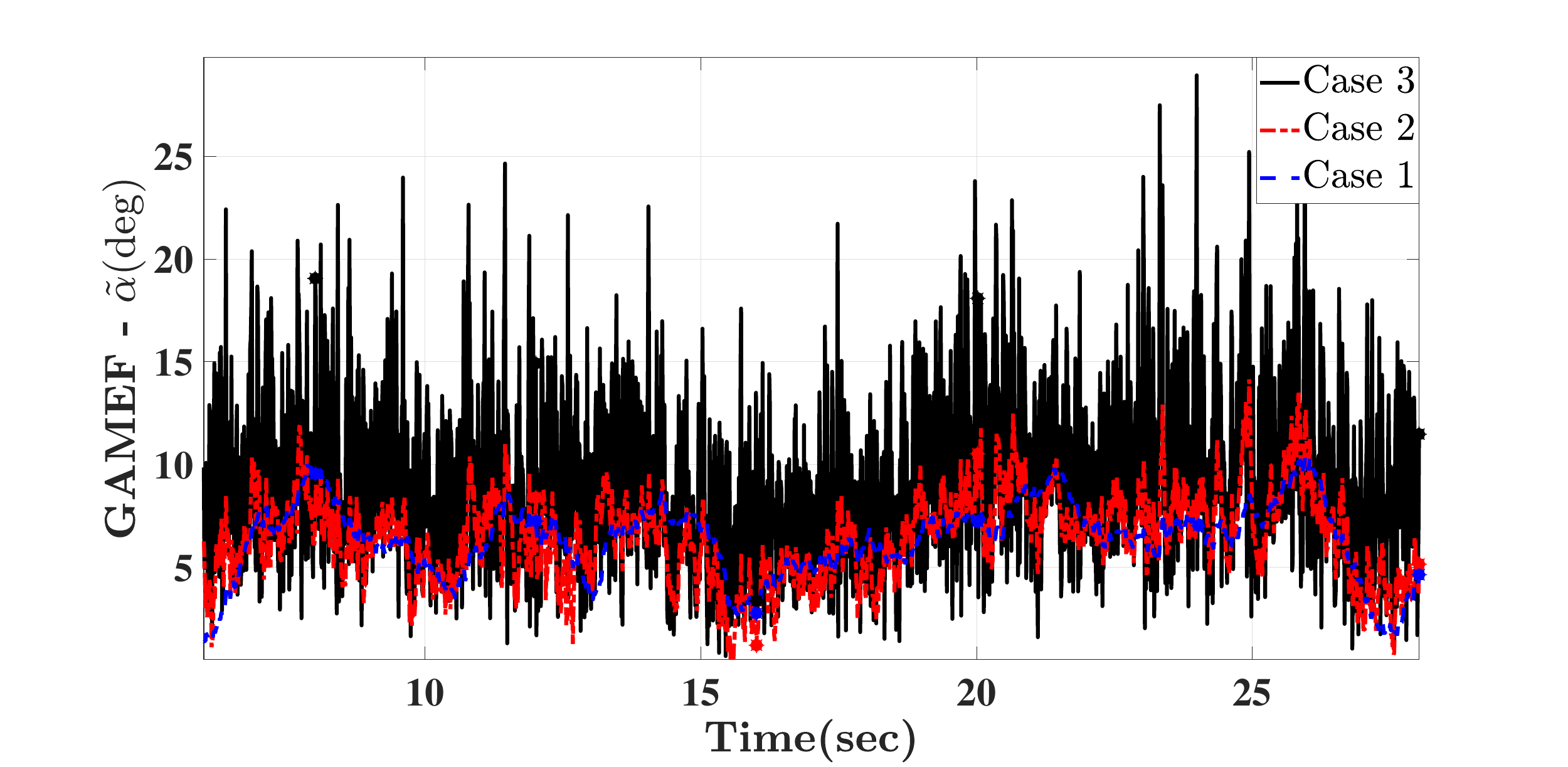}
	\begin{centering}
		\includegraphics[scale=0.3]{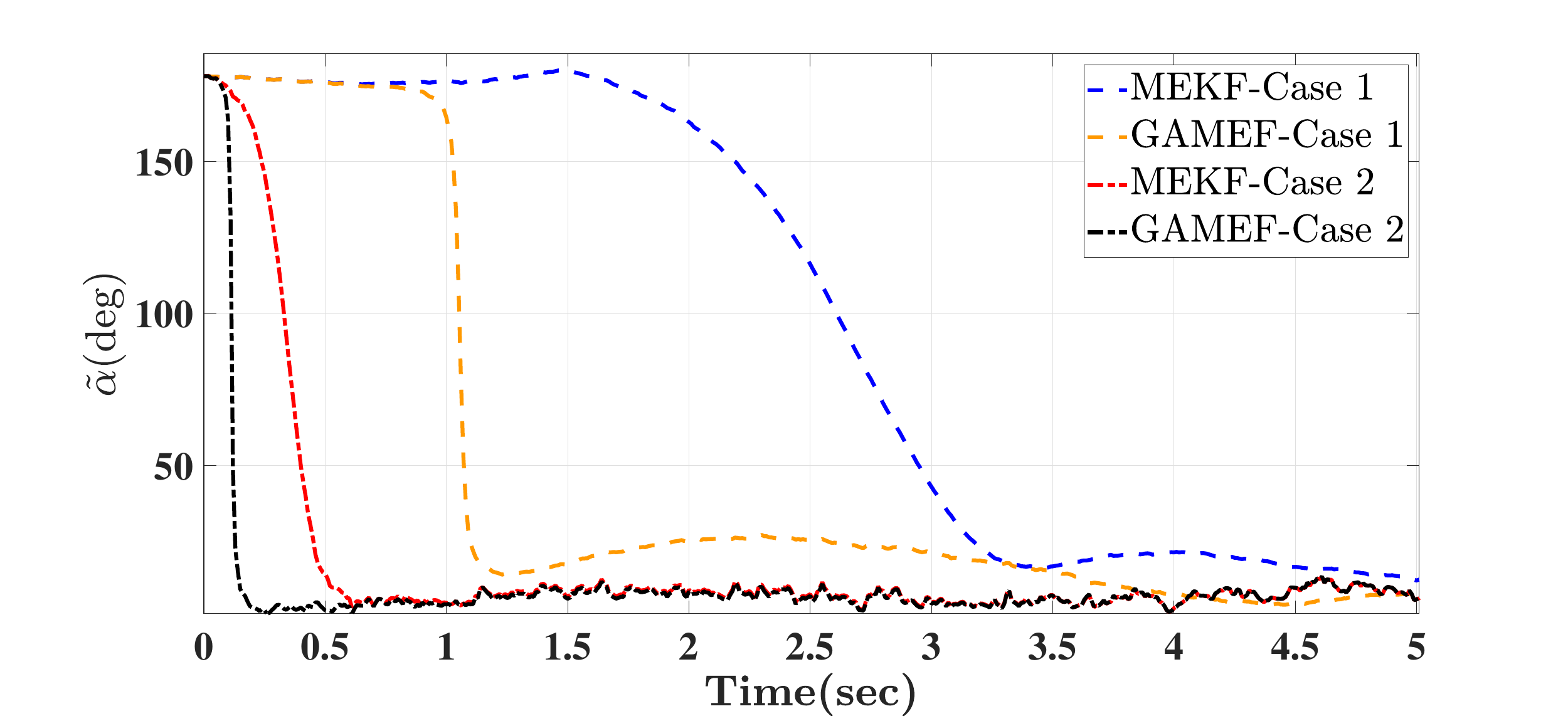}
		\par\end{centering}
	\centering{}\includegraphics[scale=0.3]{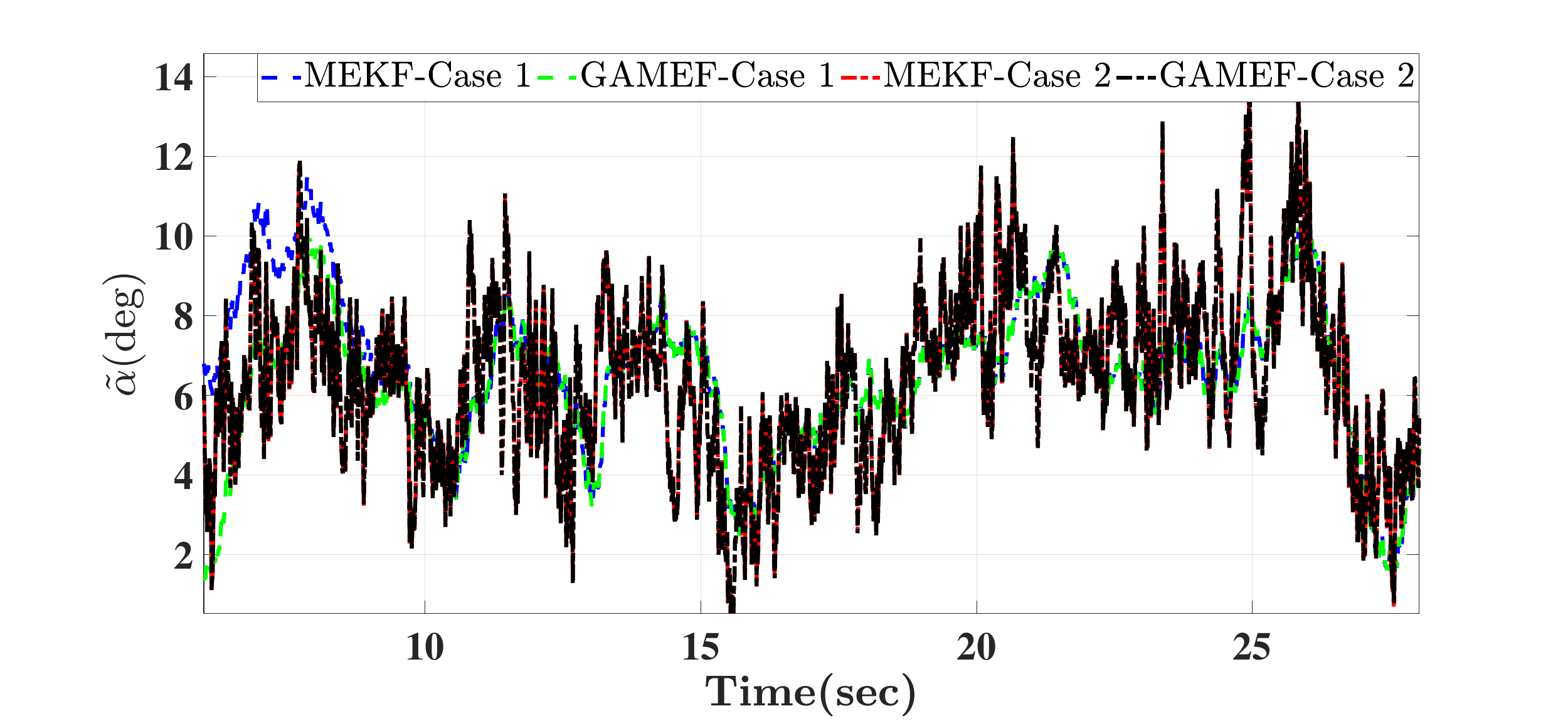}\caption{Tracking error ($\tilde{\alpha}$) of Gaussian attitude filters: MEKF
		and GAMEF.}
	\label{fig:Comp_Simu_Gaus2} 
\end{figure}

\begin{figure}[h!]
	\includegraphics[scale=0.2]{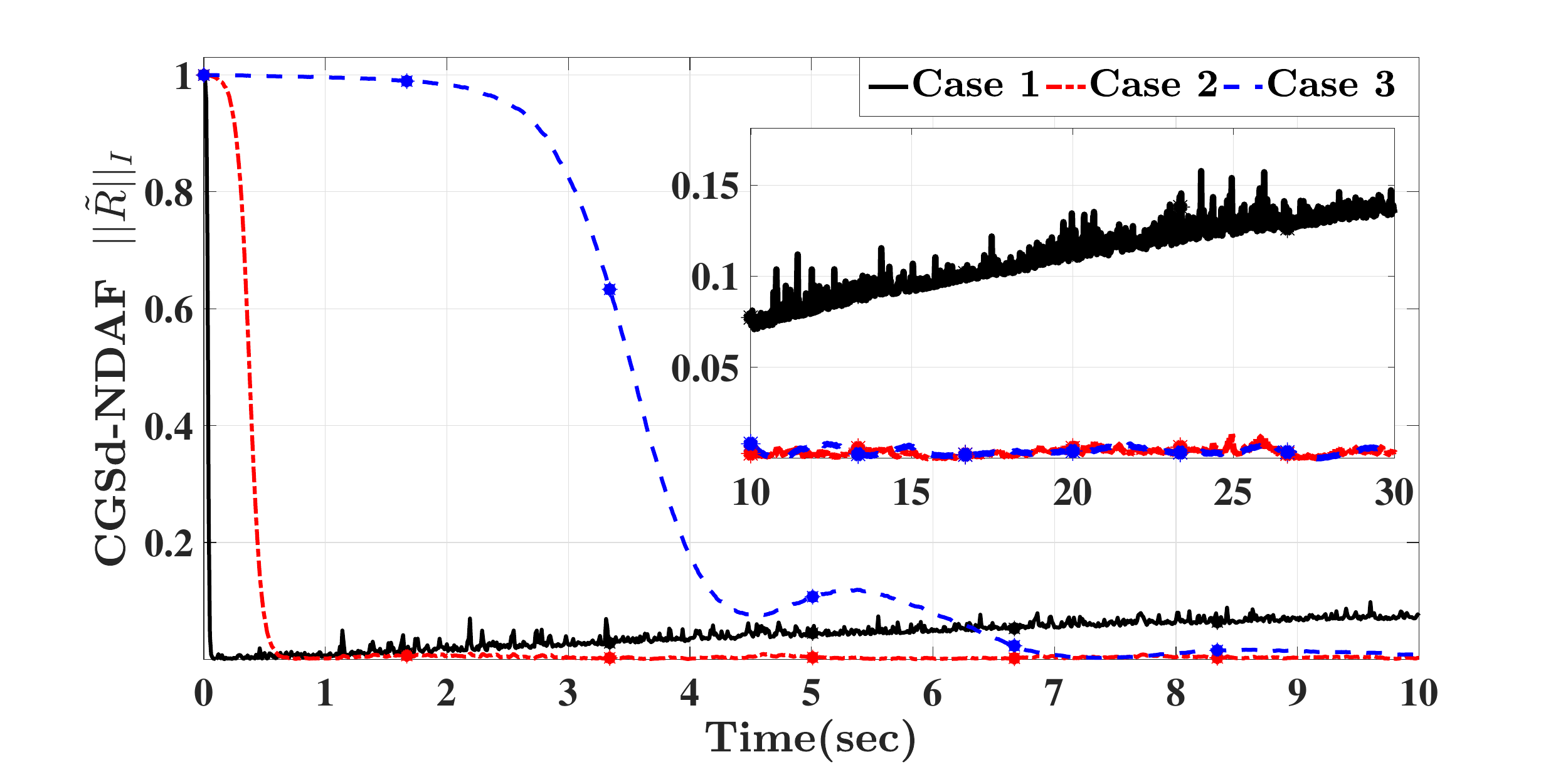}\includegraphics[scale=0.2]{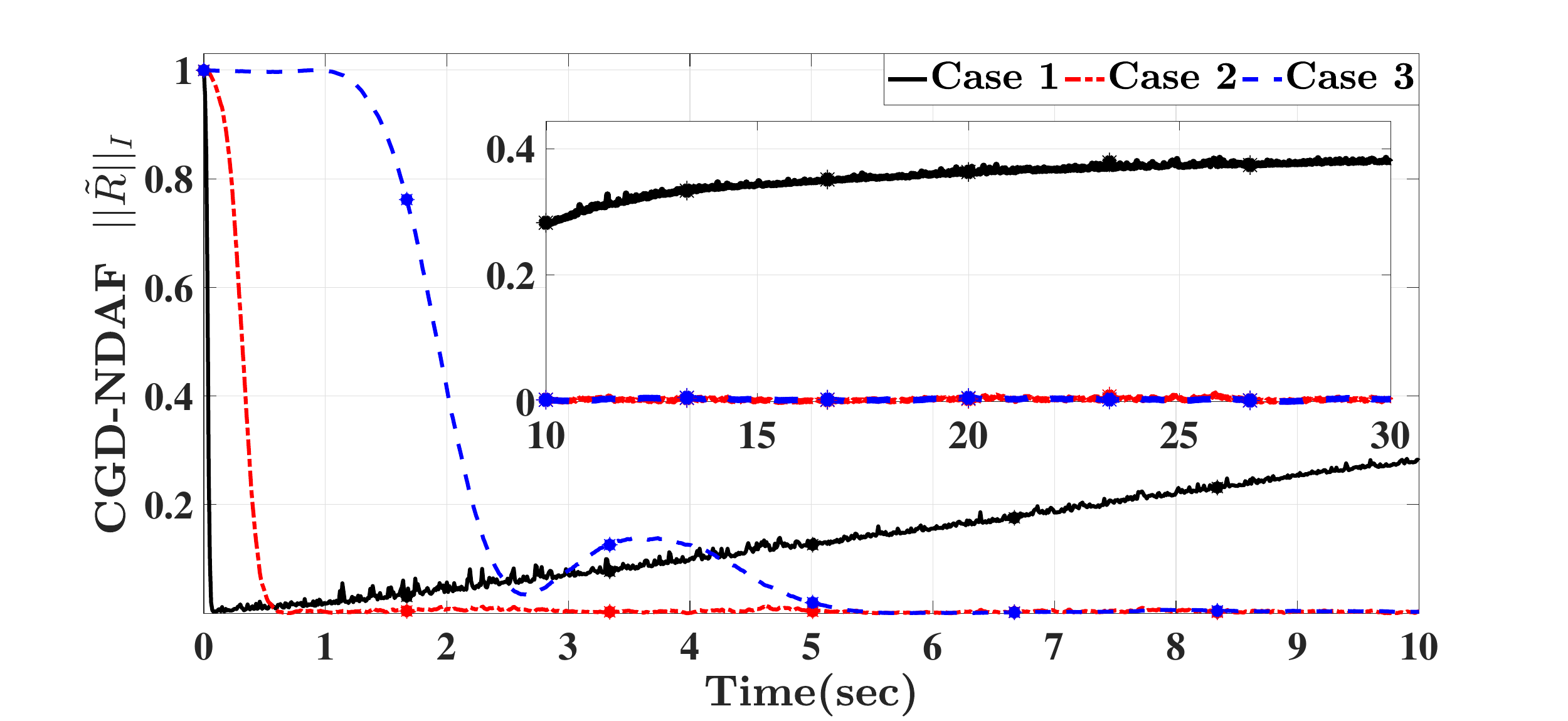}
	
	\includegraphics[scale=0.2]{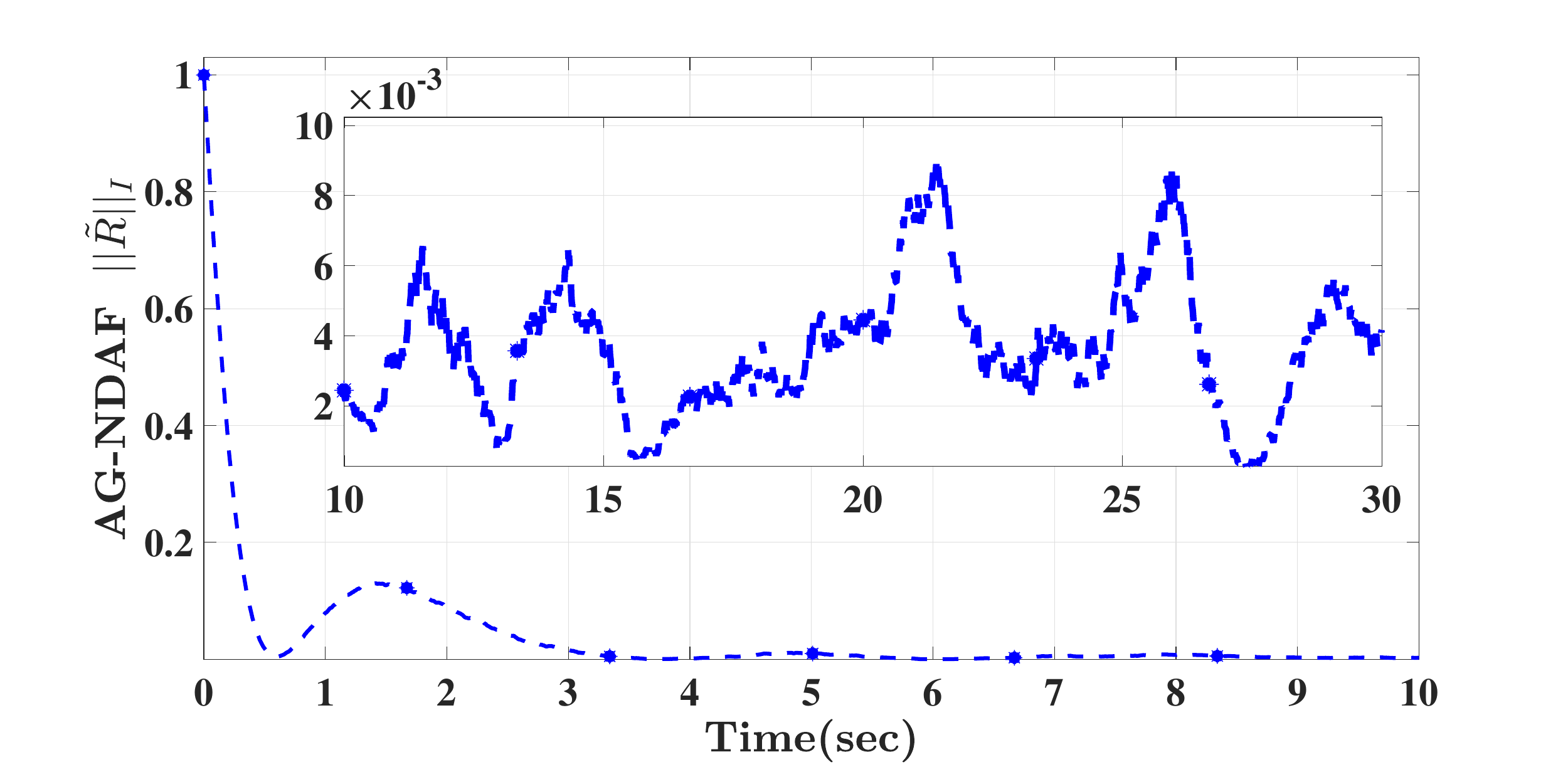}\includegraphics[scale=0.2]{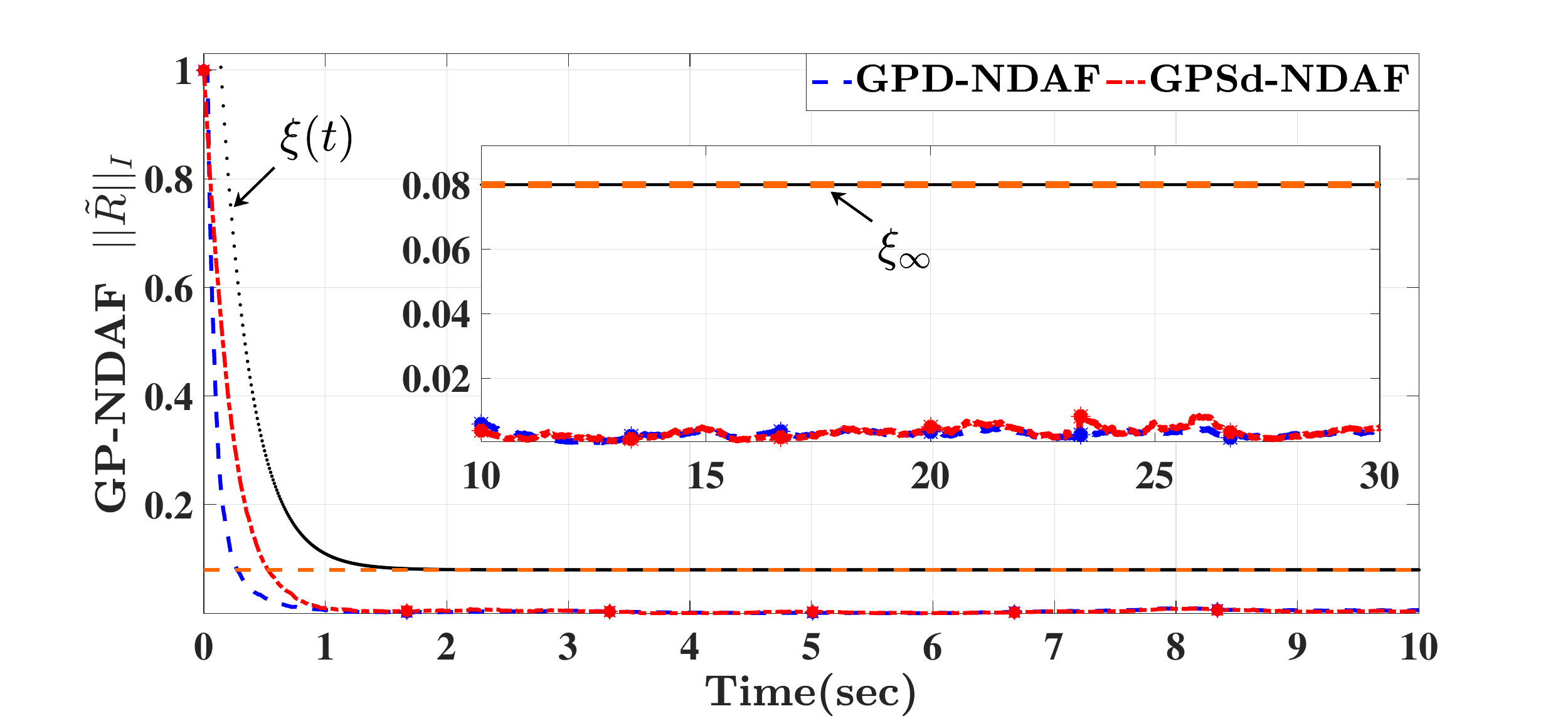}
	
	\includegraphics[scale=0.2]{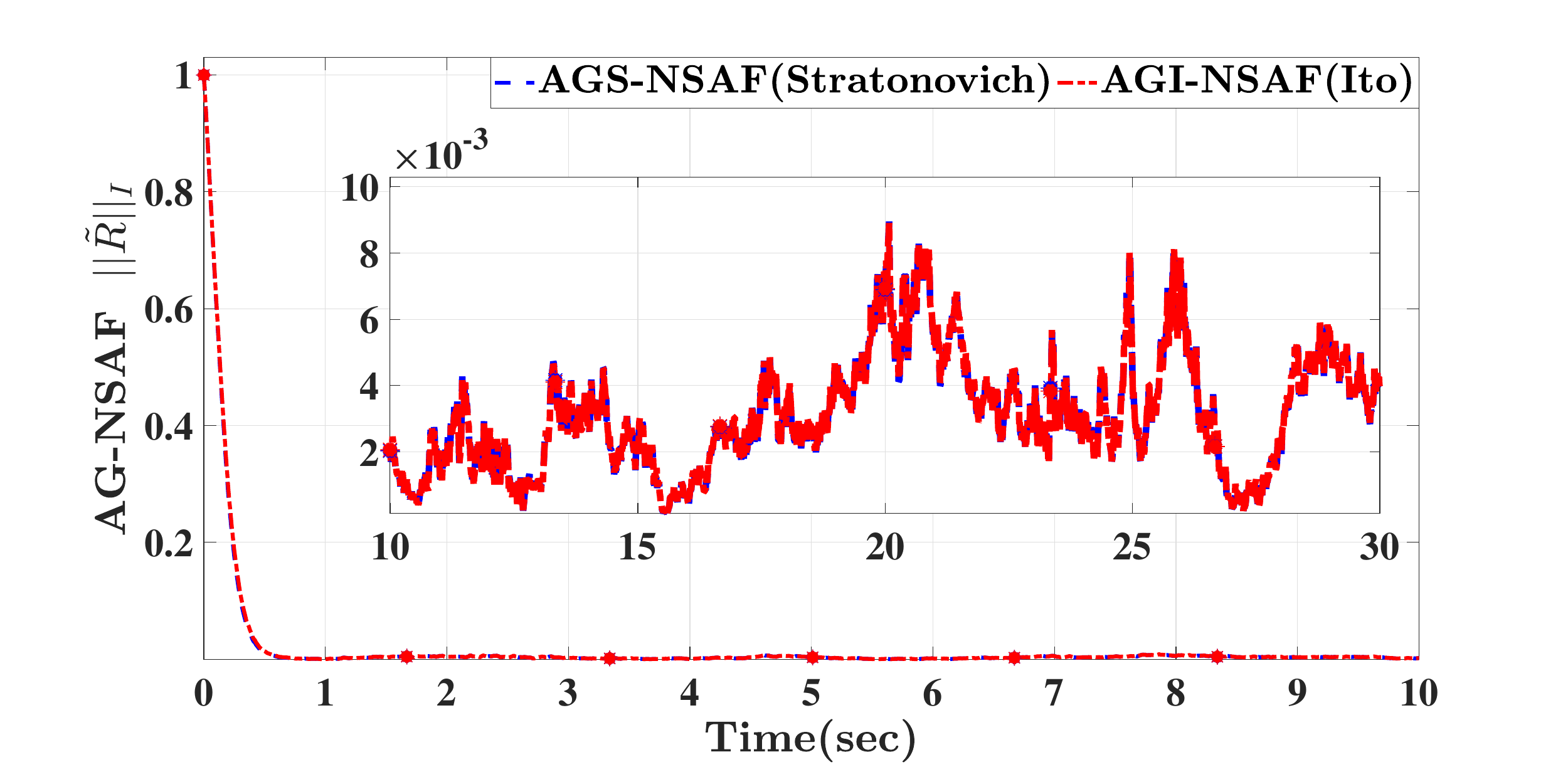}\includegraphics[scale=0.2]{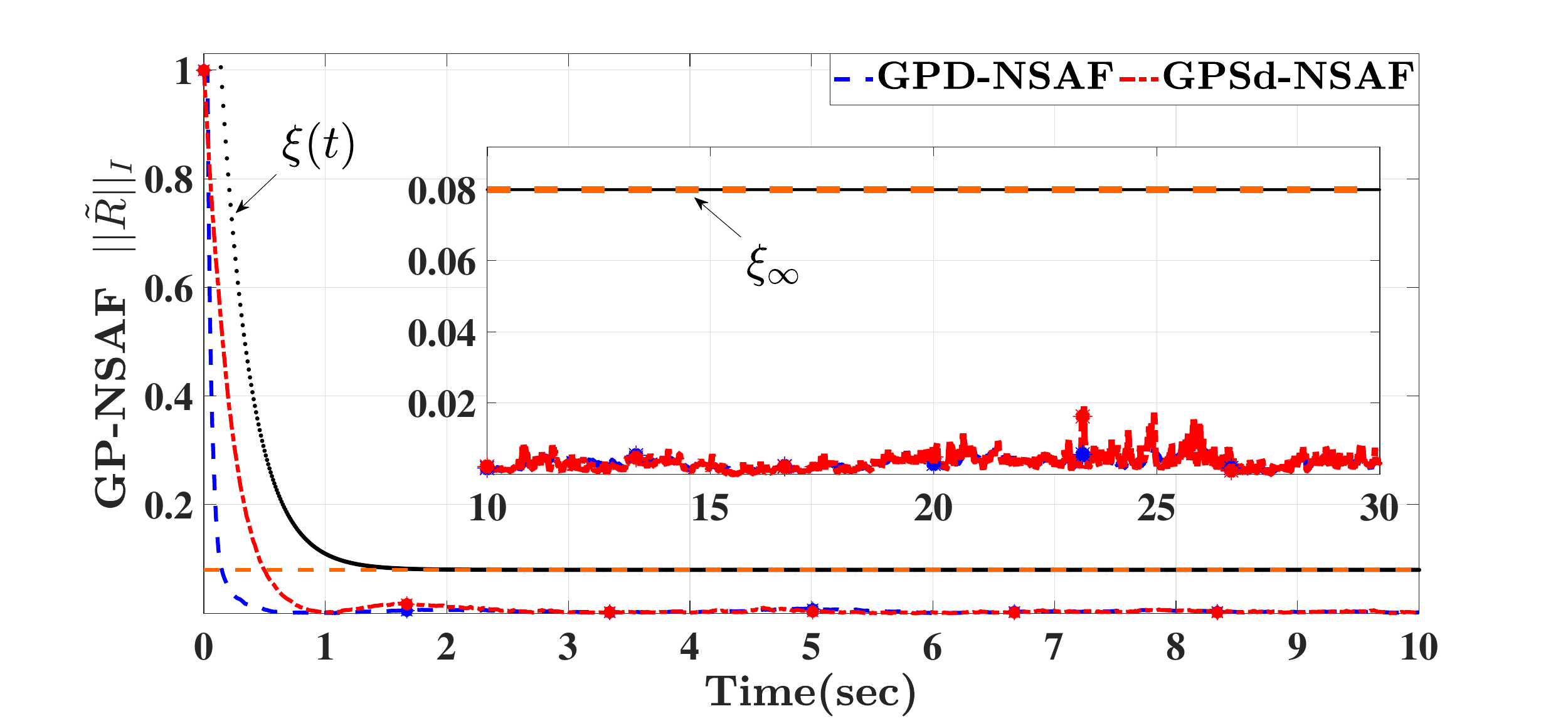}
	\centering{}\includegraphics[scale=0.4]{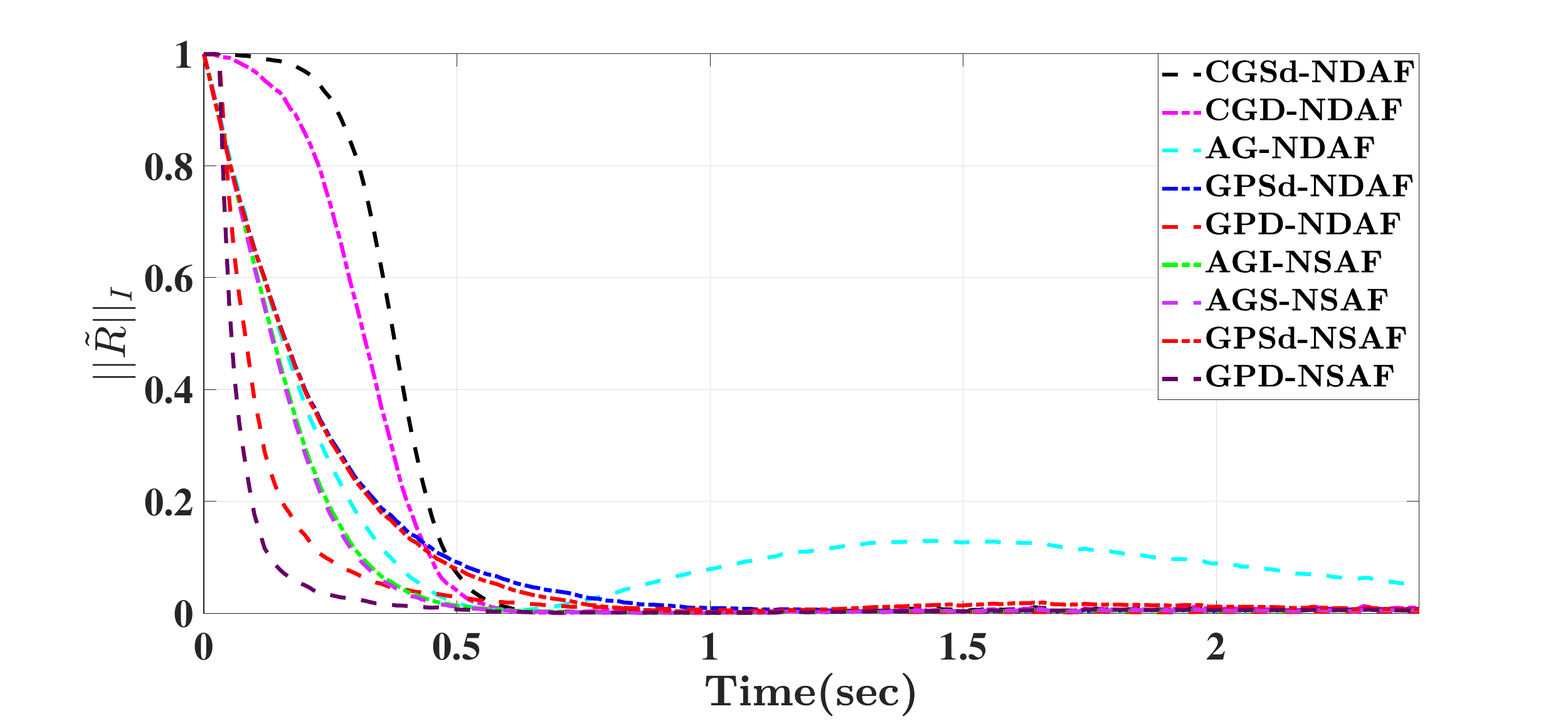}\caption{Tracking error ($||\tilde{R}||_{I}$) of nonlinear attitude filters.}
	\label{fig:Comp_Simu_Non1} 
\end{figure}

\begin{figure}[h!]
	\includegraphics[scale=0.2]{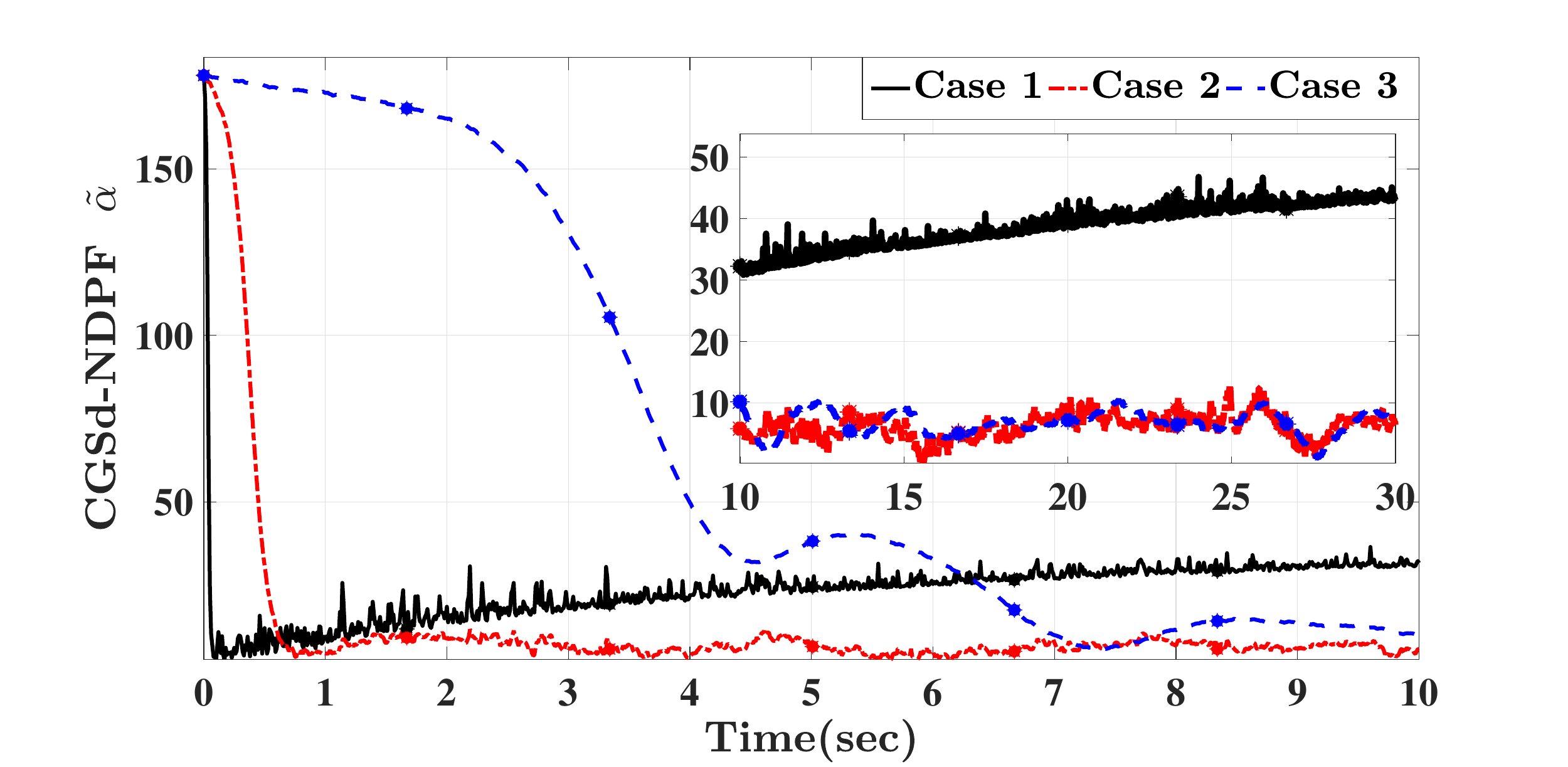}\includegraphics[scale=0.2]{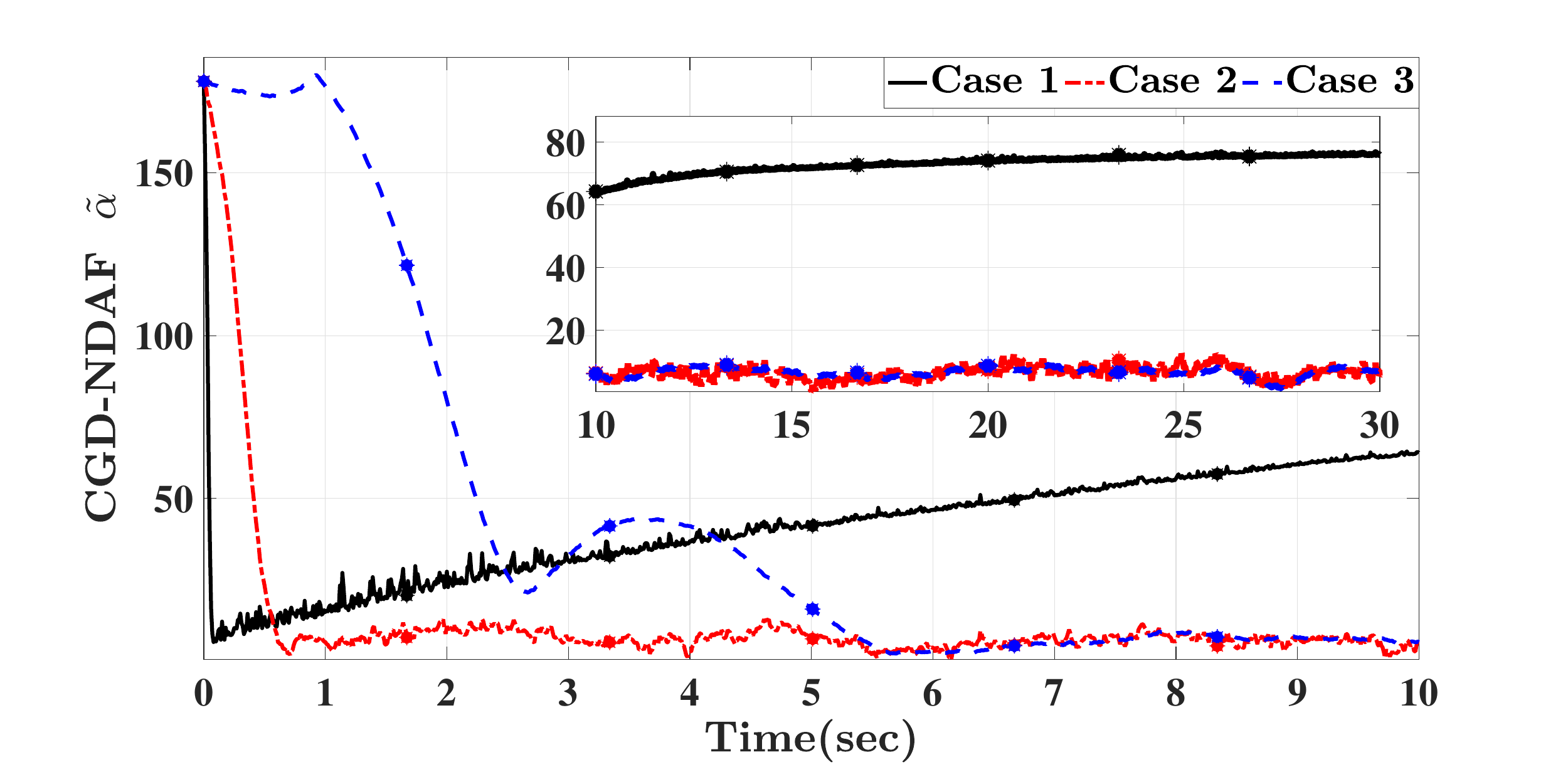}
	
	\includegraphics[scale=0.2]{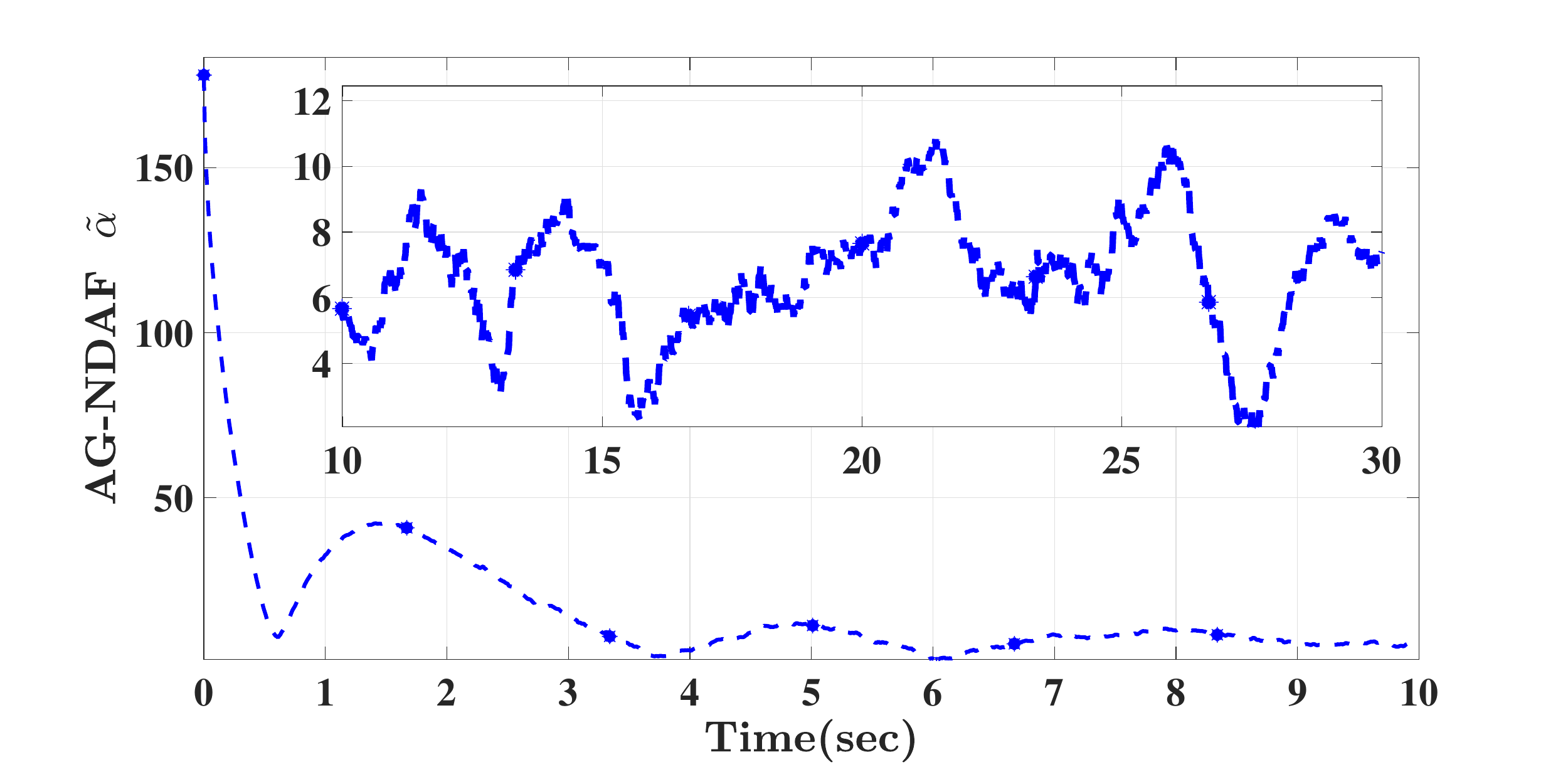}\includegraphics[scale=0.2]{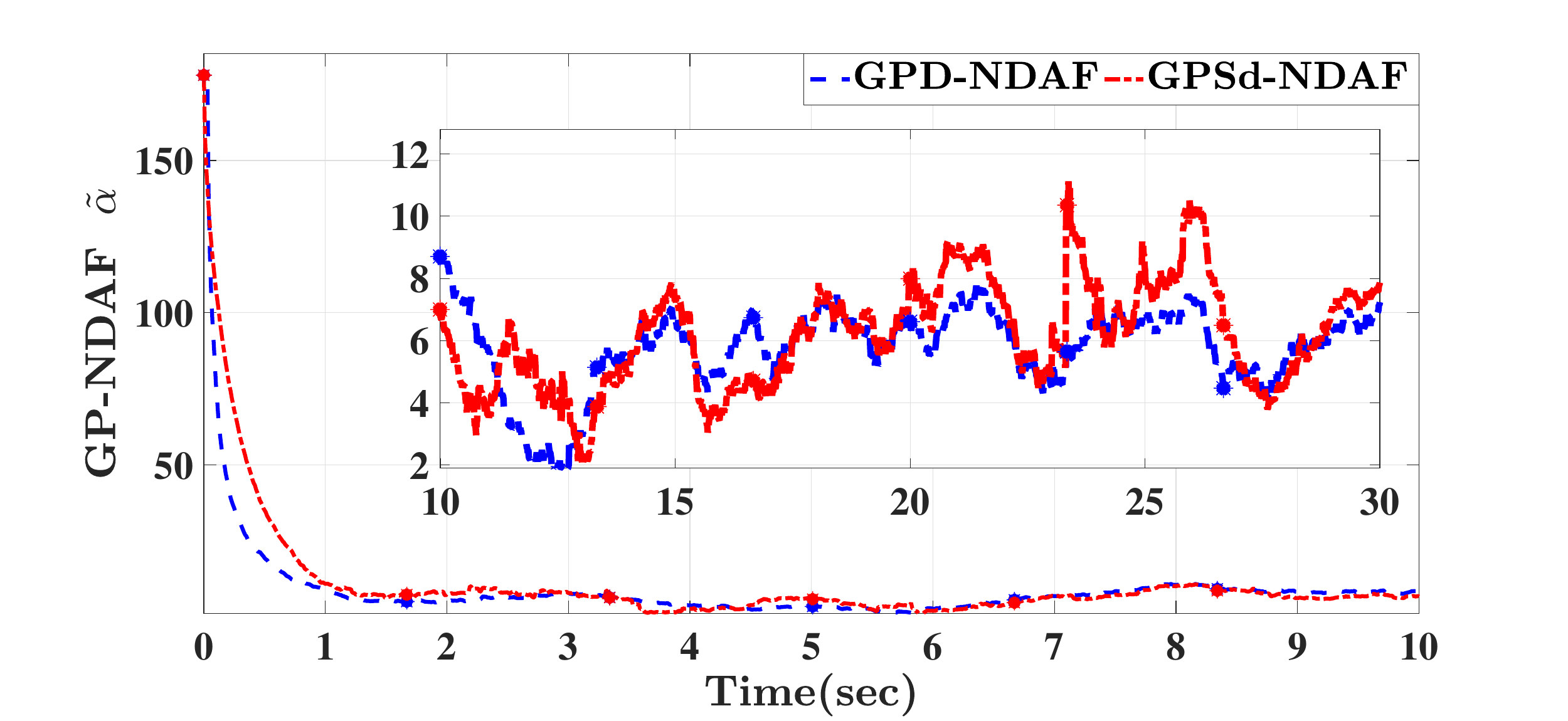}
	
	\includegraphics[scale=0.2]{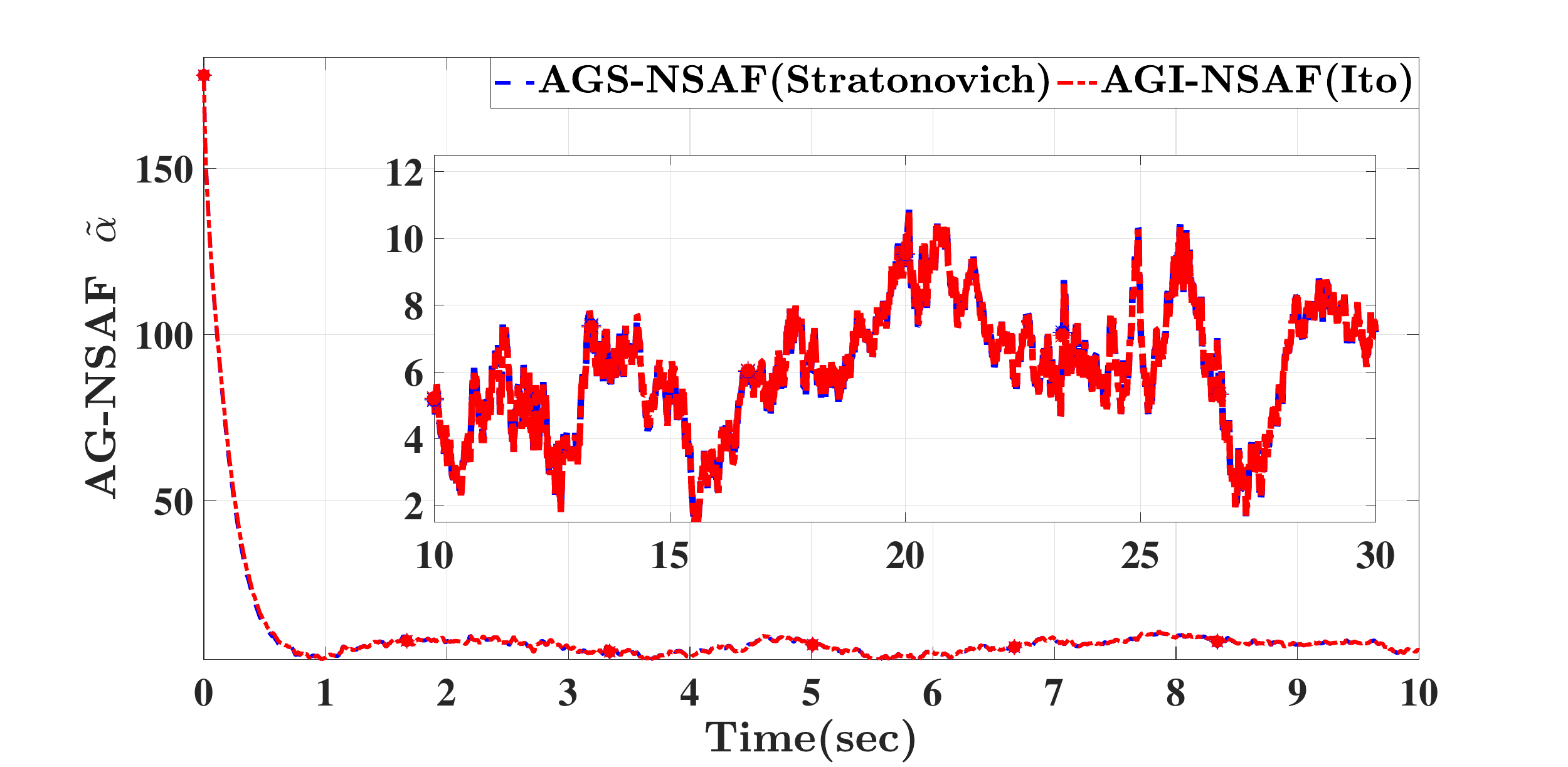}\includegraphics[scale=0.2]{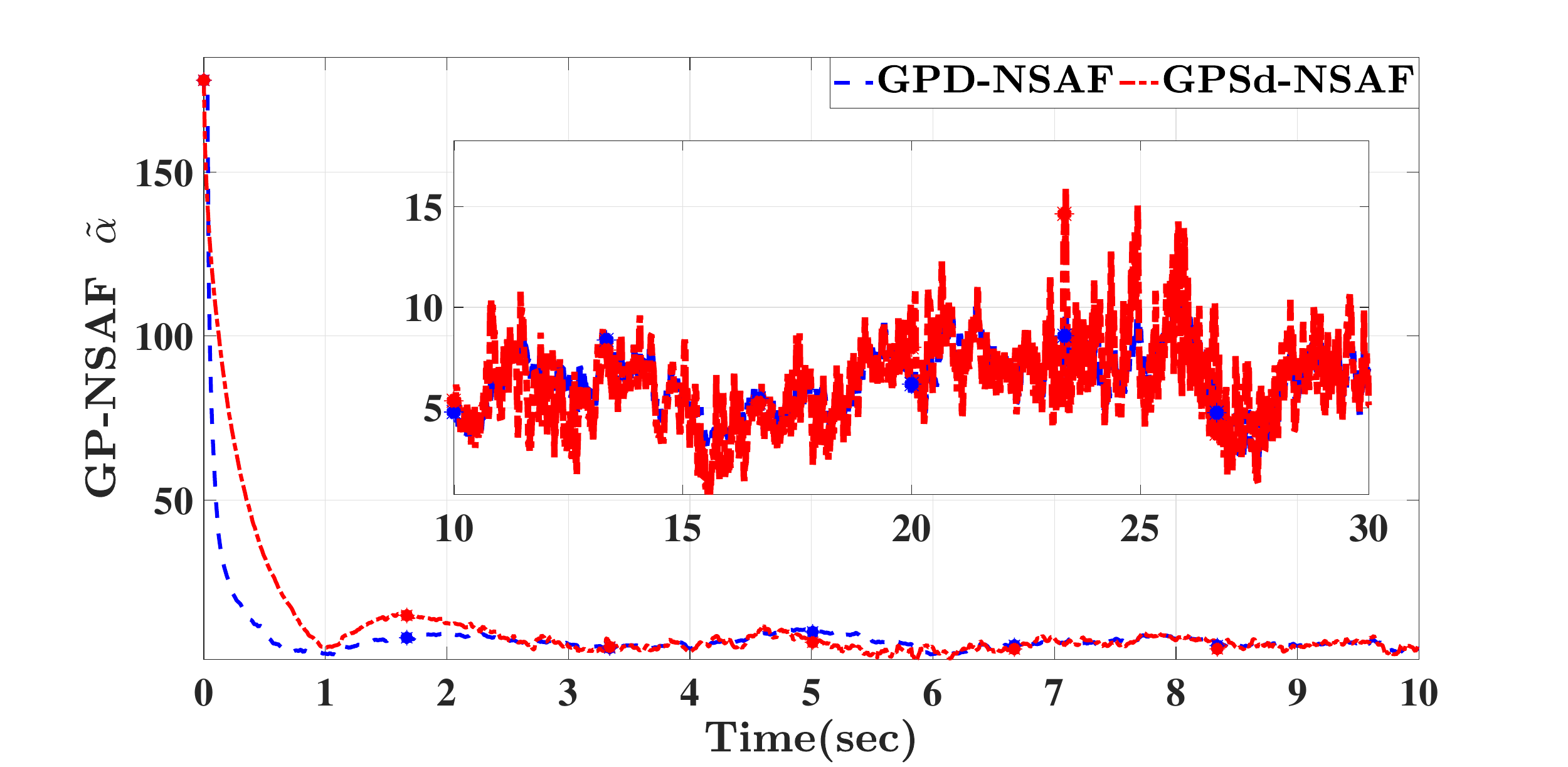}
	\centering{}\includegraphics[scale=0.4]{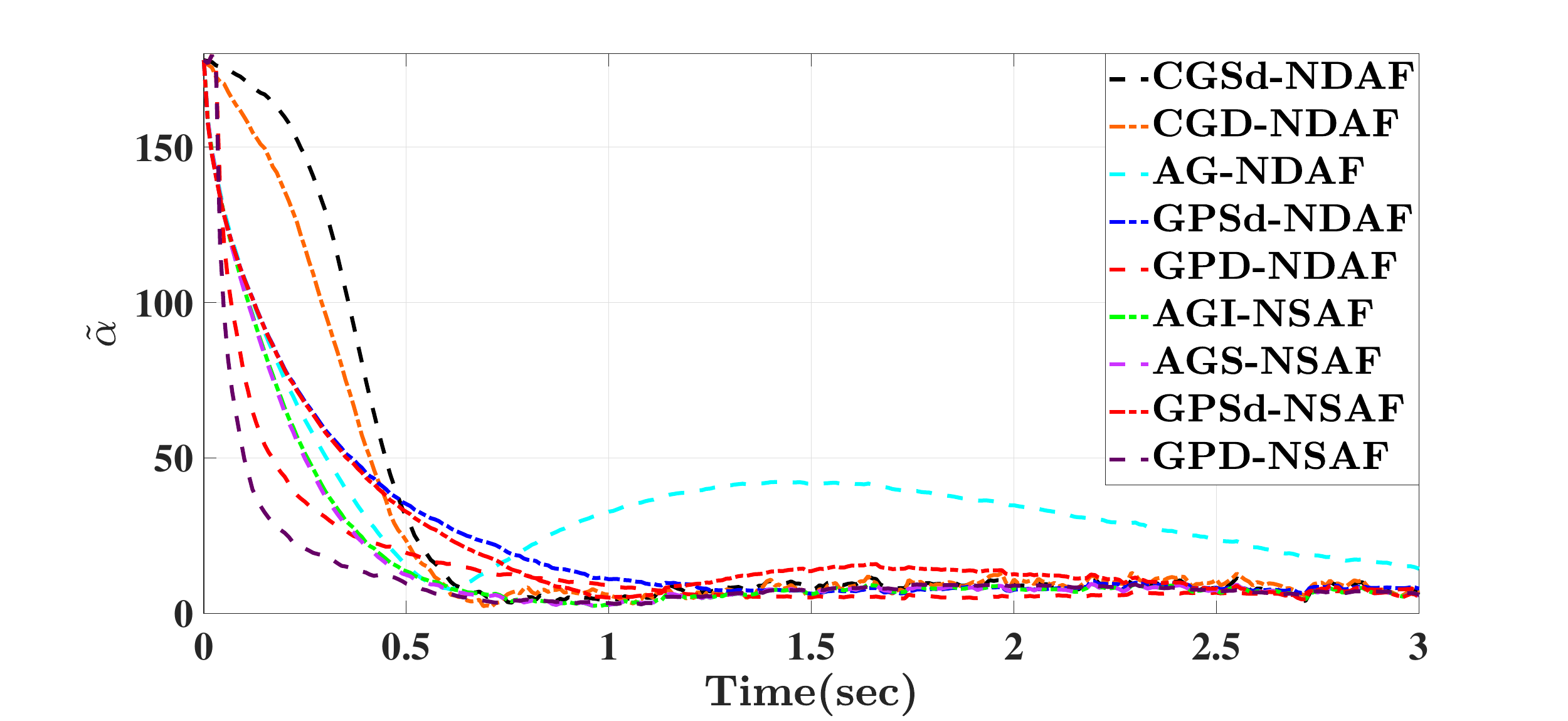}\caption{Tracking error ($\tilde{\alpha}$) of nonlinear attitude filters.}
	\label{fig:Comp_Simu_Non2} 
\end{figure}

Figure \ref{fig:Comp_Simu_Gaus1} and \ref{fig:Comp_Simu_Gaus2} demonstrate
the superiority of Gaussian attitude filters over the determination
algorithms in terms of tracking performance. It can be noticed that
the design parameters in Case 1 and Case 2 of MEKF and GAMEF provide
slower tracking performance with less oscillatory behavior in the
steady-state. In contrast, Case 3 of MEKF and GAMEF offers faster
tracking performance with higher oscillation in the steady-state.
This can be confirmed through the statistical results listed in Table
\ref{tab:Comp_Simu_Gauss}. However, MEKF requires less computational
power in comparison with GAMEF. Figure \ref{fig:Comp_Simu_Non1} and
\ref{fig:Comp_Simu_Non2} illustrate faster tracking performance of
CG-NDAF (Case 3), AG-NDAF, GP-NDAF, AG-NSAF and GP-NSAF, in comparison
with CG-NDAF (Case 1) and CG-NDAF (Case 2). Despite fast tracking
performance, the main weakness of CG-NDAF (Case 3) shows unstable
behavior. Also, CG-NDAF (Case 1), CG-NDAF (Case 2), AG-NDAF and AG-NSAF
cannot demonstrate guaranteed measures of transient and steady-state
error. It becomes apparent that the only two filters that have the
advantage of guaranteed performance of transient and steady-state
error are GP-NDAF and GP-NSAF. The side-by-side statistical comparison
of the nonlinear attitude filters in Figure \ref{fig:Comp_Simu_Non1}
and \ref{fig:Comp_Simu_Non2} can be found in Table \ref{tab:Comp_Simu_Non}.

\begin{table}[h!]
	\caption{\textcolor{blue}{\label{tab:Comp_Simu_Gauss} }Statistical analysis
		of Gaussian attitude filters (continuous time) of $||\tilde{R}||_{I}$
		and $\tilde{\alpha}$ steady-state performance: MEKF and GAMEF.}
	
	\centering{}{\footnotesize{}}%
	\begin{tabular}{l|c|c|c|c|c|c|c|c}
		\hline 
		\multicolumn{9}{c}{{\footnotesize{}Output data (Mean, STD and $\left\Vert \,\cdot\,\right\Vert _{\infty}$)
				of $||\tilde{R}||_{I}$ and $\tilde{\alpha}$ over the period (8-30
				sec)}}\tabularnewline
		\hline 
		\hline 
		\textbf{\footnotesize{}Filter} & {\footnotesize{}Mean ($||\tilde{R}||_{I}$)} & {\footnotesize{}STD ($||\tilde{R}||_{I}$)} & {\footnotesize{}$\left\Vert ||\tilde{R}||_{I}\right\Vert _{\infty}$} & {\footnotesize{}Mean ($\tilde{\alpha}$)} & {\footnotesize{}STD ($\tilde{\alpha}$)} & {\footnotesize{}$\left\Vert \tilde{\alpha}\right\Vert _{\infty}$} & {\footnotesize{}Transient} & {\footnotesize{}Overall}\tabularnewline
		\hline 
		{\footnotesize{}MEKF (Case1)} & {\footnotesize{}$0.0034$} & {\footnotesize{}$0.0016$} & {\footnotesize{}$0.0089$} & {\footnotesize{}$6.4955$} & {\footnotesize{}$1.6980$} & {\footnotesize{}$10.8532$} & {\footnotesize{}Very slow} & {\footnotesize{}Stable}\tabularnewline
		\hline 
		{\footnotesize{}MEKF (Case2)} & {\footnotesize{}$0.0035$} & {\footnotesize{}$0.0021$} & {\footnotesize{}$0.0150$} & {\footnotesize{}$6.4816$} & {\footnotesize{}$2.0886$} & {\footnotesize{}$14.0934$} & {\footnotesize{}Slow} & {\footnotesize{}Stable}\tabularnewline
		\hline 
		{\footnotesize{}MEKF (Case3)} & {\footnotesize{}$0.0075$} & {\footnotesize{}$0.0067$} & {\footnotesize{}$0.0616$} & {\footnotesize{}$9.0868$} & {\footnotesize{}$4.0262$} & {\footnotesize{}$28.7459$} & {\footnotesize{}Fast} & {\footnotesize{}Stable}\tabularnewline
		\hline 
		{\footnotesize{}GAMEF (Case1)} & {\footnotesize{}$0.0034$} & {\footnotesize{}$0.0016$} & {\footnotesize{}$0.0081$} & {\footnotesize{}$6.4402$} & {\footnotesize{}$1.6575$} & {\footnotesize{}$10.3092$} & {\footnotesize{}Very slow} & {\footnotesize{}Stable}\tabularnewline
		\hline 
		{\footnotesize{}GAMEF (Case2)} & {\footnotesize{}$0.0035$} & {\footnotesize{}$0.0021$} & {\footnotesize{}$0.0152$} & {\footnotesize{}$6.4843$} & {\footnotesize{}$2.0912$} & {\footnotesize{}$14.1452$} & {\footnotesize{}Slow} & {\footnotesize{}Stable}\tabularnewline
		\hline 
		{\footnotesize{}GAMEF (Case3)} & {\footnotesize{}$0.0075$} & {\footnotesize{}$0.0068$} & {\footnotesize{}$0.0624$} & {\footnotesize{}$9.1070$} & {\footnotesize{}$4.0463$} & {\footnotesize{}$28.9409$} & {\footnotesize{}Fast} & {\footnotesize{}Stable}\tabularnewline
		\hline 
	\end{tabular}{\footnotesize\par}
\end{table}

\newpage

\begin{table}[h!]
	\caption{\textcolor{blue}{\label{tab:Comp_Simu_Non} }Statistical analysis
		of nonlinear attitude filters (continuous time) of $||\tilde{R}||_{I}$
		and $\tilde{\alpha}$ steady-state performance.}
	
	\centering{}{\footnotesize{}}%
	\begin{tabular}{l|c|c|c|c|c|c|c|c}
		\hline 
		\multicolumn{9}{c}{{\footnotesize{}Output data (Mean, STD and $\left\Vert \,\cdot\,\right\Vert _{\infty}$)
				of $||\tilde{R}||_{I}$ and $\tilde{\alpha}$ over the period (8-30
				sec)}}\tabularnewline
		\hline 
		\hline 
		\textbf{\footnotesize{}Filter} & {\footnotesize{}Mean ($||\tilde{R}||_{I}$)} & {\footnotesize{}STD ($||\tilde{R}||_{I}$)} & {\footnotesize{}$\left\Vert ||\tilde{R}||_{I}\right\Vert _{\infty}$} & {\footnotesize{}Mean ($\tilde{\alpha}$)} & {\footnotesize{}STD ($\tilde{\alpha}$)} & {\footnotesize{}$\left\Vert \tilde{\alpha}\right\Vert _{\infty}$} & {\footnotesize{}Transient} & {\footnotesize{}Overall}\tabularnewline
		\hline 
		{\footnotesize{}CGSd-NDAF (Case1)} & {\footnotesize{}$0.0046$} & {\footnotesize{}$0.0033$} & {\footnotesize{}$0.0170$} & {\footnotesize{}$7.3376$} & {\footnotesize{}$2.6319$} & {\footnotesize{}$14.9764$} & {\footnotesize{}Very slow} & {\footnotesize{}Stable}\tabularnewline
		\hline 
		{\footnotesize{}CGSd-NDAF (Case2)} & {\footnotesize{}$0.0033$} & {\footnotesize{}$0.0019$} & {\footnotesize{}$0.0131$} & {\footnotesize{}$6.3162$} & {\footnotesize{}$1.9338$} & {\footnotesize{}$13.1318$} & {\footnotesize{}Slow} & {\footnotesize{}Stable}\tabularnewline
		\hline 
		{\footnotesize{}CGSd-NDAF (Case3)} & {\footnotesize{}$0.1076$} & {\footnotesize{}$0.0215$} & {\footnotesize{}$0.1576$} & {\footnotesize{}$38.1047$} & {\footnotesize{}$4.0626$} & {\footnotesize{}$46.7879$} & {\footnotesize{}Fast} & \textcolor{red}{\footnotesize{}Unstable}\tabularnewline
		\hline 
		{\footnotesize{}CGD-NDAF (Case1)} & {\footnotesize{}$0.0035$} & {\footnotesize{}$0.0013$} & {\footnotesize{}$0.0064$} & {\footnotesize{}$6.6584$} & {\footnotesize{}$1.4671$} & {\footnotesize{}$9.1931$} & {\footnotesize{}Very slow} & {\footnotesize{}Stable}\tabularnewline
		\hline 
		{\footnotesize{}CGD-NDAF (Case2)} & {\footnotesize{}$0.0035$} & {\footnotesize{}$0.0021$} & {\footnotesize{}$0.0130$} & {\footnotesize{}$6.4735$} & {\footnotesize{}$2.0619$} & {\footnotesize{}$13.1090$} & {\footnotesize{}Slow} & {\footnotesize{}Stable}\tabularnewline
		\hline 
		{\footnotesize{}CGD-NDAF (Case3)} & {\footnotesize{}$0.3459$} & {\footnotesize{}$0.0375$} & {\footnotesize{}$0.3852$} & {\footnotesize{}$71.9788$} & {\footnotesize{}$4.6348$} & {\footnotesize{}$76.7300$} & {\footnotesize{}Fast} & \textcolor{red}{\footnotesize{}Unstable}\tabularnewline
		\hline 
		{\footnotesize{}AG-NDAF} & {\footnotesize{}$0.0037$} & {\footnotesize{}$0.0018$} & {\footnotesize{}$0.0089$} & {\footnotesize{}$6.7044$} & {\footnotesize{}$1.8282$} & {\footnotesize{}$10.8265$} & {\footnotesize{}Fast} & {\footnotesize{}Stable}\tabularnewline
		\hline 
		{\footnotesize{}GPSd-NDAF} & {\footnotesize{}$0.0033$} & {\footnotesize{}$0.0017$} & {\footnotesize{}$0.0094$} & {\footnotesize{}$6.3198$} & {\footnotesize{}$1.7286$} & {\footnotesize{}$11.1217$} & {\footnotesize{}Guaranteed} & {\footnotesize{}Stable}\tabularnewline
		\hline 
		{\footnotesize{}GPD-NDAF} & {\footnotesize{}$0.0030$} & {\footnotesize{}$0.0014$} & {\footnotesize{}$0.0091$} & {\footnotesize{}$6.0537$} & {\footnotesize{}$1.5424$} & {\footnotesize{}$10.9516$} & {\footnotesize{}Guaranteed} & {\footnotesize{}Stable}\tabularnewline
		\hline 
		{\footnotesize{}AGI-NSAF} & {\footnotesize{}$0.0032$} & {\footnotesize{}$0.0017$} & {\footnotesize{}$0.0089$} & {\footnotesize{}$6.2420$} & {\footnotesize{}$1.8141$} & {\footnotesize{}$10.8308$} & {\footnotesize{}Fast} & {\footnotesize{}Stable}\tabularnewline
		\hline 
		{\footnotesize{}AGS-NSAF} & {\footnotesize{}$0.0032$} & {\footnotesize{}$0.0017$} & {\footnotesize{}$0.0090$} & {\footnotesize{}$6.2356$} & {\footnotesize{}$1.8140$} & {\footnotesize{}$10.8576$} & {\footnotesize{}Fast} & {\footnotesize{}Stable}\tabularnewline
		\hline 
		{\footnotesize{}GPSd-NSAF} & {\footnotesize{}$0.0033$} & {\footnotesize{}$0.0022$} & {\footnotesize{}$0.0191$} & {\footnotesize{}$6.2854$} & {\footnotesize{}$2.0817$} & \textbf{\footnotesize{}$15.8970$} & {\footnotesize{}Guaranteed} & {\footnotesize{}Stable}\tabularnewline
		\hline 
		{\footnotesize{}GPD-NSAF} & {\footnotesize{}$0.0032$} & {\footnotesize{}$0.0015$} & {\footnotesize{}$0.009$} & {\footnotesize{}$6.3104$} & {\footnotesize{}$1.5301$} & \textbf{\footnotesize{}$10.9040$} & {\footnotesize{}Guaranteed} & {\footnotesize{}Stable}\tabularnewline
		\hline 
	\end{tabular}{\footnotesize\par}
\end{table}

\noindent\makebox[1\linewidth]{%
	\rule{0.4\textwidth}{0.5pt}%
}

\subsubsection{Discrete Nonlinear Filters Results}

\begin{figure}
	\begin{centering}
		\includegraphics[scale=0.22]{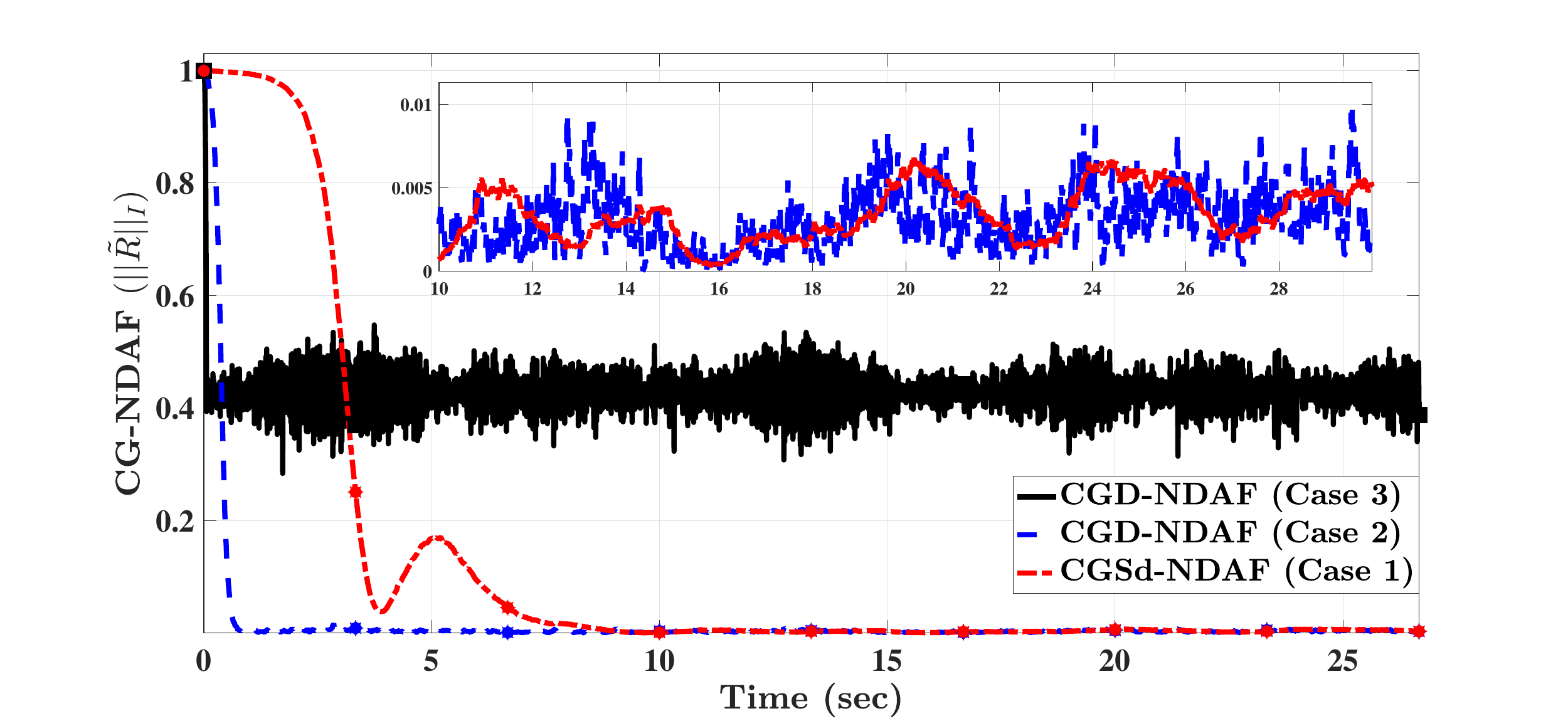}
		\par\end{centering}
	\begin{centering}
		\includegraphics[scale=0.22]{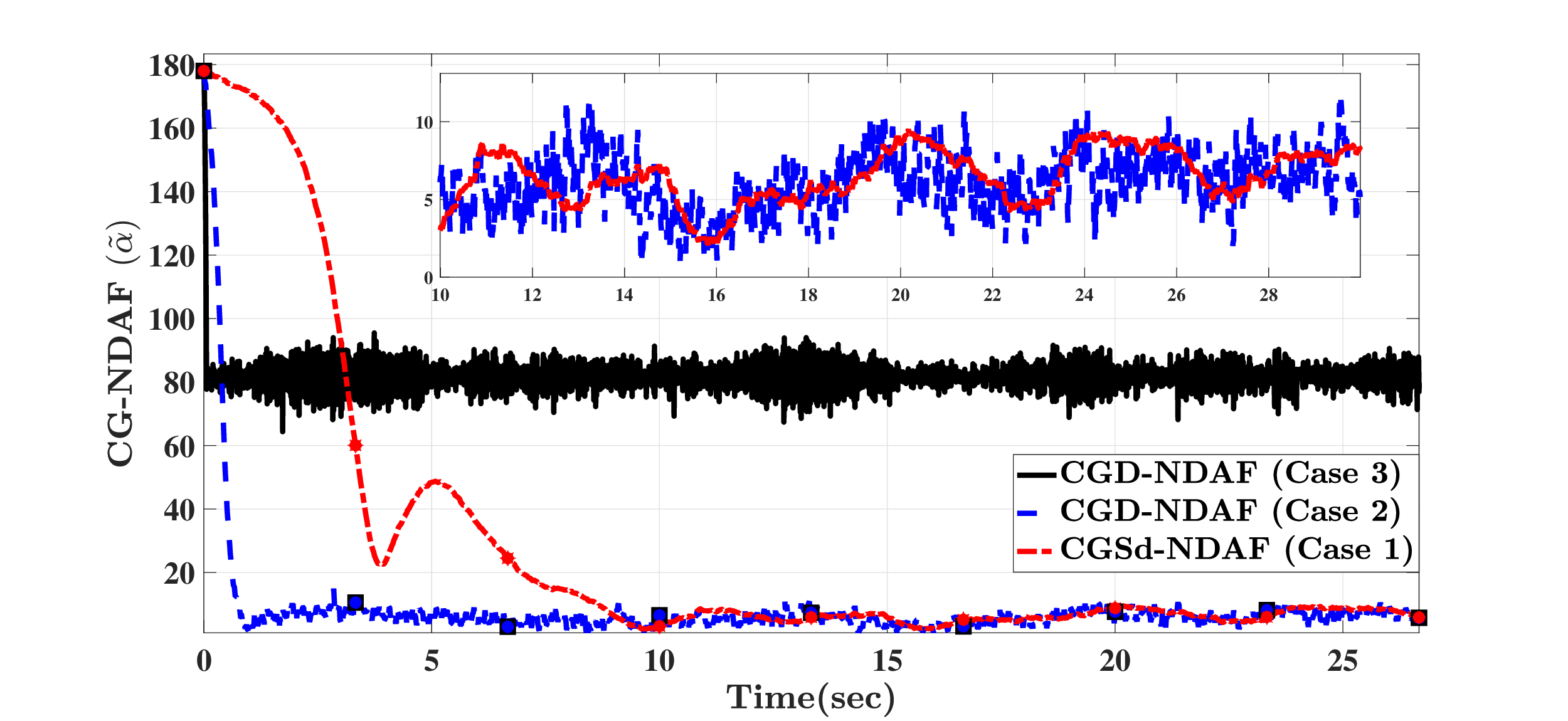}
		\par\end{centering}
	\includegraphics[scale=0.2]{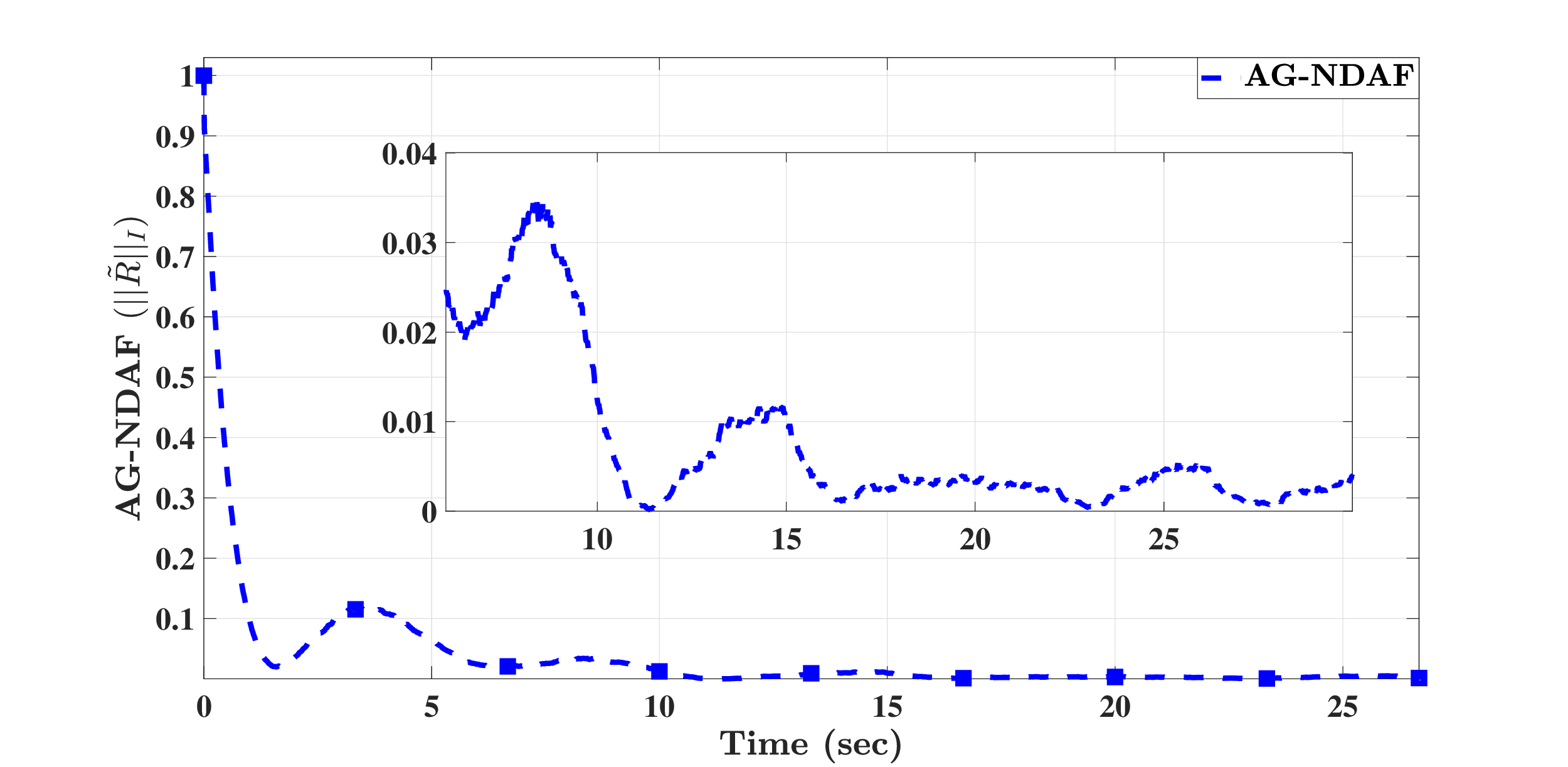}\includegraphics[scale=0.2]{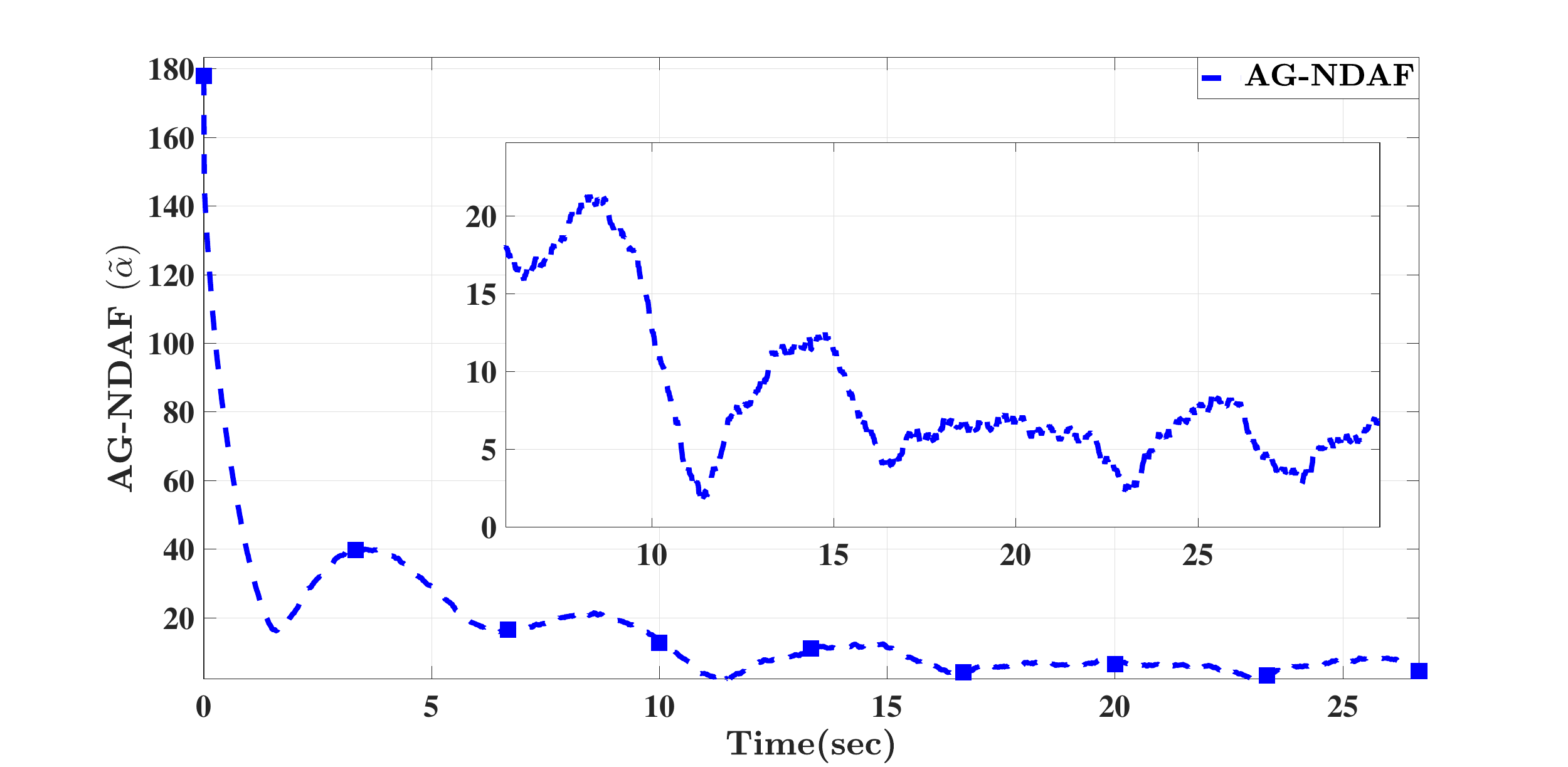}
	
	\includegraphics[scale=0.2]{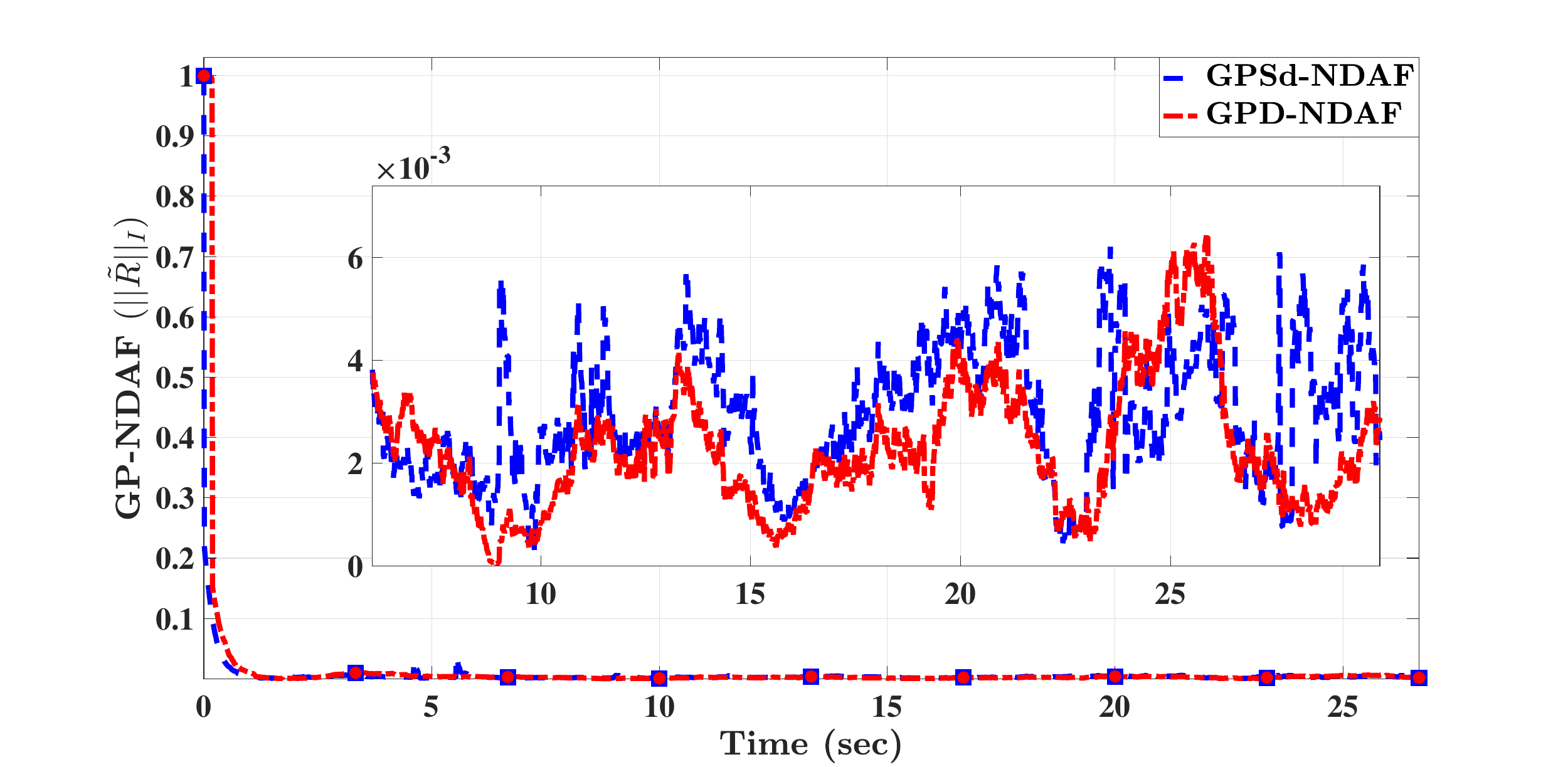}\includegraphics[scale=0.2]{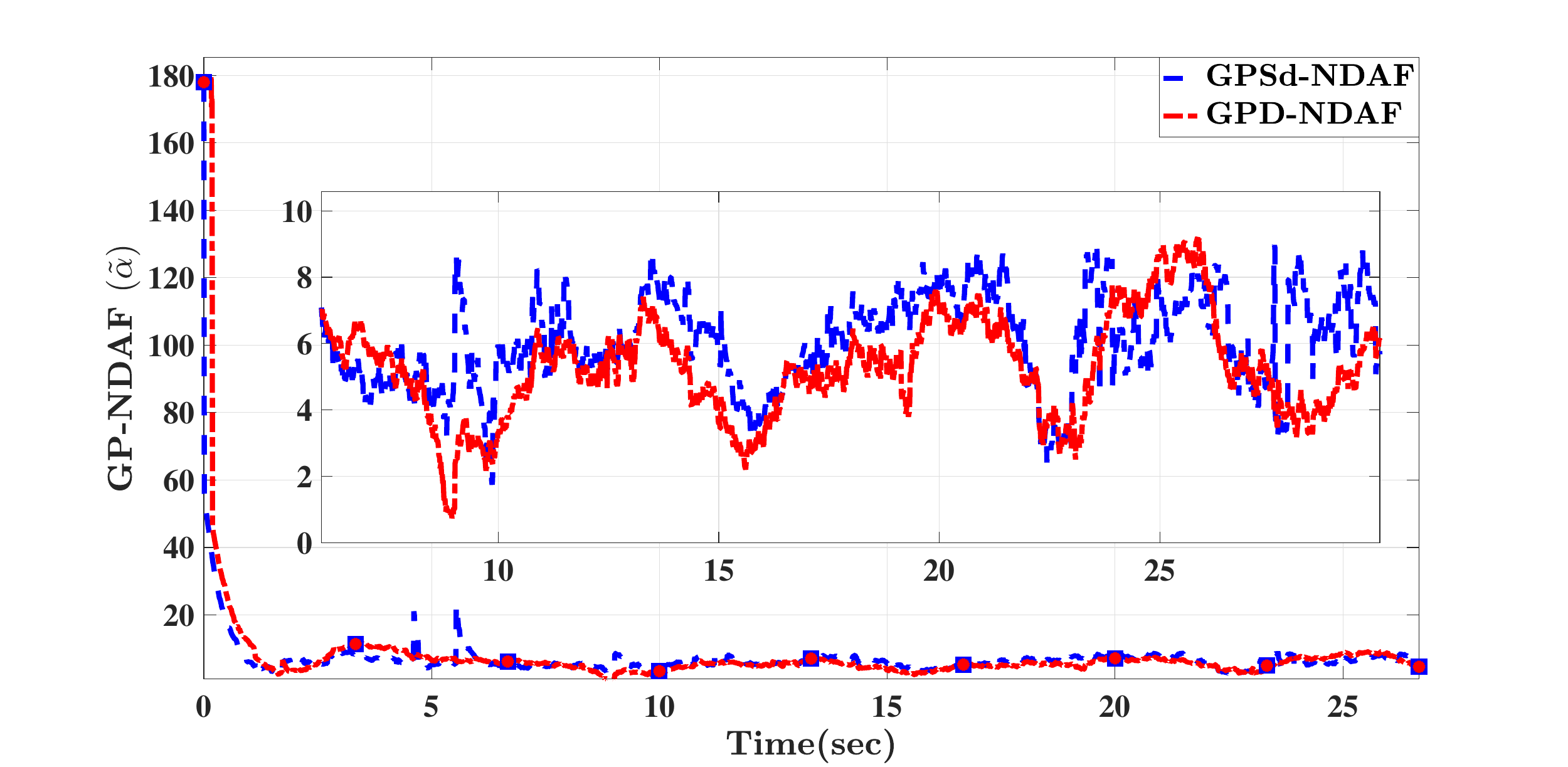}
	
	\includegraphics[scale=0.2]{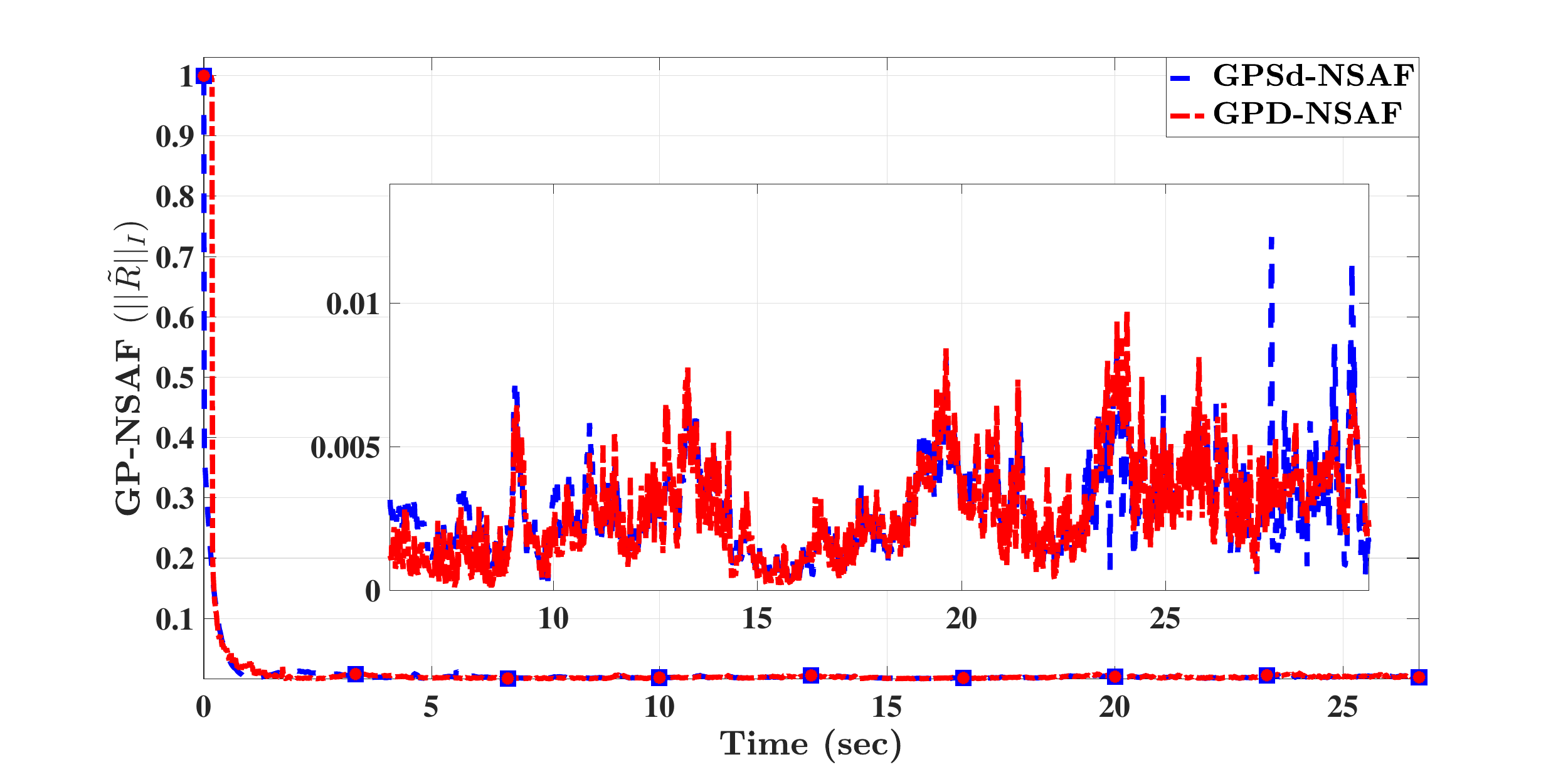}\includegraphics[scale=0.2]{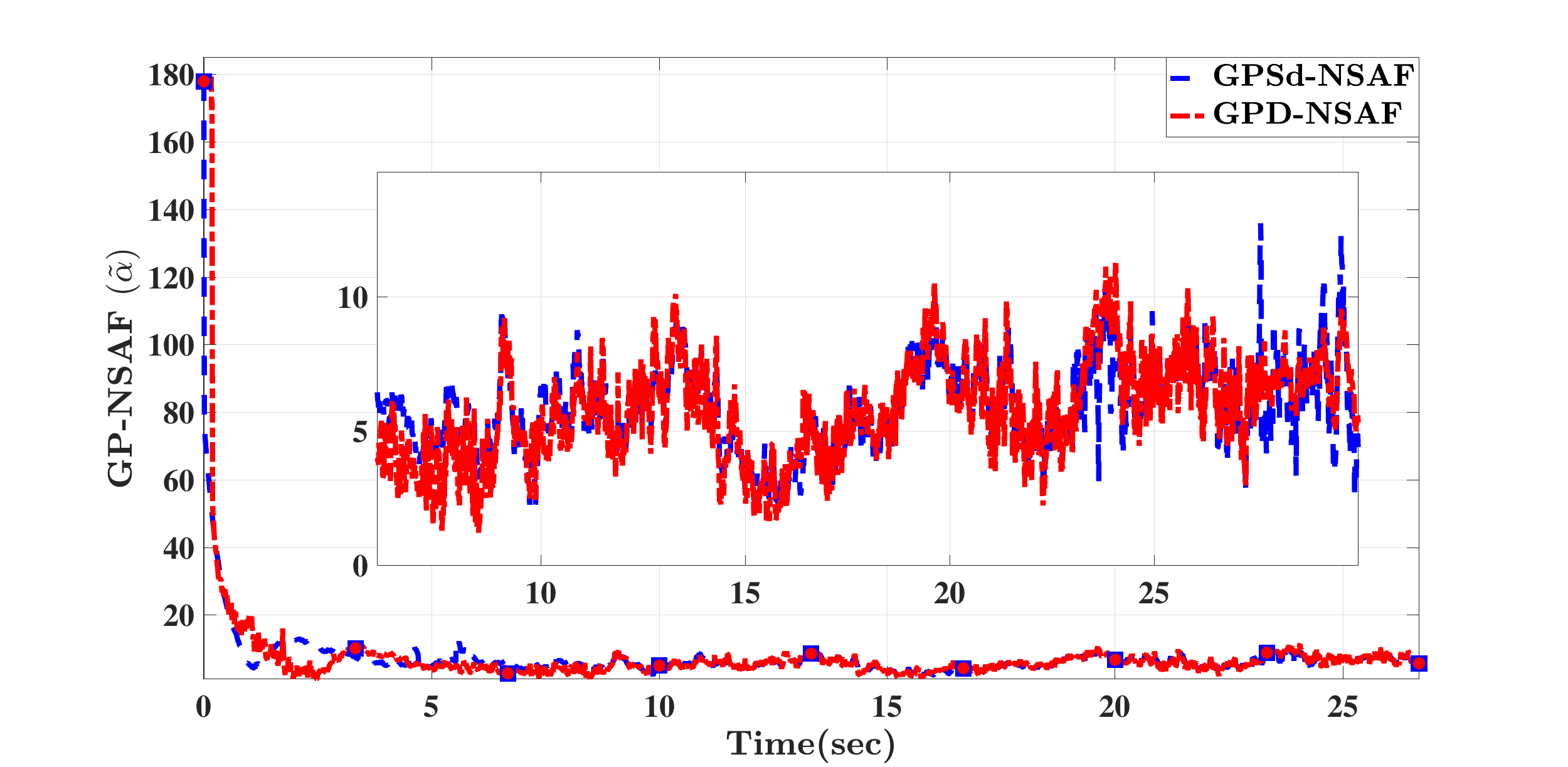}
	\centering{}\caption{Tracking error of nonlinear discrete attitude filters: $||\tilde{R}[k]||_{I}$
		and $\tilde{\alpha}[k]$.}
	\label{fig:Comp_Simu_Disc_Non} 
\end{figure}

This part contains a brief comparison between nonlinear discrete attitude
filters whose detailed descriptions can be found in the \nameref{sec:Appendix}.
The filters to be discussed are CG-NDAF (Equation \eqref{eq:Comp_Non_CGSd_NDAF-1}
and \eqref{eq:Comp_Non_CGD_NDAF-1}), AG-NDAF (Equation \eqref{eq:Comp_Non_AGNDAF-1}),
GP-NDAF (Equation \eqref{eq:Comp_Non_GPSd_NDAF-1} and \eqref{eq:Comp_Non_GPD_NDAF-1}),
and GP-NSAF (Equation \eqref{eq:Comp_Non_GPSd_NDAF-1-1} and \eqref{eq:Comp_Non_GPD_NDAF-1-1}).
The sampling time $\Delta t$ is set to 0.01 seconds. Consider the
measurement of the true angular velocity to be given similar to \eqref{eq:Comp_Simu_Ang}.
Also, let the body-frame measurements be as in \eqref{eq:Comp_Simu_Vec}
and their normalized values as in \eqref{eq:Comp_Simu_Vec2}. Figure
\ref{fig:Comp_Simu_Ang_Vec} shows the true angular velocity and the
normalized values of body-frame vectors plotted against angular velocity
measurements and the normalized values of body-frame vectorial measurements,
respectively. Figure \ref{fig:Comp_Simu_Ang_Vec} illustrates high
values of noise and bias components corrupting the measurement process.
As illustrated in Figure \ref{fig:Comp_Simu_Disc_Non}, CGD-NDAF as
well as CGSd-NDAF showed stable performance with slower transient
tracking response of $||\tilde{R}[k]||_{I}$ and $\tilde{\alpha}[k]$
for Case 1 and 2. However, For Case 3 CGD-NDAF showed fast transient
response with poor values of steady state error of $||\tilde{R}[k]||_{I}$
and $\tilde{\alpha}[k]$. AG-NDAF demonstrated fast tracking performance
with more oscillatory response in the steady-state. GPSd-NDAF and
GPD-NDAF displayed fast transient response with stable performance
in the steady-state. Similarly, GPSd-NSAF and GPD-NSAF exhibited fast
tracking performance with less oscillation in the steady-state.

\newpage

\section{Conclusion\label{sec:Comp_Conclusion}}

In conclusion, let us briefly summarize the history of development
of the attitude determination and estimation methods over the past
few decades. TRIAD algorithm is one of the earliest and simplest methods
of attitude determination for two given simultaneous observations.
SVD and QUEST displaced TRIAD and became more popular methods of attitude
determinations as they allow for the case of two or more simultaneous
observations. The family of Kalman filters was a pioneer of providing
a reasonable estimate of the true attitude, in particular the multiplicative
extended Kalman filter (MEKF). Nonlinear attitude filters were proposed
to mimic the nonlinear nature of the attitude dynamics and to provide
better results than Gaussian attitude filters. In fact, among other
advantages over the Gaussian attitude filter, nonlinear attitude filters
are simpler in derivation and require less computational power. A
brief survey of attitude determination algorithms, Gaussian attitude
filters, and nonlinear attitude filters is presented in this paper.
The output performance of each category is illustrated through the
simulation results for the purposes of validation and comparison.

\noindent\makebox[1\linewidth]{%
	\rule{0.8\textwidth}{1.4pt}%
}

\section*{Appendix\label{sec:Appendix}}
\begin{center}
	\textbf{\large{}DISCRETE: GAUSSIAN AND NONLINEAR ATTITUDE FILTERS}{\large\par}
	\par\end{center}

The Appendix contains the discrete designs of Gaussian attitude filters
(KF, MEKF and GAMEF) and nonlinear attitude filters (CG-NDAF, AG-NDAF,
GP-NDAF, AG-NSAF and GP-NSAF) presented in Section \ref{sec:Comp_Gaussian}
and \ref{sec:Comp_Nonlinear}. $\Delta t$ denotes the sampling time
which is assumed to be sufficiently small. Also, for any $x\in\mathbb{R}^{n\times m}$,
$x\left[k\right]$ refers to the value of $x$ at sample $k$. 

\subsection{Discrete KF\label{subsec:KF-1}}

For $\Omega_{m}=\Omega+\mathcal{Q}_{\omega}\omega$, recall the attitude
problem in \eqref{eq:Comp_Q_SS_Noise}
\[
\begin{cases}
\dot{Q} & =\frac{1}{2}\Gamma\left(\Omega_{m}-\mathcal{Q}_{\omega}\omega\right)Q\\
\mathcal{Y} & =\frac{1}{2}\sum_{i}^{n}\Xi\left(Q\right)\mathcal{Q}_{v\left(i\right)}\omega_{i}^{\mathcal{B}}
\end{cases}
\]
For simplicity, let $\omega_{i}^{\mathcal{B}}=\omega_{i}^{\mathcal{B}}\left[k\right]$,
$\omega=\omega\left[k\right]$, $\Omega=\Omega\left[k\right]$, $\Omega_{m}=\Omega_{m}\left[k\right]$,
$\upsilon_{i}^{\mathcal{B}}=\upsilon_{i}^{\mathcal{B}}\left[k\right]$
and $\upsilon_{i}^{\mathcal{I}}=\upsilon_{i}^{\mathcal{I}}\left[k\right]$.
The discrete form of \eqref{eq:Comp_Q_SS_Noise} is as follows: 
\[
\begin{cases}
Q\left[k+1\right] & =\exp\left(\frac{1}{2}\Gamma\left(\Omega_{m}\left[k\right]\right)\Delta t\right)Q\left[k\right]-\frac{1}{2}\Xi\left(Q\left[k\right]\right)\mathcal{Q}_{\omega}\omega\left[k\right]\Delta t\\
\mathcal{Y}\left[k\right] & =\left[\begin{array}{cc}
0 & -\left(\upsilon_{i}^{\mathcal{B}}-\upsilon_{i}^{\mathcal{I}}\right)^{\top}\\
\upsilon_{i}^{\mathcal{B}}-\upsilon_{i}^{\mathcal{I}} & -\left[\upsilon_{i}^{\mathcal{B}}+\upsilon_{i}^{\mathcal{I}}\right]_{\times}
\end{array}\right]Q\left[k\right]\\
& =\frac{1}{2}\Xi\left(Q\left[k\right]\right)\mathcal{Q}_{v\left(i\right)}\omega_{i}^{\mathcal{B}}
\end{cases}
\]
One can easily obtain the covariance by
\begin{align*}
\mathcal{Q}_{\epsilon}\left[k\right] & =\mathbb{E}\left[\left(\mathcal{Q}_{\omega}\omega\left[k\right]\right)\left(\mathcal{Q}_{\omega}\omega\left[k\right]\right)^{\top}\right]=\mathcal{Q}_{\omega}^{2}\\
\mathcal{Q}_{q}\left[k\right] & =\left(\frac{\Delta t}{2}\right)^{2}\Xi\left(Q\left[k\right]\right)\mathcal{Q}_{\omega}^{2}\Xi\left(Q\left[k\right]\right)^{\top}\\
\mathcal{R}_{\epsilon} & =\mathbb{E}\left[\sum_{i}^{n}\left(\mathcal{Q}_{v\left(i\right)}\omega_{i}^{\mathcal{B}}\right)\left(\mathcal{Q}_{v\left(i\right)}\omega_{i}^{\mathcal{B}}\right)^{\top}\right]=\sum_{i}^{n}\mathcal{Q}_{v\left(i\right)}^{2}\\
\mathcal{R}_{q}\left[k\right] & =\Xi\left(Q\left[k\right]\right)\mathcal{R}_{\epsilon}\Xi\left(Q\left[k\right]\right)^{\top}
\end{align*}
The discrete form of a basic attitude KF can be represented in two
steps. The prediction step:
\begin{equation}
\begin{cases}
\hat{Q}\left[0\right] & =\left[\begin{array}{cc}
1 & 0_{1\times3}\end{array}\right]^{\top}\\
\Psi\left[k\right] & =\exp\left(\frac{1}{2}\Gamma\left(\Omega_{m}\left[k\right]\right)\Delta t\right)\\
\hat{Q}\left[\left.k+1\right|k\right] & =\Psi\left[k\right]\hat{Q}\left[k\right]\\
\mathcal{Q}_{q}\left[k\right] & =\left(\frac{\Delta t}{2}\right)^{2}\Xi\left(\hat{Q}\left[k\right]\right)\mathcal{Q}_{\epsilon}\left[k\right]\Xi\left(\hat{Q}\left[k\right]\right)^{\top}\\
P\left[\left.k+1\right|k\right] & =\Psi\left[k\right]P\left[k\right]\Psi\left[k\right]^{\top}+\mathcal{Q}_{q}\left[k\right]
\end{cases}\label{eq:KF-1}
\end{equation}
and the correction step:
\begin{equation}
\begin{cases}
\mathcal{H}\left[k\right] & =\left[\begin{array}{cc}
0 & -\left(\upsilon_{i}^{\mathcal{B}}-\upsilon_{i}^{\mathcal{I}}\right)^{\top}\\
\left(\upsilon_{i}^{\mathcal{B}}-\upsilon_{i}^{\mathcal{I}}\right) & -\left[\upsilon_{i}^{\mathcal{B}}+\upsilon_{i}^{\mathcal{I}}\right]_{\times}
\end{array}\right]\\
\mathcal{R}_{q}\left[k+1\right] & =\frac{1}{4}\Xi\left(\hat{Q}\left[\left.k+1\right|k\right]\right)\mathcal{R}_{\epsilon}\Xi\left(\hat{Q}\left[\left.k+1\right|k\right]\right)^{\top}+\alpha\mathbf{I}_{4}\\
S\left[\left.k+1\right|k\right] & =\mathcal{H}\left[k\right]P\left[\left.k+1\right|k\right]\mathcal{H}\left[k\right]^{\top}+R_{q}\left[k+1\right]\\
K\left[k+1\right] & =P\left[\left.k+1\right|k\right]\mathcal{H}\left[k\right]^{\top}S\left[\left.k+1\right|k\right]^{-1}\\
\hat{Q}\left[k+1\right] & =\left(\mathbf{I}_{4}-K\left[k+1\right]\mathcal{H}\left[k\right]\right)\hat{Q}\left[\left.k+1\right|k\right]\\
P\left[k+1\right] & =\left(\mathbf{I}_{4}-K\left[k+1\right]\mathcal{H}\left[k\right]\right)P\left[\left.k+1\right|k\right]\left(\mathbf{I}_{4}-K\left[k+1\right]\mathcal{H}\left[k\right]\right)^{\top}\\
& \hspace{1em}+K\left[k+1\right]R_{q}\left[k+1\right]K\left[k+1\right]^{\top}\\
\hat{Q}\left[k+1\right] & =\hat{Q}\left[k+1\right]/\left\Vert \hat{Q}\left[k+1\right]\right\Vert \\
& \text{Go to prediction step Equation (93)}
\end{cases}\label{eq:KF-2}
\end{equation}
where $\alpha$ is a small positive constant. The basic attitude Kalman
filter in \eqref{eq:KF-1} and \eqref{eq:KF-2} can be modified to
account for bias compensation \cite{choukroun2006novel}. The modified
attitude Kalman filter proposed in \cite{choukroun2006novel} is given
in the following two steps. Prediction step:
\begin{equation}
\begin{cases}
\hat{x}\left[0\right] & =\left[\begin{array}{cc}
1 & 0_{1\times6}\end{array}\right]^{\top}\\
\hat{x}\left[k\right] & =\left[\begin{array}{cc}
\hat{Q}\left[k\right]^{\top} & \hat{b}\left[k\right]^{\top}\end{array}\right]^{\top}\\
\hat{\Omega}\left[k\right] & =\Omega_{m}\left[k\right]-\hat{b}\left[k\right]\\
\Psi\left[k\right] & =\exp\left(\frac{1}{2}\Gamma\left(\hat{\Omega}\left[k\right]\right)\Delta t\right)\\
\hat{x}\left[\left.k+1\right|k\right] & =\left[\begin{array}{cc}
\Psi\left[k\right] & 0_{4\times3}\\
0_{3\times4} & \mathbf{I}_{3}
\end{array}\right]\hat{x}\left[k\right]\\
\Psi\left[k\right] & =\left[\begin{array}{cc}
\Psi\left[k\right] & -\frac{\Delta t}{2}\Xi\left(\hat{Q}\left[k\right]\right)\\
0_{3\times4} & \mathbf{I}_{3}
\end{array}\right]\\
\hat{M}\left[k\right] & =\hat{Q}\left[k\right]\hat{Q}\left[k\right]^{\top}+P_{Q}\left[k\right]\\
P_{w}\left[k\right] & =\left[\begin{array}{cc}
\left(\sigma_{1}^{2}+\sigma_{2}^{2}\Delta t\right)\left({\rm Tr}\left\{ \hat{M}\left[k\right]\right\} \mathbf{I}_{4}-\hat{M}\left[k\right]\right) & 0_{4\times3}\\
0_{3\times4} & \sigma_{3}^{2}\Delta t\mathbf{I}_{3}
\end{array}\right]\\
P\left[\left.k+1\right|k\right] & =\Psi\left[k\right]P\left[k\right]\Psi\left[k\right]^{\top}+P_{w}\left[k\right]
\end{cases}\label{eq:KF-1-1}
\end{equation}
Correction step:
\begin{equation}
\begin{cases}
\mathcal{H}\left[k\right] & =\left[\begin{array}{cc}
0 & -\left(\upsilon_{i}^{\mathcal{B}}\left[k\right]-\upsilon_{i}^{\mathcal{I}}\left[k\right]\right)^{\top}\\
\left(\upsilon_{i}^{\mathcal{B}}\left[k\right]-\upsilon_{i}^{\mathcal{I}}\left[k\right]\right) & -\left[\upsilon_{i}^{\mathcal{B}}\left[k\right]+\upsilon_{i}^{\mathcal{I}}\left[k\right]\right]_{\times}
\end{array}\right]\\
\bar{\mathcal{H}}\left[k\right] & =\left[\begin{array}{cc}
\mathcal{H}\left[k\right] & 0_{4\times3}\end{array}\right]\\
\hat{M}\left[\left.k+1\right|k\right] & =\hat{Q}\left[\left.k+1\right|k\right]\hat{Q}\left[\left.k+1\right|k\right]^{\top}+P_{Q}\left[\left.k+1\right|k\right]\\
P_{v}\left[k+1\right] & =\frac{1}{4}\rho\left({\rm Tr}\left\{ \hat{M}\left[\left.k+1\right|k\right]\right\} \mathbf{I}_{4}-\hat{M}\left[\left.k+1\right|k\right]-\Gamma\left(\upsilon_{i}^{\mathcal{B}}\right)\hat{M}\left[\left.k+1\right|k\right]\Gamma\left(\upsilon_{i}^{\mathcal{B}}\right)^{\top}\right)\\
S\left[\left.k+1\right|k\right] & =\mathcal{H}\left[k\right]P_{Q}\left[\left.k+1\right|k\right]\mathcal{H}\left[k\right]^{\top}+P_{v}\left[k+1\right]\\
K\left[k+1\right] & =P\left[\left.k+1\right|k\right]\bar{\mathcal{H}}\left[k\right]^{\top}S\left[\left.k+1\right|k\right]^{-1}\\
\hat{x}\left[k+1\right] & =\left(\mathbf{I}_{7}-K\left[k+1\right]\bar{\mathcal{H}}\left[k\right]\right)\hat{x}\left[\left.k+1\right|k\right]\\
P\left[k+1\right] & =\left(\mathbf{I}_{7}-K\left[k+1\right]\bar{\mathcal{H}}\left[k\right]\right)P\left[\left.k+1\right|k\right]\left(\mathbf{I}_{7}-K\left[k+1\right]\bar{\mathcal{H}}\left[k\right]\right)^{\top}\\
& \hspace{1em}+K\left[k+1\right]P_{v}\left[k+1\right]K\left[k+1\right]^{\top}\\
\hat{Q}\left[k+1\right] & =\hat{Q}\left[k+1\right]/\left\Vert \hat{Q}\left[k+1\right]\right\Vert \\
& \text{Go to prediction step Equation (95)}
\end{cases}\label{eq:KF-2-1}
\end{equation}
where $\mathcal{Q}_{\omega}={\rm diag}\left\{ \sigma_{1},\sigma_{2},\sigma_{3}\right\} $,
$\mathcal{Q}_{\epsilon}\left[k\right]=\eta\mathbf{I}_{3}$, and $\mathcal{R}_{\epsilon}\left[k\right]=\epsilon\mathbf{I}_{3}$
with $\eta$ and $\epsilon$ being positive constants.

\noindent\makebox[1\linewidth]{%
	\rule{0.8\textwidth}{0.5pt}%
}

\subsection{Discrete MEKF\label{subsec:MEKF-1}}

The discrete form of MEKF in \eqref{eq:MEKF} and \eqref{eq:MEKF-1}
is as follows: 
\begin{equation}
\begin{cases}
\left[\begin{array}{c}
0\\
\hat{\upsilon}_{i}^{\mathcal{B}}\left[k\right]
\end{array}\right] & =\hat{Q}\left[k\right]^{-1}\odot\left[\begin{array}{c}
0\\
\upsilon_{i}^{\mathcal{I}}\left[k\right]
\end{array}\right]\odot\hat{Q}\left[k\right]\\
\hat{Q}\left[k+1\right] & =\exp\left(\frac{1}{2}\Gamma\left(\Omega_{m}\left[k\right]-\hat{b}\left[k\right]+P_{a}\left[k\right]W\left[k\right]\right)\Delta t\right)\hat{Q}\left[k\right]\\
W\left[k\right] & =\sum_{i=1}^{n}\hat{\upsilon}_{i}^{\mathcal{B}}\left[k\right]\times\mathcal{\bar{Q}}_{v\left(i\right)}^{-1}\left(\hat{\upsilon}_{i}^{\mathcal{B}}\left[k\right]-\upsilon_{i}^{\mathcal{B}}\left[k\right]\right)
\end{cases}\label{eq:MEKF-2}
\end{equation}
with
\begin{equation}
\begin{cases}
\hat{b}\left[k+1\right] & =\hat{b}\left[k\right]+P_{c}^{\top}\left[k\right]W\left[k\right]\Delta t\\
S\left[k\right] & =\sum_{i=1}^{n}\left[\hat{\upsilon}_{i}^{\mathcal{B}}\left[k\right]\right]_{\times}\mathcal{\bar{Q}}_{v\left(i\right)}^{-1}\left[\hat{\upsilon}_{i}^{\mathcal{B}}\left[k\right]\right]_{\times}\\
P_{a}\left[k+1\right] & =P_{a}\left[k\right]+\left(\mathcal{\bar{Q}}_{\omega}+2\boldsymbol{\mathcal{P}}_{s}\left(P_{a}\left[\Omega_{m}\left[k\right]-\hat{b}\left[k\right]\right]_{\times}-P_{c}\left[k\right]\right)-P_{a}\left[k\right]S\left[k\right]P_{a}\left[k\right]\right)\Delta t\\
P_{b}\left[k+1\right] & =P_{b}\left[k\right]+\left(\mathcal{\bar{Q}}_{b}-P_{c}\left[k\right]S\left[k\right]P_{c}\left[k\right]\right)\Delta t\\
P_{c}\left[k+1\right] & =P_{c}\left[k\right]-\left(\left[\Omega_{m}\left[k\right]-\hat{b}\left[k\right]\right]_{\times}P_{c}\left[k\right]+P_{a}\left[k\right]S\left[k\right]P_{c}\left[k\right]+P_{b}\left[k\right]\right)\Delta t
\end{cases}\label{eq:MEKF-1-1}
\end{equation}

\noindent\makebox[1\linewidth]{%
	\rule{0.8\textwidth}{0.5pt}%
}

\subsection{Discrete GAMEF\label{subsec:GAMEF-1}}

The discrete form of GAMEF in \eqref{eq:GAMEF1} and \eqref{eq:GAMEF2}
is as follows: 
\begin{equation}
\begin{cases}
\left[\begin{array}{c}
0\\
\hat{\upsilon}_{i}^{\mathcal{B}}\left[k\right]
\end{array}\right] & =\hat{Q}\left[k\right]^{-1}\odot\left[\begin{array}{c}
0\\
\upsilon_{i}^{\mathcal{I}}\left[k\right]
\end{array}\right]\odot\hat{Q}\left[k\right]\\
\hat{Q}\left[k+1\right] & =\exp\left(\frac{1}{2}\Gamma\left(\Omega_{m}\left[k\right]-\hat{b}\left[k\right]+P_{a}\left[k\right]W\left[k\right]\right)\Delta t\right)\hat{Q}\left[k\right]\\
W\left[k\right] & =\sum_{i=1}^{n}\hat{\upsilon}_{i}^{\mathcal{B}}\left[k\right]\times\mathcal{\bar{Q}}_{v\left(i\right)}^{-1}\left(\hat{\upsilon}_{i}^{\mathcal{B}}\left[k\right]-\upsilon_{i}^{\mathcal{B}}\left[k\right]\right)
\end{cases}\label{eq:GAMEF1-1}
\end{equation}
where
\begin{equation}
\begin{cases}
\hat{b}\left[k+1\right] & =\hat{b}\left[k\right]+P_{c}^{\top}\left[k\right]W\left[k\right]\Delta t\\
S\left[k\right] & =\sum_{i=1}^{n}\left[\hat{\upsilon}_{i}^{\mathcal{B}}\left[k\right]\right]_{\times}\mathcal{\bar{Q}}_{v\left(i\right)}^{-1}\left[\hat{\upsilon}_{i}^{\mathcal{B}}\left[k\right]\right]_{\times}\\
C\left[k\right] & =\sum_{i=1}^{n}\boldsymbol{\mathcal{P}}_{s}\left(\mathcal{\bar{Q}}_{v\left(i\right)}^{-1}\left(\hat{\upsilon}_{i}^{\mathcal{B}}\left[k\right]-\upsilon_{i}^{\mathcal{B}}\left[k\right]\right)\left(\hat{\upsilon}_{i}^{\mathcal{B}}\left[k\right]\right)^{\top}\right)\\
E\left[k\right] & ={\rm Tr}\left\{ C\left[k\right]\right\} \mathbf{I}_{3}-C\left[k\right]\\
P_{a}\left[k+1\right] & =\left(2\boldsymbol{\mathcal{P}}_{s}\left(P_{a}\left[k\right]\left[\Omega_{m}\left[k\right]-\hat{b}\left[k\right]-\frac{1}{2}P_{a}\left[k\right]W\left[k\right]\right]_{\times}-P_{c}\left[k\right]\right)+P_{a}\left[k\right]\left(E\left[k\right]-S\left[k\right]\right)P_{a}\left[k\right]\right)\Delta t\\
& \hspace{1em}\mathcal{\bar{Q}}_{\omega}\Delta t+P_{a}\left[k\right]\\
P_{b}\left[k+1\right] & =P_{b}\left[k\right]+\left(\mathcal{\bar{Q}}_{b}\left[k\right]+P_{c}\left[k\right]\left(E\left[k\right]-S\left[k\right]\right)P_{c}\left[k\right]\right)\Delta t\\
P_{c}\left[k+1\right] & =P_{c}\left[k\right]-\left(\left[\Omega_{m}\left[k\right]-\hat{b}\left[k\right]-\frac{1}{2}P_{a}\left[k\right]W\left[k\right]\right]_{\times}P_{c}\left[k\right]-P_{a}\left[k\right]\left(E\left[k\right]-S\left[k\right]\right)P_{c}\left[k\right]+P_{b}\left[k\right]\right)\Delta t
\end{cases}\label{eq:GAMEF2-1}
\end{equation}

\noindent\makebox[1\linewidth]{%
	\rule{0.8\textwidth}{0.5pt}%
}

\subsection{Discrete CG-NDAF\label{subsec:CGNDAF-1}}

Before we introduce the nonlinear filters in discrete form, let us
recall \eqref{eq:Comp_VEX_VM}, \eqref{eq:Comp_RI_VM}, and \eqref{eq:Comp_Gamma_VM}
and present them in sampling form

\begin{equation}
\begin{cases}
\mathbf{vex}\left(\boldsymbol{\mathcal{P}}_{a}\left(M^{\mathcal{B}}\left[k\right]\tilde{R}\left[k\right]\right)\right) & =\sum_{i=1}^{n}\frac{s_{i}}{2}\hat{\upsilon}_{i}^{\mathcal{B}}\left[k\right]\times\upsilon_{i}^{\mathcal{B}}\left[k\right]\\
||M^{\mathcal{B}}\left[k\right]\tilde{R}\left[k\right]||_{I} & =\frac{1}{4}\sum_{i=1}^{n}s_{i}\left(1-\left(\hat{\upsilon}_{i}^{\mathcal{B}}\left[k\right]\right)^{\top}\upsilon_{i}^{\mathcal{B}}\left[k\right]\right)\\
\boldsymbol{\Upsilon}\left(M^{\mathcal{B}}\left[k\right],\tilde{R}\left[k\right]\right) & ={\rm Tr}\left\{ \left(\sum_{i=1}^{n}s_{i}\upsilon_{i}^{\mathcal{B}}\left[k\right]\left(\upsilon_{i}^{\mathcal{B}}\left[k\right]\right)^{\top}\right)^{-1}\sum_{i=1}^{n}s_{i}\upsilon_{i}^{\mathcal{B}}\left[k\right]\left(\hat{\upsilon}_{i}^{\mathcal{B}}\left[k\right]\right)^{\top}\right\} 
\end{cases}\label{eq:Comp_VEX_VM_Discrete}
\end{equation}

\subsubsection{Semi-direct Filter}

The discrete form of CGSd-NDAF in \eqref{eq:Comp_Non_CGSd_NDAF} is
as follows: 
\begin{equation}
\begin{cases}
\hat{R}\left[k+1\right] & =\hat{R}\left[k\right]\exp\left(\left[\Omega_{m}\left[k\right]-\hat{b}\left[k\right]-k_{w}W\left[k\right]\right]_{\times}\Delta t\right)\\
W & =\mathbf{vex}\left(\boldsymbol{\mathcal{P}}_{a}\left(\tilde{R}\left[k\right]\right)\right),\hspace{1em}\tilde{R}\left[k\right]=R_{y}^{\top}\left[k\right]\hat{R}\left[k\right]\\
\hat{b}\left[k+1\right] & =\hat{b}\left[k\right]+\gamma W\left[k\right]\Delta t
\end{cases}\label{eq:Comp_Non_CGSd_NDAF-1}
\end{equation}

\noindent\makebox[1\linewidth]{%
	\rule{0.4\textwidth}{0.2pt}%
}

\subsubsection{Direct Filter}

The discrete form of CGD-NDAF in \eqref{eq:Comp_Non_CGD_NDAF} is
as follows:

\begin{equation}
\begin{cases}
\hat{R}\left[k+1\right] & =\hat{R}\left[k\right]\exp\left(\left[\Omega_{m}\left[k\right]-\hat{b}\left[k\right]-k_{w}W\left[k\right]\right]_{\times}\Delta t\right)\\
W\left[k\right] & =\mathbf{vex}\left(\boldsymbol{\mathcal{P}}_{a}\left(M^{\mathcal{B}}\left[k\right]\tilde{R}\left[k\right]\right)\right)\\
\hat{b}\left[k+1\right] & =\hat{b}\left[k\right]+\gamma W\left[k\right]\Delta t
\end{cases}\label{eq:Comp_Non_CGD_NDAF-1}
\end{equation}
with $\mathbf{vex}\left(\boldsymbol{\mathcal{P}}_{a}\left(M^{\mathcal{B}}\left[k\right]\tilde{R}\left[k\right]\right)\right)$
being obtained through vectorial measurements as in \eqref{eq:Comp_VEX_VM_Discrete}. 

\noindent\makebox[1\linewidth]{%
	\rule{0.8\textwidth}{0.5pt}%
}

\subsection{Discrete AG-NDAF\label{subsec:AGNDAF-1}}

The discrete form of AG-NDAF in \eqref{eq:Comp_Non_AGNDAF} is as
follows: 
\begin{equation}
\begin{cases}
\hat{R}\left[k+1\right] & =\hat{R}\left[k\right]\exp\left(\left[\Omega_{m}\left[k\right]-\hat{b}\left[k\right]-k_{w}W\left[k\right]\right]_{\times}\Delta t\right)\\
W\left[k\right] & =\frac{1}{1+{\rm Tr}\{\tilde{R}\left[k\right]\}}\mathbf{vex}\left(\boldsymbol{\mathcal{P}}_{a}\left(\tilde{R}\left[k\right]\right)\right),\hspace{1em}\tilde{R}\left[k\right]=R_{y}^{\top}\left[k\right]\hat{R}\left[k\right]\\
\hat{b}\left[k+1\right] & =\hat{b}\left[k\right]+\gamma W\left[k\right]\Delta t
\end{cases}\label{eq:Comp_Non_AGNDAF-1}
\end{equation}
\noindent\makebox[1\linewidth]{%
	\rule{0.8\textwidth}{0.5pt}%
}

\subsection{Discrete GP-NDAF\label{subsec:GPNDAF-1}}

\subsubsection{Semi-direct Filter}

The discrete form of GPSd-NDAF in \eqref{eq:Comp_Non_GPSd_NDAF} is
as follows: 
\begin{equation}
\begin{cases}
\hat{R}\left[k+1\right] & =\hat{R}\left[k\right]\exp\left(\left[\Omega_{m}\left[k\right]-\hat{b}\left[k\right]-k_{w}W\left[k\right]\right]_{\times}\Delta t\right)\\
W\left[k\right] & =2\frac{k_{w}\mu\left[k\right]\mathcal{E}\left[k\right]-\bar{\xi}_{d}\left[k\right]/4\xi\left[k\right]}{1-||\tilde{R}\left[k\right]||_{I}}\mathbf{vex}\left(\boldsymbol{\mathcal{P}}_{a}\left(\tilde{R}\left[k\right]\right)\right)\\
\hat{b}\left[k+1\right] & =\hat{b}\left[k\right]+\frac{\gamma}{2}\mu\left[k\right]\mathcal{E}\left[k\right]\mathbf{vex}\left(\boldsymbol{\mathcal{P}}_{a}\left(\tilde{R}\left[k\right]\right)\right)\Delta t,\quad\tilde{R}\left[k\right]=R_{y}^{\top}\left[k\right]\hat{R}\left[k\right]
\end{cases}\label{eq:Comp_Non_GPSd_NDAF-1}
\end{equation}
where

\begin{equation}
\begin{cases}
\xi\left[k\right] & =\left(\xi_{0}-\xi_{\infty}\right)\exp\left(-\ell k\right)+\xi_{\infty}\\
\bar{\xi}_{d}\left[k\right] & =\frac{\xi\left[k\right]-\xi\left[k-1\right]}{\Delta t}\\
\mathcal{E}\left[k\right] & =\frac{1}{2}\text{ln}\frac{\underline{\delta}+||\tilde{R}\left[k\right]||_{I}/\xi\left[k\right]}{\bar{\delta}-||\tilde{R}\left[k\right]||_{I}/\xi\left[k\right]}\\
\mu\left[k\right] & =\frac{1/2}{\underline{\delta}\xi\left[k\right]+||\tilde{R}\left[k\right]||_{I}}+\frac{1/2}{\bar{\delta}\xi\left[k\right]-||\tilde{R}\left[k\right]||_{I}}
\end{cases}\label{eq:Comp_Non_PPF_NDAF-1-1}
\end{equation}

\noindent\makebox[1\linewidth]{%
	\rule{0.4\textwidth}{0.2pt}%
}

\subsubsection{Direct Filter}

The discrete form of GPD-NDAF in \eqref{eq:Comp_Non_GPD_NDAF} is
as follows:

\begin{equation}
\begin{cases}
\hat{R}\left[k+1\right] & =\hat{R}\left[k\right]\exp\left(\left[\Omega_{m}\left[k\right]-\hat{b}\left[k\right]-k_{w}W\left[k\right]\right]_{\times}\Delta t\right)\\
W\left[k\right] & =\frac{4}{\underline{\lambda}}\frac{k_{w}\mu\left[k\right]\mathcal{E}\left[k\right]-\bar{\xi}_{d}\left[k\right]/\xi\left[k\right]}{1+\boldsymbol{\Upsilon}\left(M^{\mathcal{B}}\left[k\right],\tilde{R}\left[k\right]\right)}\mathbf{vex}\left(\boldsymbol{\mathcal{P}}_{a}\left(M^{\mathcal{B}}\left[k\right]\tilde{R}\left[k\right]\right)\right)\\
\hat{b}\left[k+1\right] & =\hat{b}\left[k\right]+\frac{\gamma}{2}\mu\left[k\right]\mathcal{E}\left[k\right]\mathbf{vex}\left(\boldsymbol{\mathcal{P}}_{a}\left(M^{\mathcal{B}}\left[k\right]\tilde{R}\left[k\right]\right)\right)\Delta t
\end{cases}\label{eq:Comp_Non_GPD_NDAF-1}
\end{equation}
where

\begin{equation}
\begin{cases}
\xi\left[k\right] & =\left(\xi_{0}-\xi_{\infty}\right)\exp\left(-\ell k\right)+\xi_{\infty}\\
\bar{\xi}_{d}\left[k\right] & =\frac{\xi\left[k\right]-\xi\left[k-1\right]}{\Delta t}\\
\mathcal{E}\left[k\right] & =\frac{1}{2}\text{ln}\frac{\underline{\delta}+||M^{\mathcal{B}}\left[k\right]\tilde{R}\left[k\right]||_{I}/\xi\left[k\right]}{\bar{\delta}-||M^{\mathcal{B}}\left[k\right]\tilde{R}\left[k\right]||_{I}/\xi\left[k\right]}\\
\mu\left[k\right] & =\frac{1/2}{\underline{\delta}\xi\left[k\right]+||M^{\mathcal{B}}\left[k\right]\tilde{R}\left[k\right]||_{I}}+\frac{1/2}{\bar{\delta}\xi\left[k\right]-||M^{\mathcal{B}}\left[k\right]\tilde{R}\left[k\right]||_{I}}
\end{cases}\label{eq:Comp_Non_PPF_NDAF-1}
\end{equation}
with $\mathbf{vex}\left(\boldsymbol{\mathcal{P}}_{a}\left(M^{\mathcal{B}}\left[k\right]\tilde{R}\left[k\right]\right)\right)$,
$||M^{\mathcal{B}}\left[k\right]\tilde{R}\left[k\right]||_{I}$, and
$\boldsymbol{\Upsilon}\left(M^{\mathcal{B}}\left[k\right],\tilde{R}\left[k\right]\right)$
being obtained through vectorial measurements as in \eqref{eq:Comp_VEX_VM_Discrete}.

\noindent\makebox[1\linewidth]{%
	\rule{0.8\textwidth}{0.5pt}%
}

\subsection{Discrete GP-NSAF\label{subsec:GPNSAF-1}}

\subsubsection{Semi-direct Filter}

The discrete form of GPSd-NDAF in \eqref{eq:Comp_Non_GPSd_NDSF} is
as follows: 
\begin{equation}
\begin{cases}
\hat{R}\left[k+1\right] & =\hat{R}\left[k\right]\exp\left(\left[\Omega_{m}\left[k\right]-\hat{b}\left[k\right]-W\left[k\right]\right]_{\times}\Delta t\right)\\
W\left[k\right] & =2\frac{\mathcal{E}\left[k\right]+2}{\mathcal{E}\left[k\right]+1}\mu\left[k\right]{\rm diag}\left(\boldsymbol{\Psi}(\left[k\right])\right)\hat{\sigma}\left[k\right]+2\frac{k_{w}\mu\left[k\right]\left(\mathcal{E}\left[k\right]+1\right)-\bar{\xi}_{d}\left[k\right]/4\xi\left[k\right]}{1-||\tilde{R}\left[k\right]||_{I}}\boldsymbol{\Psi}(\tilde{R}\left[k\right])\\
\hat{b}\left[k+1\right] & =\hat{b}\left[k\right]+\gamma_{1}\left(\mathcal{E}\left[k\right]+1\right)\exp\left(\mathcal{E}\left[k\right]\right)\mu\left[k\right]\boldsymbol{\Psi}(\tilde{R}\left[k\right])\Delta t\\
\hat{\sigma}\left[k+1\right] & =\hat{\sigma}\left[k\right]+\gamma_{2}\mathcal{E}\left[k\right]\left(\mathcal{E}\left[k\right]+2\right)\exp\left(\mathcal{E}\left[k\right]\right)\mu^{2}\left[k\right]{\rm diag}\left(\boldsymbol{\Psi}(\tilde{R}\left[k\right])\right)\boldsymbol{\Psi}(\tilde{R}\left[k\right])\Delta t
\end{cases}\label{eq:Comp_Non_GPSd_NDAF-1-1}
\end{equation}
where

\begin{equation}
\begin{cases}
\xi\left[k\right] & =\left(\xi_{0}-\xi_{\infty}\right)\exp\left(-\ell k\right)+\xi_{\infty}\\
\bar{\xi}_{d}\left[k\right] & =\frac{\xi\left[k\right]-\xi\left[k-1\right]}{\Delta t}\\
\mathcal{E}\left[k\right] & =\frac{1}{2}\text{ln}\frac{\underline{\delta}+||\tilde{R}\left[k\right]||_{I}/\xi\left[k\right]}{\bar{\delta}-||\tilde{R}\left[k\right]||_{I}/\xi\left[k\right]}\\
\mu\left[k\right] & =\frac{\exp\left(2\mathcal{E}\left[k\right]\right)+\exp\left(-2\mathcal{E}\left[k\right]\right)+2}{8\xi\left[k\right]\bar{\delta}}
\end{cases}\label{eq:Comp_Non_PPF_NDAF-1-1-1}
\end{equation}
where $\boldsymbol{\Psi}(\tilde{R})=\mathbf{vex}(\boldsymbol{\mathcal{P}}_{a}(\tilde{R}))$.

\noindent\makebox[1\linewidth]{%
	\rule{0.4\textwidth}{0.2pt}%
}

\subsubsection{Direct Filter}

The discrete form of GPD-NSAF in \eqref{eq:Comp_Non_GPD_NDSF} is
as follows:

\begin{equation}
\begin{cases}
\hat{R}\left[k+1\right] & =\hat{R}\left[k\right]\exp\left(\left[\Omega_{m}\left[k\right]-\hat{b}\left[k\right]-W\left[k\right]\right]_{\times}\Delta t\right)\\
W\left[k\right] & =2\frac{\mathcal{E}\left[k\right]+2}{\mathcal{E}\left[k\right]+1}\mu\left[k\right]{\rm diag}\left(\boldsymbol{\Psi}(M^{\mathcal{B}}\left[k\right]\tilde{R}\left[k\right])\right)\hat{\sigma}\left[k\right]+\frac{4}{\underline{\lambda}}\frac{k_{w}\mu\left[k\right]\mathcal{E}\left[k\right]-\bar{\xi}_{d}\left[k\right]/\xi\left[k\right]}{1+\boldsymbol{\Upsilon}\left(M^{\mathcal{B}}\left[k\right],\tilde{R}\left[k\right]\right)}\boldsymbol{\Psi}(M^{\mathcal{B}}\left[k\right]\tilde{R}\left[k\right])\\
\hat{b}\left[k+1\right] & =\hat{b}\left[k\right]+\gamma_{1}\mu\left[k\right]\left(\mathcal{E}\left[k\right]+1\right)\exp\left(\mathcal{E}\left[k\right]\right)\boldsymbol{\Psi}(M^{\mathcal{B}}\left[k\right]\tilde{R}\left[k\right])\Delta t\\
\hat{\sigma}\left[k+1\right] & =\hat{\sigma}\left[k\right]+\gamma_{2}\left(\mathcal{E}\left[k\right]+2\right)\exp\left(\mathcal{E}\left[k\right]\right)\mu^{2}\left[k\right]{\rm diag}\left(\boldsymbol{\Psi}(M^{\mathcal{B}}\left[k\right]\tilde{R}\left[k\right])\right)\boldsymbol{\Psi}(M^{\mathcal{B}}\left[k\right]\tilde{R}\left[k\right])\Delta t
\end{cases}\label{eq:Comp_Non_GPD_NDAF-1-1}
\end{equation}
where

\begin{equation}
\begin{cases}
\xi\left[k\right] & =\left(\xi_{0}-\xi_{\infty}\right)\exp\left(-\ell k\right)+\xi_{\infty}\\
\bar{\xi}_{d}\left[k\right] & =\frac{\xi\left[k\right]-\xi\left[k-1\right]}{\Delta t}\\
\mathcal{E}\left[k\right] & =\frac{1}{2}\text{ln}\frac{\underline{\delta}+||M^{\mathcal{B}}\left[k\right]\tilde{R}\left[k\right]||_{I}/\xi\left[k\right]}{\bar{\delta}-||M^{\mathcal{B}}\left[k\right]\tilde{R}\left[k\right]||_{I}/\xi\left[k\right]}\\
\mu\left[k\right] & =\frac{\exp\left(2\mathcal{E}\left[k\right]\right)+\exp\left(-2\mathcal{E}\left[k\right]\right)+2}{8\xi\left[k\right]\bar{\delta}}
\end{cases}\label{eq:Comp_Non_PPF_NDAF-1-2}
\end{equation}
with $\boldsymbol{\Psi}(M^{\mathcal{B}}\tilde{R})=\mathbf{vex}(\boldsymbol{\mathcal{P}}_{a}(M^{\mathcal{B}}\tilde{R}))$,
$\mathbf{vex}\left(\boldsymbol{\mathcal{P}}_{a}\left(M^{\mathcal{B}}\left[k\right]\tilde{R}\left[k\right]\right)\right)$,
$||M^{\mathcal{B}}\left[k\right]\tilde{R}\left[k\right]||_{I}$, and
$\boldsymbol{\Upsilon}\left(M^{\mathcal{B}}\left[k\right],\tilde{R}\left[k\right]\right)$
being obtained through vectorial measurements as in \eqref{eq:Comp_VEX_VM_Discrete}. 

\noindent\makebox[1\linewidth]{%
	\rule{0.8\textwidth}{1.4pt}%
}

\section*{Acknowledgment}

The author would like to thank \textbf{Maria Shaposhnikova }for proofreading
the article.

\bibliographystyle{IEEEtran}
\bibliography{bib_Overview_SO3}

\end{document}